\documentstyle[epsfig,psfig]{aipproc}
\begin{document}
\title{   Magnetic and orbital ordering in cuprates and manganites }
\author{}
\author { Andrzej M. Ole\'{s} }
\address{ Institute of Physics, Jagellonian University, Reymonta 4,
          PL-30059 Krak\'ow, Poland }
\author { Mario Cuoco and Natalia B. Perkins }
\address{ Dipartimento di Scienze Fisiche "E.R. Caianiello",
          Universita di Salerno, Via S. Allende, I-84081 Baronissi, Italy }

\maketitle

\begin{abstract}
We address the role played by orbital degeneracy in strongly
correlated transition metal compounds. The mechanisms of magnetic
and orbital interactions due to double exchange (DE) and
superexchange (SE) are presented. Specifically, we study the
effective spin-orbital models derived for the $d^9$ ions as in
KCuF$_3$, and for the $d^4$ ions as in LaMnO$_3$, for spins
$S=1/2$ and $S=2$, respectively. The magnetic and orbital ordering
in the undoped compounds is determined by the SE interactions that
are inherently frustrated, carrying both antiferromagnetic (AF)
and ferromagnetic (FM) channels due to low-spin and high-spin
excited states, respectively. As a result, the classical phase
diagrams consist of several magnetic phases which all have
different orbital ordering: either the same orbitals ($x^2-y^2$ or
$3z^2-r^2$) are occupied, or two different linear combinations of
$e_g$ orbitals stagger, leading either to G-AF or to A-AF order.
These phases become unstable near orbital degeneracy, leading to a
new mechanism of spin liquid. The model for $d^4$ Mn$^{3+}$ ions
in collosal magnetoresistance compounds provides an explanation of
the observed A-AF phase, with the orbital order stabilized
additionally by the Jahn-Teller effect. Possible extensions of the
model to the doped compounds are discussed both for the insulating
polaronic regime and for the metallic phase. It is shown that the
spin waves are well described by SE in the insulating regime,
while they are explained by DE for degenerate $e_g$ orbitals in
the metallic FM regime. Orbital excitations contribute to the hole
dynamics in FM planes of LaMnO$_3$, characterized by new
quasiparticles reminiscent of the $t$-$J$ model, and a large
redistribution of spectral weight with respect to mean-field
treatments. Finally, we point out some open problems in the
present understanding of doped manganites.
\end{abstract}

\newpage
\section{ Correlated transition-metal oxides with orbital degeneracy }

Theory of strongly correlated electrons is one of the most
challenging and fascinating fields of modern condensed matter. The
correlated electrons are responsible for such phenomena as
magnetic ordering in transition metals, heavy-fermion behavior,
mixed valence, and metal-insulator transitions
\cite{Ima98,Ful91,Faz99}. They play also a prominent role in
transition metal oxides, where they trigger such phenomena as
superconductivity with high transition temperatures in cuprates
and collosal magnetoresistance (CMR) in manganites. At present,
most of the current studies of strongly correlated electrons deal
with models of nondegenerate orbitals, such as the Hubbard model,
Kondo lattice model, Anderson model, and the like. Strong electron
correlations lead in such situations to new effective models which
act only in a part of the Hilbert space and describe the
low-energy excitations. A classical example is the $t$-$J$ model
which follows from the Hubbard model \cite{Cha77,Cha77a}, and
describes a competition between the magnetic superexchange and
kinetic energy of holes doped into an antiferromagnetic (AF) Mott
insulator.

\begin{figure}[t]
\centerline{\psfig{figure=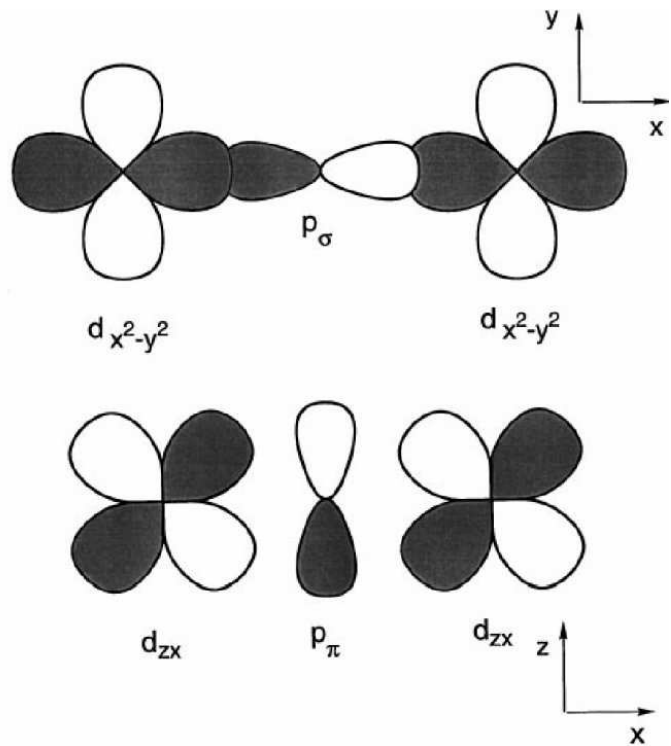,width=9cm}}
\smallskip
\caption{ Examples of configurations for transition-metal $3d$
orbitals which are bridged by ligand $2p$ orbitals in transition
metal oxides (after Ref. \protect\cite{Ima98}). } \label{fig:tmo}
\end{figure}

The realistic models of correlated electrons are, however, more
complex than the Hubbard or Kondo lattice model. Transition metal
oxides crystallize in a three-dimensional (3D) perovskite
structure, where the oxygen ions occupy bridge positions between
transition metal ions, as in LaMnO$_3$, or in similar structures
with two-dimesional (2D) planes built by transition metal and
oxygen ions, as in CuO$_2$ planes of high temperature
superconductors. The oxygen ligand $2p$ orbitals play thereby a
fundamental role in these systems, and determine both the
electronic structure and actual interactions between the electrons
(holes) which occupy correlated $3d$ orbitals of transition metal
ions. The bands in transition metal oxides are built either by
$p_{\sigma}$ or by $p_{\pi}$ oxygen orbitals which hybridize with
the respective $3d$ orbitals of either $e_g$ or $t_{2g}$ symmetry.
Taking an example shown in Fig. \ref{fig:tmo}, it is clear that
the overlap between the $p_{\sigma}$ orbitals and $d_{x^2-y^2}$
orbitals is larger than that between the $p_{\pi}$ orbitals and
the corresponding $d$ orbitals of $t_{2g}$ symmetry. Therefore,
the $t_{2g}$ and $p_{\pi}$ states are filled in the cuprates, and
the relevant model Hamiltonians known as {\it charge transfer\/}
models include frequently only the $e_g$ orbitals of transition
metal ions and the $p_{\sigma}$ oxygen orbitals between them.

There are two crucial parameters which decide about the physical
properties of a transition metal oxide, provided the $d-p$
hybridization elements are much smaller than the value of the
on-site Coulomb interaction $U$. The latter parameter has to be
compared with the splitting between the $3d$ and $2p$ orbitals,
given by the so-called charge-transfer energy,
$\Delta=|\varepsilon_p-\varepsilon_d|$, where $\varepsilon_d$ and
$\varepsilon_p$ are the energies of an electron (hole) in these
states, respectively. These systems are called Mott-Hubbard
insulators (MHI) when $U<\Delta$, and it is in this limit that the
Hubbard model would apply directly for the description of a
metal-insulator transition. In the opposite case, one deals
instead with charge-transfer insulators, as introduced by Zaanen,
Sawatzky and Allen fifteen years ago \cite{Zaa85,Zaa90}. Both
classes of correlated (in contrast to band) insulators have quite
different spectral properties, but in the strongly correlated
regime the charge-transfer insulators resemble MHI, with a
charge-transfer energy $\Delta$ playing a role of the effective
$U$ \cite{Mei93}.

\begin{figure}[t]
\centerline{\psfig{figure=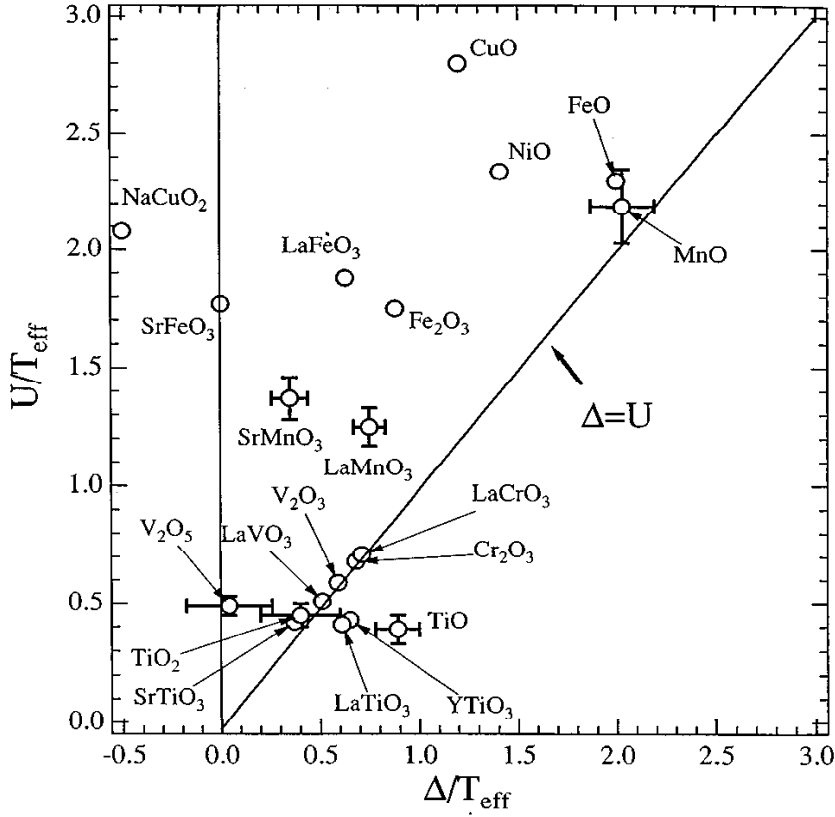,width=14.5cm}}
\smallskip
\caption{ Zaanen-Sawatzky-Allen phase diagram for $3d$ transition
metal oxides (after Ref. \protect\cite{Boc96}). } \label{fig:zsa}
\end{figure}

In reality, however, many oxides are found close to the above
qualitative boarder line between Mott-Hubbard and charge-transfer
systems (Fig. \ref{fig:zsa}), and one might expect that the only
relevant description has to be based on the charge-transfer models
which include explicitly both $d$ and $p$ orbitals. Nevertheless,
a reduction of such models to the effective simpler Hamiltonians
dealing only with correlated $d$-like orbitals is possible, and
examples of such mapping procedure have been discussed in the
literature \cite{Zha88,Fei96,Fei96a,Fei96b}. Unfortunately, there
is no general method which works in every case, but the principle
of the mapping procedure is clear, at least in perturbation
theory. We will follow this idea in the present paper and
concentrate ourselves on such simpler models which describe
interactions between $3d$ electrons, determined by the {\it
effective\/} hopping between transition metal ions which follows
from intermediate processes involving charge-transfer excitations
at the $2p$ oxygen orbitals \cite{Zaa88}. It will be clear from
what follows that while this simplification is allowed, there is
in general no way to reduce these models any further to those of
nondegenerate $d$ orbitals, at least not for the oxides with a
single electron or hole occupying (almost) degenerate $e_g$
orbitals.

We concentrate ourselves on a class of insulating strongly
correlated transition metal compounds, where the crystal field
leaves the $3d$ orbitals of $e_g$ symmetry explicitly degenerate
and thus the type of occupied orbitals is not known {\it a
priori\/}, while the effective magnetic interactions between the
spins of neighboring transition metal ions are determined by
orbitals which are occupied in the ground state
\cite{Geh75,Kug82,Zaa93,Kho97}. The most interesting situation
occurs when $e_g$ orbitals are partly occupied, which results in
rather strong magnetic interactions, accompanied by strong
Jahn-Teller (JT) effect. Typical examples of such ions are:
Cu$^{2+}$ ($d^9$ configuration, one hole in $e_g$-orbitals)
\cite{Kug73}, low-spin Ni$^{3+}$ ($d^7$ configuration, one
electron in $e_g$-orbitals) \cite{Gar92,Med97,Rod98}, as well as
Mn$^{3+}$ \cite{Ram97} and Cr$^{2+}$ ions (high-spin $d^4$
configuration with one $e_g$ electron). The situation encountered
for $d^9$ (or $d^7$) transition metal ions is simpler, as the
$t_{2g}$ orbitals are filled. The effective interactions may then
be derived by considering only $e_g$ orbital degrees of freedom
and spins $s=1/2$ at every site, and were first considered by
Kugel and Khomskii more than two decades ago \cite{Kug73}. In the
case of $d^4$ configuration one needs instead to consider larger
spins $S=2$ which interact with each other, due to virtual
excitation processes which involve either $e_g$ or $t_{2g}$
electrons \cite{Fei99}. Finally, the early transition-metal
compounds with $d^1$ or $d^2$ ions give also some interesting
examples of degenerate $t_{2g}$ orbitals
\cite{Cas78,Bao97,Bao97a,Pen97,Ezh99,Kei00}. In general, the
magnetic superexchange and the coupling to the lattice are weaker
in such cases due to a weaker hybridization between $3d$ and $2p$
orbitals (Fig. \ref{fig:tmo}). Moreover, this problem is somewhat
different due to the symmetry of the orbitals involved, and we
will not discuss it here.
The collective behavior of $e_g$ electrons follows from their
interactions. The models of interacting electrons in degenerate
$3d$ states are usually limited to the leading on-site part of
electron-electron interaction given by the Coulomb and exchange
elements, $U$ and $J_H$, respectively. The model Hamiltonian which
includes these interactions is of the form,
\begin{eqnarray}
\label{huj} &H_{int}&=(U+2J_H)
          \sum_{i\alpha}n_{i\alpha  \uparrow}n_{i\alpha\downarrow}
 + (U-J_H)\sum_{i,\alpha<\beta,\sigma}
          n_{i\alpha\sigma}n_{i\beta\sigma}
 + U      \sum_{i,\alpha<\beta,\sigma}
          n_{i\alpha\sigma}n_{i\beta\bar{\sigma}}        \nonumber \\
& &-J_H\sum_{i,\alpha<\beta,\sigma}
          d^{\dagger}_{i\alpha\sigma}d^{}_{i\alpha\bar{\sigma}}
          d^{\dagger}_{i\beta\bar{\sigma}}d^{}_{i\beta\sigma}
 + J_H\sum_{i,\alpha<\beta}
      ( d^{\dagger}_{i\alpha\uparrow}d^{\dagger}_{i\alpha\downarrow}
        d^{       }_{i\beta\downarrow}d^{       }_{i\beta\uparrow}
      + d^{\dagger}_{i\beta\uparrow}d^{\dagger}_{i\beta\downarrow}
        d^{       }_{i\alpha\downarrow}d^{       }_{i\alpha\uparrow}),
\end{eqnarray}
where summations over $\alpha<\beta$ guarantee that every pair of
different states interacts only once, and we neglected the
anisotropy of the interorbital interactions. It is important to
realize that precisely for this reason the multiplet structure of
transition metal ions \cite{Gri71} cannot be characterized by two
quantities such as $U$ and $J_H$, but one needs instead three
independent parameters, usually chosen as Racah parameters $A$,
$B$, and $C$. The commonly used relation between these parameters
and the Slater parameters $F_0$, $F_2$, and $F_4$ are given in
Table II of Ref. \cite{Ima98}. We also emphasize that the last
term  which describes the hopping of double occupancies between
different orbitals has the same amplitude as the spin exchange and
is $\propto J_H$. Such terms are frequently neglected in the
Hubbard-like models which thus cannot reproduce the correct
multiplet structure and give uncontrolled errors when
superexchange is derived from them.

Early applications of the model Hamiltonian (\ref{huj}) were
devoted to the understanding of magnetic states of transition
metals \cite{Ful91,Faz99,Ole83,Ole84a,Ole84b}. More recently, the
Hamiltonian (\ref{huj}) has been used to improve the local density
approximation (LDA) scheme for determining the electronic
structure of correlated transition metal oxides by including the
electron-electron interactions in the Hartree-Fock (HF)
approximation which gives the so-called LDA+U method \cite{Ani91}.
If the electron-electron interactions $U$ and $J_H$ are treated in
the HF approximation, they generate local potentials which act on
different local states $|i\alpha\sigma\rangle$, and allow thus for
the ground states with anisotropic distributions of charge and
magnetization over five $d$ orbitals. Such corrections improve the
gap values in Mott-Hubbard and charge-transfer insulators, and
become particularly important in the cuprates and manganites with
partial filling of $e_g$ orbitals.

The consequences of local potentials which follow from the Coulomb
and exchange terms in Eq. (\ref{huj}) are well seen on the example
of KCuF$_3$, one of the compounds which exhibits the degeneracy of
$e_g$ orbitals. We start out with the observation that according
to LDA KCuF$_3$ would be an undistorted perovskite, as the energy
increases if the lattice distortion is made (see Sec. III.C). The
reason is that the band structure of KCuF$_3$ determined by LDA
would give a band metal with a Fermi-surface which is not
susceptible to a band JT instability. LDA+U yields instead a
drastically different picture: it allows both the orbitals and the
spins to polarize which results in an energy gain of order of the
band gap, i.e., of the order of 1 eV and reproduces the observed
orbital ordering \cite{Lie95}. The orbital- and spin polarization
is nearly complete and the situation is close to the
strong-coupling limit underlying the spin-orbital model of Sec.
III.A.

Observed orbital ordering could also be obtained in manganites
using the LDA+U approach \cite{Ani97}. As a remarkable success of
this method, the orbital ordering which corresponds to the
so-called CE phase with the orbital ordering accompanied by the
charge ordering was obtained for Pr$_{1/2}$Ca$_{1/2}$MnO$_3$
\cite{Ani97}. In the undoped PrMnO$_3$ one finds that orbitals
alternate between two sublattices in $(a,b)$ planes, as also
expected following more qualitative arguments \cite{Kho97},
allowing thus for the ferromagnetic (FM) coupling between spins
within the $(a,b)$ planes. In contrast, the orbitals almost repeat
themselves along the $c$-axis, suggesting that the effective
magnetic interactions should be AF. However, when the
superexchange constants are determined using the band structure
calculations \cite{Sol96}, they do not agree with the experimental
data \cite{Hir96,Hir96a}. Not only the FM exchange constants are
larger by a factor close to four, but even the sign of the AF
superexchange along the $c$-axis {\it cannot be reproduced\/}.
Contrary to the suggestions made \cite{Sol96}, this result cannot
be corrected by effective interactions between further neighbors,
as the crystal structure and the momentum dependence of the
spin-waves in LaMnO$_3$ indicate that only nearest neighbor
interactions should contribute in the effective spin model
\cite{Hir96,Hir96a}, and represents one of the spectacular
examples how the electronic structure calculations fail in
strongly correlated systems. Therefore, it is necessary to study
the effective models which describe the low-energy sector of
excited states and treat more accurately the strong electron
correlations, as presented in this article.
\begin{figure}
\centerline{\psfig{figure=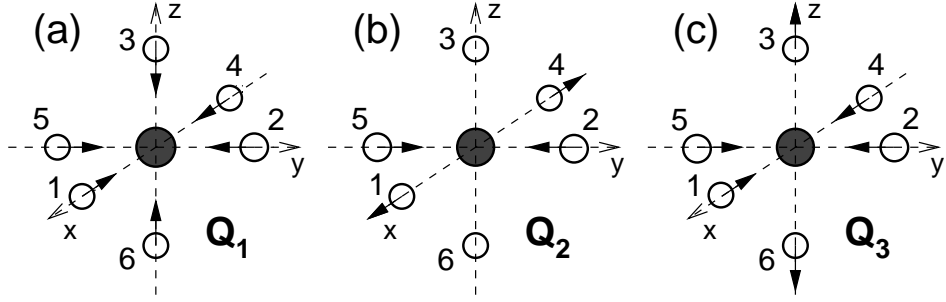,width=4cm,angle=270}}
\vspace{0.2in} \caption{ Different local modes for an MnO$_6$
octahedron: (a) breathing mode $Q_1$, (b) JT mode active in
$(a,b)$ planes, and (c) JT tetragonal distortion $Q_3$. Filled and
empty circles show Mn and O ions, respectively. }
\label{fig:jtmodes}
\end{figure}
It is impossible to discuss the magnetic and orbital states of
cuprates and manganites without paying attention to the lattice
distortions. When the cubic crystal distorts, the energies of
$e_g$ orbitals change due to the coupling to the lattice.
Depending on the type of distortion, the energy of one or the
other $e_g$ orbital will be lower. Therefore, particular lattice
distortions alone might stabilize orbital ordering. In spite of
some other views presented in the literature, we shall argue below
that this is not the case for the cuprates, and the orbital
ordering observed in KCuF$_3$ follows from the electronic
interactions between strongly correlated electrons. We shall
discuss this problem in particular for manganites (in Sec. V),
where we argue that the electronic interactions alone determine
the observed type of magnetic ordering, which is however
additionally stabilized by the orbital interactions which follow
from the JT effect \cite{Fei99}.

Let us recall first the single-ion (noncooperative) JT effect. It
was realized long ago by Kanamori that local JT effect leads to
the symmetry lowering for Cu$^{2+}$ and Mn$^{3+}$ ions with a
single hole (electron) at octahedral sites \cite{Kan60}. He
considered a single-site problem of an ion surrounded by six
neighbors which may be distorted from their initial symmetric
positions which satisfy the octahedral symmetry (see Fig.
\ref{fig:jtmodes}). The normal modes may be written as follows:
\begin{eqnarray}
\label{jtmodes} Q_1&=&\frac{1}{\sqrt{3}}(x_1-x_4+y_2-y_5+z_3-z_6),
\\ Q_2&=&\frac{1}{\sqrt{2}}(x_1-x_4-y_2+y_5),  \\
Q_3&=&\frac{1}{\sqrt{6}}(2z_3-2z_6-x_1+x_4-y_2+y_5),
\end{eqnarray}
where $x_i$, $y_i$ and $z_i$ are the coordinates of atom $i$. In
contrast to the breathing mode $Q_1$, where all the neighbors move
towards/away from the central site and the $e_g$ orbitals do not
split, the other two normal modes ($Q_2$ and $Q_3$) remove orbital
degeneracy and favor the occupancy of either $|x\rangle\equiv
|x^2-y^2\rangle$ or $|z\rangle\equiv |3z^2-r^2\rangle$ orbital.

Following Kanamori \cite{Kan60}, we write the effective
Hamiltonian in the form,
\begin{eqnarray}
\label{hjth0} H_{JT}&=&H_0+H_1+H_2,   \\ \label{hjt}
H_0&=&\beta\lambda\sum_i Q_{1i}, \\ \label{hjth1}
H_1&=&\lambda\sqrt{C}\sum_{i\alpha\beta\sigma\sigma'}
c_{i\alpha\sigma}^{\dagger}
(Q_{2i}\sigma_i^x+Q_{3i}\sigma_i^z)_{\alpha\beta}c_{i\beta\sigma'}^{},
\\ \label{hjth2} H_2&=&\case{1}{2}C\sum_i(Q_{2i}^2+Q_{3i}^2),
\end{eqnarray}
where $c_{i\alpha\sigma}^{\dagger}$ is a creation operator for an
$e_g$ electron in orbital $\alpha$ with spin $\sigma$,
$\sigma_i^x$ and $\sigma_i^z$ are the Pauli matrices, and
$\lambda$ and $\beta$ are the parameters which depend on the
system. The ions at different sites are independent and one may
solve just a single-site problem, assuming an ansatz for the
orbital state,
\begin{eqnarray}
\label{newstate} |i\theta\sigma\rangle=\cos\theta
|iz\sigma\rangle+\sin\theta |ix\sigma\rangle .
\end{eqnarray}
Using the uniform angle $\theta_i=\theta$, the classical
distortions and the coordinates and the orbital state are given as
follows:
\begin{eqnarray}
\label{jtminq0} Q_{2i}^0&=&\frac{\lambda}{\sqrt{C}}\sin 2\theta,
\hskip 1cm Q_{3i}^0=\frac{\lambda}{\sqrt{C}}\cos 2\theta, \\
\label{jtmine} \langle \sigma_i^z\rangle&=&\cos 2\theta, \hskip
1.4cm \langle \sigma_i^x\rangle=\sin 2\theta.
\end{eqnarray}
As easily recognized from Eqs. (\ref{jtminq0}) and (\ref{jtmine}),
the orbital state follows the lattice distortions and one finds
the energy minimum given by $-\lambda^2/2$, showing that the
lowest state cannot be determined uniquely. The situation changes
when anharmonic terms are included which lead to the energy
contribution of the form,
\begin{equation}
\label{hani} H_A=-A\sum_i\cos 6\theta_i,
\end{equation}
with $A>0$. This term favors directional orbitals, and tetragonal
distortions with the elongated tetragonal axis is the most stable
structure. One finds identical energy for three different
distortions, corresponding to $3z^2-r^2$ orbital at $\theta=0$,
and to $3x^2-r^2$ ($3y^2-r^2$) orbital at $\theta=\pm \pi/3$,
respectively.

The above energy contributions occur for the sites occupied by a
single $e_g$ electron, as for instance at Mn$^{3+}$ ions.
Important deformations of the lattice occur as well around
Mn$^{4+}$ ions, when the $e_g$ electron is absent. In this case
the breathing mode $Q_1$ becomes active, and the respective energy
contribution takes the form,
\begin{equation}
\label{hhole} H_{hole}=\beta\lambda\sum_i(1-n_i)Q_{1i},
\end{equation}
and typically $\beta\gg 1$.

Although the tendency towards directional orbitals might be
considered to be generic for the present systems, such states
cannot occur independently of each other in a crystal, as the
lattice distortions are correlated. Therefore, a more realistic
description requires a coupling between the oxygen distortions
realized around different manganese sites. We discuss this problem
on the example of LaMnO$_3$, which has been studied in more detail
only recently \cite{Mil96,Rod96}. If oxygens around a given
Mn$^{3+}$ ion are distorted, there are also distortions of common
oxygen atoms around the neighboring Mn$^{3+}$ ions, and in this
way the orbital angles are coupled to each other. This is called
cooperative JT effect, in contrast to the noncooperative one
\cite{Kan60} which concerns single sites. The total energy which
follows from the coupling to the lattice was derived by Millis
\cite{Mil96}:
\begin{eqnarray}
\label{jtminq} H_{lat}&=&-|E_0|\sum_i[n_i^2+\beta^2(1-n_i)^2+A\cos
6\theta_i]  \nonumber \\ & &+\kappa\sum_{\langle ij\rangle}n_in_j
\cos 2(\theta_i+\psi_{\alpha})\cos 2(\theta_j+\psi_{\alpha})
\nonumber
\\ &&+ 2\beta\kappa\sum_{\langle ij\rangle}(1-n_i)n_j \cos
2(\theta_j+\psi_{\alpha}) +\beta^2\kappa\sum_{\langle
ij\rangle}(1-n_i)(1-n_j).
\end{eqnarray}
It includes the on-site terms, and the intersite couplings along
the bonds, making the JT effect cooperative. A particular tendency
for occupying the orbitals of a given type is expressed by the
terms $\cos 2(\theta_i+\psi_{\alpha})$, with the angle
$\psi_{\alpha}$ depending on the bond direction as follows:
$\psi_c=0$, $\psi_a=-\pi/3$, and $\psi_b=\pi/3$, for the bonds
$\langle ij\rangle$ along the $c$, $a$, and $b$-axis,
respectively. The coupling constants $E_0$ and $\kappa$ depend on
the coefficient $\lambda$ introduced in Eq. (\ref{hjth1}) and are
functions of the respective force constants which describe the
coupling between the manganese and oxygens ions, and between the
pairs of oxygen ions, respectively. For the purpose of these
lectures we will treat them as phenomenological parameters, but an
interested reader may find explicit expressions and more technical
details in Ref. \cite{Mil96}.
\begin{figure}
\centerline{\psfig{figure=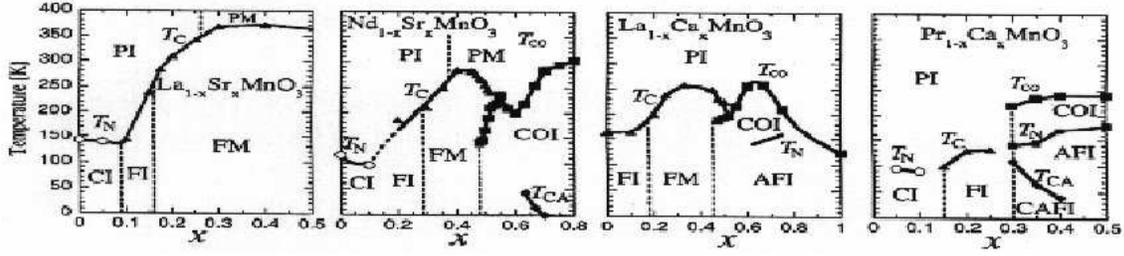,width=15cm}}
\smallskip
\caption{ Phase diagrams of A$_{1-x}$B$_x$MnO$_3$ compounds (after
Ref. \protect\cite{Ima98}). Various ordered phases are labeled as
follows: FI (PI) -- ferromagnetic (paramagnetic) insulator, FM
(PM) -- ferromagnetic (paramagnetic) metal, CI -- spin-canted
insulating, COI -- charge-ordered insulating, AFI --
antiferromagnetic insulating, CAFI -- canted antiferromagnetic
insulating. } \label{fig:expphd}
\end{figure}

Recent extensive research on the CMR manganites has generated a
huge number of papers in the scientific literature. The interest
is motivated by very spectacular experimental properties, with
typically several magnetic phases stable in different doping
regimes, most of them insulating, but one metallic FM phase
\cite{Ima98,Ram97}. With increasing temperature a transition from
the FM metallic phase occurs either to a paramagnetic metal, or at
lower doping to a paramagnetic insulator. In the latter case a
large change of the resistivity accompanies the phase transition,
and the transition temperature is strongly modified by the
external magnetic field, giving rise to the phenomenon of CMR
\cite{Jin94}. Examples of magnetic phase diagrams for
representative distorted perovskites R$_{1-x}$A$_x$MnO$_3$ are
shown in Fig. \ref{fig:expphd}. The FM metallic phase is found in
first three compounds, while in Pr$_{1-x}$Ca$_x$MnO$_3$ all
magnetic phases are insulating. As the average ionic radius of the
perovskite A site increases from (La,Sr) to (Pr,Ca) through
(Nd,Sr) or (La,Ca), orthorhombic distortion of the GdFeO$_3$ type
\cite{Ima98} increases, resulting in the decrease of the decrease
of the one-electron bandwidth $W$. When the bandwidth gets
reduced, the balance between the double-exchange (DE)
\cite{Zen51,And55} and other interactions changes and such
instabilities as JT type distortions, charge and/or orbital
ordering may occur. Moreover, the AF superexchange may play an
important role and stabilize the AF order in a broader doping
regime.

In order to understand the phase diagrams of manganites, one needs
to consider four different kinds of degrees of freedom: charge,
spin, orbital, and lattice. Therefore, the models which treat
doped manganites in a realistic way are rather sophisticated. A
somewhat simpler situation occurs in the undoped LaMnO$_3$ as the
charge fluctuations are suppressed by large on-site Coulomb
interactions and one may study effective magnetic and orbital
interactions, as we present in Sec. V.A. This problem may be also
approached in a phenomenological way by postulating model
Hamiltonians which contain such essential terms as the Hund's rule
exchange interaction between $e_g$ and $t_{2g}$ electrons, the AF
interactions between the $t_{2g}$ core spins, and the coupling to
the lattice. As an example, we present the Hamiltonian of the
degenerate Kondo lattice with the coupling to local distortions of
MnO$_6$ octahedra \cite{Hot99},
\begin{eqnarray}
\label{jtminq} H_{}&=&\sum_{ij\alpha\beta\sigma}
  t_{\alpha\beta}c_{i\alpha\sigma}^{\dagger}c_{j\beta\sigma}^{}
  -J_H\sum_{i\alpha\sigma\sigma'}{\vec S}_i\cdot
  c_{i\alpha\sigma}^{\dagger}{\vec\sigma}_{\sigma\sigma'}^{}c_{i\beta\sigma}^{}
  +J'\sum_{\langle ij\rangle}{\vec S}_i\cdot {\vec S}_j          \nonumber \\
&+&\lambda\sum_{i\alpha\beta\sigma}c_{i\alpha\sigma}^{\dagger}
(Q_{1i}\sigma^0_i+Q_{2i}\sigma^x_i+Q_{3i}\sigma^z_i)_{\alpha\beta}
  c_{i\beta\sigma}^{}
+\frac{1}{2}\sum_i(\beta Q_{1i}^2+Q_{2i}^2+Q_{3i}^2),
\label{hamhotta}
\end{eqnarray}
where $c_{i\alpha\sigma}^{\dagger}$ is the creation operator of an
$e_g$ electron with spin $\sigma$ in the $d_{x^2-y^2}$
($d_{3z^2-r^2}$) orbital at site $i$. The hopping elements
$t^{\alpha\beta}_{ij}$ between nearest neighbors follow from the
Slater-Koster rules \cite{Sla54}. ${\vec S}_i$ is the localized
spin of $t_{2g}$ electrons, and $\sigma^x_i$ and $\sigma^z_i$ are
Pauli matrices. $\lambda$ stands for the dimensionless
electron-phonon coupling constant. Different distortions $Q_{1i}$,
$Q_{2i}$, and $Q_{3i}$ are the breathing mode and two JT modes
shown in Fig. \ref{fig:jtmodes}. Hotta {\it et al.\/} \cite{Hot99}
took into account the cooperative nature of the JT phonons by
introducing the coupling between the neighboring Mn ions in the
normal coordinates for distortions of MnO$_6$ octahedra
\cite{Mil96,All99}. The important parameter is the ratio of the
vibrational energies for manganite breathing ($\omega_{b}$) and JT
($\omega_{\rm JT}$) modes, $\beta=(\omega_{b}/\omega_{\rm JT})^2$.
\begin{figure}[tbp]
\centerline{\psfig{figure=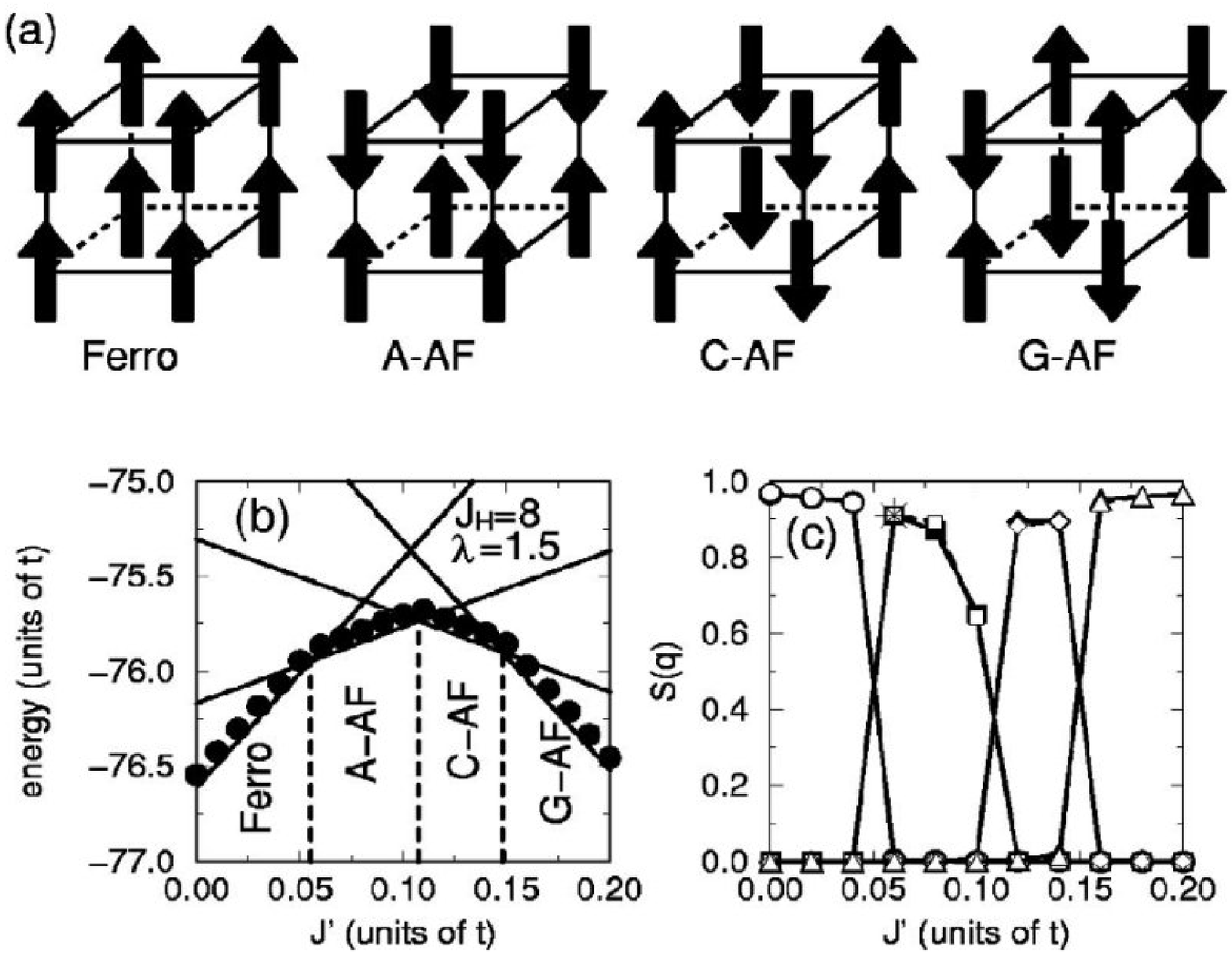,width=10cm}}
\centerline{\psfig{figure=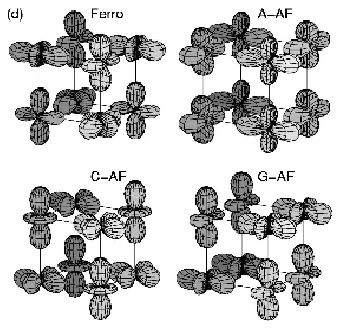,width=10cm}}
\smallskip
\caption{ Types of magnetic order (a) and orbital order (d) in
perovskite structures: Ferro, A-AF, C-AF, and G-AF, as obtained in
MC simulations by Hotta {\it et al.\/} \protect\cite{Hot99} using
a model of $e_g$ electrons coupled to the core $t_{2g}$ spins and
to JT lattice distortions (\protect\ref{hamhotta}). Quantitative
results obtained as functions of the AF coupling $J'$ between
$t_{2g}$ spins for $J_H=8$ and $\lambda=1.5$ are shown by two
middle panels: (b) the total energy for a $2\times 2\times 2$
lattice; (c) the spin-spin structure factor $S({\vec q})$ obtained
for $4\times 4\times 2$ (solid symbols) and $4\times 4\times 4$
(open symbols) clusters. The circles, squares, diamonds, and
triangles indicate $S({\vec q})$ for ${\vec q}=(0,0,0)$, ${\vec
q}=(\pi,0,0)$, ${\vec q}=(\pi,\pi,0)$, and ${\vec
q}=(\pi,\pi,\pi)$, respectively. } \label{fig:mo+oo}
\end{figure}
The calculations performed by the relaxation technique and by
Monte-Carlo \cite{Yun98} for finite 3D clusters are summarized in
Fig. \ref{fig:mo+oo}. Depending on the parameters, four different
magnetic phases are found [Fig. \ref{fig:mo+oo}(a)]: FM phase
(Ferro), so-called A-AF phase with staggered FM planes, C-AF phase
with staggered FM chains, and, finally, the G-AF phase which is
the 3D N\'eel order. They are identified by investigating the
magnetic structure factor $S({\vec q})=\frac{1}{N}\sum_{ij}
\exp[i{\vec q}({\vec R}_i-{\vec R}_j)\langle {\vec S}_i{\vec
S}_j\rangle$ [Fig. \ref{fig:mo+oo}(c)]. It is straightforward to
understand that the ground state is FM at $J'=0$. In this case the
lowest energy gain may be obtained from the combination of the
kinetic energy of $e_g$ electrons with the local Hund's rule
$\propto J_H$. Increasing $J'$ increases the tendency towards the
AF order and leads finally to the G-AF phase in the range of large
$J'/t>0.15$ [Fig. \ref{fig:mo+oo}(b)].

Hotta {\it et al.\/} \cite{Hot99} found that various magnetic
orderings are accompanied by the orbital orderings shown in Fig.
\ref{fig:mo+oo}(d)]. The shape of the occupied orbital arrangment
is not easy to understand, however. It follows from the
cooperative JT effect and expresses a compromise between the
orbital and magnetic energies. The overall picture might seem
appealing, but it is questionable whether the JT effect is the
dominating mechanism that determines the magnetic and orbital
ordering in manganites and related compounds which are known to be
primarily MHI \cite{Ima98}, i.e., the on-site Coulomb interaction
$U$ is the largest parameter, typically $5<U<10$ eV, which
dominates the hybridization. Although it has been argued that the
large Coulomb interaction will not change the main results shown
in Fig. \ref{fig:mo+oo} \cite{Hot99}, we do not think this
conclusion is allowed. In fact, in the absence of large Coulomb
interaction $U$ the magnetic interactions are dominated by DE
\cite{And55} and the system is FM at $J'=0$ as new effective
interactions arise in the presence of large $U$, and they easily
might change the delicate balance between different magnetic and
orbital ordered phases. We shall discuss the problem of magnetic
and orbital ordering in detail in Secs. III and V and show that
the electronic interactions alone give a dominating contribution
to magnetic interactions.

\begin{figure}
\centerline{\psfig{figure=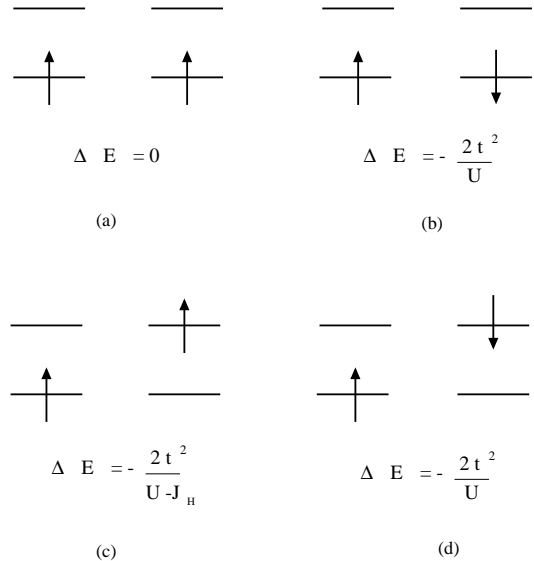,width=7cm}}
\smallskip
\caption{ The various configurations of nearest neighbors for
doubly-degenerate
    orbitals. The kinetic exchange is always AF in a nondegenerate model (b),
    while the processes which involve differently occupied orbitals on both
    sites may be either FM (c) or AF (d). $U$ and $J_H$ are the on-site Coulomb
    and Hund's rule exchange interactions (after Ref. \protect\cite{Kho97}). } \label{fig:se}
\end{figure}

Magnetism in transition metals and in their compounds is known to
be due to intraatomic Coulomb interaction \cite{Sla36}. The
simplest model which takes into account the Coulomb interaction is
that due to Hubbard (see Sec. II.A). It describes electrons in a
narrow and nondegenerate tight-binding band and allows for
repulsion $U$ between electrons only when they are at the same
site. This model has been studied intensively since Hubbard
proposed it, especially in connection to the occurrence of
magnetism \cite{Faz99}. Anderson has shown in 1959 \cite{And59}
that the Hubbard Hamiltonian is equivalent to a Heisenberg
Hamiltonian with an AF superexchange interaction given in terms of
the hopping amplitude and the Coulomb interaction, if $U$ is
large. Indeed, for two neighboring ions an extra delocalization
process $d^1_id^1_j\rightarrow d^{0}_id^2_j$ is only possible for
antiparallel arrangment of neighboring spins, decreasing the
energy and favoring this configuration (Fig. \ref{fig:se}).
Therefore, if each ion has only one nondegenerate orbital, the
superexchange is AF, as explained in Sec. II.A.

The question of whether the AF correlations might evolve by
changing the electron concentration into ferromagnetism is still
controversial. A rigorous proof of Nagaoka \cite{Nag66} of the
existence of ferromagnetism applies only in a very special case --
in the limit of infinite Coulomb repulsion $U$ when one hole or
one extra electron is added to the half-filled band ($n=1$), in a
lattice of particular symmetry. However, the occurrence of
ferromagnetism comes in a more natural way if one takes into
account the orbital degeneracy as Van Vleck has emphasized
\cite{Vlec53}. In the case of two-fold orbital degeneracy,
applying similar arguments as those used by Anderson \cite{And59},
one ends up with a richer structure of effective interactions when
the processes $d^1_id^1_j\rightarrow d^{0}_id^2_j$ are analyzed.
For the occupancy of $n=1$ electron per atom one finds four
possible situations as depicted in Fig. \ref{fig:se}: (a) same
orbital -- same spin, (b) same orbital -- different spin, (c)
different orbital -- same spin, (d) different orbital -- different
spin. This problem has been studied already in the seventies, but
mostly starting from simplified model Hamiltonians
\cite{Cyr75,Cyr75a,Cyr75b,Spa80}. In order to study the
qualitative effects, the simplest case with only diagonal hopping
and equal intra- and interorbital Coulomb elements $U$ in Eq.
(\ref{huj}) has been usually assumed.

As a qualitative new effect due to the Hund's rule exchange
$J_H>0$, {\it ferromagnetic superexchange\/} becomes possible, if
the excitation involves the high-spin state with two parallel
electrons [Fig. \ref{fig:se}(c)]. Although the processes which
contribute to the superexchange for nondegenerate orbitals [Fig.
\ref{fig:se}(b)] are also present, and there are more AF terms,
the FM term has the largest coefficient due to the structure of
Coulomb interactions (\ref{huj}). Therefore, one might expect that
under certain conditions such terms could promote ferromagnetism.
While this is not easy and happens only for rather extreme
parameters in the doubly degenerate Hubbard model with isotropic
but diagonal hopping elements \cite{Cyr75,Cyr75a,Cyr75b}, it has
been recognized in these early works that the orbital ordering may
accompany the magnetic ordering, and orbital superlattice favors
the appearance of magnetism at zero temperature. Indeed, the onset
of magnetic long-range order (LRO) is obtained for such values of
parameters that the usual Stoner criterion is not yet fulfilled.
Furthermore, the studies at finite temperature revealed that the
orbital order is more stable than the magnetic one. Therefore, two
phase transitions are expected in general: at the lower
temperature the ferromagnetism disappears and, at the higher one,
the orbital order \cite{Cyr75,Cyr75a,Cyr75b}.

In order to understand the behavior of CMR manganites, it is
necessary to include the orbital degrees of freedom for partly
occupied $e_g$ orbitals. The motivation comes both from theory and
experiment. For quite long time it was believed that the FM state
in manganites can be understood by the DE model
\cite{Zen51,And55}. In fact, it provides not more than a
qualitative explanation why the doped manganites should have a
regime of FM metallic state (Fig. \ref{fig:expphd}). However, if
one calculates the Curie temperature $T_C$ using the DE model the
values are overestimated by a factor of the order of five
\cite{Mil95}. Also the experimental dependence of the resistivity
in the metallic phase \cite{Uru95} cannot be reproduced within the
DE model \cite{Mil95}. Finally, in a FM metal one expects a large
Drude peak and no incoherent part in the optical conductivity. The
experimental result is quite different -- most of the intensity is
incoherent at low temperatures, and only a small Drude peak
appears which absorbs not more than 20 \% of the total spectral
weight \cite{Oki97,Oki95}. All these results demonstrate the
importance of orbital degrees of freedom which should be treated
on equal footing as the spins of $e_g$ electrons.
The orbital degeneracy leads therefore to a new type of models in
the theory of magnetism: {\it spin-orbital models}. They act in
the extended space and describe the (super)exchange interactions
between spins, between orbitals, and simultaneous spin-and-orbital
couplings. In order to address realistic situations encountered in
cuprates and manganites, such models cannot rely on the degenerate
Hubbard model \cite{Cyr75,Cyr75a,Cyr75b}, but have to include the
anisotropy in the hopping elements \cite{Sla54}, nonconservation
of the orbital quantum number, and realistic energetic structure
of the excited states \cite{Gri71}. Once such models are derived,
as we present in Secs. III and V, their phase diagrams may be
studied using the mean-field (MF)
approximation\cite{Fei99,Fei97,Ole00}. It turns out that their
phase diagrams show an unusual competitions between classical
(magnetic and orbital) ordering of different type, in particular
close to the degeneracy of $e_g$ orbitals. Therefore, two
interesting questions occur for such orbitally degenerate MHI:
 (i) Which {\em classical\/} states with magnetic LRO do exist in the
       neighborhood of orbital degeneracy?
(ii) Are those forms of classical order
       always stable against {\em quantum\/} (Gaussian) fluctuations?
We will show that the orbitally degenerate MHI represent a class
of systems in which spin disorder occurs due to frustration of
{\em spin and orbital\/} superexchange couplings. This frustration
mechanism is different from that which operates in quantum
antiferromagnets, and suppresses the magnetic LRO in the ground
state {\it even in three dimensions\/}.

We organized the remaining chapters of this article as follows. In
order to clarify the basic magnetic interactions in strongly
correlated oxides, we start with the superexchange and DE in Sec.
II. Next we introduce and analyze on the classical level the
simplest spin-orbital model for spins $S=1/2$ (Sec. III) which
applies to cuprates, and present its collective modes. The model
exhibits an interesting frustration of magnetic interactions, and
the classical phases are destabilized by quantum fluctuations
(Sec. IV). Therefore, we discuss the problem of quantum disorder
in low dimensional spin models and in an idealized spin-orbital
model with SU(4) symmetry in one dimension. The physics of
manganites is richer than that of cuprates close to the degeneracy
of $e_g$ orbitals, and additional interactions between orbital
variables occur due to the coupling to the lattice (Sec. V). The
understanding of various phase transitions shown in Fig.
\ref{fig:expphd} remains an outstanding problem which requires to
study simultaneously the coupling to the lattice responsible for
the insulating behavior of doped systems, and the DE model at
orbital degeneracy. This latter problem is very actual and was
addressed for the first time only last year \cite{Bri99} (Sec.
VI). The spin and orbital interactions lead in general to new type
of effective $t$-$J$ models \cite{Zaa93}, and we analyze
spin-waves in FM phase and give an example of the hole which
dresses by orbital excitations and compare this situation with the
hole motion in the $t$-$J$ model (Sec. VII). Among many open
questions for doped manganites, we selected a few such as the
CE-phase, stripes, orbital ordering and phase separation (Sec.
VIII). In our opinion, they are crucial to understand the
complexity of the experimental phase diagrams. We give a brief
summary and our conclusions in Sec. IX.


%
\section{Magnetic interactions for nondegenerate orbitals }

\subsection{Superexchange and $t$-$J$ model}

Before discussing the consequences of $e_g$ orbital  degeneracy in
cuprates and manganites, we review shortly the basic magnetic
interactions in the models of nondegenerate orbitals -- the
superexchange and the DE. The main idea to derive the
superexchange is the notion of the Mott-Hubbard insulator at $n=1$
in which the charge fluctuations are suppressed and the electrons
localize, occupying the states of the lower Hubbard band.
Therefore, part of degrees of freedom is integrated out and one
may study an effective model which captures the essential features
of the low-energy excitations.

Before deriving the effective superexchange model for degenerate
orbitals, we analyze shortly the nondegenerate orbitals filled by
one electron per atom ~($n$=1). This situation plays a fundamental
role in strongly correlated systems and elucidates the general
principle of introducing magnetic (and orbital) interactions in a
Mott-Hubbard insulator by integrating out charge fluctuations and
reducing the problem. The starting point is the Hubbard
Hamiltonian,
\begin{eqnarray}
\label{hamhub} H&=&H_t+H_U,\\ H_t&=&-t\sum_{\langle ij \rangle
\sigma} \left(c^{\dag}_{i\sigma} c_{j\sigma}+H.c.\right),\\
H_U&=&U\sum_i n_{i\uparrow}n_{i\downarrow},
\end{eqnarray}
where $U$ and $t$ are standing for the on-site screened Coulomb
repulsion, and for the hopping amplitude between nearest
neighbors, respectively, and the summation runs over the bonds
$\langle ij \rangle$ between nearest neighbors. Furthermore, the
operator $c^{\dag}_{i\sigma}$ creates an electron with spin
$\sigma$ at site $i$, and
$n_{i\sigma}=c^{\dag}_{i\sigma}c_{i\sigma}$ is the electron
density operator. Recall that the interaction term in the Hubbard
model can be reexpressed in the following way:
\begin{equation}
H_{U}=-\frac{2}{3}U\sum_{i}\left( \vec{S}_{i}\right) ^{2},
\end{equation}
so that it forces the spin $\vec{S}_{i}$ to be maximal if $U$
becomes infinitely large, i.e., doubly occupied sites are
forbidden. Only $\mid\uparrow\rangle $ and $\mid\downarrow\rangle
$ states are kept in this large $U$ limit at half-filling. Taking
the atomic limit ($t=0$), the interaction part of the Hamiltonian
(\ref{hamhub}) has infinitely many ($2^N$, where $N$ is the number
of sites) degenerate eigenstates, given by different spin
configurations. In order to lift this large degeneracy we will
keep the effects of fluctuations induced by the kinetic energy
term to leading order in an expansion in $(t/U)$. As usually, this
problem has to be solved in degenerate perturbation theory.

Suppose we begin with an arbitrary configuration which can be
labeled by the local $z$th components of the spins $S_i^z$. In the
expansion in powers of $(t/U)$, one includes contributions from
intermediate states in which one site will become doubly occupied
and, at the same time, the other site becomes empty \cite{Cha77}.
The energy of the excited state is $U$ above that of the
degenerate ground state manifold. The squared transition matrix
element is $t^{2}$ and the combinatorial factor of two has to be
included since this process can occur in two different ways. Hence
we expect that the relevant parameter of the effective spin
Hamiltonian should be $2t^2/U$. Also, the final state after the
double occupancy dissociates has to be either the same as the
initial state, or it may differ at most by a spin exchange. The
candidate for the effective Hamiltonian is, of course, the quantum
Heisenberg antiferromagnet \cite{Cha77,And59,Spa80}, since we know
that the spin-spin interaction follows from a possibility of
permuting the electrons on a lattice.

The formal derivation of the effective Heisenberg model can be
performed in a few different equivalent ways: (i) by means of a
canonical transformation \cite{Cha77}, (ii) with Schrieffer-Wolff
procedure \cite{Sch66}, and (iii) with Brillouin-Wigner
perturbation approach \cite{Baym74}. The first method is the most
transparent to use away from the half-filling, where it leads to
the $t$-$J$ model, known as a minimal model to describe the
electronic states in high temperature superconductors
\cite{Dag94}. We will not repeat here the details of the
derivation of the $t$-$J$ model \cite{Cha77} as it belongs already
to the textbook material \cite{Ful91,Faz99}. The common result of
all these procedures at $n=1$ is the removal of degeneracy within
the second order perturbation, and the effective Hamiltonian,
given by the following expression:
\begin{eqnarray}
\langle \phi|H_{eff}|\psi \rangle=-\langle \phi |H_t
\frac{1-P_0}{H_U}H_t|\psi \rangle=-\sum_{n\neq\{0\}}\langle \phi
|H_t|n\rangle \frac{1}{U} \langle n|H_t|\psi \rangle,
\end{eqnarray}
where $|\phi\rangle,|\psi\rangle$ denote states in the subspace
without double occupancies, with a projection operator P$_0$. The
states $|n\rangle$ are configurations with one doubly occupied
site, and each term in the sum can be represented by a retraceable
exchange path. Thereby we assume that $n\leq 1$; the case of $n>1$
may be treated by the same method after performing a particle-hole
transformation. Since the total spin per two sites is conserved in
the excitation process
$\mid\uparrow\rangle_{i}\mid\downarrow\rangle_{j}\rightarrow
|0\rangle_{i}\mid\uparrow\downarrow\rangle_{j}$, we can express
the operators which connect the initial and final states of this
transition by means of the projection operators for the singlet
and for the triplet state on the bond $\langle ij\rangle$,
respectively:
\begin{equation}
Q_S(i,j)=\left(-\vec{S_i}\cdot\vec{S_j}
               +\frac{1}{4}\right), \hskip 1cm
Q_T(i,j)=\left( \vec{S_i}\cdot\vec{S_j}
               +\frac{3}{4}\right).
\label{projst}
\end{equation}
The excitation energy associated to the process
$\mid\uparrow\rangle_{i}\mid\downarrow\rangle_{j}\rightarrow
|0\rangle_{i}\mid\uparrow\downarrow\rangle_{j}$ which creates a
singlet at site $j$ is $\varepsilon _{S}=U$, (if we start from a
triplet configuration, virtual processes are blocked due to the
Pauli principle). Taking into account that the double occupancy
may be created either at site $i$ or at site $j$, the effective
Hamiltonian can be expressed in the following way,
\begin{equation}
H_{eff}=-\frac{4t^{2}}{U}\sum_{\langle ij\rangle}Q_S(i,j)
       =J\sum_{\langle ij\rangle}
       \left(\vec{S_i}\cdot\vec{S_j}-\frac{1}{4}\right).
\end{equation}
Thus, one finds the AF Heisenberg model with the superexchange
constant \\$J$=$4t^{2}/U$.

\subsection{ Kondo lattice model -- double exchange }

The early theoretical studies of manganites were concentrated on
the models introduced in order to understand the FM phase which
occurs in the doped materials. The basic understanding of the
tendency towards FM order follows from the so-called {\it double
exchange\/} model \cite{Zen51,And55} -- it explains that electrons
in a partially filled band maximize their kinetic energy when
their spins are aligned with the localized spins which order
ferromagnetically. In fact, this phenomenon is quite reminiscent
of the Nagaoka state in the Hubbard model\cite{Nag66}. However, in
spite of this qualitative explanation of the existence of
ferromagnetism, several features of the experimental phase
diagrams of manganites remain unclear especially at low
temperature, where one has to go beyond the DE model. In order to
understand the reasons of its shortcomings, let us present briefly
the main consequences of the DE model.

The Kondo lattice Hamiltonian with {\it ferromagnetic\/}
spin-fermion coupling can be defined as follows,
\begin{equation}
\label{kondol} H = -t\sum_{\langle
{ij}\rangle\sigma}(c^\dagger_{i\sigma}c_{j\sigma}^{}+H.c.) - J_H
\sum_{i\alpha\beta}{\vec S}_{\bf i}\cdot
  { c^{\dagger}_{i\alpha} {\vec\sigma}_{\alpha\beta}^{}c_{i\beta}^{} },
\end{equation}
where $c_{i\sigma}^\dagger$ is a creation operator for an electron
at site $i$ with spin $\sigma$, and ${\vec S}_i$ is the total spin
of the $t_{2g}$ electrons $S=3/2$, assumed to be localized. The
first term describes the kinetic energy of electrons in a
nondegenerate band due to the electron transfer between
nearest-neighbor Mn-ions, $J_H>0$ is the FM Hund's coupling
between the itinerant electron and the $t_{2g}$ core spin
($S=3/2$). The average electronic density of $e_g$ electrons,
denoted by $n=\langle n_i\rangle$, is adjusted using a chemical
potential $\mu$.

Let us consider a bond in the perovskite structure formed by two
Mn atoms with an oxygen atom in between. In the ionic
configuration the $2p$ shell of the O$^{2-}$ ion is completely
filled. In order to treat the problem semiclassically
\cite{And55}, we assume that the Mn ions have rather large spins
$S_1$ and $S_2$, so that one could assign to them definite
directions in space, and a definite angle relative to each other.
If an itinerant $e_g$ electron is on the site $i=1$, it has two
states, of energies $E_1=\pm JS$, if the electron spin is parallel
and antiparallel to the spin ${\vec S}_1$, respectively. On atom
$i=2$ it also has similar two states, but defined with respect to
the direction of spin ${\vec S}_2$. As the electron spin direction
is conserved in the hopping process, the final state has to be
projected on the new local axis. This is equivalent to rotating
the transfer matrix between these two sites in such a way that its
elements refer correctly to the projected spin components in the
rotated basis.

Let us label the two electronic spin functions referring to the
direction of ${\vec S}_1$ by $\alpha$ and $\beta$, and those
referring to ${\vec S}_2$ by $\alpha'$ and $\beta'$. The energies
of the eigenstates on atom $i=1$ are:
\begin{equation}
 E(d_1\alpha) = -J_HS,  \hskip 1cm E(d_1\beta)=J_H(S+1),
\end{equation}
for FM coupling $J_H>0$. The energies of the eigenstates on atom
$i=2$ are given by similar expressions:
\begin{equation}
 E(d_2\alpha')=-J_HS, \hskip 1cm E(d_2\beta')=J_H(S+1).
\end{equation}
Here $d_1$ and $d_2$ describe the spin function of an electron
localized at atom 1 and atom 2, respectively.

The transformation which expresses $\alpha$ and $\beta$ in terms
of $\alpha'$ and $\beta'$ is of the form,
\begin{eqnarray}
\alpha &=& \cos (\theta/2)\alpha '+ \sin (\theta /2)\beta', \\
\beta &=&- \sin (\theta/2)\alpha '+ \cos (\theta /2)\beta',
\end{eqnarray}
where $\theta$ is the angle between the spins
$\overrightarrow{S_1}$ and $\overrightarrow{S_2}$. By considering
the Hamiltonian (\ref{kondol}) for two sites, one can write the
secular equation which has the following four solutions
\cite{And55}:
\begin{equation}
E=1/2J\pm \Big([J(S+1/2)\pm
t\cos(\theta/2)]^2+t^2\sin^2(\theta/2)\Big)^{1/2}.
\end{equation}
The energies depend on the angle $\theta$ between both spins, and
in the semiclassical case
\begin{equation}
\cos\left(\frac{\theta}{2}\right) = \frac{|\vec S_1+\vec
S_2|}{2S}.
\end{equation}
In the absence of any other interaction, the lowest energy is
obtained for the aligned spins, at $\theta=0$.

The existence of phase separation and ferromagnetism in the ground
state of the FM Kondo model can also be studied in the limit of
$d=\infty$. The dynamical mean field theory (DMFT) \cite{Geo96}
leads to a self-consistent equation which can be solved
iteratively starting from a random spin configuration, and as a
function of temperature and electron density three solutions have
been found with AF, FM, and paramagnetic character. We refer an
interested reader to Ref. \cite{Fur98} for more technical details.

\begin{figure}[t]
\centerline{\psfig{figure=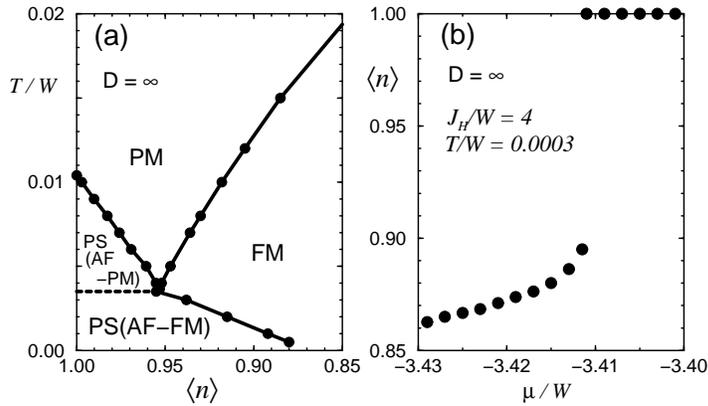,width=7cm,angle=270}}
\smallskip
\caption{ (a) Phase diagram in the $d=\infty$ limit working at
    $J_H/W = 4.0$. The ``PS(AF-PM)'' region denotes phase separation
    (PS) between a hole-poor antiferromagnetic (AF) region, and
    a hole-rich paramagnetic (PM) region. The rest of the notation
    is standard; (b) Density $\langle n \rangle$ vs $\mu/W$ obtained
    in the $d=\infty$ limit, $J_H/W=4.0$, and $T/W=0.0003$. The
    discontinuity in the density is clear (after Ref. \protect\cite{Fur98}). }
    \label{fig:dmft}
\end{figure}

The presence of ferromagnetism at finite doping and
antiferromagnetism at half-filling are quite clear from Fig.
\ref{fig:dmft}. Close to half-filling and at low temperature, the
density $n=\langle n_i\rangle$ was found to be discontinuous as a
function of $\mu$, in excellent agreement with the results
obtained by other numerical calculations. The phase separation
observed in Fig. \ref{fig:dmft}(a) occurs between AF and FM
regions. However, we note that at higher temperature the phase
separation occurs between hole-poor AF and hole-rich paramagnetic
regions [Fig. \ref{fig:dmft}(b)].

\begin{figure}[t]
\centerline{\psfig{figure=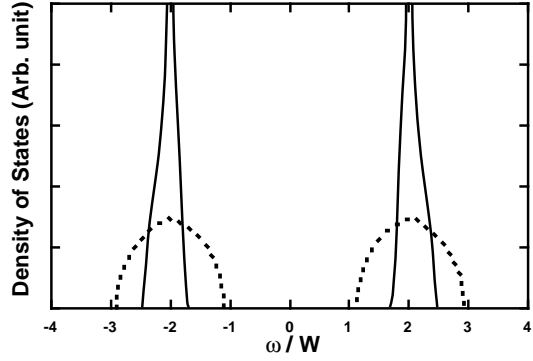,width=7cm}}
\smallskip
\caption{ Density of states in the $d=\infty$ limit corresponding
to the AF
    (solid line) and FM (dotted line) solutions, found at $J_H/W = 2.0$
    and $T/W = 0.005$ (after Ref. \protect\cite{Fur98}). } \label{fig:dosdmft}
\end{figure}

In order to illustrate the consequences of the DMFT treatment of
the Kondo lattice (DE) model (\ref{kondol}) we reproduce in Fig.
\ref{fig:dosdmft} the density of states $A(\omega)$ obtained for
the AF and FM phases at low temperature (for details of the
calculation see Ref. \cite{Fur98}). The critical value of the
chemical potential where the AF and FM phases coexist is
$\mu_c\simeq -1.40W$. In the both cases the density of states
splits into upper and lower bands due to the large Hund's coupling
$J_H$. The band splitting in the FM phase is due to the
half-metalicity of the system. The width of the upper and lower
bands is wider for the FM phase, which causes a narrower gaped
region centered at $\omega\simeq 0$.

Following Furukawa~\cite{Fur96}, we calculate the spin excitation
spectrum of the DE model, and compare the results with recent data
of the neutron inelastic scattering experiments. We use the
spin-wave approximation in the ground state, which has been
introduced by Kubo and Ohata \cite{Kub72}. Expanding the spin
operators in terms of boson operators in the FM state,
\begin{equation}
S_i^+ \simeq \sqrt{2S} a_i, \hskip 1cm S_i^- \simeq \sqrt{2S}
a_i^{\dagger}, \hskip 1cm S_i^z = S - a_i^+ a_i, \label{defSWope}
\end{equation}
the lowest-order effective Hamiltonian can be written as follows:
\begin{eqnarray}
 H &=&\sum_{\vec k}\left[
   (\varepsilon_{\vec k}-J_H)c_{{\vec k}\uparrow}^{\dagger}c_{{\vec k}\uparrow}
+(\varepsilon_{\vec k}+J_H)c_{{\vec
k}\downarrow}^{\dagger}c_{{\vec k}\downarrow}
                \right]  \nonumber\\
 &+ & J_H \sqrt{\frac{2}{SN}} \sum_{{\vec k}{\vec q}}
    \left(a_{\vec q}^{\dagger}c_{{\vec k}\uparrow}^{\dagger}
                              c_{{\vec k}+{\vec q},\downarrow}
   +a_{\vec q}c_{{\vec k}+{\vec q},\downarrow}^{\dagger} c_{{\vec k}\uparrow}
                \right)
   +\frac{J_H}{SN}\sum_{{\vec k} {\vec q}_1 {\vec q}_2,\sigma}
                \sigma a_{{\vec q}_1}^{\dagger} a_{{\vec q}_2}
       c_{{\vec k}-{\vec q}_1\sigma}^{\dagger} c_{{\vec k}-{\vec q}_2\sigma}.
\label{HSW}
\end{eqnarray}
The first line in Eq. (\ref{HSW}) describes the electron band
split by the exchange interaction with the core spins, while the
second line stands for the coupling between the electrons and spin
excitations (electron-magnon interaction).

Let us consider the lowest order terms of the $1/S$ expansion at
$T=0$, assuming that $J_H$ is finite but sufficiently large to
polarize completely the electronic band, i.e., $n_\uparrow=n$ and
$n_\downarrow=0$, at $T=0$. The electron concentration is given by
$n=1-x$. For a simple cubic 3D lattice with nearest-neighbor
hopping $t$ one finds,
\begin{equation}
\varepsilon_{\vec k}=-2t\left(\cos k_x+\cos k_y+\cos k_z\right)
                    =-6t\gamma_{\vec k},
\label{defDispRelFermicubic}
\end{equation}
where $\gamma_{\vec k}=\frac{1}{z}\sum_{\vec{\delta}} \exp(i{\vec
{k}}\cdot{\vec{\delta}})$. For perovskite manganites, estimates of
the electron bandwidth and the on-site Hund's coupling being a few
{\rm eV} has been made by the first-principle calculations.
However, one might consider that $t$ and $J_H$ in the DE model are
effective parameters which could be strongly renormalized from the
bare values due to other interactions present in the real systems.
Such effective parameters could be determined from a comparison
with experiments.

\begin{figure}[t]
\centerline{\psfig{figure=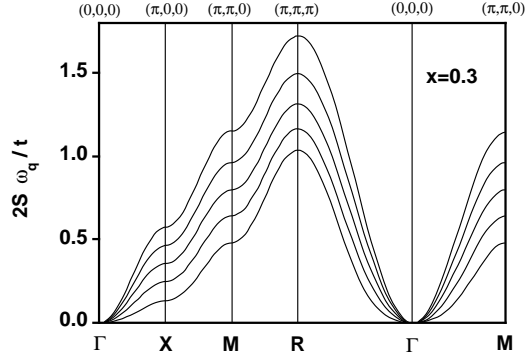,width=7cm}}
\smallskip
\caption{Spin wave dispersion in the metallic FM phase calculated
using the DE model by Furukawa \protect\cite{Fur96}. Different
curves are obtained for $J_H/t=\infty$, $24$, $12$ and $6$ from
top to bottom.} \label{fig:de}
\end{figure}

The spin-wave self-energy in the lowest order of $1/S$ expansion
is given by
\begin{equation}
  \Pi({\vec q},\omega) = \frac{1}{SN}
  \sum_{\vec k}\left(f_{{\vec k}\uparrow}-f_{{\vec k}+{\vec q}\downarrow}\right)
        \times\left(J_H + \frac{2J_H^2}
   {\omega+\varepsilon_{\vec k}-\varepsilon_{{\vec k}+{\vec q}}-2J_H}\right),
\label{fmlaPi}
\end{equation}
where $f_{{\vec k}\sigma}$ is the Fermi distribution function. We
have $f_{{\vec k}\downarrow}=0$, if the system is fully polarized.
The spin-wave dispersion relation $\omega_q$ is now obtained
self-consistently as a solution of the equation $\omega_{\vec
q}=\Pi({\vec q},\omega_{\vec q})$. Since $\Pi({\vec
q},\omega_{\vec q})\propto 1/S$, the lowest order of $1/S$
expansion gives $\omega_{\vec q}=\Pi({\vec q},0)$. Therefore, the
spin-wave dispersion is described as
\begin{eqnarray}
    \omega_{\vec q} = \frac{1}{2S} \frac{1}{N}
    \sum_{\vec k} f_{{\vec k}\uparrow}
          \frac{J_H(\varepsilon_{{\vec k}+{\vec q}}-\varepsilon_{\vec k} )}
             { J_H+(\varepsilon_{{\vec k}+{\vec q}}-\varepsilon_{\vec k} )/2 }.
\label{SWdispreldef}
\end{eqnarray}
In Fig. \ref{fig:de}, we show the spin-wave dispersion relation at
$x=0.3$ for various values of $J_H/t$. As the value of $J_H$
becomes comparable with the electron bandwidth, the softening of
the spin-wave dispersion is observed since the effective coupling
between spins becomes weak. At $J_H\to\infty$, we have
\begin{equation}
 \omega_{\vec q} \simeq  \frac{1}{2SN}\sum_{\vec k}
    (\varepsilon_{{\vec k}+{\vec q}}-\varepsilon_{\vec k})f_{{\vec k}\uparrow}=
    \frac{1}{2}W_{sw} (1 - \gamma_{\bf {\vec q}}),
\label{SW, allBZ, nn}
\end{equation}
where $W_{sw} $ is the spin-wave bandwidth given by the kinetic
energy of electrons moving in a polarized band,
\begin{equation}
  W_{sw} = \frac{6t}{SN}\sum_{\vec k} f_{{\vec k}\uparrow}\cos k_x.
\end{equation}

The dispersion relation (\ref{SW, allBZ, nn}) is identical with
that given by a FM Heisenberg model with nearest-neighbor spin
exchange $J_{eff}=W_{sw}/12$. The above correspondence can be
understood as follows. We consider a perfectly polarized FM state
at $T=0$ and then flip a spin at site $i_0$. In the case of the
strong coupling limit $J_H\gg t$, where electrons with spins
antiparallel to the localized spin on the same site are
disfavored, the electron at site $i_0$ is localized because it has
different spin orientation from that of the localized spins at
neighboring sites. Therefore, in this limit the effective
spin-spin interaction is short-ranged and the DE model in the
strong-coupling limit is mapped onto the Heisenberg model with
short-range interactions.

\section{ Spin-orbital model in cuprates }
\label{sec:somd9}

\subsection{ Superexchange for degenerate $\lowercase{e_g}$ orbitals }

Our aim is to construct the effective low-energy Hamiltonian for a
3D perovskite-like lattice, assuming the situation as in the
cuprates, i.e., $d^9$ configuration with single occupancy of one
hole in $e_g$ orbitals. This situation was considered already by
Kugel and Khomskii \cite{Kug73}; here we present a more recent
derivation which uses a correct multiplet structure of the excited
$d^8$ states \cite{Ole00}. From a general point of view, one
should approach the problem starting from the charge-transfer
multiband model which contains the hybridization between the $d$
orbitals of transition metal ions and the $2p$ orbitals of oxygen
ions. Yet, if the Coulomb interaction at the $d$ orbital and the
energies required for the electron transfer from the $3d$ to the
$2p$ orbital levels are large compared to the other parameters
involved, then it is possible to integrate out the oxygen degrees
of freedom and to deal instead with a simpler model which
describes electrons (holes) in a $d$ band.

We derive the superexchange in a similar fashion as in Sec. II.A
for the case of degenerate orbitals. Having in mind the strongly
correlated late transition metal oxides, we consider specifically
the case of the $e_{g}$ orbitals, defined by the local basis:
$x^{2}-y^{2}\equiv |x\rangle $ and $(3z^{2}-r^{2})/\sqrt{3}\equiv
|z\rangle$. Although we focus here on the case of the $d^{9}$
configuration, though the presented analysis can be easily
generalized to the low-spin $d^{7}$ configuration with a single
electron; in the case of the early transition metal oxides the
$d^{1}$ case would involve the $t_{2g}$ orbitals occupied by a
single electron instead
\cite{Cas78,Bao97,Bao97a,Pen97,Ezh99,Kei00}.

We take as a starting point the following Hamiltonian which describes $d$%
-holes on transition metal ions,
\begin{equation}
\label{hband} H_{e_g} = H_{t} + H_{int} + H_z,
\end{equation}
and includes the kinetic energy $H_{kin}$, and the
electron-electron interactions $H_{int}$, restricted now to the
subspace of the $e_g$ orbitals (the $t_{2g}$ orbitals are filled
by electrons, do not couple to $e_g$ orbitals by the hoppings via
oxygens, and hence can be neglected). The last term $H_z$
describes the crystal-field splitting of the $e_g$ orbitals.

Due to the shape of the two $e_{g}$ orbitals $|x\rangle $ and
$|z\rangle $, their hybridization with oxygen orbitals is unequal
in the three cubic directions \cite{Zaa93}, so that the effective
hopping elements are direction dependent and different for
$|x\rangle $ and $|z\rangle $. The only nonvanishing hopping in
the $c$-direction connects two $|z\rangle $ orbitals, while the
elements in the $(a,b)$ planes fulfill the Slater-Koster relations
\cite{Sla54}, as presented in Ref. \cite{Zaa93}. Taking the
hopping $t$ along the $c$-axis as a unit, the kinetic energy is
given by,
\begin{eqnarray}
\label{hkin} H_t&=&{\case{1}{4}}t\sum_{\langle
ij\rangle{\parallel},\sigma}\left[
    3d_{ix\sigma}^{\dagger}d_{jx\sigma}^{}
   +\pm\sqrt{3}
  (d_{iz\sigma}^{\dagger}d_{jx\sigma}^{}+d_{ix\sigma}^{\dagger}d_{jz\sigma}^{})
   +d_{iz\sigma}^{\dagger}d_{jz\sigma}^{}+H.c.\right]           \nonumber \\
  &&\hskip 1.3cm+t\sqrt{\beta }\sum_{\langle ij\rangle{\perp},\sigma}
   \left(d_{iz\sigma}^{\dagger}d_{jz\sigma}+H.c.\right),
\end{eqnarray}
where $d_{ix\sigma}^{\dagger}$ ($d_{iz\sigma}^{\dagger}$) creates
a hole in $|x\rangle$ ($|z\rangle$) orbital with spin $\sigma$.
The sums run over the bonds between nearest neighbors oriented
along the cubic axes: $\langle ij\rangle{\parallel}$ within the
$(a,b)$-planes, and $\langle ij\rangle{\perp}$ along the $c$-axis
[perpendicular to $(a,b)$-planes], respectively, and $\beta=1$ in
a cubic system. The $x-z$ hopping in the $(a,b)$ planes depends on
the phases of the $x^{2}-y^{2}$ orbitals along $a$- and $b$-axis,
respectively, included in the factors $\pm\sqrt{3}$ in Eq.
(\ref{hkin}).
\begin{figure}[t]
\centerline{\psfig{figure=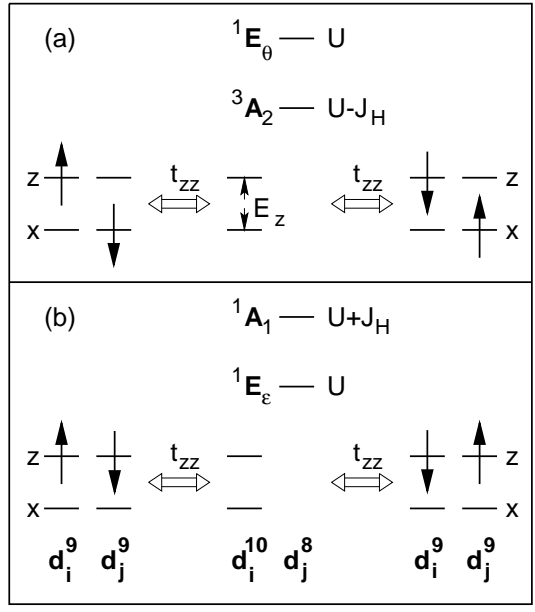,width=7cm}} \vspace{0.2in}
\caption {Virtual transitions $d_i^9d_j^9\rightarrow
d_i^{10}d_j^8$ which lead to a spin-flip and generate effective
interactions for a bond $\langle ij\rangle\parallel c$-axis, with
the excitation energies at $E_z=0$. For two holes in different
orbitals (a), either the triplet $^3A_2$ or the interorbital
singlet $^1E_{\theta}$ occurs as an intermediate $d^8$
configuration, while if both holes are in $|z\rangle$ orbitals
(b), two other singlets, $^1E_{\epsilon}$ and $^1A_1$, with double
occupancy of $|z\rangle$ orbital, contribute. The latter processes
are possible either from $i$ to $j$ or from $j$ to $i$ ~(after
Ref. \protect\cite{Ole00}).} \label{fig:d9se}
\end{figure}
The electron-electron interactions are described by the on-site
terms, which we write in the following form,
\begin{eqnarray}
\label{hint} H_{int}&=&(U+\case{1}{2}J_H)
          \sum_{i\alpha}n_{i\alpha  \uparrow}n_{i\alpha\downarrow}
 + (U-J_H)\sum_{i\sigma}n_{ix\sigma}n_{iz\sigma}
 +(U-\case{1}{2}J_H)\sum_{i\sigma}n_{ix\sigma}n_{iz\bar{\sigma}} \nonumber \\
&&-\case{1}{2}J_H \sum_{i\sigma}d^{\dagger}_{ix
\sigma}d^{}_{ix\bar{\sigma}}
                               d^{\dagger}_{iz\bar{\sigma}}d^{}_{iz \sigma}
 + \case{1}{2}J_H\sum_{i}
            ( d^{\dagger}_{ix  \uparrow}d^{\dagger}_{ix\downarrow}
              d^{       }_{iz\downarrow}d^{       }_{iz  \uparrow}
            + d^{\dagger}_{iz  \uparrow}d^{\dagger}_{iz\downarrow}
              d^{       }_{ix\downarrow}d^{       }_{ix  \uparrow} ),
\end{eqnarray}
with $U$ and $J_{H}$ standing for the Coulomb and Hund's rule
exchange interaction, respectively, and $\alpha =x,z$. Moreover,
we have used the simplified notation $\bar{\sigma}=-\sigma $. For
convenience, $U$ has been defined as the average
$d^9_id^9_j\rightarrow d^{10}_id^8_j$ excitation energy of the
$d^8$ configuration, which coincides with the energy of the
central $|^1E\rangle$ doublet. Therefore, $U$ is here not the
interorbital Coulomb element. The interaction element $J_H$ stands
for the singlet-triplet splitting in the $d^8$ spectrum (Fig.
\ref{fig:d9se}) and is just twice as big as the exchange element
$K_{xz}$ usually used in quantum chemistry \cite{Gri71,Gra92}.
With the present definition of $J_H$, the interorbital interaction
between holes of opposite (equal) spins is $U-J_H/2$ ($U-J_H$),
respectively. This Hamiltonian (\ref{hint}) describes correctly
the multiplet structure of $d^{8}$ (and $d^{2}$) ions
\cite{Gri71}, and is rotationally invariant in the orbital space
\cite{Ole83}. The wave functions have been assumed to be real
which gives the same element $J_{H}/2$ for the exchange
interaction and for the pair hopping term between the $e_{g}$
orbitals, $|x\rangle $ and $|z\rangle $.

The last term in Eq. (\ref{hband}) stands for the crystal field
which lifts the degeneracy of the two $e_g$ orbitals and breaks
the symmetry in the orbital space,
\begin{equation}
\label{hz}
H_{z}=\sum_{i\sigma}(\varepsilon_xn_{ix\sigma}+\varepsilon_zn_{iz\sigma})
     =-\case{1}{2}E_z\sum_{i\sigma}(n_{ix\sigma}-n_{iz\sigma}),
\end{equation}
if $\varepsilon_x\neq \varepsilon_z$ (and neglecting a constant
term $\propto \varepsilon_x+\varepsilon_z$). Here $\varepsilon_x$
and $\varepsilon_z$ are the energies of a hole at $|x\rangle$ and
$|z\rangle$ orbitals, respectively, and
\begin{equation}
\label{ez} E_z=\varepsilon_z-\varepsilon_x.
\end{equation}
Its effect is like that of a magnetic field in the orbital space,
and together with the parameter $\beta$ in $H_{kin}$ (\ref{hkin})
quantifies the deviation in the electronic structure from the
ideal cubic local point group.

In the atomic limit, i.e., at $t=0$, one finds at $E_z=0$ a highly
degenerate problem, with orbital degeneracy next to spin
degeneracy. All four basis states per site, with a hole occupying
either orbital, $|x\rangle$ or $|z\rangle$, and either spin state,
$\sigma=\uparrow$ or $\sigma=\downarrow$, have the same energy.
Therefore, the system of $N$ $d^9$ ions has a degeneracy $4^N$,
which is, however, removed by the effective interactions between
each pair of nearest neighbor ions $\{i,j\}$ that originate from
virtual transitions to the excited states,
$d^9_id^9_j\rightleftharpoons d^{10}_id^8_j$, due to hole hopping
$t\neq 0$. Hence, we derive the effective spin-orbital model
following Kugel and Khomskii \cite{Kug73}, starting from the
Hamiltonian in the atomic limit, $H_{at}=H_{int}+H_z$, and
treating $H_{kin}$ as a perturbation. However, we report here the
study which includes the \emph{full multiplet structure\/} of the
excited states within the $d^8$ configuration which gives
corrections of the order of $J_H$ compared with the earlier
results of Refs. \cite{Kug73} and \cite{Kug82}.

Knowing the multiplet structure of the $d^8$ intermediate states,
the derivation of the effective Hamiltonian can be done in various
ways. The most straightforward but lengthy procedure is a
generalization of the canonical transformation method used earlier
for the Hubbard \cite{Cha77} and the three-band\cite{Zaa88} model.
A significantly shorter derivation is possible, however, using the
cubic symmetry and starting with the interactions along the
$c$-axis. Here the derivation simplifies tremendously as one finds
only effective interactions which result from the hopping of holes
between the directional $|z\rangle$ orbitals, as shown in Fig.
\ref {fig:d9se}. Next the interactions in the remaining directions
can be generated by the appropriate rotations to the other cubic
axes $a$ and $b$, and by applying the symmetry rules for the
hopping elements between the $e_g$ orbitals \cite{Sla54}.

Following the above argument, the derivation of the effective
interactions between two $d^{9}$ ions at sites $i$ and $j$ takes
the simplest form for a bond $\langle ij\rangle $ oriented along
the $c$-axis. In that case due to the vanishing hopping from/to
$|x\rangle$ orbital, the orbital occupancies in the initial and
final $ d_{i}^{9}d_{j}^{9}$ states have to be identical (apart
from a possible simultaneous and opposite spin flip at both
sites), i.e., the $z$th component of the pseudospin $T^z$ is
conserved. The possible initial states are described by a direct
product of the total spin state, either a triplet ($S=1$) or a
singlet ($S=0$), and the orbital configuration, given by one of
four possibilities: $|x_ix_j\rangle$, $|x_iz_j\rangle$,
$|z_{i}x_{j}\rangle$, or $|z_{i}z_{j}\rangle$. Moreover, the
effective interaction vanishes if the holes occupy the
$|x_{i}x_{j}\rangle$ configuration. The total spin per two sites
is conserved in the $d_{i}^{9}d_{j}^{9}\rightarrow
d_{i}^{10}d_{j}^{8}$ excitation process, and therefore the spin
dependence of the resulting second order Hamiltonian can be
expressed in terms of the projection operators on the total spin
states, defined for a given bond $\langle ij\rangle$ by Eq.
(\ref{projst}).

Depending on whether the initial state is $|z_ix_j\rangle$ or
$|z_iz_j\rangle$, the intermediate $d_{i}^{10}d_{j}^{8}$
configuration resulting from the hole-hop
$|z_{i}\rangle\rightarrow |z_{j}\rangle$, involves different
$d^{8}$ excited states: either the interorbital states, the
triplet $^3A_2$ and the singlet $^1E_{\theta}$ (for
$|z_ix_j\rangle$), or the two singlets built from the states with
doubly occupied orbitals, $^1E_{\varepsilon}$ and $^1A_1$ (for
$|z_iz_j\rangle$). Of course, since the wave function has to be
antisymmetric, the spins have to be opposite in the latter case,
while in the former case also parallel spin configurations
contribute in the triplet channel. The eigenstates within the
$e_{g}$ subspace are:
\begin{description}
\item[(i)] triplet: $|^3\!A_2\rangle=\{|z\!\uparrow\rangle|x\!\uparrow\rangle$,
$\frac{1}{\sqrt 2}(|z\!\uparrow\rangle|x\!\downarrow\rangle+
|z\!\downarrow\rangle|x\!\uparrow\rangle)$,
$|z\!\downarrow\rangle|x\!\downarrow\rangle\}$,
\item[(ii)] interorbital singlet $|^1E_{\epsilon}\rangle =
\frac{1}{\sqrt 2}(|z\!\uparrow\rangle|x\!\downarrow\rangle-
|z\downarrow\rangle|x\uparrow\rangle$),
\item[(iii)] bonding $|^1\!E_{\theta}\rangle$ and antibonding $|^1\!A_1\rangle$
             singlets:

 $|^1\!E_{\theta}\rangle=\frac{1}{\sqrt 2}
 (|z\!\uparrow\rangle|z\!\downarrow\rangle+
  |x\!\uparrow\rangle|x\!\downarrow\rangle)$, and

 $|^1\!A_1\rangle=\frac{1}{\sqrt 2}
 (|z\!\uparrow\rangle|z\!\downarrow\rangle-
  |x\!\uparrow\rangle|x\!\downarrow\rangle)$, with double occupancies of
 both orbitals.
\end{description}

The energies of the states $|^{3}A_{2}\rangle $ and
$|^{1}E_{\epsilon}\rangle$ are straightforwardly obtained using
${\vec{S}}_{ix}\cdot {\vec{S}}_{iz}=+1/4$ and ${\vec{S}}_{ix}\cdot
{\vec{S}}_{iz}=-3/4$, for $S=1$ and $S=0 $ states, respectively.
The remaining two singlet energies are found by diagonalizing a
$2\times 2$ problem in the subspace of doubly occupied states.
Hence, the resulting excitation energies which correspond to the
{\it local excitations\/}
 $d_{i}^{9}d_{j}^{9}\rightarrow d_{i}^{10}d_{j}^{8}$ on a given bond
$\langle ij\rangle$ are,
\begin{eqnarray}
\label{specd81} \varepsilon(^{3}A_{2}) &=&U-J_{H},           \\
\label{specd82} \varepsilon(^{1}E_{\epsilon })&=&U,          \\
\label{specd83} \varepsilon(^{1}E_{\theta})&=&U+\frac{1}{2}J_{H}
-\frac{1}{2}J_{H}\left[1+(E_{z}/J_{H})^{2}\right] ^{1/2},
\\ \label{specd84} \varepsilon(^{1}A_{1})&=&U+\frac{1}{2}J_{H}
            +\frac{1}{2}J_{H}\left[ 1+(E_{z}/J_{H})^{2}\right]^{1/2},
\end{eqnarray}
At $E_{z}=0$ it consists of equidistant states, with a distance of
$J_H$ between the triplet $|^{3}A_{2}\rangle$ and the degenerate
singlets $|^1E_{\theta}\rangle$ and $|^{1}E_{\epsilon}\rangle$
(which form of course an orbital doublet), as well as between the
above singlets and the highest energy singlet $|^1A_1\rangle$.
Note that when the pair hopping term $\propto J_H$ is neglected in
Hamiltonian (\ref{hint}), the spectrum is incorrect, with
$\varepsilon(^1E_{\theta})=\varepsilon(^1A_{1})=U+J_H/2$.

At this point we have all the elements for deriving the effective
spin-orbital model. Hence, its general form is given by the
formula which includes all possible virtual transitions to the
excited $d^{8}_id^{10}_j$ configurations,
\begin{equation}
\label{allv} H_{\langle ij\rangle }=-\sum_{n,\alpha\beta
}\frac{t^{2}}{\varepsilon _{n}} Q_S(i,j) P_{i\alpha }P_{j\beta },
\end{equation}
where $t$ stands for the $z-z$ hopping along the $c$-axis,
$Q_S(i,j)$ is one of the projection operators on the total spin
        state (\ref{projst}), either $S=0$ or $S=1$, and $P_{i\alpha}$
is the projection operator on the orbital state $\alpha $ at site
$i$, while $\varepsilon_n$ stands for the excitation energies
given by Eqs. (\ref{specd81})--(\ref{specd84}). The orbital
projection operators on $|x\rangle $ and $|z\rangle$ orbital in
the initial and final state of the $d^9$ configuration at site $i$
are, respectively,
\begin{eqnarray}
\label{orbproject} P_{ix} &=&|ix\rangle \langle
ix|=\case{1}{2}+\tau_i^c,    \\ P_{iz} &=&|iz\rangle \langle
iz|=\case{1}{2}-\tau_i^c,
\end{eqnarray}
where $\tau_i^c$ stands for the $z$th component of pseudospin and
is given by
\begin{equation}
\label{orbop0} \tau _{i}^{c} =\case{1}{2}\sigma_{i}^{z}.
\end{equation}
The interaction terms along the bonds $\langle ij\rangle\parallel
(a,b)$ are represented by the projection operators similar to
$P_{ix}$ and $P_{iz}$, with $\tau_i^c$ replaced by the orbital
operators $\tau_i^a$ and $\tau_i^b$ which are expressed in terms
of the Pauli matrices as follows:
\begin{equation}
\label{orbop1} \tau_i^a
=-\case{1}{4}(\sigma_{i}^{z}-\sqrt{3}\sigma_{i}^{x}),  \hskip 1cm
\tau_i^b =-\case{1}{4}(\sigma_{i}^{z}+\sqrt{3}\sigma_{i}^{x}).
\end{equation}
Here, the $\sigma_i^{\alpha}$ are Pauli matrices acting on the
orbital pseudospins:
\begin{equation}
\label{pseudospins} |x\rangle =\left(
\begin{array}{c}
1 \\0 \end{array} \right) ,\hskip 1.5cm |z\rangle =\left(
\begin{array}{c}
0 \\1 \end{array} \right).
\end{equation}

Expanding Eq. (\ref{allv}) for a bond $\langle ij\rangle$ along
the $c$-direction, one finds
\begin{eqnarray}
 \nonumber H_{\langle ij\rangle }&=&-\frac{t^{2}}{\varepsilon
(^{3}A_{2})}\left( \vec{S}_{i}\cdot \vec{S}_{j}+\frac{3}{4}\right)
\left( P_{ix}P_{jz}+P_{iz}P_{jx}\right)  +\frac{t^{2}}{\varepsilon
(^{1}E_{\epsilon })}\left( \vec{S}_{i}\cdot
\vec{S}_{j}-\frac{1}{4}\right) \times  \\  && \left(
P_{ix}P_{jz}+P_{iz}P_{jx}\right)
+\!\left[ \frac{t^{2}}{\varepsilon (^{1}E_{\theta })}+\frac{t^{2}}{%
\varepsilon (^{1}A_{1})}\right] \!\left( \vec{S}_{i}\cdot
\vec{S}_{j} -\frac{1}{4}\right) 2P_{iz}P_{jz}. \label{fullij}
\end{eqnarray}
As one can see, the magnetic interactions in the first two terms
in Eq. (\ref{fullij}) cancel each other in the limit of
$\eta\rightarrow 0$, while the last term favors AF spin
orientation independently of $\eta$. We recognize that Hamiltonian
(\ref{fullij}) describes the superexchange along the bond $\langle
ij\rangle\parallel c$, with the superexchange constant of
$4t^{2}/U$ \cite{And59,Cha77}. However, the hopping in the other
directions $\langle ij\rangle\parallel (a,b)$ is reduced and thus
we define for convenience $J=t^{2}/U$ as the energy unit. For
simplifying the form (\ref{fullij}) we use an expansion of the
excitation energies $\varepsilon _{n}$ in the denominators for
small $J_{H}$, and introduce
\begin{equation}
\label{eta} \eta =J_{H}/U
\end{equation}
as a parameter which quantifies the Hund's rule exchange. Using
the explicit form of the orbital projection operators
$P_{i\alpha}$ (\ref{orbproject}) this results in the following
form of the effective Hamiltonian for the bond $\langle
ij\rangle\parallel c$,
\begin{eqnarray}
\label{expij} H_{\langle ij\rangle } &=&J\left[ (1+\eta )\left(
\vec{S}_i\cdot\vec{S}_j+\frac{3}{4}\right)
        -\left( \vec{S}_i\cdot\vec{S}_j-\frac{1}{4}\right) \right]
        \times \left( P_{ix}P_{jz}+P_{iz}P_{jx}\right)         \nonumber \\
&+&4J\left( 1-\frac{1}{2}\eta \right) \left(
 \vec{S}_{i}\cdot \vec{S}_{j}-\frac{1}{4}\right) P_{iz}P_{jz} ,
\end{eqnarray}
which may be represented explicitly by the orbital operators
$\tau_i^c$ and $\tau_j^c$ in the following way,
\begin{eqnarray}
\label{effij} H_{\langle ij\rangle } &=&J\left( 4\vec{S}_{i}\cdot
\vec{S}_{j}+1\right) \left( \tau _{i}^{c}-\frac{1}{2}\right)
\left( \tau _{j}^{c}-\frac{1}{2} \right) +\tau _{i}^{c}+\tau
_{j}^{c}-1                       \nonumber \\ &+&J\eta \left(
\vec{S}_i\cdot \vec{S}_j\right) (\tau _i^c+\tau_j^c-1)
+\frac{1}{2}J\eta\left[ \left( \tau_{i}^{c}-\frac{1}{2}\right)
\left( \tau _{j}^{c}-\frac{1}{2}\right) +3\left( \tau _{i}^{c}\tau
_{j}^{c}- \frac{1}{4}\right) \right] .
\end{eqnarray}
The first line represents the AF superexchange interactions
$\propto J$, while the second line describes the weaker FM
interactions $\propto J\eta$, which originate from the multiplet
splittings of the $d^{8}$ excited states.

It is straightforward to verify that the above form of the
effective Hamiltonian simplifies in the limit of occupied
$|z\rangle $ orbitals to
\begin{equation}
H_{\langle ij\rangle }=4J\left( 1-\frac{1}{2}\eta \right) \left( \vec{S}%
_{i}\cdot \vec{S}_{j}-\frac{1}{4}\right) ,
\end{equation}
and one recognizes the same constant $-\frac{1}{4}$, and the same
superexchange interaction $4J=4t^{2}/U$ as in the $t-J$ model at
half-filling \cite{Cha77}. However, the effective superexchange is
somewhat reduced by the factor $(1-\frac{1}{2}\eta )$ in the
presence of the Hund's rule interaction, which increases the
excitation energy $\varepsilon(^1A_1)$. The effective interactions
along the bonds within the $(a,b)$ planes may be now obtained by
rotating Eq. (\ref{fullij}) with the projection operators $P_{ix}$
and $P_{iz}$ [or its simplified version (\ref{effij}) with the
orbital operators $\tau_i^c$] by $\pi/2$ to the cubic axes $a$ and
$b$, which generates the orbital operators $\tau_{i}^{a}$ and
$\tau _{i}^{b}$ (\ref {orbop1}), respectively \cite{Ole00}. This
results in a nontrivial coupling between the orbital and spin
degrees of freedom.

Following the above procedure, we have derived the effective
Hamiltonian $H$ in spin-orbital space,
\begin{equation}
H(d^9)={H}_{J}+{H}_{\tau}, \label{somcu}
\end{equation}
where the superexchange part ${H}_{J}$ can be most generally
written as follows (a simplified form was discussed recently in
Refs. \cite{Fei97} and \cite{Ole00}),
\begin{eqnarray}
{H}_{J}&=&\sum_{\langle ij\rangle
}\left\{-\frac{t^{2}}{\varepsilon (^{3}A_{2})}\left(
\vec{S}_{i}\cdot \vec{S}_{j}+\frac{3}{4}\right) {\cal P}_{\langle
ij\rangle }^{\zeta \xi }+\frac{t^{2}}{\varepsilon (^{1}E_{\epsilon
})}\left( \vec{S}_{i}\cdot \vec{S}_{j}-\frac{1}{4}\right) {\cal
P}_{\langle ij\rangle }^{\zeta \xi }\right.         \nonumber \\ &
&\hskip 0.7cm \left. +\left[ \frac{t^{2}}{\varepsilon
(^{1}E_{\theta })} +\frac{t^{2}}{\varepsilon(^{1}A_{1})}\right]
\left( \vec{S}_{i}\cdot\vec{S}_{j}-\frac{1}{4}\right) {\cal
P}_{\langle ij\rangle }^{\zeta\zeta }\right\}, \label{somj}
\end{eqnarray}
and the crystal-field term (\ref{htau}) we rewrite now in the
form,
\begin{equation}
\label{htau} {H}_{\tau} = - E_z \sum_i \tau^c_i.
\end{equation}
In general, the energies of the two orbital states, $|x\rangle$
and $|z\rangle$, are different, and thus the complete effective
Hamiltonian of the $d^9$ model (\ref{somcu}) includes as well the
crystal-field term. It acts as a "magnetic field" for the orbital
pseudospins, and is loosely associated with an uniaxial pressure
along the $c$-axis.

The operators $\vec{S}_{i}$ in Eq. (\ref{somj}) refer to a spin
$S=1/2$ at site $i$, while ${P}_{\langle ij\rangle
}^{\alpha\beta}$ are projection operators on the orbital states
for each bond,
\begin{eqnarray}
\label{projbond} {\cal P}_{\langle ij\rangle }^{\zeta \xi } &=&
 (\frac{1}{2}+\tau_{i}^{\alpha })(\frac{1}{2}-\tau _{j}^{\alpha })
+(\frac{1}{2}-\tau_{i}^{\alpha })(\frac{1}{2}+\tau _{j}^{\alpha
}),   \\ \label{projbond1} {\cal P}_{\langle ij\rangle
}^{\zeta\zeta} &=& 2(\frac{1}{2}-\tau _{i}^{\alpha
})(\frac{1}{2}-\tau_{j}^{\alpha }),
\end{eqnarray}
where $\alpha=a,b,c$ refers to the cubic axes, respectively. The
individual projection operators on the orbital state which is
parallel (perpendicular) to the bond direction are:
\begin{equation}
\label{projind} P_{i\zeta}=\case{1}{2}-\tau_i^{\alpha }, \hskip
1.5cm P_{i\xi  }=\case{1}{2}+\tau_i^{\alpha },
\end{equation}
and are constructed with the orbital operators (\ref{orbop0}) and
(\ref{orbop1}) associated with the three cubic axes. The global
operators (\ref{projbond}) and (\ref{projbond1}) select orbitals
that are either parallel ($P_{i\zeta}$) to the direction of the
bond $\langle ij\rangle$ on site $i$, and perpendicular
($P_{j\xi}$) on the other site $j$, as in ${\cal P}_{\langle
ij\rangle}^{\zeta\xi}$, or parallel on both sites, as in ${\cal
P}_{\langle ij\rangle}^{\zeta\zeta}$, respectively. Hence, we find
a Heisenberg Hamiltonian for the spins, coupled into an orbital
problem. While the spin problem is described by the continuous
symmetry group $SU(2)$, the orbital problem is clock-model like,
i.e., there are three directional orbitals: $3x^{2}-r^{2}$,
$3y^{2}-r^{2}$, and $3z^{2}-r^{2}$, but they are not independent,
and transform into each other by appropriate cubic rotations. In
general, the occupied orbital state at a given site $i$ may be
expressed by the following transformation of bond basis
$\{|z\rangle,|x\rangle\}$ with an assigned angle $\theta$
(\ref{newstate}). In order to give an idea of the possible orbital
configuration one can get by changing $\theta$, we have summarized
the results obtained for a few representative angles in Table
\ref{tabtheta}.

\smallskip
\begin{table}
\caption{ Orbital configuration for a few representative values of
the orbital rotation angle $\theta$ [see Eq.
(\protect{\ref{newstate}})] for the site $|i\rangle$. }
\vspace{0.15in}
\begin{tabular}{ccccc}
\\
   $\theta$   & \hspace{1in}   & $|i\rangle$ & \\
   \\
\hline
\\
      0    &  \hspace{1in}  & $\case{1}{\sqrt{3}}(3z^2-r^2)\equiv |z\rangle$ &
               \\
  $\case{\pi}{6}$ & \hspace{1in}  &  $z^2-y^2$   &      \\
  $\case{\pi}{4}$  & \hspace{1in} &
  $\case{1}{\sqrt{6}}\left[2z^2+(\sqrt{3}-1)x^2-(\sqrt{3}+1)y^2\right]$ &
   \\
  $\case{\pi}{3}$  & \hspace{1in} &  -$\case{1}{\sqrt{3}}(3y^2-r^2)$ &     \\
  $\case{\pi}{2}$  & \hspace{1in} &  $x^2-y^2\equiv |x\rangle$ &
                       \\
 $\case{3\pi}{4}$  & \hspace{1in} &
  $\case{1}{\sqrt{6}}\left[2z^2-(\sqrt{3}+1)x^2+(\sqrt{3}-1)y^2\right]$ &
   \\
   \\
\end{tabular}
\label{tabtheta}
\end{table}
The $d^9$ spin-orbital model (\ref{somcu})-(\ref{htau}) depends
thus on two parameters: (i) the crystal field splitting $E_z$
(\ref{ez}), and (ii) the Hund's rule exchange $J_H$ (\ref{eta}).
While the first two terms in (\ref{somj}) cancel for the magnetic
interactions in the limit of $\eta\rightarrow 0$, the last term
favors AF spin orientation. Using again $\eta$ (\ref{eta}) as an
expansion parameter which quantifies the Hund's rule exchange, one
finds the following form of the effective exchange Hamiltonian in
the $d^{9}$ model (\ref{somcu}) \cite{Fei97},
\begin{eqnarray}
\nonumber {H}_{J}&\!\simeq \!&J\sum_{\langle ij\rangle }\left[
2\left( {\vec{S}}_{i}\cdot {\vec{S}}_{j}-\frac{1}{4}\right)
P_{\langle ij\rangle}^{\zeta\zeta}-P_{\langle ij\rangle
}^{\zeta\xi}\right]
\\\label{somjexp} &-&\!J\eta \sum_{\langle ij\rangle }\left[
{\vec{S}}_{i}\cdot {\vec{S}}_{j}\left( {P}_{\langle ij\rangle
}^{\zeta \zeta }\! +{P}_{\langle ij\rangle }^{\zeta \xi }\!\right)
+\frac{3}{4}{P}_{\langle ij\rangle }^{\zeta \xi
}-\frac{1}{4}{P}_{\langle ij\rangle }^{\zeta \zeta }\right] .
\end{eqnarray}

The first term in Eq. (\ref{somjexp}) describes the AF
superexchange $\propto J=t^2/U$ (where $t$ is the hopping between
$|\zeta\rangle$ orbitals along the $\langle ij\rangle$ bond), and
is the leading interaction term obtained when the splittings
between different excited $d^8$ states $\propto J_H$ are
neglected. As we show below, in spite of the AF superexchange
$\propto J$, \textit{no \textrm{LRO} can stabilize in a system
described by the spin-orbital model (\ref{somcu}) in the limit
$\eta\to 0$ at orbital degeneracy $(E_z=0)$} because of the
presence of the frustrating orbital interactions $\propto
P_{\langle ij\rangle}^{\zeta\zeta}$ which give a highly degenerate
classical ground state. We emphasize that even in the limit of
$J_H\to 0$ the present Kugel-Khomskii model {\it does not obey\/}
SU(4) symmetry, essentially because of the directionality of the
$e_g$ orbitals. Therefore, such an idealized SU(4)-symmetric model
(see Sec. IV.A) does not correspond to the realistic situation of
degenerate $e_g$ orbitals and is expected to give different
answers concerning the interplay of spin and orbital ordering in
cubic crystals.

Taking into account the multiplet splittings, we obtain [see Eq.
(\ref {somjexp})] again a Heisenberg-like Hamiltonian for the
spins coupled into an orbital problem, with a reduced interaction
$\propto J\eta $. It is evident that the new terms support FM
rather than AF spin interactions for particular orbital orderings.
This net FM superexchange originates from the virtual transitions
which involve the triplet state $|^{3}A_{2}\rangle $, having the
lowest energy and thus providing the strongest effective magnetic
coupling.

The important feature of the spin-orbital model (\ref{somcu}) is
that the \textit{actual magnetic interactions depend on the
orbital pattern\/}. This follows essentially from the hopping
matrix elements in $H_{t}$ (\ref{hkin}) being different between a
pair of $|x\rangle $ orbitals, between a pair of different
orbitals (one $|x\rangle $ and one $|z\rangle $ orbital), and
between a pair of $|z\rangle $ orbitals, respectively, and
depending on the bond direction either in the $(a,b)$ planes, or
along the $c$-axis \cite {Zaa93}. We show below that this leads to
a particular competition between magnetic and orbital
interactions, and the resulting phase diagram contains a rather
large number of classical phases, stabilized for different values
of $E_{z}$ and $J_{H}$.

\subsection{Classical phases and phase diagrams}


The simplest approach to the $d^9$ spin-orbital model as given by
Eqs. (\ref{somcu}), and (\ref{somjexp}) for getting an insight
into the competition between spin and orbital interactions is the
MF theory which is formally obtained by replacing the scalar
products $\vec{S}_i\cdot\vec{S}_j$ by the Ising terms,
$S^z_iS^z_j$. We report here the MF study of the phase diagram
after Ref. \cite{Ole00} for a distorted system with respect to the
cubic perovskite lattice. Therefore, we introduce a parameter
$\beta$ which controls the anisotropy along the $c$-axis and leads
to the different exchange constants in $(a,b)$ planes
($J_a=J_b=J$), and along $c$-direction ($J_c=J\beta$):
\begin{eqnarray}
{\cal H}_{\rm MF} &\! \simeq\!& \sum_{\langle ij\rangle}
J_{\alpha}\left[
      2\left( S_i^zS_j^z -\case{1}{4} \right)
       {\cal P}_{\langle ij\rangle}^{\zeta\zeta}
     -{\cal P}_{\langle ij\rangle}^{\zeta\xi  }\right] \nonumber
 \\  &-&\! \eta\sum_{\langle ij\rangle} J_{\alpha}\left[ S_i^zS_j^z
       \left(\!{\cal P}_{\langle ij\rangle}^{\zeta\zeta}
       +{\cal P}_{\langle ij\rangle}^{\zeta\xi  }\!\right)
       +\case{3}{4}{\cal P}_{\langle ij\rangle}^{\zeta\xi  }
       -\case{1}{4}{\cal P}_{\langle ij\rangle}^{\zeta\zeta}\right]
- E_z \sum_i \tau^c_i. \label{somcumf}
\end{eqnarray}
Here $\beta<1$ ($\beta>1)$ corresponds to the elongation
(compression) of the bond $\langle ij\rangle$ $\parallel c$,
respectively. The two limiting cases: $\beta=0$ and $\beta=1$,
stand for the 2D (square) lattice, and the 3D undistorted
(perovskite) lattice, respectively. At first sight the MF
Hamiltonian (\ref{somcumf}) contains a dominating AF exchange
$\propto J$ which competes with a FM one $\propto \eta J$, and
suggests that one should search for a solution with different
exchange constants along the three cubic axes. In the following we
will consider several magnetic patterns with two- and
four-sublattice 3D structures. They include the possibility of
having: the G-AF order (AF spin alternating along all three cubic
directions), A-AF or 1D-AF phase (FM interaction along two cubic
directions and AF along the third axis), and C-AF order (FM
exchange along 1D chains, and AF exchange in the directions
perpendicular to them).

Moreover, the interaction between orbital variables has also an AF
character, $\sim J\tau^{\alpha}_i\tau^{\alpha}_j$, suggesting that
it might be energetically more favorable to alternate the orbitals
in a certain regime of parameters, and pay thereby part of the
magnetic energy. This gives the main idea of the complex
frustration present in this system. Therefore, to any classical
arrangements of spins one has to find the optimal configuration of
{\it occupied orbitals\/} which minimizes the total energy. Hence,
we allow for mixed orbital states of the type as given in Eq.
(\ref{newstate}),
\begin{equation}
\label{mixing}
|i\mu\sigma\rangle=\cos\theta_i|iz\sigma\rangle+\sin\theta_i|ix\sigma\rangle,
\end{equation}
with the set of angles $\{\theta_i\}$ to be found variationally
from the minimization of the classical energy. Let us suppose that
the orbitals occupied at sites $i$ and $j$ are given by the
superposition of the states $\{|iz\sigma\rangle,|ix\sigma\rangle
\}$ (\ref{mixing}) with an angle $\theta_i$ and $\theta_j$,
respectively. One finds then the average values of the operator
projection operators $\{P_{i\alpha} \}$ for the bonds $\langle
ij\rangle\parallel c$:
\begin{eqnarray}
\label{genorbavc1} \langle P_{ix}P_{jz}+P_{iz}P_{jx}\rangle&=&
\cos^2 \theta_i \sin^2 \theta_j+\cos^2 \theta_j \sin^2 \theta_i ,
\\ \label{genorbavc2} \langle 2P_{iz}P_{jz}\rangle &=&2 \cos^2
\theta_i \cos^2 \theta_j,
\end{eqnarray}
while for the bonds $\langle ij\rangle\parallel (a,b)$ they are:
\begin{eqnarray}
\label{genorbavab1} \langle
P_{i\xi}P_{j\zeta}+P_{i\zeta}P_{j\xi}\rangle&=& \frac{1}{8}\left[
4-2 \cos2(\theta_j-\theta_i)+ \cos2(\theta_j+\theta_i)-\sqrt 3
\sin2(\theta_j+\theta_i)\right] ,
\\ \label{genorbavab2} \langle 2P_{i\zeta}P_{j\zeta}\rangle&=&
\frac{1}{8} (-2+\cos 2\theta_i+\sqrt 3 \sin 2\theta_i) (-2+\cos
2\theta_j+\sqrt 3 \sin 2\theta_j).
\end{eqnarray}
By means of these expressions one can easily determine the MF
energy for any orbital configuration, assuming that the spin
structure is assigned. Let us start from the MF solutions with
G-AF type of magnetic structure, that is from the 3D N\'eel state.

\begin{figure}
\centerline{\psfig{figure=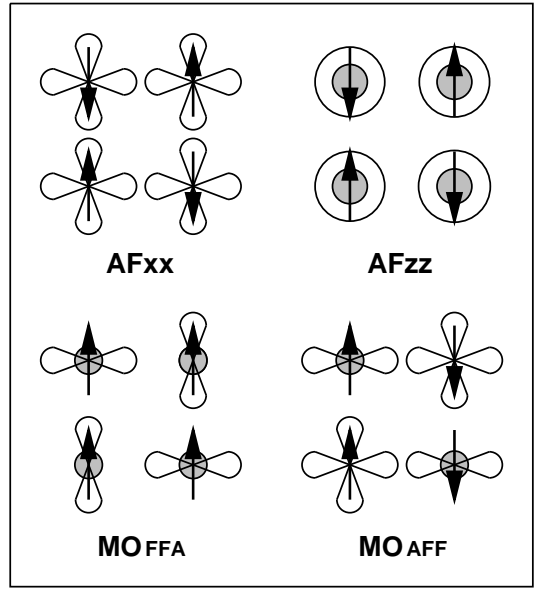,width=7cm}} \vspace{0.2in}
\caption {Schematic representation of magnetic and orbital
long-range orderings in $(a,b)$ planes for the classical phases:
AFxx, AFzz, MO{\protect{\scriptsize FFA}} and
MO{\protect{\scriptsize AFF}} phases. Grey parts of different
$e_g$ orbitals are oriented along the $c$-axis (after Ref.
\protect\cite{Ole00}).} \label{fig:d9mfa}
\end{figure}
It is clear that at large positive $E_z$, where the crystal field
strongly favors $|x\rangle$-occupancy over $|z\rangle$-occupancy,
one expects that $\theta_i=\pi/2$ in Eq. (\ref{mixing}), and the
holes occupy $|x\rangle$ orbitals at every site. In this case the
spins do not interact in the $c$-direction (see Fig.
\ref{fig:d9se}), and there is also no orbital energy contribution.
Hence, the $(a,b)$ planes will decouple magnetically, while within
each plane the superexchange is AF and equal to $9J/4$ along $a$
and $b$. These interactions stabilize a 2D antiferromagnet, called
further AFxx phase. On the contrary, if $E_z<0$ and $|E_z|$ is
large, then the holes occupy $|z\rangle$ orbitals and $\theta_i=0$
in Eq. (\ref{mixing}). By means of the expressions
(\ref{genorbavc1}) -- (\ref{genorbavab2}), we find that the spin
system has then strongly anisotropic AF superexchange, being $4J$
on the bonds $\langle ij\rangle$ along the $c$-axis, and $J/4$ on
the bonds within the $(a,b)$ planes, respectively. This 3D N\'eel
state with the holes occupying $|z\rangle$ orbitals is called AFzz
phase. The spin and orbital order in both AF phases is shown
schematically within the $(a,b)$ planes in Fig. \ref{fig:d9mfa}.
In this case the energies normalized per one site are given by:
\begin{eqnarray}
\label{mfaf} E_{\rm AFxx}&=&-3J\left(1-{\eta\over
4}\right)-{1\over 2}E_z,     \\ E_{\rm
AFzz}&=&-J\left(1+{\eta\over 4}\right)
          -2J\beta\left(1-{\eta\over 2}\right)+{1\over 2}E_z.
\end{eqnarray}
The AFxx and AFzz phases are degenerate in a cubic system
($\beta=1$) along the line $E_z=0$, while decreasing $\beta$ moves
the degeneracy point to negative values of $E_z$, given by
$E_z=-2J(1-\beta)(1-{\eta\over 2})$.

However, for intermediate values of $E_z$ one may expect to
optimize the energy by realizing mixed orbital configurations
($0<\theta<\pi/2$). In this case, guided by the observation that
the orbital interaction is AF-like, we look for solutions with
alternating orbitals at two sublattices, $A$ and $B$. The
alternation is chosen in a way to allow the orbitals being
parallel (optimizing the magnetic energy) in one direction, and
being (almost) orthogonal in the other two (optimizing the orbital
energy). Such states are realized by choosing in Eq.
(\ref{mixing}) the angles alternating between two sublattices in
particular planes: $\theta_i=+\theta$ for $i\in A$, and
$\theta_j=-\theta$ for $j\in B$, respectively,
\begin{eqnarray}
\label{orbmoffa}
|i\mu\sigma\rangle&=&\cos\theta|iz\sigma\rangle+\sin\theta|ix\sigma\rangle,
                                                        \nonumber \\
|j\mu\sigma\rangle&=&\cos\theta|jz\sigma\rangle-\sin\theta|jx\sigma\rangle.
\end{eqnarray}

Let us assume first the G-AF state. By evaluating the orbital
operators following Eqs. (\ref{genorbavc1}) -- (\ref{genorbavab2})
for this case, one finds easily the energy as a function of
$\theta$ in Eqs. (\ref{orbmoffa}),
\begin{eqnarray}
\label{enemoaaat}
E(\theta)&=&-\frac{J}{4}(1+\frac{\eta}{2})(7-4\cos^2 2\theta)
         -\frac{J}{4} (1-\frac{\eta}{2})(1-2\cos 2\theta)^2   \nonumber \\
         &-&\frac{J}{2}\beta(1+\frac{\eta}{2})(1-\cos^2 2\theta)
         -\frac{J}{2}\beta(1-\frac{\eta}{2})(1+\cos 2\theta)^2
         +\frac{1}{2}E_z\cos 2\theta.
\end{eqnarray}
This expression has a minimum at
\begin{equation}
\label{orbmoaaa} \cos 2\theta=-{{(1-\frac{\eta}{2})(1-\beta)
+\frac{1}{2}\varepsilon_z\over (2+\beta)\eta}},
\end{equation}
where $\varepsilon_z=E_z/J$, if $\eta\neq 0$, and provided that
$|\cos 2\theta|\le 1$ (a similar condition applies to all the
other states with MO considered below). So, as long as
$2J(\beta-1)-3J(\beta+1)\eta \le E_z \le
2J(\beta-1)+J(5+\beta)\eta$, there is genuine MO order, while upon
reaching the smaller (larger) boundary value for $E_z$, the
orbitals go over smoothly into $|z\rangle$ ($|x\rangle$), i.e.,
one retrieves the AFzz (AFxx) phase. Taking the magnetic ordering
in the three cubic directions $\{a,b,c\}$ as a label to classify
the classical phases with MO (\ref{orbmoffa}), we call the phase
obtained in the regime of genuine MO order MO{\scriptsize AAA},
with classical energy given by
\begin{eqnarray}
\label{enemoaaa} E_{\rm MO{\scriptsize AAA}}&=&
-\left(2+\beta+\frac{3}{4}\eta\right)J -J {\left[
(2-\eta)(1-\beta)+\varepsilon_z \right]^2\over 4(2+\beta)\eta}.
\end{eqnarray}

In a similar fashion we can get the MF solutions for other
possible spin configurations of A-AF type. Consider first the
MO{\scriptsize FFA} phase, with FM order within the $(a,b)$
planes, and AF order along the $c$-axis. The classical energy as a
function of $\theta$ is given by:
\begin{eqnarray}
 \nonumber E(\theta)&=&-\frac{J}{4}(1+\eta)(7-4\cos^2
2\theta)-\frac{J}{2}\beta(1+\frac{\eta}{2})(1-\cos^2 2\theta)
\\ \label{enemoffat} &&-\frac{J}{2}\beta(1-\frac{\eta}{2})(1+\cos 2\theta)^2
         +\frac{1}{2}E_z\cos 2\theta ,
\end{eqnarray}
with a minimum at
\begin{equation}
\label{thetamoffa} \cos
2\theta={\beta(1-\frac{\eta}{2})-\frac{1}{2}\varepsilon_z
              \over 2+(2+\beta)\eta},
\end{equation}
where again the MO exist as long as $|\cos 2\theta|\le 1$. Using
Eqs. (\ref{enemoffat}) and (\ref{thetamoffa}) one finds that the
classical energy of the MO{\scriptsize FFA} phase is given by
\begin{equation}
\label{enemoffa} E_{\rm MO{\scriptsize
FFA}}=-\frac{J}{4}\left(11-7\eta\right)
  -\frac{J}{2}{[\beta(1-\frac{\eta}{2})-\frac{1}{2}\varepsilon_z]^2
               \over 2+(2+\beta)\eta}.
\end{equation}
This solution is stable for $E_z<0$, while for $E_z>0$ the other
two degenerate phases: the MO{\scriptsize FAF} and MO{\scriptsize
AFF} phase have a lower energy, as they are characterized by a
lower hole density in $|z\rangle$ orbitals which become
unfavorable. In this case, due to the breaking of local symmetry
of the magnetic interactions within the $(a,b)$ planes, with one
direction AF and the other FM, one is forced to look for solutions
with different angles on the two sublattices \cite{Ole00}.

Finally, one may consider how the degeneracy of the AFxx phase is
removed by the interactions along the $c$-axis. One possibility is
the MO{\scriptsize AAA} phase, with the energy given above by Eq.
(\ref{enemoaaa}). If the interactions along the $c$-axis are
instead FM, one finds the classical energy of the MO{\scriptsize
AAF} phase given by
\begin{eqnarray}
\label{enemoaaf} E_{\rm MO{\scriptsize AAF}}=
-\left(2+\frac{3}{4}\eta\right)J-\frac{1}{2}\beta(1+\eta)
-J{\left( 2-\eta+\varepsilon_z\right)^2\over
2[\beta(1+\eta)+2\eta]},
\end{eqnarray}
with the mixing angle
\begin{equation}
\label{orbmoaaf}
\cos2\theta=-{{1-\frac{\eta}{2}+\frac{1}{2}\varepsilon_z
             \over \beta(1+\eta)+2\eta}}.
\end{equation}
This solution turns out to be stable with respect to the
MO{\scriptsize AAA} as long as $1+\cos 2\theta<\eta$. This means
that when the hole density in the $|z\rangle$ orbitals $\sim
\cos^2\theta$ grows smoothly from zero (at $\theta=\pi/2$) with
decreasing $E_z$, it tends to stabilize first the MO{\scriptsize
AAF} phase by FM terms $\sim J\eta\cos^2\theta$, while at higher
occupancy of $|z\rangle$ orbitals the AF interactions $\sim
J\cos^4\theta$ take over.

Thus, one obtains the classical phase diagram of the 3D
spin-orbital model (\ref{somcu}) by comparing the energies of the
six above phases for various values of two parameters, $E_z/J$ and
$J_H/U$: two AF phases with two sublattices and pure orbital
character (AFxx and AFzz), three A-AF phases with four sublattices
(MO{\scriptsize FFA} and two degenerate phases: MO{\scriptsize
AFF} and MO{\scriptsize FAF}), one C-AF phase (MO{\scriptsize
AAF}), and one G-AF phase with MO's (MO{\scriptsize AAA}). By
looking at the phase diagram one can see that the generic sequence
of classical phases at finite $\eta$ and decreasing $E_z/J$ is:
AFxx, MO{\scriptsize AAF}, MO{\scriptsize AAA}, MO{\scriptsize
AFF}, MO{\scriptsize FFA}, and AFzz, and the magnetic order is
tuned together with the gradually increasing $|z\rangle$ character
of the occupied orbitals. By making several other choices of
orbital mixing and classical magnetic order, it has been verified
that no other commensurate ordering with up to four sublattices
can be stable in the present situation. Although some other phases
have been found, they were degenerate with the above phases only
at the $M=(0,0)$ point of the phase diagram, and otherwise had
higher energies.
\begin{figure}
\centerline{\psfig{figure=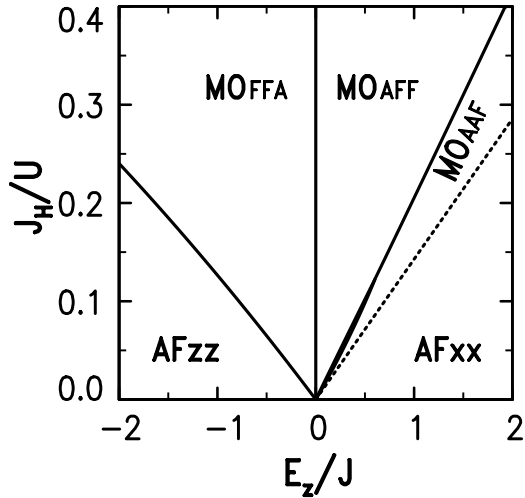,width=9cm}} \caption {Mean-field
phase diagram of the 3D spin-orbital model (\protect{\ref{somcu}})
in the $(E_z,J_H)$ plane for $\beta=1$ (after Ref.
\protect\cite{Ole00}). The lines separate the classical states
shown in Fig. \protect{\ref{fig:d9mfa}}; the transition from AFxx
to MO{\protect{\scriptsize AFF}} phase is second order (dashed
line), while all the other transitions are first order (full
lines). } \label{fig:3d}
\end{figure}

The result for cubic symmetry ($\beta=1$) is presented in Fig.
\ref{fig:3d}, where one finds all six phases, but the
MO{\scriptsize AAA} phase does stabilize only in a very restricted
range of parameters for $J_H/U<0.1$, in between AFxx and
MO{\scriptsize AFF} phases. Only the first of the above
transitions is continuous, while the other lines in Fig.
\ref{fig:3d} are associated with jumps in the magnetic and in
orbital patterns. We would like to emphasize that {\it all the
considered phases are degenerate at the $M=(0,0)$ point\/}
\cite{Fei97}. It is a multicritical point, where the orbitals may
be rotated freely when the spins are AF, and a few other states
with FM planes, and tuned to them orbital order of the MO type
gives precisely the same energy.

When $\beta\neq1$, the phase diagram changes quantitatively but
not qualitatively, with either expanded or reduced areas
corresponding to the different classical phases \cite{Ole00}. In
particular, $\beta>1$ stabilizes the MO phases [especially the
MO{\scriptsize AFF}(MO{\scriptsize FAF}) states]. On the contrary,
the MO phases are stable in a reduced range of $E_z$ for a fixed
value of $J_H/U$, if $\beta<1$. It is worth emphasizing that the
multicritical point $M$ is a common feature of the classical phase
diagram independently of the value of $\beta$. It follows from the
degenerate multiplet structure of $d^8$ ions, and its coordinate
moves along the $\eta=0$ line, according to the following
relation: $E_z=-2J(1-\beta)$. This is a clear demonstration of the
frustrated nature of the spin and orbital superexchange in the
model, whereas the crystal field term just compensates the
enhanced or suppressed magnetic interactions in the $(a,b)$
planes.

\begin{figure}
\centerline{\psfig{figure=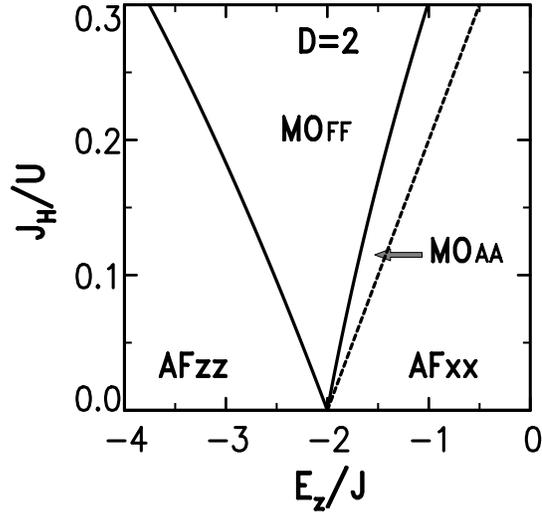,width=9cm}} \caption {Mean-field
phase diagram of the spin-orbital model (\protect{\ref{somcu}}) in
the $(E_z,J_H)$ plane in two dimensions ($\beta=0$). Full lines
separate the classical states AFxx, AFzz, and
MO{\protect{\scriptsize FF}} shown in Fig.
\protect\ref{fig:d9mfa}, while the spin order in the
MO{\protect{\scriptsize AA}} phase is AF, and the orbitals are in
between those in AFxx and MO{\protect{\scriptsize FF}} phase
(after Ref. \protect\cite{Ole00})} \label{fig:2d}
\end{figure}

A special role plays the case with $\beta=0$ which corresponds to
the 2D spin-orbital model. In this case the MO{\scriptsize AFF}
phase disappears completely while the other two phases with AF
order in the $(a,b)$ planes MO{\scriptsize AAA} and MO{\scriptsize
AAF} collapse into a single MO{\scriptsize AA} phase. The
resulting phase diagram is shown in Fig. \ref{fig:2d}. The
MO{\scriptsize FF} is still stable in a large region of the
parameter space which demonstrates that the strong AF exchange
along the $c$ axis in the corresponding 3D MO{\scriptsize FFA}
phase is not instrumental to stabilize this phase, but the orbital
energy within the FM planes is a dominating mechanism.
\begin{figure}
\centerline{\psfig{figure=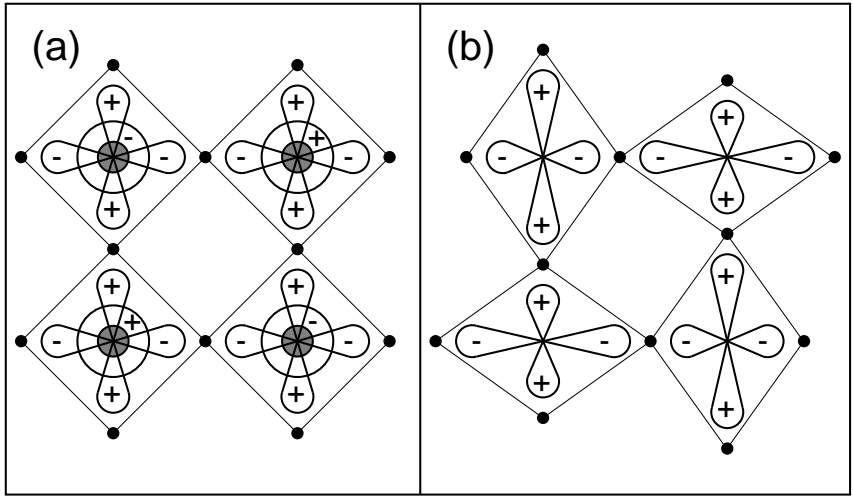,width=7cm,angle=270}}
\vspace{0.2in} \caption {Schematic representation of the mixed
orbitals in $(a,b)$ planes of the MO{\protect{\scriptsize FF}}
phase in a 2D model: (a) the orbitals with their phases, and (b)
the resulting distortion in the oxygen lattice, stabilized by the
orbital ordering (after Ref. \protect\cite{Ole00})} \label{fig:jt}
\end{figure}
It is interesting to compare the results obtained on the classical
level with some relevant physical systems. For La$_2$CuO$_4$ and
Nd$_2$CuO$_4$ the crystal field splitting is large, $E_z\simeq
0.64$ eV \cite{Gra92}, so that one falls in the region of the 2D
AFxx phase observed in neutron scattering. If on the contrary the
orbital splitting is small, the orbital ordering sets in and has
to couple strongly to the lattice. The net result is a quadrupolar
distortion as indicated in Fig. \ref{fig:jt}. This lattice
instability is again related to the question on the origin of the
orbital ordering: is it due to JT and/or to electronic mechanism?
The deformations found in KCuF$_3$ (or LaMnO$_3$) could in
principle be entirely caused by phonon-driven collective JT
effects. One might therefore attempt to neglect electron-electron
interactions, and focus on the electron-phonon coupling. In case
that the ions are characterized by a JT (orbital) degeneracy, one
can integrate out the (optical) phonons, and one finds effective
Hamiltonians with phonon mediated interactions between the
orbitals. In the specific case of $e_g$ degenerate ions in a cubic
crystal, these look quite similar to the orbital interactions in
the $d^9$ Hamiltonian, except that the spin dependent term is
absent \cite{KKphon}. Any orbital order resulting from this
Hamiltonian is now accompanied by a lattice distortion of the same
symmetry.

The size of the quadrupolar deformation in the $(a,b)$ plane of
KCuF$_3$ is actually as large as 4 \% of the lattice constant
$a_0$. It is therefore often argued that the orbital order is
clearly phonon-driven, and that the orbital interactions discussed
above are less important. Although appealing at first sight, this
argument is flawed: large displacements do not necessarily imply
that phonons are the driving mechanism. Unfortunately, the
deformations of the lattice and the orbital degrees of freedom
cannot be disentangled using general principles: they constitute
an irreducible subsector of the problem. The issue is therefore a
quantitative one, and may be answered by calculating the
electronic structure.
\begin{figure}[t]
\centerline{\psfig{figure=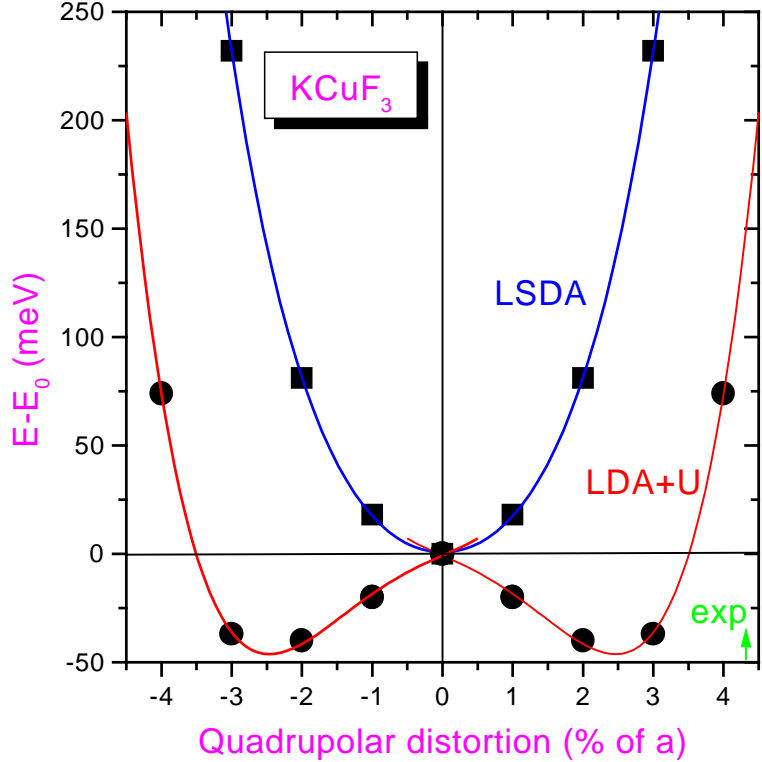,width=12cm}}
\smallskip
\caption{The dependence of the total energy of KCuF$_3$ on the
quadrupolar lattice distortion according to LSDA and LDA+U band
structure calculations (after Ref. \protect\cite{Lie95}).}
\label{fig:kcuf3}
\end{figure}

We start out with the observation that according to LDA KCuF$_3$
would be an undistorted, cubic system: the energy increases if the
distortion is switched on (see Fig. \ref{fig:kcuf3}). The reason
is that KCuF$_3$ would be a band metal according to LDA (the usual
Mott-gap problem) with a Fermi-surface which is not susceptible to
a band JT instability. Therefore, the effects of strong on-site
Coulomb interaction should be included and the LDA+U method
\cite{Ani91} is a well designed method to serve this purpose. It
is constructed to handle the physics of electronic orbital
ordering, keeping the accurate treatment of the electron-lattice
interaction of LDA intact. According to LDA+U calculations the
total energy gained by the deformation of the lattice is only a
small contribution of $\sim 50$ meV (Fig. \ref{fig:kcuf3}) to the
energies involved in the electronic orbital ordering \cite{Lie95}.
Therefore, the coupling to the lattice is here not a driving force
for the orbital and magnetic ordering, but {\it the lattice
follows the orbital state\/}.

Although the energy gained in the deformation of the lattice is
rather small, the electron-phonon coupling is quite effective in
keeping KCuF$_3$ away from the frustrated interactions associated
with the origin of the phase diagram (Fig. \ref{fig:3d}). Since
the FM interactions in the $(a,b)$ plane of KCuF$_3$ are quite
small ($J_{(a,b)}=-0.2$ meV, as compared to the `1D' AF exchange
$J_c=17.5$ meV \cite{Ten95,Ten95a,Ten95b}), one might argue that
the effective Hund's rule coupling $J\eta$ as of relevance to the
low energy theory is quite small. Such a strong anisotropy of
magnetic interactions $J_{(a,b)}$ and $J_c$ has been reproduced
recently within the {\it ab initio\/} method, but not in
unrestricted HF, demonstrating the importance of electron
correlation effects.
Although this still needs further study, it might well be that in
the absence of the electron-phonon coupling KCuF$_3$ would be
close to the origin of Fig. \ref{fig:3d}. Therefore, although
further work is needed to clarify the role played by
electron-phonon coupling, it might be that phonons are to a large
extent responsible for the stability of KCuF$_3$'s classical
ground state. In any case, one cannot rely just on the size of the
lattice deformations to resolve this issue.

\subsection{ Elementary excitations in the $d^9$ model }
\label{sec:genrpa}

The presence of the orbital degrees of freedom in the Hamiltonian
(\ref{somcu}) yields excitation spectra that are qualitatively
different from those of the quantum antiferromagnet with a single
spin-wave mode. In the present case one gets two transverse
excitations: {\em spin waves\/} and {\em spin-and-orbital waves\/}
\cite{Fei98}; and also longitudinal excitations -- {\em orbital
waves\/}, thus producing three elementary excitations for the
present spin-orbital model (\ref{somcu})
\cite{Fei97,Ole00,Fei98,Bri98}. This gives therefore the same
number of modes as found in a 1D SU(4) symmetric spin-orbital
model (see Sec. IV.A) in the Bethe ansatz method
\cite{Sut75,Fri99}. We emphasize that this feature is a
consequence of the dimension (equal to 15) of the $so(4)$ Lie
algebra of the local operators, as explained below, and is not
related to the global symmetry of the Hamiltonian. In this
chapter, we report the analysis of the realistic $d^9$
spin-orbital model for the 3D simple cubic (i.e., perovskite-like)
lattice (\ref{somcu}), using linear spin-wave (LSW) theory
\cite{Aue94,Tak89}, generalized in such a way that makes it
applicable to the present situation.

Before we introduce the excitation operators, it is convenient to
rewrite the spin-orbital model (\ref{somcu}) in a different
representation which uses a four-dimensional space:
$\{|x\!\uparrow\rangle$, $|x\!\downarrow\rangle$,
 $|x\!\downarrow\rangle\}$, $|z\!\uparrow\rangle$,
instead of a direct product of the spin and orbital subspaces.
This will demonstrate explicitly that three different elementary
excitations appear in a natural way. Hence, we introduce operators
which define purely {\it spin excitations\/} in individual
orbitals,
\begin{equation}
S^{+}_{ixx}=d^{\dagger}_{ix\uparrow}d^{}_{ix\downarrow}, \hskip
1.2cm S^{+}_{izz}=d^{\dagger}_{iz\uparrow}d^{}_{iz\downarrow},
\label{splus}
\end{equation}
and operators for simultaneous spin-flip and transfer between the
orbitals, {\it spin-and-orbital excitations\/},
\begin{equation}
K^{+}_{ixz}=d^{\dagger}_{ix\uparrow}d^{}_{iz\downarrow}, \hskip
1.0cm K^{+}_{izx}=d^{\dagger}_{iz\uparrow}d^{}_{ix\downarrow}.
\label{oplus}
\end{equation}
The corresponding operators $S^z_{i\alpha\alpha}$ and
$K^z_{i\alpha\beta}$ are defined as follows,
\begin{eqnarray}
S^{z}_{ixx}&=&\case{1}{2}(n_{ix\uparrow}-n_{ix\downarrow}), \hskip
2.3cm S^{z}_{izz}=\case{1}{2}(n_{iz\uparrow}-n_{iz\downarrow}), \\
\label{szet} K^{z}_{ixz}&=&\case{1}{2}(d^{\dagger}_{ix
\uparrow}d_{iz\uparrow}
                        -d^{\dagger}_{ix\downarrow}d_{iz\downarrow}),
\hskip 1.0cm K^{z}_{izx}=\case{1}{2}(d^{\dagger}_{iz
\uparrow}d_{ix\uparrow}
                        -d^{\dagger}_{iz\downarrow}d_{ix\downarrow}).
\label{kzet}
\end{eqnarray}
The Hamiltonian (\ref{somcu}) contains also purely orbital
interactions which can be expressed using the following {\it
orbital excitation\/} operators,
\begin{eqnarray}
\label{ozet} T_{ixz}&=&\case{1}{2}(d^{\dagger}_{ix  \uparrow}d_{iz
\uparrow}
                    +d^{\dagger}_{ix\downarrow}d_{iz\downarrow}),
\hskip 1cm T_{izx}=\case{1}{2}(d^{\dagger}_{iz  \uparrow}d_{ix
\uparrow}
                   +d^{\dagger}_{iz\downarrow}d_{ix\downarrow}),
\end{eqnarray}
while the anisotropy in the orbital space is expressed by
orbital-polarization operators,
\begin{equation}
n_{i-} = \case{1}{2}(d^{\dagger}_{ix  \uparrow}d_{ix  \uparrow}\!+
                     d^{\dagger}_{ix\downarrow}d_{ix\downarrow}\! -\!
                     d^{\dagger}_{iz  \uparrow}d_{iz  \uparrow}\! -\!
                     d^{\dagger}_{iz\downarrow}d_{iz\downarrow} ).
\label{nzet}
\end{equation}
In order to simplify the notation, we also introduce global
operators for the spin, spin-and-orbital and orbital excitations,
\begin{eqnarray}
\label{top} S^{+}_{i}&=&S^{+}_{ixx}+S^{+}_{izz},  \hskip 1cm
S^{z}_{i} = S^{z}_{ixx}+S^{z}_{izz},  \\
K^{+}_{i}&=&K^{+}_{ixz}+K^{+}_{izx},  \hskip 1cm K^{z}_{i} =
K^{z}_{ixz}+K^{z}_{izx},  \\ T_{i}&=&T    _{ixz}+T    _{izx} .
\end{eqnarray}

The number of collective modes in a particular phase may be
determined as follows. The $so(4)$ Lie algebra consists of three
Cartan operators, i.e., operators diagonal on the local basis of
the symmetry-broken phase under consideration (e.g. $S^z_{ixx}$,
$S^z_{izz}$, and $n_{i-}$ in the AFxx phase), plus twelve
nondiagonal operators turning the eigenstates into one another
(like $S^{+}_{ixx}$ and $S^{+}_{izz}$ in the AFxx phase). Out of
those twelve operators, six connect two excited states (like
$S^{+}_{izz}$ in the AFxx phase), and are physically irrelevant
(in the lowest order), because they give only rise to the
so-called 'ghost' modes, the modes for which the spectral function
vanishes identically at $T=0$. The remaining six operators connect
the local ground state with excited states, three of them
describing an excitation and three a deexcitation, and only these
six operators are physically relevant. Out of the three
excitations (deexcitations), two are transverse, i.e., change the
spin, and one is longitudinal, i.e., does not affect the spin. For
a classical phase with $L$ sublattices one therefore expects $4L$
transverse and $2L$ longitudinal modes. Because of time-reversal
invariance they all occur in pairs with opposite frequencies, $\pm
\omega^{(n)}_{\vec k}$.

Finally, the SU(2) spin invariance of the Hamiltonian guarantees
that the transverse operators raising the spin are decoupled from
those lowering the spin, and that both sets of operators are
described by equivalent equations of motion, so that the
transverse modes are pairwise degenerate. Such a simplification
does not occur in the longitudinal sector. So, in conclusion, in
an $L$-sublattice phase there are $L$ doubly-degenerate
positive-frequency transverse modes and $L$ nondegenerate
positive-frequency longitudinal modes, accompanied by the same
number of negative-frequency modes. This may be compared with the
well-known situation in the quantum antiferromagnet \cite{Aue94},
where there is, with only spin excitation operators involved, only
one (not two) doubly-degenerate positive-frequency (transverse)
mode and the corresponding negative-frequency mode in the
two-sublattice N\'{e}el state.

For the actual evaluation it is convenient to decompose the
superexchange terms (\ref{somj}) in the spin-orbital Hamiltonian
(\ref{somcu}),
\begin{equation}
{\cal H}_J={\cal H}_{\parallel}+{\cal H}_{\perp}, \label{somlong}
\end{equation}
into two parts which depend on the bond direction:

(i) for the bonds $\langle ij\rangle\parallel (a,b)$,
\begin{eqnarray}
\label{hpara} {\cal H}_{\parallel}&=&\case{1}{4}J\sum_{\langle
ij\rangle{\parallel}}
  \left[(1-\case{1}{2}\eta)
       (3{\vec S}_{ixx}+{\vec S}_{izz}+\lambda_{ij}\sqrt{3}{\vec K}_{i})
 \cdot (3{\vec S}_{jxx}+{\vec S}_{jzz}+\lambda_{ij}\sqrt{3}{\vec K}_{j})
  \right.                                            \nonumber \\
 & &\left. \hskip 1.1cm -2\eta {\vec S}_{i}\cdot {\vec S}_{j}
+(1+2\eta)(n_{i-}+\pm\sqrt{3}T_{i})(n_{j-}+\pm\sqrt{3}T_{j})
      -(3+\eta)\right],
\end{eqnarray}

(ii) for the bonds $\langle ij\rangle\perp (a,b)$, i.e., along the
$c$-axis,
\begin{eqnarray}
\nonumber {\cal H}_{\perp}&=&J\sum_{\langle ij\rangle{\perp}}
  [(4-2\eta){\vec S}_{izz}\cdot{\vec S}_{jzz}
  -\eta ({\vec S}_{ixx}\cdot{\vec S}_{jzz}
  + {\vec S}_{izz}\cdot{\vec S}_{jxx})
 \\ \label{hperp}  &&\hskip 1cm +(1+2\eta)n_{i-}n_{j-}-\case{1}{4}(3+\eta) ].
\end{eqnarray}
Here and in the following paragraphs we consider a 3D cubic model
with $\beta=1$. We note that the orbital interactions
(\ref{oplus}) are quite different in $H_{\parallel}$ and
$H_{\perp}$; propagating composite spin-and-orbital excitations
are possible only within the $(a,b)$ planes, where they are
coupled to the spin excitations, while in the $c$-direction only
pure spin excitations and pure spin-and-orbital excitations occur,
which are decoupled from one another. This apparent breaking of
symmetry between $H_{\parallel}$ and $H_{\perp}$ is a consequence
of the choice of basis as $|x\rangle$ and $|z\rangle$ orbitals.

In the following, we report briefly the results obtained for
transverse and longitudinal excitations in the various
symmetry-broken classical states of the spin-orbital model
(\ref{somcu}). The transverse excitations, i.e., spin-waves and
spin-and-orbital-waves, are calculated using the spin-rising
operators which make a transition to a state realized in a
classical phase at a given site $i$. As an example we use the AFxx
phase to illustrate the formalism and calculation method with the
excitation operators:
\begin{equation}
S^+_{ixx}=d^{\dagger}_{ix\uparrow}d^{}_{ix\downarrow},    \hskip
1cm K^+_{ixz}=d^{\dagger}_{ix\uparrow}d^{}_{iz\downarrow}.
\label{excopt}
\end{equation}
The longitudinal excitations (without spin-flip) are most
conveniently obtained starting from spin-dependent orbital
excitation operators,
\begin{equation}
T_{ixz\sigma}=d^{\dagger}_{ix\sigma}d^{}_{iz\sigma},      \hskip
1cm T_{izx\sigma}=d^{\dagger}_{iz\sigma}d^{}_{ix\sigma},
\label{topera}
\end{equation}
as these excitations conserve the spin component and we ask a
question whether such a longitudinal excitation may propagate
coherently in a given symmetry-broken classical state.

The nature and dispersion of elementary excitations in the
spin-orbital model (\ref{somcu}) can be conveniently studied in
the leading order of the $1/S$ expansion using the Green function
formalism. The starting point are the equations of motion for the
Green functions generated by the excitation operators
(\ref{excopt}) written in the energy representation
\cite{Zub60,Hal72},
\begin{eqnarray}
\label{gfafxx1} E\langle\langle     S_{ixx}^+|...\rangle\rangle
&=& {1\over 2\pi}\langle [S_{ixx}^+,...]\rangle +
 \langle\langle [S_{ixx}^+,H]|...\rangle\rangle,          \\
\label{gfafxx2} E\langle\langle     K_{ixz}^+|...\rangle\rangle
&=& {1\over 2\pi}\langle [K_{ixz}^+,...]\rangle +
 \langle\langle [K_{ixz}^+,H]|...\rangle\rangle,
\end{eqnarray}
where the average of the commutator on the right hand side, e.g.
$\langle [S_{ixx}^+,S_{jxx}^-]\rangle$, is evaluated in the
classical ground state. We note, however, that equivalent results
for the AFxx and AFzz phases can be obtained using instead an
expansion around a classical saddle point with Schwinger bosons
\cite{Aue94}.

The equations of motion have been derived for the Green functions
generated by the set of operators
$\{S_{ixx}^+,K_{ixz}^+,S_{jxx}^+,K_{jxz}^+\}$, where $i\in A$ and
$j\in B$, and used the random-phase approximation (RPA) for
spinlike operators which linearizes the equations of motion by a
decoupling procedure \cite{Zub60,Hal72}. Thereby, the operators
which have nonzero expectation values in the considered classical
state give finite contributions, e.g. for the first spin-flip
Green function one uses,
\begin{equation}
\langle\langle S_{ixx}^+S_{mxx}^z|...\rangle\rangle\simeq \langle
S_{mxx}^z\rangle \langle\langle S_{ixx}^+|...\rangle\rangle,
\label{rpas}
\end{equation}
and a similar formula for the mixed spin-and-orbital excitation
described by $\langle\langle K_{ixz}^+|...\rangle\rangle$,
\begin{equation}
\langle\langle K_{ixz}^+S_{mxx}^z|...\rangle\rangle\simeq \langle
S_{mxx}^z\rangle \langle\langle K_{ixz}^+|...\rangle\rangle .
\label{rpak}
\end{equation}
In the present case of the AFxx phase one uses the respective
N\'eel state average values,
\begin{eqnarray}
\langle S_{ixx}^z\rangle&=&-\langle S_{jxx}^z\rangle=\case{1}{2},
\\
   \langle n_{i-}\rangle&=& \langle n_{j-}\rangle=\case{1}{2},
\label{avx}
\end{eqnarray}
where $i\in A$ and $j\in B$, and $A$ and $B$ are the two
sublattices in a 2D lattice of the AFxx phase, and all the
remaining averages vanish. It is crucial that the decoupled
operators have different site indices, and thus the decoupling
procedure preserves the local commutation rules. Instead, if one
uses products of spin and orbital operators, e.g.,
$K_{ixz}^+=S_{ixx}^+\sigma_i^+$, one is tempted to decouple these
operators locally \cite{Cas78,Kha97} which would violate the
algebraic structure of the $so(4)$ Lie algebra.

The translational invariance of the N\'eel state implies that the
transformed Green functions are diagonal in the reduced Brillouin
zone (BZ). As in the Heisenberg antiferromagnet, the Fourier
transformed functions are defined for the Green functions which
describe the spin dynamics on a given sublattice, either $A$ or
$B$. For instance, the pure spin-flip Green functions are
transformed as follows,
\begin{eqnarray}
\langle\langle S_{{\vec k}xx}^+|...\rangle\rangle_A&=&
\frac{1}{\sqrt{N}}\sum_{i\in A}e^{i{\vec k}{\vec R}_i}
\langle\langle S_{ixx}^+|...\rangle\rangle_A,  \nonumber \\
\langle\langle S_{{\vec k}xx}^+|...\rangle\rangle_B&=&
\frac{1}{\sqrt{N}}\sum_{j\in B}e^{i{\vec k}{\vec R}_j}
\langle\langle S_{jxx}^+|...\rangle\rangle_B, \label{fourier}
\end{eqnarray}
where $N$ is the number of sites in one sublattice. Hence, the
problem of finding the elementary excitations of the considered
spin-orbital model (\ref{somcu}) reduces to the diagonalization of
the following $4\times 4$ dynamical matrix at each ${\vec
k}$-point:

\begin{equation}
\label{gfeq} \left(\begin{array}{cccc}
 \lambda_{\alpha}-\overline{\omega}_{\vec k} & 0 & Q_{\alpha\vec k}
                                            & P_{\alpha\vec k}   \\
    0 & \tau_{\alpha}-\overline{\omega}_{\vec k} & P_{\alpha\vec k}
                                            & R_{\vec k}         \\
-Q_{\alpha\vec k} & -P_{\alpha\vec k}
                  & -\lambda_{\alpha}-\overline{\omega}_{\vec k}&0 \\
      -P_{\alpha\vec k} & -R_{\vec k} & 0
                        & -\tau_{\alpha}-\overline{\omega}_{\vec k}
\end{array} \right)
\left( \begin{array}{c}
 \langle\langle S^+_{{\vec k}xx}|\cdots\rangle\rangle_A   \\
 \langle\langle K^+_{{\vec k}xz}|\cdots\rangle\rangle_A   \\
 \langle\langle S^-_{{\vec k}xx}|\cdots\rangle\rangle_B   \\
 \langle\langle K^-_{{\vec k}xz}|\cdots\rangle\rangle_B
\end{array} \right) = 0,
\end{equation}
The symmetric positive and negative eigenvalues $\pm\omega_{\vec
k}^{(n)}$, with $n=1,2$, solved from the matrix in Eq.
(\ref{gfeq}) may be written in the following form for the AFxx
phase,
\begin{eqnarray}
\label{afsw} [\omega_{\vec k}^{(n)}]^2&=&
J^2\left(\lambda_{x}^2+\tau_{x}^2 -Q_{x\vec k}^2-R_{\vec k}^2
-2P_{x\vec k}^2\right) \pm J^2\left[ (\lambda_{x}^2-\tau_{x}^2)^2
-2(\lambda_{x}^2-\tau_{x}^2)(Q_{x\vec k}^2-R_{\vec k}^2) \right.
                                                           \nonumber \\
&-&\left. 4(\lambda_{x}-\tau_{x})^2P_{x\vec k}^2 +(Q_{x\vec
k}^2+R_{\vec k}^2+2P_{x\vec k}^2)^2
 -  4(Q_{x\vec k}R_{\vec k}-P_{x\vec k}^2)^2\right]^{1/2}.
\end{eqnarray}
Here the quantities $\lambda_x$ and $\tau_x$ play the role of
local potentials and follow from the model parameters, $E_z$ and
$J_H$,
\begin{eqnarray}
\label{afxxlambda} \lambda_x&=&\case{9}{2}-3\eta,
\\ \label{afxxtau}
\tau_x&=&\case{7}{2}-4\eta-2-\eta+\varepsilon_z.
\end{eqnarray}
The remaining terms are ${\vec k}$-dependent, and depend on
\begin{eqnarray}
\label{gammap} \gamma_{+}(\vec k)&=&\case{1}{2}(\cos k_x+\cos
k_y),                 \\ \label{gammam} \gamma_{-}(\vec
k)&=&\case{1}{2}(\cos k_x-\cos k_y),                 \\
\label{gammaz} \gamma_{z}(\vec k)&=&\cos k_z.
\end{eqnarray}
The quantities $Q_{x\vec k}$ and $P_{x\vec k}$ for the AFxx phase
take the form,
\begin{eqnarray}
\label{afxxq} Q_{x\vec k}&=&(\case{9}{2}-3\eta)\gamma_{+}(\vec k),
\\ \label{afxxp} P_{x\vec
k}&=&\case{1}{2}\sqrt{3}(3-\eta)\gamma_{-}(\vec k),
\end{eqnarray}
while the last dispersive term,
\begin{equation}
R_{\vec k}=\case{3}{2}\gamma_{+}(\vec k) , \label{rdef2}
\end{equation}
carries no index and remains identical for both G-AF phases (AFxx
and AFzz). We emphasize that the coupling between the spin-wave
and spin-and-orbital-wave excitations occurs due to the terms
$\propto P_{x\vec k}$, as seen from Eq. (\ref{gfeq}). It vanishes
in the planes of $k_x=\pm k_y$, but otherwise plays an important
role, as discussed in Sec. \ref{sec:somd9}. In the limit of large
$E_z\to\infty$, Eq. (\ref{afsw}) reproduces the spin-wave
excitations for a 2D antiferromagnet with an AF superexchange
interaction of $J(\frac{9}{4}-\frac{3}{2}\eta)$, as given between
the occupied $|x\rangle$ orbitals,
\begin{equation}
\label{limitafxx} \omega_{\vec
k}^{(1)}=J\left(\case{9}{2}-3\eta\right)
                       [1-\gamma_{+}^2(\vec k)]^{1/2},
\end{equation}
while the dispersion of the high-energy spin-and-orbital
excitation, $\omega_{\vec k}^{(2)}\simeq E_z$, becomes negligible.
As explained above, both modes are doubly degenerate.

Consider now the orbital (excitonic) excitations generated by the
orbital-flip operators (\ref{ozet}). They are found by considering
the equations of motion,
\begin{equation}
\label{gfafxxl1} E\langle\langle T_{i\alpha\beta \sigma
}|...\rangle\rangle = {1\over 2\pi}\langle [T_{i\alpha\beta \sigma
},...]\rangle + \langle\langle [T_{i\alpha\beta \sigma
},H]|...\rangle\rangle,
\end{equation}
where spin $\sigma$ corresponds to the occupied state in the
symmetry-broken N\'eel state. By making a Fourier transformations
as for the transverse operators (\ref{fourier}), one may show that
only two operators per sublattice suffice to describe the modes in
an antiferromagnet. The structure of the respective RPA dynamical
matrix is given by
\begin{equation}
\label{gfeqor} \left(\begin{array}{cccc}
 u_{\alpha}-\overline{\zeta}_{\vec k} & 0 & +\rho_{\alpha\vec k}
                                          & +\rho_{\alpha\vec k} \\
0 & -u_{\alpha}-\overline{\zeta}_{\vec k} & -\rho_{\alpha\vec k}
                                          & -\rho_{\alpha\vec k} \\
-\rho_{\alpha\vec k} & -\rho_{\alpha\vec k}
                     & -u_{\alpha}-\overline{\zeta}_{\vec k} & 0 \\
+\rho_{\alpha\vec k} & +\rho_{\alpha\vec k} & 0
                     & u_{\alpha}-\overline{\zeta}_{\vec k}
\end{array} \right)
\left( \begin{array}{c}
 \langle\langle T_{{\vec k}xz\uparrow  }|\cdots\rangle\rangle_A   \\
 \langle\langle T_{{\vec k}zx\uparrow  }|\cdots\rangle\rangle_A   \\
 \langle\langle T_{{\vec k}xz\downarrow}|\cdots\rangle\rangle_B   \\
 \langle\langle T_{{\vec k}zx\downarrow}|\cdots\rangle\rangle_B
\end{array} \right) = 0,
\end{equation}
with
\begin{eqnarray}
\label{afxxu} u_x&=&\varepsilon_z-3\eta,  \\ \label{afxxrho}
\rho_{x\vec k}&=&\case{3}{2}\eta\gamma_{+}(\vec k),
\end{eqnarray}
and one finds two, in general nondegenerate, positive-frequency
modes,
\begin{equation}
\label{genorb} \zeta_{\vec k}= J\;\left[u_{\alpha}
                        (u_{\alpha}\pm 2\rho_{\alpha\vec k})\right]^{1/2}.
\end{equation}

\begin{figure}
\centerline{\psfig{figure=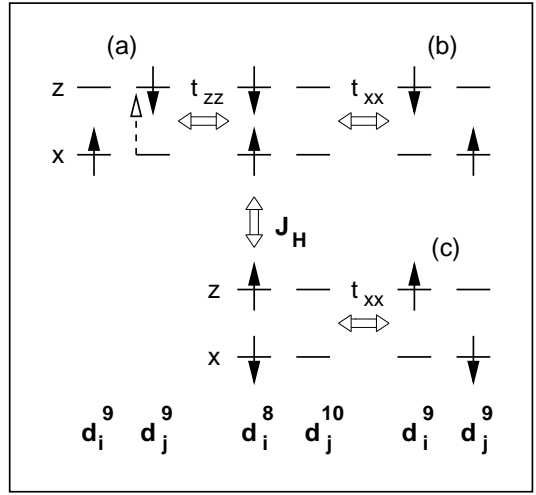,width=7cm}} \vspace{0.2in}
\caption {Schematic propagation of the orbital (excitonic)
excitation (a). If $J_H=0$, an orbital excitation can propagate
only to state (b) and is accompanied by a spin-flip (top), while
$J_H>0$ allows also the spin-flip in the intermediate $d_i^8$
state, and thus the propagation without spin-flip (c) becomes
possible (bottom) (after Ref. \protect\cite{Ole00}).}
\label{fig:orbex}
\end{figure}

It is important to realize that the propagation of longitudinal
excitations, being equivalent to a finite dispersion of
longitudinal modes, becomes possible only at $\eta>0$. This
follows from the multiplet structure of the excited $d^8$ states,
which allows a spin-flip between the orbitals in the
$|^1E_{\theta}\rangle$ and in the $S^z=0$ component of the
$|^3A_2\rangle$-state only if $J_H\neq 0$, as illustrated in Fig.
\ref{fig:orbex}. The processes $\sim t_{xz}$ are not shown, as
they would also lead to a final state given in Fig.
\ref{fig:orbex}(b), i.e., to a propagation of a spin-and-orbital
excitation which was already considered above. In contrast, the
relevant longitudinal orbital excitation in the symmetry-broken
state implies that the exciton has the same spin as imposed by the
N\'eel state of the background; this state is shown in Fig.
\ref{fig:orbex}(c). Therefore, in a perfect N\'eel state without
FM interactions due to $\eta\neq 0$, only local orbital
excitations are possible. These local excitations cost no energy
in the limit of $\varepsilon_z\to 0$ which demonstrates again the
frustration of magnetic interactions at the classical degeneracy
point, $\varepsilon_z=\eta=0$.
%
\begin{figure}
\centerline{\psfig{figure=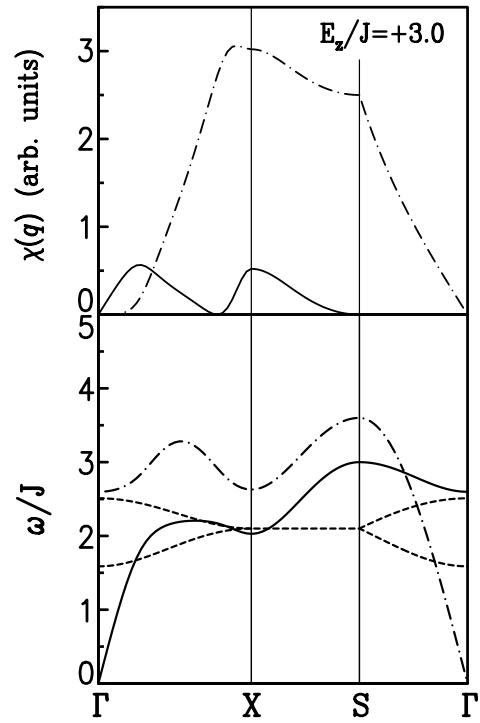,width=9cm}}
\smallskip
\caption{ Spin-wave and spin-and-orbital-wave transverse
excitations (full lines) and longitudinal excitations (dashed
lines) in AFxx phase (bottom), and neutron intensities of the
transverse excitations (top). Parameters: $E_z/J=3.0$ and
$J_H/U=0.3$ (after Ref. \protect\cite{Ole00}). }
\label{fig:modesxx}
\end{figure}
An example of the excitation spectra is presented in Fig.
\ref{fig:modesxx} for the main directions in the 2D BZ, with
$X=(\pi,0)$ and $S=(\pi/2,\pi/2)$. Near the $\Gamma$ point one
finds a (doubly-degenerate) Goldstone mode $\omega_{\vec k}^{(1)}$
with dispersion $\sim k$ at ${\vec k}\to 0$, as in the Heisenberg
antiferromagnet, and a second (doubly-degenerate) transverse mode
at higher energy, $\omega_{\vec k}^{(2)}\simeq \omega_0+ak^2$.
Near the $\Gamma$ point the Goldstone mode is essentially purely
spin-wave, the second mode is purely spin-and-orbital wave. With
increasing ${\vec k}$ these modes start to mix due to the
$P_{x{\vec k}}$ term along the $\Gamma-X$ direction. This is best
illustrated by the intensity measured in the neutron scattering
experiments, which see only the spin-wave component in each
transverse mode. Indeed, the intensity $\chi({\vec q})$ is
transferred from one mode to the other along the $\Gamma-X$
direction in the 2D BZ (Fig. \ref{fig:modesxx}), demonstrating
that indeed the lowest (highest) mode is predominantly
spin-wave-like (spin-and-orbital-wave-like) before the
anticrossing point, while this is reversed after the anticrossing
of the two modes. Thus, we make here a specific prediction that
{\it two spin-wave-like modes could be measurable in certain parts
of the 2D BZ}, in particular in the vicinity of the anticrossing,
if only an AFxx phase was realized for parameters not too distant
from the classical degeneracy point.

Unfortunately, for the realistic parameters for the cuprates
\cite{Gra92}, one finds $E_z/J\simeq 10$ which makes the
spin-and-orbital excitation and the changes of the spin-wave
dispersion hardly visible in neutron spectroscopy. The orbital
(longitudinal) excitations are found for the parameters of Fig.
\ref{fig:modesxx} at a finite energy, being of the same order of
magnitude as the energy of the spin-and-orbital excitation,
$\omega^{(2)}_{\vec k}$. The weak dispersion of these modes
follows from the spin-flip processes in the {\em excited\/}
states, as explained in Fig. \ref{fig:orbex} and discussed above.
We emphasize that the orbital mode has a gap and {\it does not
couple\/} to any spin excitation. At the classical degeneracy
point $M$ the orbital mode falls to zero energy and is
dispersionless, expressing that the orbital can be changed locally
without any cost in energy.

The transverse excitations in the AFzz phase are determined by
considering the complementary set of Green functions to that given
by Eqs. (\ref{gfafxx1}) and (\ref{gfafxx2}), with the excitation
operators $S_{izz}^+$ and $K_{izx}^+$. After deriving the RPA
equations, one finds the final form of the equations of motion by
performing a Fourier transformation and using the following
nonvanishing expectation values,
\begin{eqnarray}
\langle S_{izz}^z\rangle&=&-\langle S_{jzz}^z\rangle=\case{1}{2},
\\
   \langle n_{i-}\rangle&=& \langle n_{j-}\rangle=-\case{1}{2},
\label{avz}
\end{eqnarray}
in the AFzz phase, with $i\in A$ and $j\in B$. This leads again to
the general form (\ref{gfeq}), with the elements $\lambda_x$,
$\tau_x$, $Q_{x\vec k}$, and $P_{x\vec k}$ now replaced by,
\begin{eqnarray}
\label{afzzlambda} \lambda_z&=&\case{1}{2}-\eta+2(2-\eta), \\
\label{afzztau}
\tau_z&=&-\case{1}{2}-\eta+2(1-2\eta)-\varepsilon_z,  \\
\label{afzzq} Q_{z\vec k}&=&(\case{1}{2}-\eta)\gamma_{+}(\vec k)
            +2(2-\eta)\gamma_{z}(\vec k), \\
\label{afzzp} P_{z\vec
k}&=&\case{1}{2}\sqrt{3}(1-\eta)\gamma_{-}(\vec k) .
\end{eqnarray}
Thus, the transverse excitations have the same form (\ref{afsw})
as in the AFxx phase, but the above quantities
(\ref{afzzlambda})--(\ref{afzzp}) have to be used.

In the limit of large $E_z\to-\infty$ one finds the spin-wave for
a 3D anisotropic antiferromagnet with strong superexchange equal
to $2J(2-\eta)$ along the $c$-axis, and weak superexchange
$\frac{1}{4}J(1-2\eta)$ within the $(a,b)$-planes,
\begin{equation}
\label{limitafzz} \omega_{\vec k}^{(1)}=J\left\{
       \left[(\case{1}{2}-\eta)+2(2-\eta)\right]^2
      -\left[(\case{1}{2}-\eta)\gamma_{+}(\vec k)
            +2(2-\eta)\gamma_{z}\right]^2\right\}^{1/2},
\end{equation}
while the spin-and-orbital excitation, $\omega_{\vec
k}^{(2)}\simeq -E_z$, is dispersionless. Again, both these
transverse modes are doubly degenerate.

The representative excitation spectrum for the AFzz phase may be
found in Ref. \cite{Ole00}. One finds again a Goldstone mode
$\omega_{\vec k}^{(1)}$ at the $\Gamma$ point which is
spin-wave-like, accompanied by a finite energy spin-and-orbital
mode $\omega_{\vec k}^{(2)}$. The first one is linear, while the
second changes quadratically with increasing ${\vec k}$. The
dispersion in the $\Gamma-X$ direction is, however, only $\sim
0.7J$, while in the AFxx phase a large dispersion of $\sim 2.5J$
was found. This demonstrates the large difference between the
superexchange in the $(a,b)$-planes in these two AF phases. Here
one should bear in mind, that in a strongly anisotropic
antiferromagnet, such as the AFzz phase, the dispersion of the
spin-wave mode in the $(k_x,k_y)$ plane is roughly given by
$(zJ_{ab}J_c)^{1/2}S$, so actually enhanced by
$(J_c/zJ_{ab})^{1/2}$ compared with the planar exchange constant.

The (longitudinal) orbital excitations in the AFzz phase are found
using the equations of motion of the form (\ref{gfafxxl1}) which
lead to Eq. (\ref{gfeqor}) with,
\begin{eqnarray}
\label{afzzu} u_z&=&-\varepsilon_z-3\eta, \\ \label{afzzrho}
\rho_{z\vec k}&=&-\case{3}{2}\eta\gamma_{+}(\vec k),
\end{eqnarray}
and we find again zero-energy nondispersive modes at
$\varepsilon_z=\eta=0$. The orbital excitation is found at the
$X=(\pi,0,0)$ and $L=(\pi/2,\pi/2,\pi/2)$ points at the same
energy as that of a {\it local\/} excitation from $|z\rangle$ to
$|x\rangle$ orbital. It depends only on the energy difference
between the orbitals, and has a weak dispersion $\sim J\eta$ due
to the same mechanism as described above for the AFxx phase (Fig.
\ref{fig:orbex}).

The excitation operators which couple to the local states in the
MO{\scriptsize FFA} phase with mixed orbitals are linear
combinations of the operators considered above. It is therefore
convenient to make a unitary transformation of the Hamiltonian
(\ref{somcu}) to new orbitals defined as follows for $i\in A$ or
$i\in D$ sublattice,
\begin{equation}
\label{newstatei} \left( \begin{array}{c}
 |i\mu\rangle   \\
 |i\nu\rangle
\end{array} \right) =
\left(\begin{array}{cc}
 \cos\theta &  \sin\theta \\
-\sin\theta &  \cos\theta
\end{array} \right)
\left( \begin{array}{c}
 |iz\rangle   \\
 |ix\rangle
\end{array} \right) ,
\end{equation}
and for $j\in B$ or $j\in C$ sublattice,
\begin{equation}
\label{newstatej} \left( \begin{array}{c}
 |j\mu\rangle   \\
 |j\nu\rangle
\end{array} \right) =
\left(\begin{array}{cc}
 \cos\theta & -\sin\theta \\
 \sin\theta &  \cos\theta
\end{array} \right)
\left( \begin{array}{c}
 |jz\rangle   \\
 |jx\rangle
\end{array} \right) .
\end{equation}
With these definitions and by choosing the angle $\theta$ at the
value which minimizes the classical energy (\ref{thetamoffa}), we
guarantee that $|i\mu\rangle$ and $|j\mu\rangle$, respectively,
are the orbital states realized in the classical MO{\scriptsize
FFA} phase at each site, which is G-type with respect to the
orbital ordering, while $|i\nu\rangle$ and $|j\nu\rangle$ are the
excited states, so that one can readily define the excitation
operators pertinent to the symmetry-broken ground state of this
phase.

\begin{figure}[t]
\centerline{\psfig{figure= 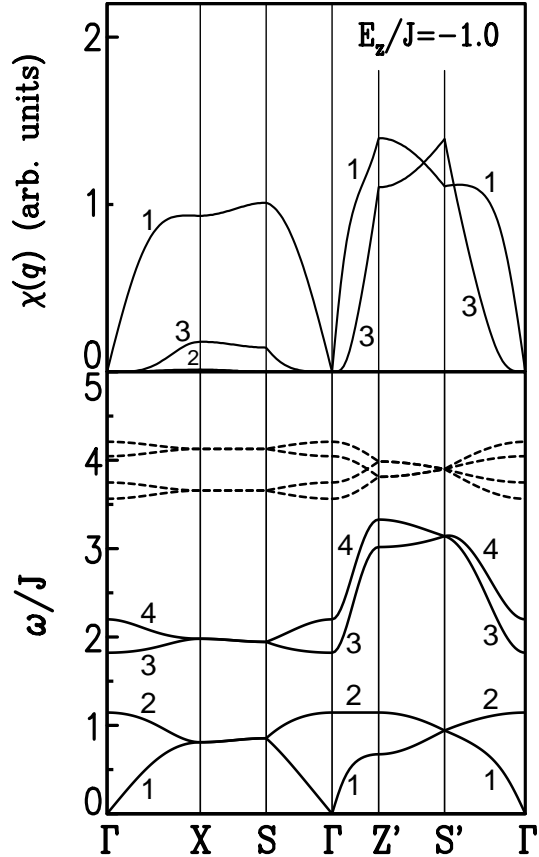,width=7cm}}
\smallskip
\caption {The same as in Fig. \protect\ref{fig:modesxx}, but for
the MO{\protect{\scriptsize FFA}} phase, as obtained for
$E_z/J=-1.0$ and $J_H/U=0.3$. Different modes are labelled by the
increasing indices $i=1,\dots,4$ with increasing energy (after
Ref. \protect\cite{Ole00}).} \label{fig:modesffa}
\end{figure}
Thus the spin ${\cal S}_{i\mu\mu}^+$, spin-and-orbital ${\cal
K}_{i\mu\nu}^+$, and orbital ${\cal T}_{i\mu\nu\sigma}$ operators
are defined in terms of the new reference orbital states
$\{|\mu\rangle,|\nu\rangle\}$, and fulfill the same commutation
rules as the non-transformed operators: $S_{i\alpha\alpha}^+$
$K_{i\alpha\beta}^+$, and $T_{i\alpha\beta\sigma}$, respectively.
To simplify the notation we also introduce total spin ${\cal
S}_i^{+}$ and spin-and-orbital ${\cal K}_i^{+}$ operators, as
explained above. The Hamiltonian (\ref{somcu}) has to be
transformed by the inverse transformations to those given by Eqs.
(\ref{newstatei}) and (\ref{newstatej}) \cite{Ole00}. Hence, the
transverse excitations may be found starting from the relevant
raising operators that lead to the local state
$|i\mu\uparrow\rangle$ realized in one of the sublattices,
analogous to those introduced for the AFxx phase (\ref{excopt}),
i.e., the set $\{{\cal S}_{i\mu\mu}^+,{\cal K}_{i\mu\nu}^+, {\cal
S}_{j\mu\mu}^+,{\cal K}_{j\mu\nu}^+, {\cal S}_{k\mu\mu}^+,{\cal
K}_{k\mu\nu}^+, {\cal S}_{l\mu\mu}^+,{\cal K}_{l\mu\nu}^+\}$,
where $i\in A$, $j\in B$, $k\in C$, and $l\in D$; they lead as
usual to the orbitals $\{|i\mu\rangle, |j\mu\rangle\}$
(\ref{orbmoffa}) realized in the MO{\scriptsize FFA} phase. We
applied the same RPA procedure as explained above for the AFxx and
AFzz phase in order to determine the Green function equations in
the ${\vec k}$-space. The longitudinal excitations can be obtained
from operators ${\cal T}_{i\mu\nu\sigma}$ similar to those used in
the AFxx and AFzz phases (\ref{topera}), taking $\sigma=\uparrow$
for the $(a,b)$ planes occupied with $\uparrow$-spins, and
$\sigma=\downarrow$ for the $(a,b)$ planes occupied with
$\downarrow$-spins. As expected, there are four doubly-degenerate
positive-frequency transverse modes, and four non-degenerate
positive-frequency longitudinal modes, consistent with the
MO{\scriptsize FFA} phase having four sublattices.

An example of the transverse and longitudinal modes in the
MO{\scriptsize FFA} phase is presented in Fig. \ref{fig:modesffa}.
The modes are shown in the respective BZ which corresponds to the
magnetic unit cell of the MO{\scriptsize FFA} phase: The 2D part
along $\Gamma-X-S-\Gamma$ resembles the modes in the AFxx phase
(compare Fig. \ref{fig:modesxx}), reflecting the orbital
alternation, while the AF coupling along the $c$-axis results in
the folding of the zone along the $\Gamma-Z$ direction, with
$Z'=(0,0,\pi/2)$ and $S'=(\pi/2,\pi/2,\pi/2)$. One finds one
Goldstone mode, and three other finite-energy modes at the
$\Gamma$ point. If no AF coupling along the $c$-axis is present,
similar positive-energy modes describe the excitation spectrum in
the MO{\scriptsize FF} phase in the 2D part of the BZ (in the
region of stability shown in Fig. \ref{fig:2d}), and the symmetric
negative-frequency modes carry then no weight. In contrast, due to
the strong AF interactions in the MO{\scriptsize FFA} phase, the
negative modes give a large energy renormalization due to quantum
fluctuations, as discussed in more detail in Sec. IV.B.

The spin-wave and spin-and-orbital-wave excitations are well
separated along the $\Gamma-X-S-\Gamma$ path, with a gap of $\sim
0.5J$, as the FM interactions $\propto J\eta$ are considerably
weaker than the orbital interactions which are $\propto J$.
Therefore, the neutron intensity $\chi({\vec q})$ is found mainly
as originating from the lowest energy mode, $\omega_{\vec
k}^{(1)}$, with a small admixture of the higher-energy
spin-and-orbital excitation, $\omega_{\vec k}^{(3)}$. The magnetic
interactions are considerably stronger along the $c$-axis; the
modes mix and the higher-energy excitations, $\omega_{\vec
k}^{(n)}$ with $n=3,4$, have a larger dispersion in the remaining
directions with $k_z\neq 0$. Strong mixing of the modes in this
part of the BZ is also visible in the intensity distribution, with
the modes $n=1$ and $n=3$ contributing with comparable intensities
(Fig. \ref{fig:modesffa}). The fact that modes labeled as 2 and 4
have zero intensity is due to the path $\Gamma-Z'-S'-\Gamma$ being
in the high-symmetry BZ plane where $k_x=k_y$ so that
$\gamma_{-}(\vec{k})=0$. Then modes 2 and 4 have equal amplitude
but are exactly out-of-phase between $A$ and $B$ sites as well as
between $C$ and $D$ sites, and so their neutron intensities
vanish, and only the companion in-phase modes 1 and 3 are
observable by neutrons. Unlike in the AF phases, the purely
orbital excitation is here energetically separated from the spin
wave and spin-and-orbital wave. The dispersion is quite small and
decreases with $\eta$.

Interestingly, although the order in the $(a,b)$ planes is FM, the
energy of the Goldstone mode increases {\it linearly in all three
directions\/} with increasing ${\vec k}$, and the slopes are
proportional to the respective exchange interactions. This
behavior is a manifestation of the A-AF spin order; a
qualitatively similar spectrum is found experimentally in
LaMnO$_3$ \cite{Hir96,Hir96a}, where, however, the excitation
spectra correspond to large spins $S=2$ of Mn$^{3+}$ ions. The
rather small dispersion of the spin-wave part at low energies is
due to small values of the exchange constants for the actual
optimal orientation of orbitals found at $J_H/U=0.3$. We note,
however, that the AF interactions along the $c$-axis are much
stronger at $J_H\to 0$ than in the present case. The AF structure
along the $c$-axis may be easily recognized from the symmetric
spin-wave mode in the $\Gamma-Z$ direction with respect to
$Z'=(0,0,\pi/2)$, while this mode increases all the way from the
$\Gamma$ to the $X$ point. Unfortunately, no experimental
verification of these spectra is possible at present, as the spin
excitations measured in neutron scattering for KCuF$_3$ are
consistent with the Bethe ansatz and thus suggest a spin-liquid
ground state with strong 1D AF correlations instead of the A-AF
phase with magnetic LRO \cite{Ten95,Ten95a,Ten95b}.

The elementary excitations in the MO{\scriptsize AFF} phase may be
obtained using a similar scheme to that described here for the
MO{\scriptsize FFA} phase. In this case, the transverse
excitations have a similar dependence on the ${\vec k}$-vector to
those found in the MO{\scriptsize FFA} phase, but the value of the
crystal-field $E_z$ is effectively smaller by a factor of two in
comparison with the MO{\scriptsize FFA} phase. This asymmetry is a
consequence of the choice of $|x\rangle$ and $|z\rangle$ states as
orbital basis to which $E_z$ refers. Most importantly, one finds
that the classical phases are all stable on the RPA level in the
regions of their stability in the phase diagram of Fig.
\ref{fig:3d}. However, there are characteristic low-frequency
modes which follow from the mixing between the spin wave and
spin-and-orbital wave modes, and these modes are responsible for
enhanced quantum fluctuations (sec. IV.B).




\section{Spin liquid due to orbital fluctuations}

\subsection{Idealized case: $SU(4)$ model}

Motivated by the recent studies of the strongly correlated systems
with orbital degeneracy, and by the role played by the orbital
degree of freedom in spin systems, the SU(4) symmetric
spin-orbital model has attracted a lot of interest
\cite{Fri99,Yam98,Li98,Li98a,Li98b,Jos99,Mil99}. It describes the
localized electrons in the twofold degenerate Hubbard model with
the diagonal isotropic hopping $t$ between the same type of
orbitals for the filling of one electron per site (the filling by
one hole is equivalent by the particle-hole transformation). As
the off-diagonal hopping elements are absent, the $z$th component
of pseudospin is conserved, unlike in the $d^9$ model of Sec. III.
If the Coulomb interaction $U$ is large compared with $t$, $U\gg
t$, the effective Hamiltonian may be derived in a similar way as
shown in Secs. II.A and III.A, and one finds anisotropic orbital
interactions with extra terms $\propto T_i^zT_j^z$, while the spin
interactions are as usually SU(2) symmetric \cite{Yam98}. The
highly symmetric SU(4) model follows in the limit of vanishing
Hund's rule exchange, $J_H\to 0$, when the spectrum of the excited
states collapses into a spin triplet multiplied by orbital
singlets, and an orbital triplet multiplied by three spin
singlets, all at the same energy $U$. Taking the projection
operators on the spin triplet and spin singlet (\ref{projst}), and
introducing similar operators for the orbital pseudospin states,
one finds the Hamiltonian of the form,
\begin{equation}
\label{dersu4} H=2J\sum_{\langle ij\rangle}
           \left[\left(\vec{S}_i\cdot\vec{S}_{j}+\frac{3}{4}\right)
                 \left(\vec{T}_i\cdot\vec{T}_{j}-\frac{1}{4}\right)
+\left(\vec{S}_i\cdot\vec{S}_{j}-\frac{1}{4}\right)
           \left(\vec{T}_i\cdot\vec{T}_{j}+\frac{3}{4}\right)\right],
\end{equation}
where $\vec S_i$ and $\vec T_i$ are spin $S=1/2$ and pseudospin
$T=1/2$, corresponding to spin and orbital degrees of freedom,
respectively, and we defined the energy unit for the superexchange
interaction $J=2t^2/U$. An interesting observation is that a pure
spin $\sim \vec{S}_i\cdot\vec{S}_{j}$ interaction has here a
prefactor $J=2t^2/U$ which is by a factor of two smaller than in
the $t$-$J$ model (see Sec. II.A). This reduction follows from the
competition between the orbital triplet and orbital singlets in
the present case. The derived expression (\ref{dersu4}) explains
the physical origin of a fully symmetric Hamiltonian in spin and
pseudospin (orbital) space, usually written as,
\begin{equation}\label{hamsu4}
H=J\sum_{\langle ij\rangle}
          \left(2\vec{S}_i\cdot\vec{S}_{j}+\frac{1}{2}\right)
          \left(2\vec{T}_i\cdot\vec{T}_{j}+\frac{1}{2}\right).
\end{equation}
It was noticed out only very recently
\cite{Yam98,Li98,Li98a,Li98b}, that the Hamiltonian (\ref{hamsu4})
has not only the obvious SU(2)$\times$SU(2) symmetry, but the full
symmetry of Eq. (\ref{hamsu4}) obeys even the higher symmetry
group SU(4). It is worth pointing out that SU($N$) symmetric
models in one dimension were studied by Affleck, using conformal
field theory \cite{Aff86}. He showed that any 1D system of SU($N$)
symmetry is critical. In this case, the critical exponents and
zero temperature correlations at the very low energy scale are
equivalent to $N-1$ free massless bosons. These general results
naturally also applies to the case with $N=4$. We note also that a
different SU(4) symmetric model has been introduced by Santoro
{\it et al.\/} \cite{San99}, which has a different low-energy
physics from the present Hamiltonian (\ref{hamsu4}). In realistic
materials, however, this high SU(4) symmetry is practically always
broken by an anisotropic hybridization \cite{Zaa93}, or by the JT
effect \cite{Kug82}, as we have emphasized in other Sections.

The advantage of the high SU(4) symmetry is that the rigorous
analysis of this model is possible in one dimension. The SU(4)
model (\ref{hamsu4}) belongs to a class of models which are
exactly soluble in one dimension by the Bethe ansatz \cite{Bet31}.
The Bethe ansatz solution obtained by Sutherland gives the exact
ground state energy and three branches of low-energy gapless
excitations \cite{Sut75}, having all a common velocity $v=\pi
J/2$. The physical interpretation of these branches is not
straightforward though.

The essential complexity comes from the large local degeneracy.
For a single bond, the ground state is six-fold degenerate: either
spin triplet multiplied by any of the three orbital singlets, or
any of spin singlets multiplied by the orbital triplet. It is
rotationally invariant not only in $\vec{S}$-space, but also in
$\vec{T}$-space. Furthermore, it has an interchange symmetry
between spin and orbital operators. In such a case, the standard
mean-field approach \cite{Cas78} that leads to FM correlations for
one type of variables and AF correlations for the others, is not
reliable and more powerful methods have to be applied. The first
investigation of the thermodynamic properties of the SU(4) model
(\ref{hamsu4}) has been performed by Frischmuth, Mila, and Troyer
\cite{Fri99} by means of the continuous time quantum Monte Carlo
(QMC) loop algorithm \cite{Eve93,Bea96}, adapted to spin-orbital
models. The ground state energy for a chain of $L=100$ with
periodic boundary condition amounts to
$\epsilon_0(L=100)=-0.8253(1)J$, and is in perfect agreement with
the Bethe Ansatz result for the infinite chain, $-0.8251189\ldots
J$ \cite{Sut75}. In contrast, if the MF decoupling is made, one
finds the energy of $-0.3863J$ \cite{Fri99} which demonstrates
that the MF method cannot be used even for qualitative insight
into the nature of the ground state.

\begin{figure}[t]
\centerline{\psfig{figure= 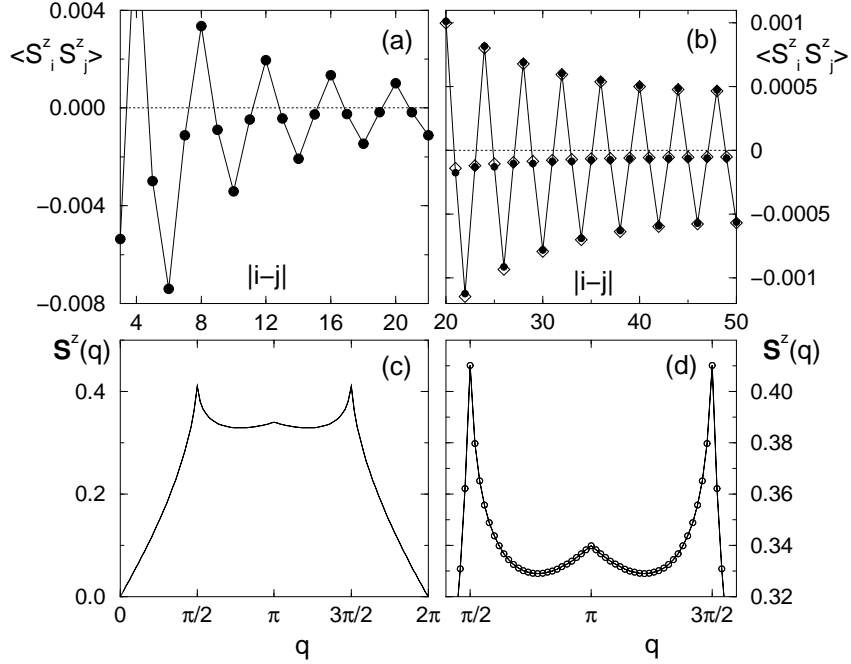,width=11cm}}
\smallskip
\caption {(a) QMC results for the correlation function
    $w_{ij}\equiv\langle S_i^z S_j^z\rangle$ (\protect\ref{symmetry}) (solid points) as a
    function of $|i-j|$ for a SU(4) chain of length $L=100$ with PBC which
    is predominantly in the ground state. The
    correlations for $|i-j|=1,\,2$ and 4 (which are out of the plot
    range) are -0.07168(1), -0.04011(1) and 0.008261(4), respectively.
    Part.~(b) shows the correlations $w_{ij}$ at large distances
    $|i-j|$ and the fit to the QMC data
    (open diamonds). The statistical error bars of the QMC
    calculations are much smaller than the symbols. Parts (c) and (d) show
    the Fourier transform ${\cal S}^z(k)$ of $w_{ij}$ on two different
    scales (after Ref. \protect\cite{Fri99}).
} \label{fig:su4}
\end{figure}

The structure of the ground state becomes more transparent when
the correlation functions are investigated. The QMC study of
Frischmuth {\it et al.\/} \cite{Fri99} gives the zero-temperature
correlation function $w_{ij}(T=0)\equiv\langle S_i^z
S_j^z\rangle(T=0)$ as a function of distance $|i-j|$ along a 1D
chain (for $L=100$). We reproduce their results in Fig.
\ref{fig:su4}. Due to the SU(4) symmetry, all the following
correlations are equal:
\begin{equation}
\label{symmetry}
 w_{ij}=\langle S_i^\alpha S_j^\alpha \rangle
 = \langle T_i^\alpha T_j^\alpha \rangle
 = \langle 4 S_i^\alpha S_j^\alpha T_i^\beta T_j^\beta \rangle,
\end{equation}
independent of the indices $\alpha,\,\beta=x,y,z$. This relation
is valid for zero as well as for finite temperatures, and is
easily violated by any MF decoupling. While the first equality
also holds for an arbitrary SU(2)$\times$SU(2) symmetric model
with exchange symmetry of the $\vec{S}$ and $\vec{T}$-variable,
the second one is a special property of the SU(4) symmetric model.
We observe that the correlation function $w_{ij}$ exhibits a clear
four-site periodicity (Fig. \ref{fig:su4}). Its sign is positive
if $|i-j|=4N$ with $N$ integer, and negative otherwise. The reason
for this behavior is the tendency for every four neighboring sites
to form an SU(4) singlet \cite{Bos00}.
Looking at the results for $w_{ij}$, it can be concluded that they
correspond to a disordered state and the two dominant modes are
those with $k=\pi/2$ (positive prefactor) and $k=0$ (negative
prefactor) \cite{Fri99}. This is also reflected in the Fourier
transform ${\cal S}^z(k)$ of the correlation function $w_{ij}$,
having a characteristic cusp structure at $k=0,\,\pi/2$ and $\pi$
(see Fig. \ref{fig:su4}). While the cusps at $k=0$ and $\pi/2$ are
quite sharp, the one at $k=\pi$, however, is not so pronounced,
indicating that the $k=\pi$ mode is the least dominant mode in the
correlation function of all the three modes.

\begin{figure}[t]
\centerline{\psfig{figure= 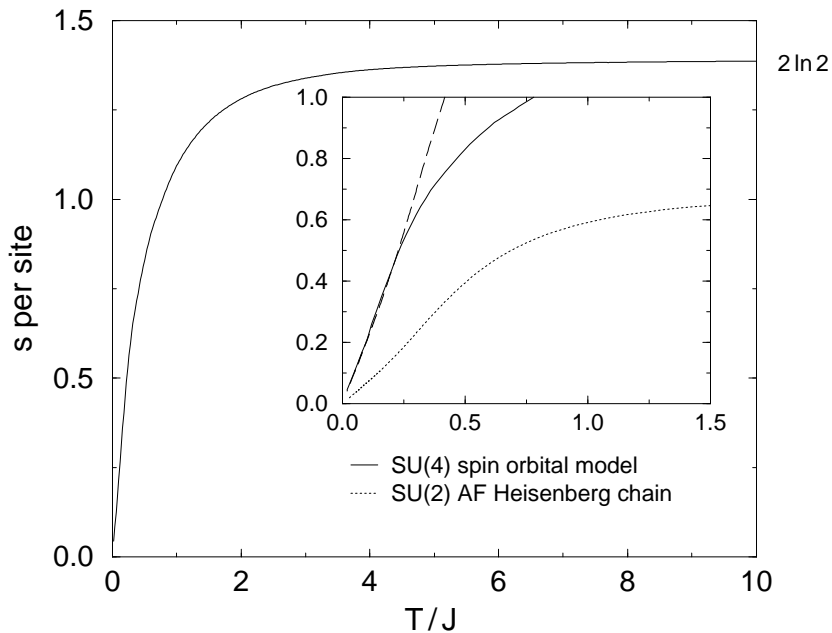,width=11.5cm}}
\smallskip
\caption {Temperature dependence of the entropy $s$ per site for
the 1D SU(4)
 spin-orbital model (\ref{hamsu4}) (solid line). In the inset
the entropy per site is shown on larger temperature scale together
with the entropy $s_{\rm HB}$ per site of a SU(2) spin-1/2 AF
Heisenberg chain (dotted line). For comparison also $3s_{\rm HB}$
is shown (dashed line) (after Ref. \protect\cite{Fri99}). }
\label{fig:su4en}
\end{figure}

The large degeneracy of the SU(4) invariant model (\ref{hamsu4})
becomes transparent in the entropy $s(T)$ per site determined by
the QMC loop algorithm \cite{Fri99}; its $T$-dependence is shown
in Fig. \ref{fig:su4en}. With increasing $T$, the entropy $s$
increases monotonically from zero towards the high temperature
value $k_B\ln 4$. This shows that the short-range correlations are
gradually lost, and the spins and orbitals are fluctuating
individually at every site at high temperatures, in some
similarity with the spin fluctuations in the spin model which
gives the high temperature value of entropy $k_B\ln 2$. At low
temperatures the entropy varies linearly in both models. However,
the slope of $s(T)$ in the spin-orbital model (\ref{hamsu4}) is
about a factor {\it three} bigger than that in the Heisenberg
antiferromagnet, as shown in the inset of Fig. \ref{fig:su4en}.
This is consistent with the statement of Affleck \cite{Aff86} that
the AF Heisenberg model is equivalent to {\em one} free massless
boson, while the SU(4) invariant spin-orbital model is equivalent
to {\em three} massless bosons. The velocity of these bosons are
all equal to $\pi J/2$ \cite{Sut75}. Therefore we expect the low
energy density of states (and hence the entropy) of these two
models to differ just by a factor of three. This analysis,
however, does not give a clear indication on the nature of the
ground state and of the corresponding excitations, and this
problem is still open.
Modifications of the SU(4) Hamiltonian in one dimension give also
interesting physics. If the anisotropy in the orbital sector is
introduced, either gapless or gaped phases are found depending on
the balance of excitations; in the gaped phase an alternation of
spin and orbital singlets gives the lowest energy \cite{Pat98}.
The ground state is known exactly and can be written as a product
of the fully polarized FM ground state for the spins and the
Jordan-Wigner Fermi sea for the local states \cite{Pat00}. On
addition to this phase, one finds two different dimerized phases,
where the spin and orbital variables are dimerized in a correlated
pattern. In one phase spin singlets and orbital singlets arise on
alternating rungs, whereas in the other phase spin singlet and
orbital triplet share the same rungs. In spite of AF interactions,
no gapless AF spin phase is found \cite{Pat00}. These results show
that spin-gap phases of this type survive anisotropies in the
orbital space and are quite generic in spin-orbital models.

Furthermore, the 1D model has also been studied in two limiting
cases: the pure XY model both in spin and in orbital space
\cite{Mil99}, and the dimerized XXZ model \cite{Mil99}. In the
pure XY case, a phase separation takes place between two phases
with free -- fermion like, gapless excitations, while in the
dimerized case the low-energy effective Hamiltonian reduces to the
1D Ising model with gaped excitations. In both cases, all the
elementary excitations involve simultaneous flips of the spin and
orbital degrees of freedom, which gives a clear indication of the
breakdown of the traditional MF theory.

The SU(4) symmetric model represents an idealized case for
investigating a spin and orbital disordered ground state in higher
dimensions. Unfortunately, the QMC method of Frischmuth {\it et
al.\/} cannot be used \cite{Fri99}, as it suffers from a severe
minus sign problem in 2D lattices. Hence, unbiased results may be
obtained only by an exact diagonalization technique. As pointed
out by Li {\it et al.\/} \cite{Li98,Li98a,Li98b}, the SU(4)
symmetric model (\ref{hamsu4}) is characterized by a strong
tendency towards a liquid ground state with resonating plaquette
singlets. This follows from a particular stability of a (spin and
orbital) singlet on a square, with the energy of $-4J$, while the
first excited state of energy $-2J$ has a degeneracy of fifty
\cite{Bos00}. The diagonalization of larger (8-site and 16-site)
clusters provides a good evidence that the spin-liquid ground
state is realized in fully symmetric SU(4) model in two
dimensions. The nature of the disordered ground state is not
completely understood, but the results obtained so far indicate
that a singlet-multiplet gap survives in the thermodynamic limit,
and low-lying SU(4) singlets might exist within this gap
\cite{Bos00}. In the view of these results it is interesting to
search for a spin-liquid ground state as well in a more realistic
situation described by the $d^9$ model. As a matter of fact, the
spin-liquid ground state as resulting from the orbital quantum
fluctuations has been predicted first in the $d^9$ model
\cite{Fei97}.

\subsection{ Quantum corrections in the $d^9$ model }
\label{sec:rpa}

The size of quantum fluctuation corrections to the classical order
parameter determines the stability of the classical phases. The
frustration of magnetic interactions might lead in spin models to
divergent quantum corrections within the LSW theory
\cite{Cha88,Cha88a,Cha88b,Cha88c,Cha88d,Chu91,Mil94,San94,Chu95}.
Before calculating these corrections in the present situation, a
generalization of the usual RPA procedure to a system with several
excitations is necessary. Here we present only the relations
needed to calculate the quantum corrections to the LRO parameter
and to the ground state energy \cite{Ole00}.
\begin{figure}
\centerline{\psfig{figure=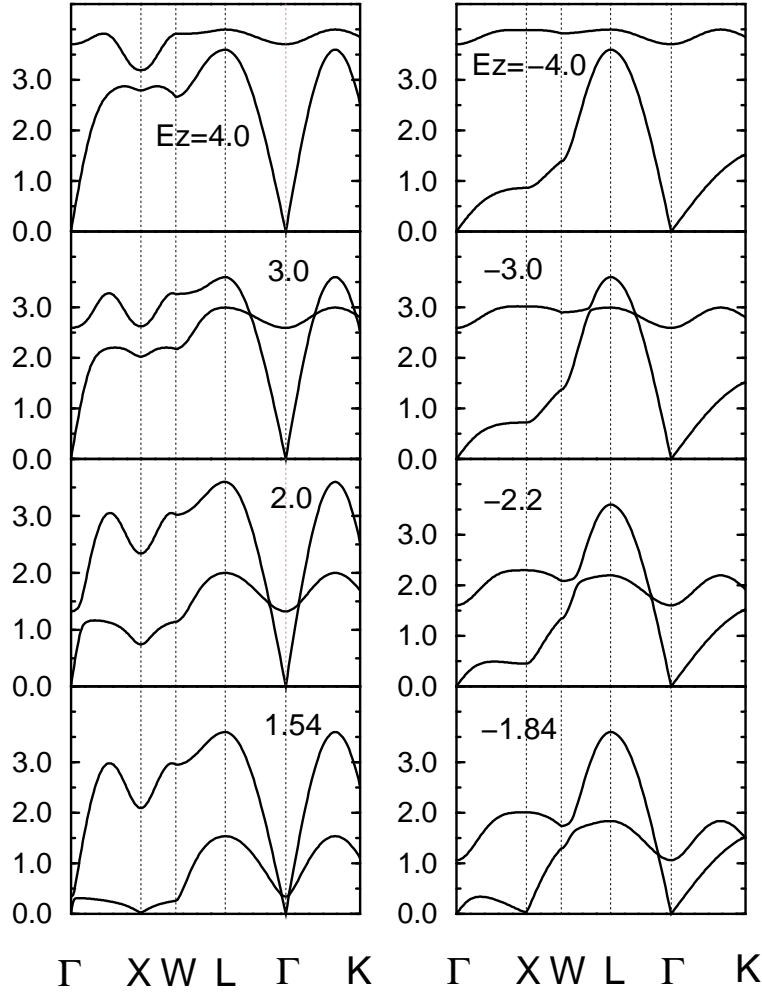,width=12cm}}
\smallskip
\caption {Spin-wave and spin-and-orbital-wave excitations in the
G-AF phases: AFxx (left) and AFzz (right), in the main directions
of the 3D BZ for a few values of $E_z$ (in the units of $J$), and
for $J_H/U=0.3$. The lower-energy mode becomes soft for
$E_z/J<1.54$ ($E_z/J>-1.84$) in the AFxx (AFzz) phase. To allow a
direct comparison, both phases are shown using the 3D BZ for a
$bcc$ lattice with the standard notation of high symmetry points:
$\Gamma=(0,0,0)$, $X=(\pi,0,0)$, $W=(\pi,\pi/2,0)$,
$L=(\pi/2,\pi/2,\pi/2)$ and $K=(3\pi/4,3\pi/4,0)$ (after Ref.
\protect\cite{Ole00}). } \label{fig:soft}
\end{figure}
For that purpose, let us introduce here the local operators
constituting the $so(4)$ Lie algebra at site $i$ as Hubbard
operators, $X_i^{\alpha\beta}=|i\alpha\rangle \langle i\beta|$.
Using the unity operator, $\sum_{\beta}X_i^{\beta\beta}={\cal{I}}
$, the diagonal operator $X_i^{\alpha\beta}$ that refers to the
state $|i\alpha\rangle$ {\em realized at site $i$ in the classical
ground state\/} of the phase under consideration may be expanded
in terms of the excitation operators,
\begin{equation}
\label{expex} X_i^{\alpha\alpha}={\cal{I}}
-\sum_{\beta\neq\alpha}X_i^{\beta\alpha}X_i^{\alpha\beta},
\end{equation}
while the diagonal operators referring to {\em excited\/} states
$|i \beta\rangle$ are expressed as
\begin{equation}
\label{expexb}
X_i^{\beta\beta}=X_i^{\beta\alpha}X_i^{\alpha\beta}.
\end{equation}
Applying these equations to the $z$-th spin component
$S^z_i=S^z_{ixx}+S^z_{izz}$ of the total spin at site $i$ in one
of the G-AF phases with pure orbital character (say AFxx for
definiteness), one finds, for $i$ in the spin-up sublattice
\cite{Ole00},
\begin{eqnarray}
\label{exps} S_i^z &=& \frac{1}{2} (X_i^{x\uparrow,x\uparrow}
                    - X_i^{x\downarrow,x\downarrow}
                    + X_i^{z\uparrow,z\uparrow}
                    - X_i^{z\downarrow,z\downarrow} ) \nonumber \\
      &=& \frac{1}{2} {\cal{I}}
          - X_i^{x\downarrow,x\uparrow} X_i^{x\uparrow,x\downarrow}
          - X_i^{z\downarrow,x\uparrow} X_i^{x\uparrow,z\downarrow}
       = \frac{1}{2} {\cal{I}}
          - S_{ixx}^- S_{ixx}^+ - K_{izx}^- K_{ixz}^+ .
\end{eqnarray}
Taking the average of Eq. (\ref{exps}) one obtains, with the MF
value $\langle S_i^z\rangle_{\rm MF}=\frac{1}{2}$,
\begin{equation}
\label{expes} \langle S_i^z\rangle_{\rm RPA}= \frac{1}{2}- \langle
S_{i}^- S_{i}^+\rangle - \langle K_{i}^-K_{i}^+\rangle =\langle
S_i^z\rangle-\delta\langle S_i^z\rangle,
\end{equation}
where the averages like $\langle S_{ixx}^-S_{izz}^+\rangle$ are
zero since they involve `ghost' modes, so that one may formally
replace $S_{ixx}^+$ by $S_{ixx}^++S_{izz}^+=S_i^+$, {\it
etcetera\/}. The first contribution $\propto\langle
S_i^-S_i^+\rangle$ is the usual renormalization due to spin waves,
while the second term $\propto\langle K_i^-K_i^+\rangle$ stands
for the reduction of $\langle S_i^z\rangle_{\rm RPA}$ due to
spin-and-orbital excitations. Both terms involve a local
excitation preceded by a deexcitation which reproduces the initial
local state. As expected, only transverse excitations contribute
to the spin renormalization. Note that, since Eq. (\ref{exps}) is
an {\em exact operator relation\/}, the present procedure
guarantees that Eq. (\ref{expes}) is a {\em conserving
approximation\/} which respects the sum rule for the occupancies
of all states, $\sum_{\beta}\langle X_i^{\beta\beta}\rangle=1$.
The generalization of Eq. (\ref{expes}) to the MO phases using the
operators ${\cal S}_i^{\pm}$ and ${\cal K}_i^{\pm}$ in the
expansion of ${\cal S}_i^z$, or to other order parameters, like
the orbital polarization (\ref{nzet}), is straightforward.

The local correlation functions which renormalize the order
parameter in Eq. (\ref{expex}) are determined in the standard way
\cite{Hal72},
\begin{equation}
\label{theorem} \langle B_i^{\dagger}A_i\rangle=
   \frac{1}{N}\sum_{\vec k}\int_{-\infty}^{+\infty}d\omega
   {\cal A}_{AB^{\dagger}}({\vec k},\omega)\frac{1}{\exp(\beta\omega)-1},
\end{equation}
where $\beta=1/k_BT$, and
\begin{equation}
\label{weight} {\cal A}_{AB^{\dagger}}({\vec k},\omega<0) =
   2{\rm Im}\langle\langle A_{\vec k}|
      B^{\dagger}_{\vec k}\rangle\rangle_{\omega-i\epsilon}
    = \sum_{\nu<0}{\cal A}_{AB^{\dagger}}^{(\nu)}({\vec k})
      \delta(\omega-\omega_{\vec k}^{(\nu)})
\end{equation}
is the respective spectral density for the negative frequencies
($\nu<0$), and ${\cal A}_{AB^{\dagger}}^{(\nu)}({\vec k})$ are the
respective spectral weights. Therefore, the correlation functions
at $T=0$ are found by summing up the total spectral weight at the
negative frequencies,
\begin{equation}
\label{average} \langle
B_i^{\dagger}A_i\rangle=\frac{1}{N}\sum_{\vec k}\sum_{\nu<0}
                                {\cal A}_{AB^{\dagger}}^{(\nu)}({\vec k}).
\end{equation}

Before discussing the renormalization of the order parameter and
the corresponding energies in RPA, we concentrate ourselves on the
behavior of the transverse excitations when the crossover lines
between the classical phases are approached. As already emphasized
in Sec. III.C, the spin-wave and spin-and-orbital-wave excitations
couple. As a consequence, the modes in all considered phases {\em
soften\/} when the transition lines between different classical
phases, or classical degeneracy point are approached. This
softening is shown for a representative value of $J_H/U=0.3$ in
Fig. \ref{fig:soft} for the two AF phases with either $|x\rangle$
or $|z\rangle$ orbitals occupied. In the AFxx phase the energy
scales of both excitations are separated for $E_z>4J$, while the
spin-and-orbital mode moves towards zero energy with decreasing
$E_z$, and finally becomes soft at the $X$ point, along
$\vec{k}=(\pi,0,k_z)$ and along equivalent lines in the BZ for
$E_z\simeq 1.54J$. A similar mode softening is found for the AFzz
phase at $E_z<0$, with the soft mode along $\Gamma-X$ and
equivalent directions in the BZ at $E_z\simeq -1.84J$. This
peculiar softening along lines and not at points in the BZ shows
that the modes behave 2D-like instead of 3D-like:
constant-frequency surfaces are cylinders contracting towards
lines, not spheres contracting towards a point.

By making an expansion of Eq. (\ref{afsw}) around the soft-mode
lines, one finds that the (positive) excitation energies are
characterized by {\em finite} masses in the perpendicular
directions:
\begin{equation}
\label{massx} \omega_{\rm AFxx}(\vec{k}) \rightarrow \Delta_x
      + B_x \left( {\bar k}_x^4+14{\bar k}_x^2k_y^2+k_y^4 \right)^{1/2},
\end{equation}
independently of $k_z$ (here ${\bar k}_x=k_x-\pi$) for the AFxx
phase, and
\begin{equation}
\label{massz} \omega_{\rm AFzz}(\vec{k}) \rightarrow \Delta_z
      + B_z \left( k_y^2 + 4k_z^2 \right),
\end{equation}
independently of $k_x$, and similarly along the $\Gamma-Y$
direction with $k_y$ replaced by  $k_x$ for the AFzz phase. As an
example we give explicit expressions for the AFxx phase at
$\eta=0$,
\begin{equation}
\label{massxx} \Delta_x= \frac{9}{2}
\frac{\varepsilon_z}{\varepsilon_z+3}, \hskip 1.2cm
     B_x=\frac{27}{16}\frac{      1      }{\varepsilon_z+3},
\end{equation}
where one finds that the gap $\Delta_x\to 0$ when $\varepsilon_z
\to 0$, i.e., upon approaching the $M=(E_z,J_H)=(0,0)$ point at
which the AF order is changed to the AFzz phase. This illustrates
a general principle: $\Delta_i\to 0$ when the crossover line to
another phase is approached, and $B_i\neq 0$ when the modes
(\ref{massx}) and (\ref{massz}) soften, making quantum fluctuation
corrections to the order parameter to diverge logarithmically,
$\langle\delta S\rangle\sim \int d^3k/\omega(\vec{k})\sim\int
d^2k/(\Delta_i+B_ik^2)\sim\ln\Delta_i$. We emphasize that for the
occurrence of this divergence not only the finiteness of the mass,
but also the 2D-like nature of the dispersion is essential. It
enables a 3D system to destabilize LRO by what are essentially 2D
fluctuations. So the divergence of the order parameter near the
crossover lines in the phase diagram and the associated
instability of the classical phases, may be regarded as another
manifestation of the effective reduction of the dimensionality
occurring in the spin-orbital model. We do not present explicitly
the softening of the longitudinal modes which also happens at the
transition lines, but has no direct relation to the stability of
classical phases.
\begin{figure}
\centerline{\psfig{figure=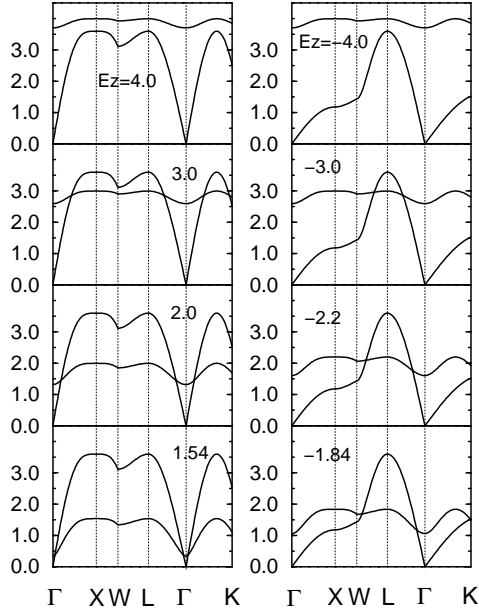,width=7cm}}
\smallskip
\caption {The same as in Fig. \protect\ref{fig:soft}, but for
decoupled spin-wave and spin-and-orbital-wave excitations in the
G-AF phases (after Ref. \protect\cite{Ole00}). }
\label{fig:softbug}
\end{figure}
A seemingly attractive way to simplify the calculation of the
transverse excitations would be to make a decoupling of the
spin-waves and spin-and-orbital-waves. However, this is equivalent
to violating the commutation rules between the spin and
spin-and-orbital operators \cite{Fei98}, and this changes the
physics. It gives the same excitation energies as Eq.
(\ref{afsw}), but with $P_{\alpha\vec k}=0$; the numerical result
is given in Fig. \ref{fig:softbug}. Of course, the spin-wave
excitation does not depend then on the orbital splitting $E_z$,
and the spin-and-orbital-wave excitation gradually approaches the
line $\omega_{\vec k}=0$ with decreasing $|E_z|$. It has a weak
dispersion which depends on $J_H$ and on the value of $|E_z|$, and
gives an instability at the $\Gamma$ point only, not at certain
lines in the BZ. The instability occurs well beyond the transition
lines in the phase diagram of Fig. \ref{fig:3d}, i.e., within the
MO{\scriptsize FFA} and MO{\scriptsize AFF} phase for $E_z<0$ and
$E_z>0$, respectively. Such spin-wave and spin-and-orbital-wave
modes give, of course, much smaller quantum corrections of the
order parameter and energy than the correct RPA spectra of Fig.
\ref{fig:soft} \cite{Fei98,Ole00}.
\begin{figure}
\centerline{\psfig{figure= 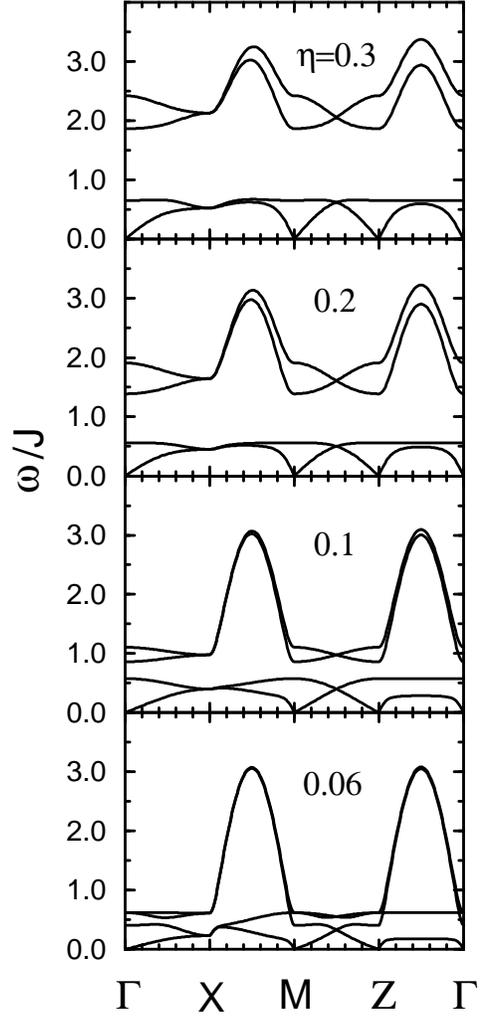,width=7cm}}
\smallskip
\caption {Transverse (full lines) and longitudinal (dashed lines)
excitations in MO{\protect{\scriptsize FFA}} phase in the main
directions of the 3D BZ for a few values of $J_H/U$, and for
$E_z/J=-0.5$. The lower-energy mode becomes soft for $J_H/U<0.06$
(after Ref. \protect\cite{Ole00}). } \label{fig:moffa}
\end{figure}

The spin waves in the MO{\scriptsize FFA} phase, stable at
$E_z<0$, soften with decreasing $\eta$ (\ref{eta}), as shown in
Fig. \ref{fig:moffa}. At large $\eta$ the spin-and-orbital waves
at high energies are well separated from the spin-wave modes. The
latter have a rather small dispersion at $J_H/U=0.3$ which follows
from relatively weak FM interactions in the $(a,b)$ planes, and AF
interactions along the $c$-axis. The modes start to mix stronger
with decreasing $\eta$, and finally the gap in the spectrum closes
below $\eta=0.1$. The mode softening occurs again along lines in
the BZ, namely along the $\Gamma-X$ direction. Unfortunately, an
analogous analytic expansion of the energies near the softening
point to those in the AFxx and AFzz phases could not be performed,
but the reported numerical results suggest a qualitatively similar
behavior to these two phases. The MO{\scriptsize AFF} phase gives
an analogous instability for $E_z>0$.

\begin{figure}
\centerline{\psfig{figure= 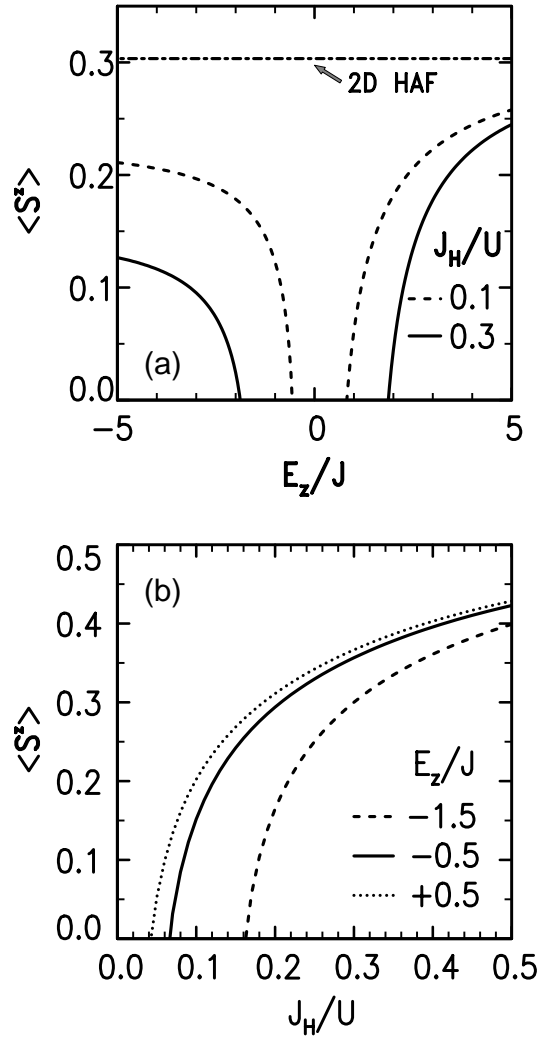,width=7cm}}
\smallskip
\caption {Renormalization of the magnetic LRO parameter $\langle
S^z_i\rangle$ by quantum fluctuations as obtained in RPA in: (a)
AFzz (left) and AFxx (right) phases as functions of $E_z/J$ for
$J_H/U=0.1$ and $0.3$; (b) MO{\protect{\scriptsize FFA}} phase as
function of $J_H/U$ for $E_z/J=0.5$, -0.5 and -1.5 (after Ref.
\protect\cite{Ole00}). } \label{fig:szreal}
\end{figure}

The soft modes in the excitation spectra give a very strong
renormalization of the order parameter $\langle S^z\rangle_{\rm
RPA}$ in RPA (\ref{expes}) near the mode softening, as shown in
Fig. \ref{fig:szreal}. The quantum corrections {\em exceed\/} the
MF values of the order parameter in the AFxx and AFzz phases in a
region which separates these two types of LRO. Although one might
expect that another classical phase with mixed orbitals and FM
planes sets in instead, and the actual instabilities where
$\delta\langle S_z\rangle\to\infty$ are found indeed beyond the
transition lines to another phase, the lines where $\delta\langle
S^z\rangle=\langle S^z\rangle$ occur still {\em before\/} the
phase boundaries in the phase diagram of Fig. \ref{fig:3d} (see
Fig. 1 of Ref. \onlinecite{Fei97}). This leaves a window where
{\em no classical order is stable\/} in between the G-AF and A-AF
spin structures.
The origin of such a strong renormalization of $\langle
S^z\rangle$ may be better understood by decomposing the quantum
corrections into individual contributions as given in Eq.
(\ref{expes}). The leading correction comes from the local spin
fluctuation expressed by $\langle S^-_iS^+_i\rangle$ and enhanced
with respect to the pure spin model, while the spin-and-orbital
fluctuation, $\langle K^-_iK^+_i\rangle$, increases rapidly when
the instability lines $\langle S^z\rangle_{\rm RPA}=0$ are
approached. Interestingly, the latter fluctuation is stronger in
the AFxx than in the AFzz phase for the same values of $J_H$ and
$|E_z|$ which demonstrates that the AFzz phase is more robust due
to the directionality of the $|z\rangle$ orbitals and the strong
AF bonds along the $c$-axis. This asymmetry is also visible in
Fig. \ref{fig:szreal}, where $\langle S^z\rangle_{\rm RPA}$
decreases somewhat faster towards zero for $E_z>0$.

\begin{figure}
\centerline{\psfig{figure= 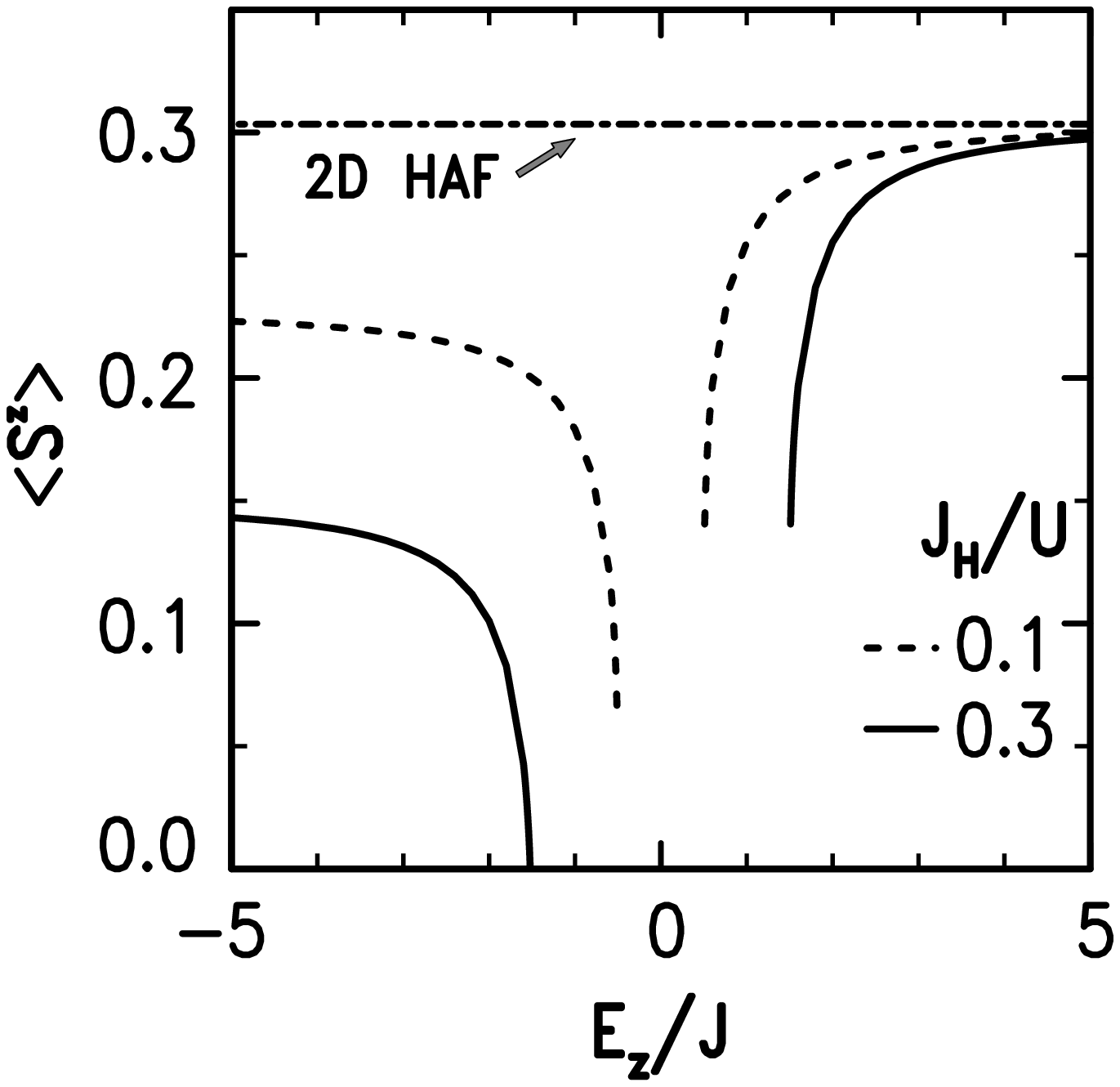,width=7cm}}
\bigskip
\caption {Renormalization of the magnetic LRO parameter $\langle
S^z_i\rangle$ by quantum fluctuations as obtained in RPA in: (a)
AFzz (left) and AFxx (right) phases as functions of $E_z/J$ for
    $J_H/U=0.1$ (dashed lines) and $0.3$ (solid lines);
(b) MO{\protect{\scriptsize FFA}} phase as functions of $J_H/U$
for $E_z/J=0.5$, -0.5 and -1.5 (after Ref. \protect\cite{Ole00}).
} \label{fig:szpoor}
\end{figure}

In both G-AF phases (AFxx and AFzz) the leading contribution to
the renormalization of $\langle S^z\rangle_{\rm RPA}$ comes from
the lower-energy mode, especially at larger values of $J_H$. In
the case of $J_H=0$ one finds, however, that the contribution from
the lower mode either stays approximately constant (in the AFxx
phase), or even decreases (in the AFzz phase) when the line of the
collapsing LRO is approached at $|E_z|\to 0$. This latter behavior
shows again that the coupling between the spin-wave and
spin-and-orbital-wave excitations is of crucial importance
\cite{Fei98}. This is further illustrated by Fig.
\ref{fig:szpoor}, which shows the renormalization of $\langle
S_z\rangle$ as obtained when spin waves and spin-and-orbital waves
are decoupled in the manner discussed above. One observes that
significant reduction of $\langle S_z \rangle$ then sets in only
very close to the actual divergence.
The reduction of $\langle S^z\rangle_{\rm RPA}$ in the
MO{\scriptsize FFA}/MO{\scriptsize AFF} phases, described by a
relation similar to Eq. (\ref{expes}), is in general weaker than
that in the G-AF phases. This is understandable, as the quantum
fluctuations contribute here only from a single AF direction,
while the FM order in the planes does not allow for excitations
which involve spin flips and stabilizes the LRO of A-AF type. For
fixed $J_H$ one finds increasing quantum corrections
$\delta\langle S^z\rangle$ when the lines of phase transitions
towards the AF phases are approached. These corrections increase
faster with increasing $|E_z|$ in the MO{\scriptsize FFA} phase,
as the increasing occupancy of the $|z\rangle$-orbital makes the
AF interaction stronger there than in the MO{\scriptsize AFF}
phase, where the occupancy of the $|x\rangle$ orbital increases
slower roughly by a factor of two. It may be again concluded
\cite{Ole00} that the collapse of the LRO in the A-AF (MO) phases
is primarily due to increasing spin fluctuations, $\langle {\cal
S}^-_i{\cal S}^+_i\rangle$, while the spin-and-orbital $\langle
{\cal K}^-_i{\cal K}^+_i\rangle$ fluctuations become of equal
importance only when the multicritical point of the Kugel-Khomskii
model $M=(E_z,J_H)=(0,0)$ is approached.

\begin{figure}
\centerline{\epsfig{figure= 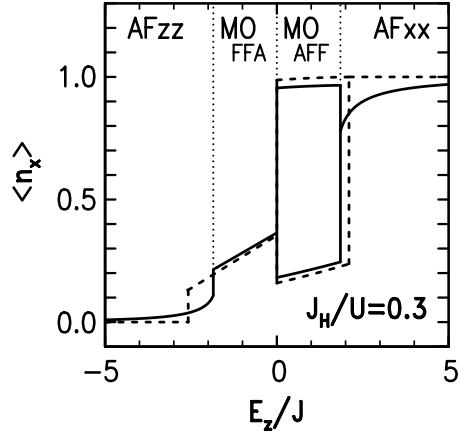,width=7cm}}
\smallskip
\caption {Average density of $|x\rangle$-holes $\langle
n_x\rangle$ as obtained in RPA for $J_H/U=0.3$ in MF approximation
(dashed lines) and with the quantum corrections calculated in RPA
(full lines). The splitting of lines for $E_z/J>0$ corresponds to
the MO{\protect{\scriptsize AFF}} phase with two different hole
densities $\langle n_x\rangle_A\neq \langle n_x\rangle_B$ on the
ions belonging to two sublattices shown in Fig.
\protect{\ref{fig:d9mfa}} (after Ref. \protect\cite{Ole00}). }
\label{fig:nxholes}
\end{figure}

The orbital polarization (\ref{nzet}) is also renormalized by the
quantum fluctuations, but this is a rather mild effect not showing
any instability (Fig. \ref{fig:nxholes}). In fact, this
renormalization involves only the spin-and-orbital and the orbital
excitation but not the spin excitation, which gives the largest
weight in the lowest transverse mode that goes soft. The value
determined in RPA is calculated from an expression similar to Eq.
(\ref{expes}), e.g. in the AFxx phase from
\begin{equation}
\label{expnx} \langle n_{ix}\rangle_{\rm RPA} = 1-4 \, \langle
T_{izx}T_{ixz}\rangle - \langle K^-_i K^+_i\rangle.
\end{equation}
The density $\langle n_{ix}\rangle_{\rm RPA}$ decreases gradually
with decreasing $E_z$, with somewhat increased quantum corrections
close to the transition lines between different classical phases.
In the MO{\scriptsize AFF} phase one finds nonequivalent
sublattices, with enhanced/reduced $\langle n_{ix}\rangle_{\rm
RPA}$ from its average value, reflecting the shape of the occupied
orbital on a given sublattice.
Finally, we note that the dominating contribution to the quantum
corrections to the energy comes from the transverse excitations.
The longitudinal excitations do not contribute at all at
$J_H/U=0$, where these modes are dispersionless. Otherwise, the
orbital excitations have always a significantly smaller dispersion
than the value of the orbital gap in the spectrum, and the
resulting quantum corrections are therefore almost negligible.

Summarizing, we have reported the case that a generic
(Kugel-Khomskii) model for the dynamics of an orbitally degenerate
MHI is characterized by a number of peculiar features. Assuming
that the ground state exhibits some particular classical spin- and
orbital order, the stability of this order can be investigated by
considering the Gaussian fluctuations around this state. In this
way we find that in various regimes of the zero-temperature
phase-diagram, classical order is defeated by the quantum
fluctuations, and we expect a qualitative phase diagram with a
region of disordered phases \cite{Fei97}. These disordered phases
with short-range spin and orbital correlations are discussed in
the next Section.


\subsection{Spin disorder: VB and RVB states}

Generally speaking, the ground state is unknown for most systems
described by the Heisenberg model with nonferromagnetic
interactions. Even those that are known in analytical form, such
as the Bethe solution in one dimension \cite{Bet31}, require
numerical computations to determine the spin correlations. Under
these circumstances, variational wave functions can give good
guesses for the ground state. In this respect, the valence bond
(VB) states and/or plaquettes valence bond (PVB) states are
variational wave functions which play a prominent role in
approximating the leading spin-spin correlations for the AF
Heisenberg model. They have been studied largely in the context of
quantum magnetism, especially in recent years when new motivation
occurred due to the research aimed at a better understanding of
the magnetic states realized in high temperature superconductors.

Their general form can be written in the following way
\cite{Faz99,Aue94}:
\begin{eqnarray}
|c_{\alpha},S\rangle=\sum_{\alpha} c_{\alpha} |\alpha \rangle,
\label{variastate}
\end{eqnarray}
where $c_{\alpha}$ are variational parameters and
\begin{eqnarray}
|\alpha \rangle = \prod_{\langle ij\rangle\epsilon \Lambda_\alpha}
    (a^{\dagger}_i b^{\dagger}_j-b^{\dagger}_i a^{\dagger}_j )|0\rangle,
\label{valence}
\end{eqnarray}
with $a_i$ and $b_i$ being the Schwinger bosons on site $i$
\cite{Aue94}. $\Lambda_\alpha$ is a particular configuration of
bonds $\langle ij\rangle$ which cover the whole lattice and are
occupied by spin singlets. The essence of a VB state is that there
are no magnetic correlations between different singlets. The
condition on $\Lambda_\alpha$ is that precisely $2S$ bonds will
emanate from each site. In some cases there will be only few
configurations to contribute in the large lattice limit. The case
were there are macroscopically many configurations and VB states
resonate between them has been called resonating valence bond
(RVB) states and was introduced by Anderson \cite{And87}.
Considering the expression given in Eq. (\ref{variastate}), it is
easy to verify that all bond operators are invariant under SU(2)
transformations. Therefore, the VB wave function
$|c_{\alpha},S\rangle$ is a singlet in the total spin. It provides
an example of spin-disordered state, the so-called {\it spin
liquid.\/}

There exists a class of special Hamiltonians for which the VB
states are the {\it exact\/} ground states. Majumdar and Ghosh
introduced the Hamiltonian for spin $S=1/2$ \cite{Maj69}:
\begin{eqnarray}
H_{\rm MG}=\frac{4K}{3} \sum_{i=1}^{N} \left( {\vec S}_i \cdot
{\vec S}_{i+1}+\frac{1}{2} {\vec S}_i \cdot {\vec S}_{i+2}
\right)+\frac{1}{2}N,
\end{eqnarray}
where $i$ stands for the sites of a 1D chain with an even number
of sites $N$, $K>0$ and one assumes periodic boundary conditions,
with ${\vec S}_{N+1}={\vec S}_1$ and ${\vec S}_{N+2}={\vec S}_2$.
It is straightforward to show that the dimer state,
\begin{eqnarray}
\label{dimersta} |d\rangle_{\pm} = \prod_{n=1}^{N/2} \left(
|\uparrow_{2n} \rangle |\downarrow_{2n \pm 1} \rangle -
|\downarrow_{2n} \rangle |\uparrow_{2n \pm 1} \rangle
\right)/\sqrt{2},
\end{eqnarray}
has energy zero, that is $H_{\rm MG} |d\rangle_{\pm}=0$, and all
the other eigenenergies are positive. As one can see the
Hamiltonian $H_{\rm MG}$ includes AF interactions between
next-nearest neighbors which frustrates the nearest neighbor
correlations. Thus, one expects the ground state to be more
disordered than that in the usual Heisenberg model. Indeed, for
the antiferromagnet with only nearest neighbor interactions the
spin-spin correlations decay as an inverse of a power of distance,
while the wave function $|d\rangle_{\pm}$ has state dimer
correlations that is they vanish beyond the nearest neighbors. The
proof is interesting since it provides a method for constructing a
more general family of Hamiltonians whose ground states are VB
states \cite{Aue94}.

The basic idea is to express $H_{MG}$ in terms of projection
operators. If we consider three spins $S=1/2$ on an arbitrary
triad of sites, the total spin $J$ may be either equal to $3/2$ or
to $1/2$. The total spin of three sites $(i-1,i,i+1)$ is given by
\begin{eqnarray}
{\vec J}_i={\vec S}_{i-1}+{\vec S}_i+{\vec S}_{i+1}
\end{eqnarray}
and its square is connected to the $H_{MG}$ via the projection
operator on a configuration of spin $1/2$ on three sites in the
following way:
\begin{eqnarray}
Q_{3/2}(i-1,i,i+1)&=&\frac{1}{3} \left(J_i^2-\frac{3}{4} \right) =
\frac{1}{2}+\frac{2}{3} \left( {\vec S}_i \cdot {\vec S}_{i-1}+
{\vec S}_{i-1} \cdot {\vec S}_{i+1}+{\vec S}_{i} \cdot {\vec
S}_{i+1} \right),
\end{eqnarray}
so that the Hamiltonian can be rewritten as
\begin{eqnarray}
H_{\rm MG}=K \sum_i Q_{3/2}(i-1,i,i+1) .
\end{eqnarray}
Of course, $Q_{3/2}(i-1,i,i+1)$ annihilates any state with $J=1/2$
at the three sites $\{i-1,i,i+1\}$. Since the dimer states
$|d\rangle_\pm$ (\ref{dimersta}) do not contain states with total
$J_z>1/2$ on any three sites and due to the rotational invariance
of the state $|d\rangle_{\pm}$ there cannot be any $J>1/2$
component in it. Therefore, each operator $Q_{3/2}(i-1,i,i+1)$
annihilates $|d\rangle_{\pm}$. Moreover, since the
$Q_{3/2}(i-1,i,i+1)$ have only positive eigenvalues,
$|d\rangle_{\pm}$ is the ground state of $H_{\rm MG}$.

In a similar fashion Affleck, Kennedy, Lieb, and Tasaki
\cite{Aff87} have constructed the Hamiltonians for the VB solids
as ground states which cover the whole lattice,
\begin{eqnarray}
H_{\rm AKLT}=\sum_{\langle ij\rangle}\sum_{J=2S-M+1}^{2S} K_J
Q_J(ij), \label{aklt}
\end{eqnarray}
where the bond projector $Q_J(ij)$ projects the total bond spin of
magnitude $J$. In this case it can be proven as before, that the
ground state is
\begin{eqnarray}
|\Omega_{\rm VBS} \rangle =\prod_{\langle ij \rangle} \left(
a^{\dagger}_i b^{\dagger}_j-b^{\dagger}_i a^{\dagger}_j
\right)^M|0 \rangle,
\end{eqnarray}
where all the nearest neighbor bonds $\langle ij\rangle$ are
included in the product, and $M=2S/z$ is an integer which is
related to the spin $S$ and to the coordination number $z$ of the
considered lattice. By construction, the smallest $S$ which allows
to construct such states is $S=1$ for a 1D lattice.

\begin{figure}
\centerline{\epsfig{figure= 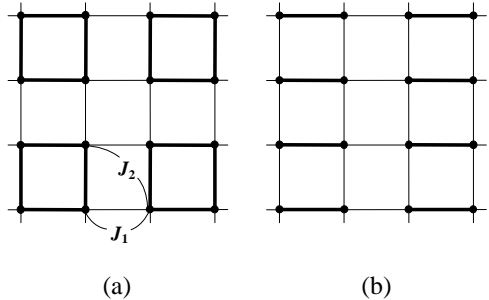,width=7cm}}
\smallskip
\caption {Disordered magnetic states for a 2D square lattice with
AF interactions $J_1$ and $J_2$: (a) plaquette-RVB, and (b)
columnar (VB) dimer ground states. The bold lines indicate the
bonds occupied by spin singlets. } \label{fig:dimer}
\end{figure}

Spontaneous spin liquid states occur in 2D lattices due to
frustrating spin interactions. The basic model which shows such a
behavior is the 2D AF $J_{1}-J_{2}$ Heisenberg model on a square
lattice \cite{Cha88,Cha88a,Cha88b,Cha88c,Cha88d,Chu91}. There are
two main reasons why this model holds a special place in the
physics of spin systems. It is one of the simplest models which
exhibits quantum transitions between long-range ordered phases and
a quantum disordered phase -- a topic of fundamental interest
\cite{Sac99}. Moreover, even though the $J_{1}-J_{2}$ model (Fig.
\ref{fig:dimer}) does not deal with charge dynamics, it can
represent a good starting point for the understanding of how
translational symmetry is broken in a purely insulating spin
background for approaching this question in spin systems with
finite doping. The $J_{1}-J_{2}$ model has been discussed in
numerous works over the last ten years and some of the important
issues that have been addressed are: (i) how is the N\'{e}el
order, present for small frustration ($J_{2}$), destroyed as
frustration increases, and (ii) is a quantum disordered phase
present in a finite window of frustration, and what is the
structure of this phase?
Spin-wave calculations, both at the non-interacting level as well
as including interactions perturbatively in powers of $1/S$ ($S$
is the spin value), have found that the magnetization decreases
with increasing frustration, ultimately vanishing at a critical
value \cite{Iga93}. These calculations however cannot predict the
structure of the phase beyond the instability point, or the
location of the phase boundary with high accuracy, since as the
magnetization decreases more and more powers of $1/S$ have to be
included (strong spin-wave interactions). Exact diagonalization
(ED) of clusters as large as $N=36$ \cite{Sch94,Sch96} have found
a finite region of quantum disordered (gaped) phase, but have
failed to determine accurately the dominant short-range
correlations or type of order (e.g. dimer, plaquette, {\it
etcetera\/}) characterize this phase. The ED calculations also
suffer from large finite-size corrections, especially for strong
frustration. An insight into the structure of the disordered phase
was possible with  the help of the large $N$ expansion technique
\cite{Sac90,Cha88b}. These authors predicted the quantum
disordered phase to be spontaneously dimerized in a particular
(columnar) configuration (Fig. \ref{fig:dimer}). High order dimer
series expansions around this configuration were performed
\cite{Gel89,Kot99}, all confirming its stability in a window of
frustration. Thus the spontaneously dimerized state has emerged as
the most probable candidate for a disordered ground state. Let us
mention in this connection that in 1D systems the
Lieb-Schultz-Mattis (LSM) theorem guarantees that a gaped phase in
a quantum spin system breaks translational symmetry \cite{Lieb61}.
Extension of the LSM theorem to two-dimensions was proposed
\cite{Aff88} but not yet proven in the most general case. The
large $N$ and dimer series results, however, seem to confirm the
validity of the LSM theorem in two dimensions as well, including,
in fact, the case of finite doping \cite{Sac99a}.

The previous case is a classical example that inclusion of
additional interactions, such as dimerization and/or frustration,
leads to increased quantum fluctuations, and ultimately to
vanishing LRO at a critical coupling. In this context, further
examples of transitions caused by local alternation of the
exchange couplings are the dimerized Heisenberg antiferromagnets
\cite{Kat93}, the two-layer Heisenberg model
\cite{Mil94,San94,Chu95} and the CaV$_4$O$_9$ lattice (1/5-
depleted square lattice) \cite{Ued96,Ued96a,Tro96,Whi96}. For
these situations, the local dimer or plaquette correlations
eventually win over the long-range N\'{e}el order, leading to a
disordered ground state.

An important issue concerning the quantum transitions mentioned
       above is their universality class. It is generally accepted that
       the effective low-energy theory for the 2D Heisenberg systems
       with a collinear (N\'{e}el) order parameter is the O(3)
       non-linear sigma model (NLSM) in 2+1 dimensions \cite{Fra91}. This
       field theory contains a single effective coupling constant $g$
       and, at $T=0$, describes the ordered N\'{e}el phase for $g<g_{c}$.
       For $g>g_{c}$ the NLSM is in a quantum disordered phase with a
       finite correlation length. However the determination  of $g_{c}$
       and the nature of the disordered phase are beyond the field
       theory formulation and depend on the specific details of the
       model. In addition, Berry phases associated with instanton
       tunneling between topologically different configurations are
       present in the NLSM \cite{Hal88}.
       In one dimension the Berry phase
       effects are known to be important, essentially leading to the
       difference between the excitations in the integer and half
       odd-integer spin chains \cite{Aff88} (Haldane conjecture). In two
       dimensions Berry phases are also present, but their role is less clear.
       If one neglects these purely quantum effects, the universality class
       of the quantum transitions in the 2D Heisenberg antiferromagnet should
be the same as that of the classical O(3) vector model in three
dimensions.
QMC simulations performed on the two-layer antiferromagnet
\cite{Sand95}, and on the CaV$_4$O$_9$ lattice \cite{Tro96}
confirm with high accuracy that the quantum transitions in the
above two models happen in the O(3) universality class.

Furthermore, the $J_1-J_2$ model, which exhibits a quantum
transition due to frustration, could represent a case of
difference from the O(3) universality class. Read and Sachdev
showed that there are two correlation lengths close to the
criticality \cite{Cha88b}, suggesting to think that there are
deviations in the universality class. On the other hand, they
showed that only one of the two is relevant implying that even
though the Berry phases are relevant in the disordered phase,
ultimately, near the critical point, their effect disappears.

As we have discussed, the main characters in this topic are
low-dimensional quantum spin systems (spin chains \cite{Whi94} and
ladders \cite{Dag96,Dag96a,Dag96b}), and it proves difficult to
achieve quantum melting of magnetic LRO in empirically relevant
systems in higher dimensions. It is therefore worth pointing out
that there is a class of systems in which quantum-melting occurs
due to a {\it unique mechanism} which operates in three
dimensions: small spin, orbital degenerate MHI, the so-called
Kugel-Khomskii systems \cite{Kug82}. There might exist already a
physical realization of such a 3D {\it quantum spin-orbital
liquid\/}: spin disorder in LiNiO$_2$\cite{Hir85,Hir85a}.

Here, we report how the orbital degeneracy operates through the
same basic mechanisms known from spin systems, to produce quantum
melting in the spin-orbital $d^9$ systems. The novelty is that
these systems tend to "self-tune" to (critical) points of high
classical degeneracy. There are interactions which may lift the
classical degeneracy, but they are usually weak. An interaction of
this kind is the electron-phonon coupling -- the degeneracy is
lifted by a change in crystal structure, the conventional
collective JT instability. As we have discussed in Sec. III.C, the
lattice has to react to the symmetry lowering in the orbital
sector, but it was recently convincingly shown, at least in the
archetypical compound KCuF$_3$, that the structural distortion is
a side effect \cite{Lie95}. The fundamental question which arises
in this context is what happens when the classical order becomes
unstable against quantum fluctuations. Although the subject is
much more general (singlet-triplet models in rare earth compounds
\cite{Hsi72}, V$_2$O$_3$ \cite{Cas78}, LaMnO$_3$ \cite{Miz96},
heavy fermions \cite{Ful91,Faz99,Cox87}), we focus here on the
simplest situation encountered in KCuF$_3$ and related systems,
described by the $d^9$ model \cite{Fei97}.

If the classical limit is as sick as explained in Sec. IV.B, what
is happening instead? {\it A priori\/} it is not easy to give an
answer to this question. There are no `off the shelf' methods to
treat quantum spin problems characterized by classical
frustration, and the situation is similar to what is found in,
e.g. $J_1-J_2-J_3$ problems
\cite{Cha88,Cha88a,Cha88b,Cha88c,Cha88d}. A first possibility is
quantum order-out-of-disorder \cite{Chu91}: quantum fluctuations
can stabilize a particular classical state over other classically
degenerate states, if this particular state is characterized by
softer excitations than any of the other candidates. Khaliullin
and Oudovenko \cite{Kha97} have suggested that this mechanism is
operative in the present spin-orbital model, where the 3D
anisotropic AFzz antiferromagnet is the one becoming stable. Their
original argument was flawed because of the decoupling procedure
they used which violates the so(4) dynamical algebra constraints
\cite{Fei98}. Nevertheless, there is yet another possibility: VB
singlet (or spin-Peierls) order, which at the least appears in a
more natural way in the present context than is the case of higher
dimensional spin-only problems, because it is favored by the
directional nature of the orbitals.

In similarity to the purely spin systems, in the presence of
orbital interactions either one particular covering of the lattice
with these `spin-dimers' might be favored (VB or spin-Peierls
state), or the ground state might become a coherent superposition
of many of such coverings (RVB state). On a cubic lattice the
difficulty is that although much energy is gained in the formation
of the singlet pairs, the bonds between the singlets are treated
poorly. Nevertheless, both in 1D spin systems (Majumdar-Ghosh
\cite{Maj69}, AKLT-systems \cite{Aff87}) and in the large $N$
limit of $SU(N)$ magnets in two dimensions \cite{Sac99}, ground
states are found characterized by spin-Peierls/VB order
\cite{Cha88b}.
\begin{figure}
\centerline{\epsfig{figure=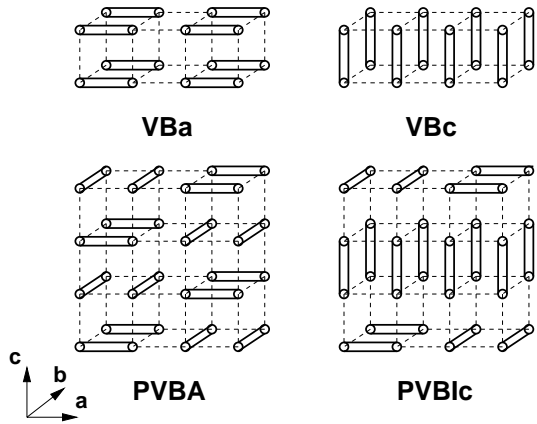,width=7cm}} \vspace{0.5in}
\caption {A variety of VB (VBa and VBc) and PVB (PVBA and PVBIc)
solids discussed in the text for the $d^9$ model (after Ref.
\protect\cite{Zaa99}). } \label{fig:vb}
\end{figure}
In principle, various topologically different coverings of the 3D
lattice may be considered, in analogy with the square lattice
\cite{Kiv87}. Two obvious choices, defined by the sets
$\Lambda_{\alpha}$ of parallel singlets in Eq. (\ref{valence}),
and suggested by the tendency towards particular directional
orbitals for positive (negative) $E_z$ are:
 ($i$) singlets along the $a$-axis with orbitals close to $3x^2-1$ (VBa),
       expected to be favored for $E_z>0$, and
($ii$) singlets along the $c$-axis with orbitals $\sim |z\rangle$
(VBc),
       preferred if $E_z<0$ (see Fig. \ref{fig:vb}).
Interestingly, the immanent frustration of the magnetic
interactions in the spin-orbital model (\ref{somcu}) causes that
these phases have lower energies than the classical phases in a
broad parameter regime. This result is already quite spectacular
when one realizes that such VB phases thus appear to be better
approximations to the exact ground state than the classical phases
with magnetic LRO in {\em three dimensions}. An example is shown
in Fig. \ref{fig:vbene} for $J_H/U=0.2$; as expected, VBa (VBc)
has a lower energy for $E_z>0$ ($E_z<0$), and both phases are
degenerate at $E_z=0$. We have recognized that the exceptional
stability of these (nonresonating) VB states is due to a unique
mechanism involving the orbital sector. Unlike in the spin system
with a simple Heisenberg exchange, the bonds not occupied by the
singlets {\em contribute orbital energy} and this is optimized
when singlets in orthogonal directions are connected (which is not
the case in the VBa and VBc phases; see further below).
\begin{figure}
\centerline{\epsfig{figure=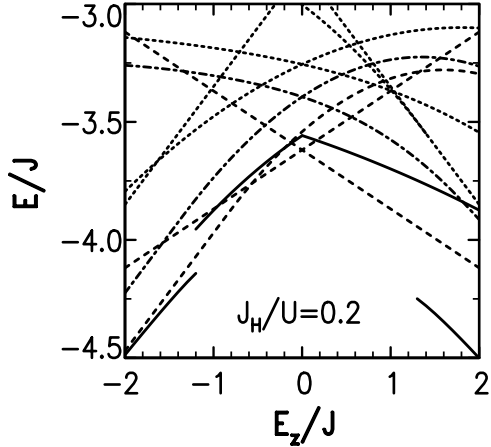,width=7cm}}
\smallskip
\caption {Energies of the various ordered and disordered phases
for the $d^9$ model (\protect\ref{somcu})--(\protect\ref{htau}) as
functions of $E_z/J$, for $J_H/U=0.2$. The energies of classical
phases were calculated within MF (short-dashed lines) and RPA
approach (full lines), while the energies of VBa and VBc states
(see Fig. \protect\ref{fig:vb}) are given by dashed-dotted lines.
The energies of disordered states (RVBc, PVBI, and PVBA) stable
close to the orbital degeneracy are given by the long-dashed lines
(after Ref. \protect\cite{Foz00}). } \label{fig:vbene}
\end{figure}
Further improvement of energy may be expected by including the
leading quantum fluctuations in the VB (VBa and VBc) states. In
the case of the VBa phase this leads directly to constructing the
PRVB states, with the plaquette wave functions of the type,
\begin{equation}
\label{plaq} |\Psi_{\Box}\rangle = {1\over\sqrt{2|1+S|}}
(|\Psi_a\rangle +e^{i\phi}|\Psi_b\rangle),
\end{equation}
where the two components are composed of the singlet pairs along
$a$ and $b$-axis, $|\Psi_a\rangle$ and $|\Psi_b\rangle$,
respectively, and $S=e^{i\phi}\langle\Psi_a|\Psi_b\rangle$ is the
overlap. It is straightforward to show that this Ansatz gives an
exact energy $E_0=-2J_0$ for a plaquette occupied by spins
$S=1/2$, assuming the AF interaction $J_0$. The wave function
$|\Psi_{\Box}\rangle$ turns out to be identical with the exact
wave function of an isolated plaquette \cite{Isk87}, if we make
the choice $\phi=0$ in Eq. (\ref{plaq}). When averaged over the
square lattice, this gives an energy of $-J_0/2$ per site, the
same energy as that of the classical N\'eel state. However, in the
present case the resonance between the two singlet structures is
much weaker, as the optimized orbitals are either $\parallel a$,
or $\parallel b$ in the two components of the plaquette wave
function (\ref{plaq}). This results in a much smaller overlap
$S=1/32$ and typically very small offdiagonal energy elements,
$\langle\Psi_a|H|\Psi_b\rangle$. As a result, the improvement due
to this resonance is marginal, and it turns out to be better to
optimize the singlet distribution over the lattice with respect to
the energy contributions which originate from the bonds not
occupied by the singlets.

\begin{figure}
\centerline{\epsfig{figure=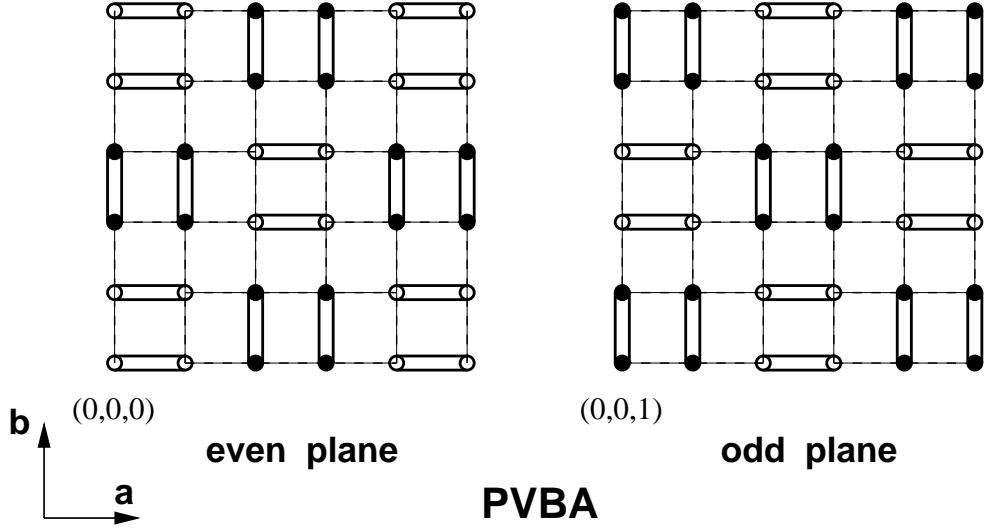,width=7cm,angle=270}}
\vspace{0.5in} \caption {Schematic representation of the PVBA
state, stable in a range of $J_H/U$ and for $E_z/J>0$. The open
and full circles refer to the directional $|\zeta\rangle$ orbitals
parallel to the bonds occupied by the singlets along $a$ and $b$
axis, respectively (after Ref. \protect\cite{Foz00}). }
 \label{fig:pvba}
\end{figure}

There are three different possible energy contributions for these
bonds depending on whether they connect:
  ($i$) two singlets along a single ($a$, $b$, or $c$) line,
 ($ii$) two singlets oriented along two different lines with an angle of
        $\pi/2$, or
($iii$) two parallel singlets, with the bond making itself an
angle of
        $\pi/2$ to both of them.
As the second type of the non-singlet bonds is energetically the
most favorable, more energy than in the PRVB state (\ref{plaq}) is
gained if the plaquette wave functions $|\Psi_a\rangle$ and
$|\Psi_b\rangle$ alternate and form a superlattice (see Fig.
\ref{fig:pvba}). This results in a plaquette VB (PVB) state for
each $(a,b)$ plane. As also the energy of the vertical ($\parallel
c$) bonds is optimized when these bonds connect one singlet
$\parallel a$ and another $\parallel b$, the optimal phase with a
lower energy that the quantum-corrected MO{\scriptsize AFF} phase
found for $E_z>0$ (Fig. \ref{fig:vbene}) is given by a PVB
alternating (PVBA) state, with the alternation of two along the
$c$-axis (see Figs. \ref{fig:vb} and \ref{fig:pvba}). We note that
the energy comparison between the AFxx state and the PVBA state
resembles qualitatively the situation in a 2D 1/5-depleted lattice
\cite{Ued96,Ued96a}, but the spin-disordered state is remarkably
more stable in our case, and provides an approximate Ansatz for
the ground state in {\it three dimensions\/}.
In contrast, the energy of the VBc phase can be improved by
considering the resonance between the singlets along the vertical
($\parallel c$) direction. The energy of the resulting resonating
VBc (RVBc) state could be obtained using the Bethe ansatz result
for the 1D Heisenberg antiferromagnet \cite{Bet31}, and adding the
orbital energy contributions due to the bonds $\perp c$. This
qualitatively different behavior has to do with the symmetry
breaking in the orbital space between $E_z>0$ and $E_z<0$, as the
system has to choose between two equally favorable directional
orbitals $\zeta\rangle$ ($\parallel a$ and $\parallel b$) if
$E_z>0$, which makes the PVBA phase optimal due to the
intersinglet contributions, while a single possibility
($|z\rangle$ orbitals) remains if $E_z<0$. As shown in Fig.
\ref{fig:vbene}, the energy of the RVBc phase is lower than the
energy of the MO{\scriptsize FFA} phase corrected by quantum
fluctuations.

In the crossover regime between the RVBc and PVBA phases one may
expect that some other arrangment of singlets in the cubic lattice
gives still a better energy that the two above magnetically
disordered phases. In fact, we have found a PVB interlayered
(PVBI) state, composed of single planes of the PVBA phase (Fig.
\ref{fig:pvba}) interlayered with two planes of VBc phase along
the $c$-direction (Fig. \ref{fig:vb}), to be more stable in the
crossover regime, as shown in Fig. \ref{fig:vbene}. The energy in
the PVBI state is gained both from more $|z\rangle$ orbital
character in the singlets $\parallel c$, and from the orbital
energy contributions due to the bonds not occupied by the
singlets, connecting a single layer of the PVBA phase with the
singlets $\parallel c$ in the neighboring planes. Of course, this
phase is destabilized for $E_z>0$, where the holes occupying
$|z\rangle$ orbitals have higher energy.

It is straightforward to understand that the interplay of orbital-
and spin degrees of freedom tends to stabilize VB order. Since the
orbital sector is governed by a discrete symmetry, the orbitals
tend to condense in some classical orbital order. It is precisely
for this reason that the best ground state wave functions do not
resemble the AKLT models (\ref{aklt}) with alternating spin and
orbital singlets \cite{Pat00}, but form instead composite
spin-and-orbital singlets on the same bonds, as for instance in
the PVBA phase (Fig. \ref{fig:pvba}). Different from the fully
classical phases, one now looks for orbital configurations
optimizing the energy of the spin VB configurations. The spin
energy is optimized by having directional orbitals $|\zeta\rangle$
parallel to the bond $\langle ij\rangle$ at both sites $i$ and $j$
at which the VB spin-pair also lives. This choice maximizes the
overlap between the wave functions, and thereby the binding energy
of the singlet. At the same time, this choice of orbitals
minimizes the unfavorable overlaps with spin pairs located in
directions orthogonal to $|\zeta\rangle$. The net result is that
VB states are much better variational solutions for the $d^9$
model (\ref{somcu}), than in the standard spin systems.

\begin{figure}
\centerline{\epsfig{figure=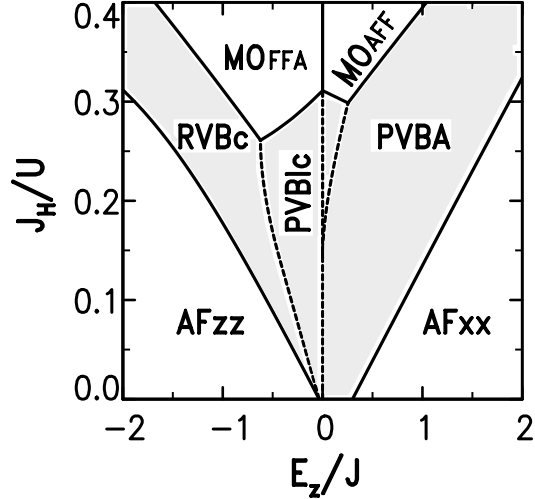,width=7cm}}
\smallskip
\caption {The same as in  Fig. \protect\ref{fig:3d}, but including
quantum fluctuations as determined in RPA. The spin liquid (RVBc,
PVBIc, PVBIa, and PVBA) takes over in the shaded region between
G-AF and A-AF phases (after Ref. \protect\cite{Fei97}). }
\label{fig:rpa}
\end{figure}

Addressing the problem of {\it spin-liquid\/} in the $d^9$ model
systematically, it has been found \cite{Fei97} that two families
of VB states are most stable: (i) The `staggered' VB states like
the PVBA and PVBIc states of Fig. \ref{fig:vb}. These states have
in common that the overlap between neighboring VB pairs is
minimized: the large lobes of the $|\zeta\rangle$ orbitals of
different pairs are never pointing to each other. (ii) The
`columnar' VB states like the VBc (or VBa) state of Fig.
\ref{fig:vb}. In the orbital sector, this is nothing else than the
AFzz state of Fig. \ref{fig:d9mfa} ($3z^2-r^2$ orbitals on every
site). Different from the AFzz state, the spin system living on
this orbital backbone is condensed in a 1D spin-Peierls state
along the $c$-direction which is characterized by strong exchange
couplings. The spins in the $a(b)$-directions stay uncorrelated,
due to the weakness of the respective exchange couplings as
compared to the VB mass gap. These considerations lead to a phase
diagram of the $d^9$ model (\ref{somcu})--(\ref{htau}) shown in
Fig. \ref{fig:rpa}. If $\eta<0.30$, the PVBA state is stable at
$E_z>0$, while a similar PVBIc interlayered state with alternating
layers of $(a,b)$-plane/$c$-axis bonds (Fig. \ref{fig:vb}), and
the RVBc state are stable at $E_z<0$. Thus, a spin liquid is
stabilized by the {\it orbital degeneracy} over the ordered MO
phases with RPA fluctuations in a broad regime. This resembles the
situation in a 2D 1/5-depleted lattice \cite{Ued96,Ued96a}, but
the present instability is much stronger and happens in {\it three
dimensions\/}.

Summarizing, there is a strong theoretical arguments supporting
the conjecture that quantum-melting might occur in orbital
degenerate MHI. Why does it not occur always (e.g., in KCuF$_3$)?
Next to the Hund's rule coupling, the electron-phonon coupling
$\lambda$ (\ref{hjth1}) may destroy quantum disorder. The lattice
will react to the orbital fluctuations, dressing them up in
analogy with polaron physics, and thereby reducing the coupling
constant. In order to quantum melt KCuF$_3$-like states, one
should therefore look for ways to reduce both the effective $J_H$
and $\lambda$. It was first suggested in Ref. \cite{Fei97} that
this situation might occur in LiNiO$_2$: although the spin-spin
interactions in the (111) planes should be very weakly FM
according to the Goodenough-Kanamori rules, magnetic LRO is absent
\cite{Hir85} and the system might represent the spin-orbital
liquid. More strikingly, LiNiO$_2$ is cubic and should undergo a
collective JT transition, which absence is actually an old
chemistry mystery! Upon electron-hole transformation, $d^7$
low-spin Ni$^{3+}$ maps on $d^9$ Cu$^{2+}$ in KCuF$_3$, but with a
difference in chemistry. While the $e_g$ hole in KCuF$_3$ is
nearly entirely localized on the Cu, the $e_g$ electron in
LiNiO$_2$ is rather strongly delocalized over the Ni and
surrounding O ions which reduces both $J_H$ and $\lambda$, and
explains the absence of classical ordering. A more precise
experimental characterization of LiNiO$_2$ is needed to decide
whether it provides an example for a quantum disorder of the kind
discussed here.

Comparing with spin models, the understanding of the disordered
ground states in spin-orbital models is much less developed. We
have discussed in this Section only these results which could be
obtained so far using simple methods based on variational wave
functions of the VB and RVB type. There are a few questions which
deserve a future study. For instance, by deriving similar rules to
those known for the spin systems \cite{Sut88}, for calculating the
overlaps between different VB-type configurations on the lattice
would simplify the analysis of these situations in which the
resonance of different configurations might play a role in
spin-orbital systems. Formulating the problem on the abstract
level, one might attempt to find parent Hamiltonians similar to
the AKLT models (\ref{aklt}) to particular dimerized states with
alternating spin and orbital singlets. However, the $e_g$ orbitals
are more classical than the pseudospins in the symmetric SU(4)
model of Sec. IV.A. Further studies should clarify to what extent
the $e_g$ orbitals exhibit really quantum behavior in realistic
situations, such as described by the $d^9$ model
(\ref{somcu})--(\ref{htau}). However, we believe that even if the
orbitals are more classical than the spins, they amplify the
quantum effects just by their fluctuations which couple to spin
fluctuations.


\section{ Spin-orbital model in manganites }

\subsection{ Superexchange and orbital interactions in LaMnO$_3$ }

As we have shown in Fig. \ref{fig:expphd}, the phase diagrams of
manganites are very rich and one would like to understand  the
microscopic origin of the experimentally observed A-AF order in
LaMnO$_3$ \cite{Wol55} in first place, and next why the magnetic
interactions change so drastically that the system becomes FM at
larger doping. Naively, the FM order follows just from the DE
model as presented in Sec. II.B. However, we have presented
arguments in Sec. I that this qualitative argument does not
explain the physics of manganites and the realistic approach has
to include the double degeneracy of $e_g$ orbitals occupied by one
or less electron per Mn ion.

As we already know from the earlier studies \cite{Kug82,Goo55} and
from Sec. III of the present paper, the magnetic order is coupled
to the orbital one. Therefore, one possible explanation of the
observed A-AF phase might be that it follows from a cooperative JT
effect \cite{Mil96} which stabilizes a particular order of the
singly occupied $e_g$ orbitals \cite{Kug73}. Although this might
sound quite attractive and attempts were made to understand the
superexchange constants observed in the A-AF phase of LaMnO$_3$ in
{\it ab initio\/} calculations \cite{Sol96,Mry97}, these studies
did not give the superexchange constants close to their
experimental values. A model of degenerate $e_g$ orbitals
\cite{Marco} is more successful, but only when the orbital
ordering is not calculated but fitted to the experiment. Here we
argue that the leading mechanism comes instead from strong Coulomb
interaction $U$ is the dominating energy scale in late transition
metal oxides which gives superexchange induced by the hopping of
$e_g$ electrons \cite{Fei99}. At the same time the coupling to the
lattice is much smaller than the Coulomb interactions which is
consistent with the existence of the insulating phase above the JT
transition \cite{Pat99}. Of course, the JT effect has to be
included in a complete model, as we will also show below, as it
leads to particular orbital interactions and stabilizes the
orbital ordering well before (when the temperature is decreased)
the magnetism sets in. Therefore, we believe that the JT effect is
of crucial importance in the doped regime where it drives the
transition from the insulating polaronic regime to a metal
\cite{MMS96}.

If transition metal ions have partly filled $e_g$ orbitals close
to orbital degeneracy, as we have discussed for the KCuF$_3$, the
strong Coulomb interactions lead to an effective low energy
Hamiltonian, where the spin and orbital degrees of freedom are
coupled \cite{Kug73,Fei97,Ole00}. As an important difference with
respect to the d$^9$ configuration (Sec. III.A), one has to
construct {\it superexchange between total spins\/} $S=2$ of
Mn$^{3+}$ ions in LaMnO$_3$ compound \cite{Fei99}. A simpler
approach in which the superexchange is considered separately for
the $e_g$ electrons with spins $s=1/2$ and for $t_{2g}$ electrons
with spins $S_t=3/2$ \cite{Ish97,Ish00} is not realistic as the
spin dynamics of $e_g$ and $t_{2g}$ spins does not occur
independently of each other and the spin waves measured by
neutrons correspond to the spin excitations of spins $S=2$
\cite{Hir96,Hir96a}. This simplified approach proposed recently by
Ishihara, Inoue and Maekawa \cite{Ish97} emphasizes correctly the
role of orbitals, but violates the SU(2) spin symmetry and
involves a Kondo coupling $K$ between $e_g$ and $t_{2g}$ spins,
which by itself is not a faithful approximation to the multiplet
structure, as we explain below. The complete effective model of
the undoped LaMnO$_3$ has to include also the $t_{2g}$-part of
superexchange and the orbital interactions induced by the JT
effect \cite{Fei99}. We will show that these terms, while
unessential qualitatively, are very important for a quantitative
understanding.

The superexchange between the $d^4$ Mn$^{3+}$ ions originates in
the large-$U$ regime from virtual ($e_g$ or $t_{2g}$) excitations,
$d_i^4d_j^4\rightleftharpoons d_i^3d_j^5$. The spin-orbital model
presented below follows from the full multiplet structure of the
manganese ions in octahedral symmetry, both in the $d^4$
($t_{2g}^3e_g$) configuration of the Mn$^{3+}$ ground state, and
in the $d^3$ ($t_{2g}^3$) and $d^5$ ($t_{2g}^3e_g^2$) virtually
excited states of Mn$^{4+}$ and Mn$^{2+}$ ions, respectively. Our
starting point is that each Mn$^{3+}$ ($d^4$) ion is in the
high-spin ($t_{2g}^3e_g$) ground state in agreement with large
Hund's rule interaction $J_H$, i.e., the high-spin ($S=2$) orbital
doublet $^5\!E$.
\begin{figure}
\centerline{\epsfig{figure=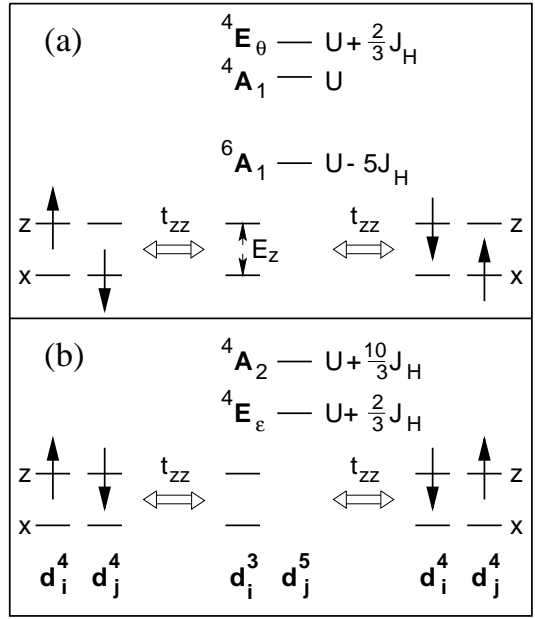,width=7cm}} \vspace{0.5in}
\caption {Virtual $d_i^4d_j^4\rightarrow d_i^3d_j^5$ excitations
 which generate effective interactions for a bond $(ij)\parallel c$-axis:
 (a) for one $|x\rangle$ and one $|z\rangle$ electron, and
 (b) for two $|z\rangle$ electrons (after Ref. \protect\cite{Fei99}).
} \label{fig:d4se}
\end{figure}
First, we analyze the strongest channel of superexchange, which
follows from the hopping of $e_g$ electrons between nearest
neighbor sites $i$ and $j$. The derivation is similar to that
analyzed in detail for the $d^9$ configuration in Sec. III.A, but
the important difference is found in the spin sector, where the
derived effective interactions have to be represented by the
superexchange terms between total spins $S=2$ per site
\cite{Fei99}. When we consider a bond oriented along the cubic
$c$-axis, only a $3z^2-r^2$ electron can hop as in the $d^9$ case
(Sec. III.A), and four $d^5$ states may be reached: the high-spin
$^6\!A_1$ state ($S=5/2$), and the lower-spin ($S=3/2$) $^4\!A_1$,
$^4\!E$, and $^4\!A_2$ states (Fig. \ref{fig:d4se}). The
$d_i^4d_j^4\rightleftharpoons d_i^3(t_{2g}^3)d_j^5(t_{2g}^3e_g^2)$
excitation energies require for their description in principle
{\em all three\/} Racah parameters, $A$, $B$ and $C$ \cite{Gri71}:
$\varepsilon(^6\!A_1)=A-8B$,  $\varepsilon(^4\!A_1)=A+ 2B+5C$,
$\varepsilon(^4\!E  )\simeq A+6B+5C$,
$\varepsilon(^4\!A_2)=A+14B+7C$. In view of the realistic values
of $B=0.107$ and $C=0.477$ eV for Mn$^{2+}$ ($d^5$) ions
\cite{Boc92}, one may use an approximate relation $C\simeq 4B$,
and write the excitation energies in terms of Coulomb, $U\equiv
A+2B+5C$, and Hund's exchange, $J_H\equiv 2B+C\simeq 0.69$ eV,
parameters:
\begin{eqnarray}
\label{excd51} \varepsilon(^6\!A_1) &=& U - 5 J_H,
\\
\label{excd52} \varepsilon(^4\!A_1) &=& U,
\\
\label{excd53} \varepsilon(^4\!E\,) &=& U + \frac{2}{3} J_H,
\\
\label{excd54} \varepsilon(^4\!A_2) &=& U + \frac{10}{3} J_H.
\end{eqnarray}
We emphasize that $U$ is here by definition {\it not\/} the
Coulomb matrix element, but the reference energy of the
$|^4\!A_1\rangle$ state; this definition is different from some
other conventions \cite{Boc92,Miz95}. A value of $U=7.3$ eV was
deduced for LaMnO$_3$ by Feiner and Ole\'s \cite{Fei99} from the
available spectroscopic data \cite{Zaa90,Boc92,Miz95}.

Using the spin algebra (Clebsch-Gordon coefficients), the
reduction of product representations in cubic site symmetry
\cite{Gri71} for the intermediate states, and making a rotation of
the terms derived for a bond $\langle ij\rangle\parallel c$ to the
bonds $\rangle ij\langle\parallel a$ and $\rangle
ij\langle\parallel b$ as in Sec. III.A, one finds a compact
expression,
\begin{eqnarray}
\label{egterm} H_J(e_g)&=&\frac{1}{16}\sum_{\langle
ij\rangle}\left\{
 - \frac{8}{5} \frac{t^2}{\varepsilon(^6\!A_1)}
   \left(\vec{S}_i\cdot\vec{S}_j+6\right)
   {\cal P}_{\langle ij\rangle}^{\zeta\xi}\right.              \nonumber \\
& &\hskip 1.0cm \left.
 + \left[\frac{t^2}{\varepsilon(^4\!E)}
   + \frac{3}{5}\frac{t^2}{\varepsilon(^4\!A_1)} \right]
   \left(\vec{S}_i\cdot\vec{S}_j-4\right)
   {\cal P}_{\langle ij\rangle}^{\zeta\xi}\right.              \nonumber \\
& &\hskip 1.0cm \left. +\left[ \frac{t^2}{\varepsilon(^4\!E)}
   + \frac{t^2}{\varepsilon(^4\!A_2)} \right]
   \left(\vec{S}_i\cdot\vec{S}_j-4\right)
   {\cal P}_{\langle ij\rangle}^{\zeta\zeta}\right\},
\end{eqnarray}
where $t$ is the hopping element along the $c$-axis, while ${\cal
P}_{\langle ij\rangle}^{\zeta\xi}$ and ${\cal P}_{\langle
ij\rangle}^{\zeta\zeta}$ are the projection operators introduced
in Eqs. (\ref{projbond}) and (\ref{projbond1}). A similar
expression was derived by Shiina, Nishitani and Shiba using the
group theory arguments \cite{Shi97}. However, this method does not
allow to determine accurately the coefficients of different terms,
and thus the balance between AF and FM interactions is different.

The terms in the first line of Eq. (\ref{egterm}) are obtained
from the virtual processes which involve different occupancies at
the sites $i$ and $j$ -- at one site the orbital along the bond
$|\zeta\rangle$ is occupied by an $e_g$ electron, while an
orthogonal orbital $|\xi\rangle$ is occupied at the other [Fig.
\ref{fig:d4se}(a)]. In agreement with the general rule as
presented in Secs. I and III.A, the processes which involve the
high-spin state $|^6\!A_1\rangle$ lead to FM superexchange, while
the low-spin state $|^4\!A_1\rangle$ gives an AF interaction. The
terms in the second line arise from the configurations with
occupied directional orbitals $|\zeta\rangle$ at both sites which
give the excited states with a double occupancy in one of the
$|\zeta\rangle$ orbitals [Fig. \ref{fig:d4se}(b)]. After
projecting this double occupancy onto the eigenstates
$|^4\!E\rangle$ and $|^4\!A_2\rangle$ one finds the AF
interactions. Thus, the structure is similar to that of Eq.
(\ref{somj}), but the coefficients are different and follow from
the multiplet structure of the excited $d^5$ states and from the
spin algebra.

A similar derivation gives the $t_{2g}$ superexchange
\cite{Fei99},
\begin{equation}
\label{t2gterm} H_J(t_{2g}) = \frac{1}{4}J_t\sum_{\langle
ij\rangle}
           \left(\vec{S}_i\cdot\vec{S}_j-4\right) ,
\end{equation}
where $J_t$ is an average of the processes which couple different
low-spin $d^5$ ($t_{2g}^4e_g$) and $d^3$ ($t_{2g}^3$) excited
states $^4T_1$ and $^4T_2$ of Mn$^{2+}$ and Mn$^{4+}$ ions,
respectively, $J_t=(J_{11}+J_{22}+J_{12}+J_{21})/4$. The
individual exchange elements,
$J_{mn}=t_{\pi}^2/\varepsilon(^4T_m,^4\!T_n)$, result from
different local $d_i^4d_j^4 \rightleftharpoons
d_i^5(t_{2g}^4e_g)d_j^3(t_{2g}^2e_g)$ excitations within an
$\langle ij\rangle$ bond, with energies
$\varepsilon(^4T_1,^4\!T_1)\simeq U+8J_H/3$,
$\varepsilon(^4T_1,^4\!T_2)\simeq U+2J_H/3$,
$\varepsilon(^4T_2,^4\!T_1)\simeq U+4J_H$,
$\varepsilon(^4T_2,^4\!T_2)\simeq U+2J_H$, expressed by the same
elements $U$ and $J_H$ as above, where the states $^4T_m$
($^4T_n$) label the symmetry of $d_i^5$ ($d_j^3$) excited
configurations, respectively. The hopping element between the
$t_{2g}$ orbitals involves two-step processes via the oxygen
orbitals in bridge positions (see Fig. 1), and thus one finds that
$t_{\pi}=t/3$. We neglected the correction terms which describe
the anisotropy of $t_{2g}$ superexchange depending on the actual
configuration of $e_g$ electrons.

As we already pointed out, the complete spin-orbital model for the
undoped manganites \cite{Fei99},
\begin{equation}
{\cal H}=H_J(e_g)+H_J(t_{2g})+H_{\rm JT}+H_{\tau}, \label{full}
\end{equation}
includes both above superexchange terms due to $e_g$ and $t_{2g}$
excitations [$H_J(e_g)$ and $H_J(t_{2g})$], the effective
interactions between orbital degrees of freedom which follow from
the JT effect ($H_{\rm JT}$), and a low-symmetry crystal field
($H_{\tau}$). The JT term may be derived from the general energy
expression (\ref{jtminq}) for the cooperative JT effect
\cite{Mil96}. The operator form of the intersite orbital
interaction $\propto\kappa$ (\ref{jtminq}),
\begin{equation}
\label{jtterm} H_{\rm JT}=\kappa\sum_{\langle ij\rangle}\left(
           {\cal P}_{\langle ij\rangle}^{\zeta\zeta}
         -2{\cal P}_{\langle ij\rangle}^{\zeta\xi  }
          +{\cal P}_{\langle ij\rangle}^{\xi  \xi  }\right)
     =2\kappa\sum_{\langle ij\rangle}\left(
           1-2{\cal P}_{\langle ij\rangle}^{\zeta\xi  }\right),
\end{equation}
favors orbital alternation, i.e., one $|\zeta\rangle$ and one
$|\xi\rangle$ as occupied orbitals at two ends of the same bond
$\langle ij\rangle$. In analogy to Eqs. (\ref{projbond}) and
(\ref{projbond1}) we define ${\cal P}_{\langle
ij\rangle}^{\xi\xi}=2P_i^{\xi}P_j^{\xi}$. The second equality in
Eq. (\ref{jtterm}) follows from the completeness relation,
 ${\cal P}_{\langle ij\rangle}^{\zeta\zeta}
+2{\cal P}_{\langle ij\rangle}^{\zeta\xi  }
 +{\cal P}_{\langle ij\rangle}^{\xi  \xi  }=2$. The coefficient $\kappa$ is
a parameter which was estimated in Ref. \cite{Fei99} from the
experimental temperature of the structural phase transition $T_s$.
Eq. (\ref{jtterm}) gives therefore the same energetically favored
orbital configurations as those involved in the superexchange
terms in Eq. (\ref{egterm}) which occur due to the
$|^6\!A_1\rangle$ and $|^4\!A_1\rangle$ excited states. The last
term is the crystal-field splitting of the $e_g$ orbitals
(\ref{htau}) which we reproduce here for completeness,
\begin{equation}
\label{ezh} H_{\tau}=-E_z\sum_{i}\tau_{i}^c.
\end{equation}

The strength of $e_g$ and $t_g$ superexchange can be estimated
fairly accurately from the basic electronic parameters for the Mn
ion as determined from spectroscopy \cite{Zaa90,Boc92,Miz95}, with
an estimated accuracy of $\sim 20\%$. Using $U=7.3$ eV and
$J_H=0.69$ eV, and taking into account that the Mn-Mn hopping
occurs in effective processes via the bridging oxygen, one finds
$t=0.41$ eV which follows from $t=t_{pd}^2/\Delta$, with Mn-O
hopping $t_{pd}=1.5$ eV and charge transfer energy $\Delta=5.5$
eV. This yields $J=t^2/U=23$ meV which we take as a measure of the
$e_g$-part of superexchange, and $J_t=2.1$ meV. The accuracy of
these parameters may be appreciated from the resulting prediction
for the N\'eel temperature of CaMnO$_3$, where a similar
derivation gives for spin $S=3/2$,
\begin{equation}
\label{t2gcamno3} \hat{H}_t= \hat{J}_t\sum_{\langle ij\rangle}
\left(\frac{4}{9}\vec{S}_i\cdot\vec{S}_j-1\right),
\end{equation}
where $\hat{J}_t\simeq J_t(1+J_H/U)$. Using the estimates from
spectroscopy, one obtains $\hat{J}_t=4.6$ meV and thus $T_N=124$ K
\cite{Fei99}, in excellent agreement with the experimental value
$T_N=110$ K \cite{Wol55}.

Here it is worthwhile to discuss the difference between the
present approach and the popular simplified model introduced by
Ishihara, Inoue, and Maekawa \cite{Ish97}. These authors also
emphasized the role of orbital variables in the
$e_g$-superexchange, but assumed that the total spin is conserved
within the $e_g$-subband, and one can thus use the same derivation
as presented in Sec. III.A, with the excitation energies adopted
to the new situation when the $e_g$ electrons interact also with
the $t_{2g}$ core states by the Hund's rule interaction $K$
between one $e_g$ and one $t_{2g}$ electron defined in the same
way as in Eq. (\ref{kondol}). Although the original notation was
somewhat more involved \cite{Ish97}, using our projection
operators (\ref{projbond}) and (\ref{projbond1}), and the spin
projection operators (\ref{projst}), their result may be rewritten
as follows,
\begin{eqnarray}
{H}_{\rm IIM}&=&\sum_{\langle ij\rangle }\left\{
-\frac{t^{2}}{\bar{U}'-\bar{J}_H} \left( \vec{S}_{i}\cdot
\vec{S}_{j}+\frac{3}{4}\right) {\cal P}_{\langle ij\rangle
}^{\zeta \xi } +\frac{t^{2}}{\bar{U}'+\bar{J}_H+K} \left(
\vec{S}_{i}\cdot \vec{S}_{j}-\frac{1}{4}\right) {\cal P}_{\langle
ij\rangle }^{\zeta \xi }\right.         \nonumber \\ & &\hskip
.8cm \left. +\frac{2t^{2}}{\bar{U}+K} \left(
\vec{S}_{i}\cdot\vec{S}_{j}-\frac{1}{4}\right) {\cal P}_{\langle
ij\rangle }^{\zeta\zeta }\right\}, \label{somiim}
\end{eqnarray}
Therefore, the multiplet structure given by Eqs.
(\ref{excd51})--(\ref{excd54}) is now replaced by:
\begin{eqnarray}
\label{exci1} \varepsilon(^6\!A_1) &\simeq & \bar{U}' - \bar{J}_H,
\\ \label{exci2}
\varepsilon(^4\!E\,) &\simeq & \bar{U}' + \bar{J}_H + K,
\\ \label{exci3} \varepsilon(^4\!A_2) &\simeq & \bar{U}  + K.
\end{eqnarray}

It is evident that the Hamiltonian (\ref{somiim}) {\em is not
equivalent\/} to the expression derived using the correct
multiplet structure as given by Eq. (\ref{egterm}) \cite{Fei99}.
First of all, note that $\bar{J}_H$ has here the same meaning as
the Hund's rule element used to describe the exchange interactions
between a pair of $d$ electrons.
Then the excitation spectrum (\ref{exci1})--(\ref{exci3}) should
be equivalent to the excitation spectrum given by Eqs.
(\ref{specd81})--(\ref{specd84}) in the absence of $t_{2g}$ spins
at $K=0$. The correspondence between the Coulomb and exchange
elements in Eqs. (\ref{somiim}) and (\ref{huj}) is given by
$\bar{U}'=U-J_H$ and $\bar{J}_H=J_H$, and the first two energies
of the $|e_g^2\rangle$ excited state agree, but the highest energy
does not. The reason is that the intraorbital Coulomb interaction
$\bar{U}$ is not an independent parameter, but
$\bar{U}=\bar{U}'+2\bar{J}_H$ \cite{Ole83}. Second, the state
$|^4\!A_2\rangle$ comes out to be doubly degenerate in Eq.
(\ref{somiim}) as the processes $\propto\bar{J}_H$ which transfer
a pair of electrons with opposite spins between different orbitals
in Eq. (\ref{huj}) were not included and thus the doublet occurs
as the highest-energy state rather than the state in the middle.
Furthermore, for the choice of parameters of Ref. \cite{Ish97}
with $\bar{U}'=5$, $\bar{U}=7$ and $\bar{J}_H=2$ eV, not only the
structure of levels is incorrect, but even the $|^4\!E\rangle$ and
$|^4\!A_2\rangle$ states have accidental degeneracy. Third, the
parameter $K$ in the case of $d_i^4d_j^4\rightleftharpoons
d_i^3(t_{2g}^3)d_j^5(t_{2g}^3e_g^2)$ excitations is not a free
parameter, but $K=3\bar{J}_H$ (the factor of three comes about due
to three $t_{2g}$ electrons which form a large spin $S=3/2$
interacting with an $e_g$ electron by the Kondo term). This last
condition is obeyed by the actual choice of parameters in Ref.
\cite{Ish97}. Finally, the state $|^4\!A_1\rangle$ has been
completely missed in Eqs. (\ref{exci1})--(\ref{exci3}). Therefore,
we conclude that the Hamiltonian (\ref{somiim}): (i) violates the
SU(2) invariance of superexchange interactions in spin space, and
(ii) does not correspond to the correct multiplet structure of
$d^5$ ions in any nontrivial limit. A common feature with Eq.
(\ref{egterm}) is that FM interactions are enhanced due to the
lowest excited $|^6\!A_1\rangle$ state, but the dependence of the
magnetic interactions on ${\bar J}_H$ is quite different, and it
gives a different answer concerning the stability of the A-AF
phase.

The model of Ishihara {\it et al.\/} \cite{Ish97} contains also a
superexchange term between core spins $t_{2g}$ which takes the
same form as that given in Eq. (\ref{t2gcamno3}). In contrast to
the present derivation of Ref. \cite{Fei99}, no formula for the
interaction $\hat{J}_t$ was derived in Ref. \cite{Ish97}, but
instead the value of $\hat{J}_t$ was fitted in order to reproduce
the stable A-AF ordering, as observed in LaMnO$_3$. This results
in a value of $\hat{J}_t$ overestimated by a factor close to seven
and having no relation to the experimental value of the N\'eel
temperature for CaMnO$_3$.

A comparison between Eqs. (\ref{egterm}) and (\ref{somiim}) is
possible only when the magnetic interactions concern the same
spins $S=2$. Therefore, we have renormalized the terms $\sim
t^2/\varepsilon_n$, where $\varepsilon_n$ is the excitation
energy, and $\hat{J}_t$ in the model of Ishihara, Inoue and
Maekawa \cite{Ish97} to act on the total spins $S=2$. After this
modification, the superexchange terms obtained from both models
may be written in the same form:
\begin{equation}
\label{hamag} {\cal H}_{SE}= \sum_{\langle ij\rangle}\left(
-J_1{\cal P}_{\langle ij\rangle }^{\zeta\xi  } +J_2{\cal
P}_{\langle ij\rangle }^{\zeta\xi  } +J_3{\cal P}_{\langle
ij\rangle }^{\zeta\zeta} + J_{AF}\right) \vec{S}_i\cdot\vec{S}_j,
\end{equation}
where the first interaction $\propto J_1$ is FM, and the remaining
interactions are AF. We present the values of different constants
in Table 2. First of all, the superexchange terms are much smaller
than the hopping integral $t$ which justifies {\it a posteriori\/}
the perturbative approach. Second, taking the multiplet structure
of Mn$^{2+}$ ions one finds a competition between the FM and AF
contributions to the $e_g$-promoted superexchange
(\protect\ref{egterm}), with the AF terms being roughly half of
the largest FM term \cite{Fei99}. As we show below, this ratio
between the FM and AF interactions gives automatically the A-AF
order as observed in LaMnO$_3$.

In contrast, the (effective) interactions $J_n$ found from Eq.
(\ref{somiim}) are not balanced due to overestimated excitation
energies $\varepsilon_n\simeq 13.0$ eV to the low-spin states
($n=2,3$). On one hand, one finds a dominating FM term $J_1=3.48$
meV, not far from $J_1=4.40$ meV found from Eq. (\ref{egterm}),
while on the other hand the AF terms are smaller by a factor close
to two. Therefore, the $e_g$ part of the superexchange alone gives
a FM state in all three directions, and one is forced to simulate
the missing AF interactions by a large superexchange term coming
from the hopping of $t_{2g}$ electrons \cite{Ish97} (see Table 2).
Another consequence of too low values of $J_2$ and $J_3$ is a
rather narrow parameter regime for the stable A-AF order, much
narrower than in the approach using the correct multiplets, where
the A-AF order is generic \cite{Fei99}.

\smallskip
\begin{table}
\hspace{0.5cm} \caption{Magnetic interactions $J_n$ in Eq.
(\protect\ref{hamag}), and the excitation energies $\varepsilon_n$
used in perturbation theory, as obtained from the $d^4$
spin-orbital model (\protect\ref{egterm}) \protect\cite{Fei99} and
from the model of Ishihara, Inoue and Maekawa
(\protect\ref{somiim}) \protect\cite{Ish97}. The excitation
energies $\varepsilon_2$ and $\varepsilon_3$ in the case of Eq.
(\protect\ref{egterm}) were averaged over two states which
contribute to $J_n$. } \hspace{0.2cm}
\begin{tabular}{cc|ccc|cc}
\hline
    &     &  \multicolumn{3}{c} {model (\ref{egterm}) \cite{Fei99}} &
             \multicolumn{2}{c} {model (\ref{somiim}) \cite{Ish97}} \\
superexchange & orbital operator
              & $\varepsilon_n$ (eV) & $J_n$ (meV) & $J_i/t$ ($10^{-2}$) &
                $\varepsilon_n$ (eV) & $J_n$ (meV) \\ \hline
$J_1$  & ${\cal P}^{\zeta\xi  }_{\langle ij\rangle}$ & 3.80 & 4.40
& 1.07 &
                                                               3.0 & 3.48 \\
$J_2$  & ${\cal P}^{\zeta\xi  }_{\langle ij\rangle}$ & 7.53 & 2.21
& 0.54 &
                                                              13.0 &  0.81 \\
$J_3$  & ${\cal P}^{\zeta\zeta}_{\langle ij\rangle}$ & 8.68 & 2.44
& 0.60 &
                                                              13.0 & 1.62 \\
$J_{AF}$ &          --           & 8.91 & 0.53 & 0.13 &  --  &
3.50 \\
\end{tabular}
\label{ishihara}
\end{table}

\begin{figure}
\centerline{\epsfig{figure=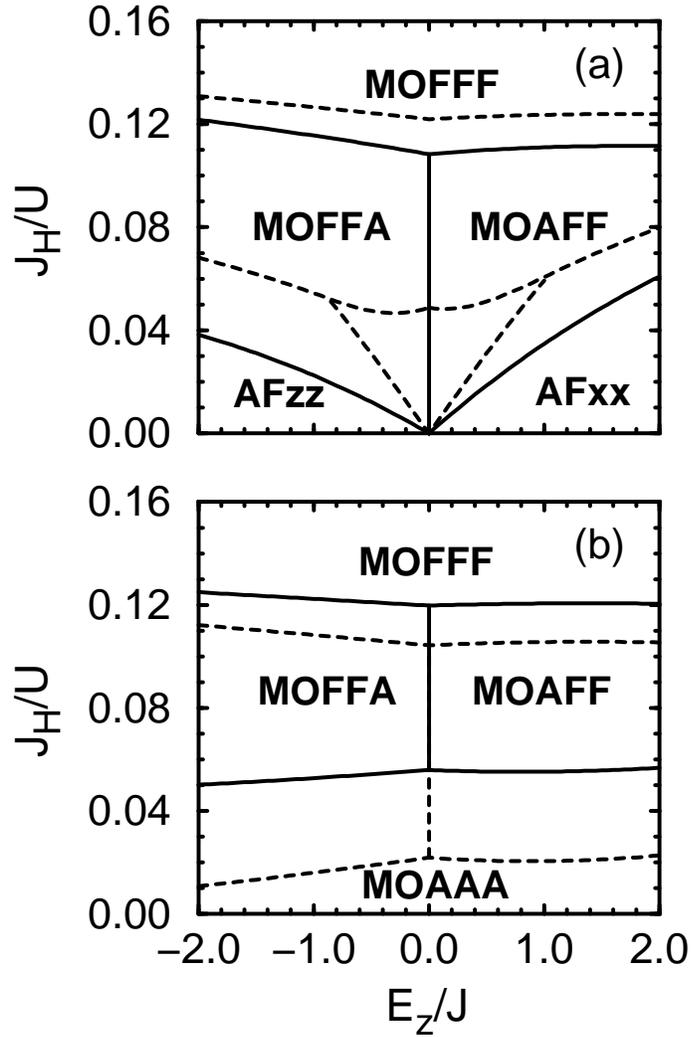,width=12cm}}
\smallskip
\caption {Classical phase diagram of the manganite model
(\protect{\ref{full}}):
 (a) no JT effect ($\kappa=0$): $J_t=0$      (full lines) and
                                $J_t=0.092J$ (dashed lines),
     with the AFxx and AFzz phases separated by a MOAAA phase;
 (b) including JT effect ($\kappa=0.5J$): $J_t=0$      (dashed lines) and
                                          $J_t=0.092J$ (full lines)
                      (after Ref. \protect\cite{Fei99}).
} \label{fig:d4mfa}
\end{figure}

The classical phases found in the present $d^4$ model (Fig.
\ref{fig:d4mfa}) are similar to those shown before in Fig.
\ref{fig:d9mfa}, but at present the spins of $e_g$ and $t_{2g}$
electrons are aligned and form total spins $S=2$. Due to the
identical interactions in the orbital sector, the MF phase diagram
of the $e_g$-part of the manganese $d^4$ model (\ref{full}),
$H=H_J(e_g)+H_{\tau}$, at $T=0$ is similar to that of the cuprate
$d^9$ spin-orbital model (\ref{somcu}) \cite{Fei97,Ole00} analyzed
in Sec. III.B: at large positive (negative) $E_z$, one finds AF
phases with either $|x\rangle$ (AFxx) or $|z\rangle$ (AFzz)
orbitals occupied, while MO phases with occupied orbitals given as
linear combinations (\ref{mixing}) of the basis states $|x\rangle$
and $|z\rangle$ are favored by increasing $J_H$. As in the $d^9$
case, the $e_g$ part of superexchange (\ref{egterm}) is frustrated
at $J_H=0$ and $E_z=0$. However, due to the large value of total
spin $S=2$, the quantum corrections are expected to be much
smaller and we may assume that the classical order is robust, at
least certainly at finite $J_H$ and in the presence of $t_{2g}$ AF
superexchange (\ref{t2gterm}).
The A-AF order is found at finite $J_H/U\simeq 0.1$ [Fig.
\ref{fig:d4mfa}(a)]. If $E_z<0$ the spin order is FM (AF) in the
$(a,b)$ planes (along the $c$-axis) in the MO{\footnotesize FFA}
phase, while at $E_z>0$ two similar phases, MO{\footnotesize AFF}
and MO{\footnotesize FAF}, are degenerate. For the parameters
appropriate for LaMnO$_3$, with $J_H/U\simeq 0.095$, one finds a
MO{\footnotesize FFA}/MO{\footnotesize AFF} ground state, i.e.,
{\em A-AF magnetic order\/}, while a FM (MO{\footnotesize FFF})
phase is stable only at $J_H/U>0.12$. The result is therefore
qualitatively similar to the classical phase diagram of the $d^9$
model (Fig. \ref{fig:3d}). We emphasize again that the region of
stability of the A-AF phase is somewhat modified by
$t_{2g}$-superexchange [Fig. \ref{fig:d4mfa}(a)], but this change
is small as $J_t\ll J$.

\begin{figure}
\centerline{\epsfig{figure=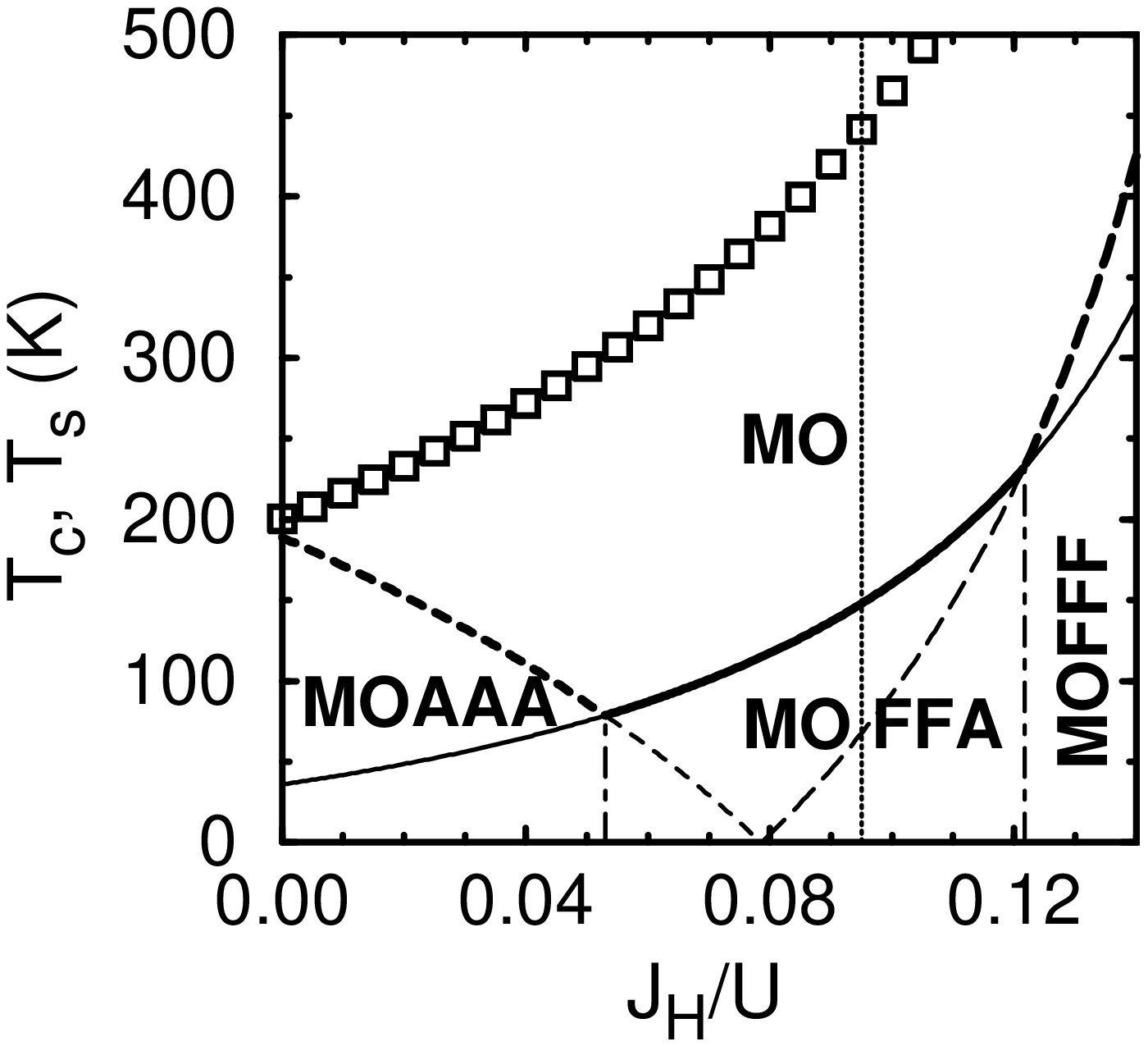,width=7cm}}
\smallskip
\caption {Magnetic transition temperatures $T_c$ ($J=23$ meV,
$E_z=0$, $J_t=0.092J$,
 $\kappa=0.5J$) for:
 MOAAA (dashed line), MOFFF (long-dashed line), and MOFFA (full line)
 phases, and $T_s$ for the structural (MO) phase transition at $\kappa=0$
 (squares). The dotted line indicates realistic $J_H/U=0.095$ for LaMnO$_3$
 (after Ref. \protect\cite{Fei99}).
 }
\label{fig:finiT}
\end{figure}

The value of $\kappa$ in Eq. (\ref{jtterm}) is the only parameter
of the manganite model (\ref{full}) which had to be determined
from experiment \cite{Fei99}. First of all, the $e_g$ part of
superexchange alone favors the alternation of orbitals, i.e.,
maximizes the average $\langle {\cal P}_{\langle ij\rangle
}^{\zeta\xi}\rangle$ and one can verify that a structural
transition follows already from this term. However, the transition
temperature obtained in this way amounts to $T_s^e\simeq 440$ K
(see Fig. \ref{fig:finiT}) and is far below the experimental value
of $T_s^{exp}\simeq 750$ K. Therefore, the structural phase
transition is induced to a larger extent by the orbital
interactions generated by the JT effect included via the H$_{\rm
JT}$ term (\ref{jtterm}). Their contribution to the mean-field
value of the transition temperature is equal
$T_s^{\kappa}=6\kappa$. Using the typical ratio between the MF
value $T_s$ and the experimental transition temperature
$T_s^{exp}$ for pseudospins 1/2, $T_s/T_s^{exp}=1.6$, it follows
from $T_s=T_s^e+T_s^{\kappa}=1200$ K that $6\kappa\simeq 760$ K.
Thus $\kappa\simeq 11$ meV which agrees with the estimate of
Millis that $\kappa>10$ meV (where he did not separate the orbital
interactions into those originating from the JT effect and from
superexchange). Therefore, the representative value $\kappa/J=0.5$
was considered in Ref. \cite{Fei99}.
Due to the structural transition which occurs in LaMnO$_3$ at
$T_s\simeq 750$ K, one has to use a MF theory for the phases with
coupled order parameters, assuming that the order parameters
corresponding to $\langle{\vec S}\rangle$,
$\langle{\vec\tau}\rangle$, and $\langle{\vec S}{\vec\tau}\rangle$
are independent variables \cite{Dit80}. When the orbital
interactions induced by the JT effect are included via H$_{\rm
JT}$ term (\ref{jtterm}), one gets with decreasing temperature
first a transition due to the largest orbital interactions to the
phase with $\langle{\vec\tau}\rangle\neq 0$. This induces
significant changes in the phase diagram close to the origin
$(E_z,J_H)=(0,0)$ [Fig. \ref{fig:d4mfa}(b)]. In fact, the JT
coupling $\kappa$ enforces alternating orbitals with
$\theta=\pi/4$ in Eqs. (\ref{orbmoffa}) which stabilizes G-AF spin
order in the MO{\footnotesize AAA} phase at small $J_H/U$, while
the pure AFxx and AFzz phases are suppressed. However, in the
physically interesting regime of larger $J_H/U$ the orbital order
is mainly driven by the $e_g$-superexchange interactions
(\ref{egterm}). Besides, from Fig. \ref{fig:d4mfa} one can see
that the A-AF phase is not affected by the JT coupling in the
physical regime of parameters for the LaMnO$_3$.

At finite temperature one may study the competition of different
magnetic phases. One finds that the same magnetic phases develop
in the presence of orbital ordering at finite temperature (Fig.
\ref{fig:finiT}) as those found independently at $T=0$. For the
experimental value of $J_H/U\simeq 0.095$ the magnetic transition
into the A-AF phase is obtained at $T_{N}^{\rm MF}\simeq 148$ K
which after reduction gives a prediction for the experimental
transition temperature of $T_{N}\simeq 106$ K, in reasonable
agreement with the actually observed value of 136 K \cite{Kaw96}.
\begin{figure}
\centerline{\epsfig{figure=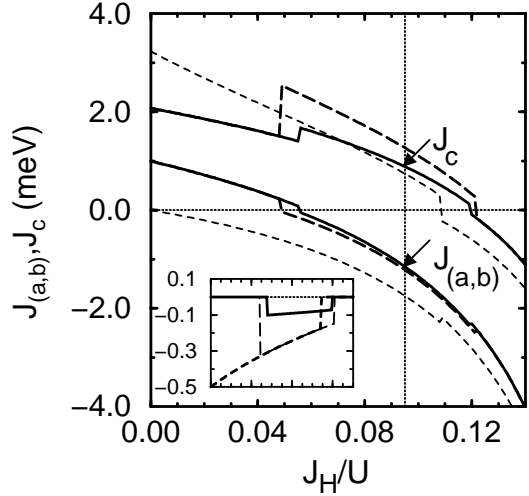,width=8.5cm}}
\smallskip
\caption {Exchange interactions $J_{(a,b)}$ and $J_c$ in the
    ground state for increasing $J_H/U$, for $J=23$ meV and:
    $J_t=0$,      $\kappa=0$    (dashed lines),
    $J_t=0.092J$, $\kappa=0$    (long-dashed lines), and
    $J_t=0.092J$, $\kappa=0.5J$ (full lines).
    The jumps in $J_{(a,b)}$ and $J_c$ correspond to phase transitions of
    Fig. \protect\ref{fig:d4mfa}. The inset gives the values of $\cos
    2\theta$ defined as in Eqs. (\ref{orbmoffa}),
    in the same range of $J_H/U$ for the optimal orbital state
    (after Ref. \protect\cite{Fei99}).
 }
\label{fig:bothj}
\end{figure}
As a remarkable success of the model (\ref{full}), the magnetic
interactions obtained for the A-AF phase are close to the
experimental values. The FM interactions in the $(a,b)$ planes
$J_{(a,b)}$, and the AF interactions along the $c$-axis $J_c$ are
found from Eqs. (\ref{egterm}) and (\ref{t2gterm}) by averaging
over the orbital operators in the actual classical ground state
(Fig. \ref{fig:bothj}). The values obtained at $J_H/U=0.095$:
$J_{(a,b)}=-1.15$ meV and $J_c=0.88$ meV \cite{Fei99} are somewhat
larger than the experimental values of -0.83 and 0.58 meV
\cite{Hir96,Hir96a}, respectively. It is interesting to realize
that although the observed A-AF phase would be obtained from the
$e_g$ superexchange alone, the ratio of the above exchange
constants is then $J_c/|J_{(a,b)}|=2.25$. Instead, if the $t_{2g}$
superexchange and the JT term are included in the present model,
the value $J_c/|J_{(a,b)}|=0.77$ is found which agrees very well
with the experimental ratio of 0.7 \cite{Hir96,Hir96a}. The
$t_{2g}$ term alone enhances the AF interactions, but does modify
the orbital ordering, and gives thus a different ratio
$J_c/|J_{(a,b)}|=1.04$. Therefore, we conclude that all the
interaction terms in Eq. (\ref{full}) are important in order to
explain the experimentally observed exchange interactions in
LaMnO$_3$. These values could not be reproduced up to now by {\it
ab initio\/} calculations performed in LSDA which fails even to
give the correct sign of $J_c$ \cite{Sol96}, while the values
somewhat  closer to experiment and to those of Ref. \cite{Fei99}
were obtained recently in the HF approximation: $J_{(a,b)}=-0.44$
meV and $J_c=0.11$ meV \cite{Kap00}.

\subsection{ Polaronic regime at low hole doping }

The spin-orbital model (\ref{full}) which includes the complete
superexchange and the orbital interactions which follow from the
JT effect provides a good starting point to describe the interplay
between magnetic and orbital interactions in doped manganites. It
reproduces very well the spin and orbital pattern observed in the
ground state of LaMnO$_3$ \cite{Fei99} which can be well
understood as mainly due to the spin and orbital interactions
contained in the $e_g$ superexchange (\ref{egterm}), with small
corrections given by the JT effect. Thus, the undoped case
involves spin, orbital and lattice degrees of freedom. When the
holes are doped into this ordered ground state, the charge degrees
of freedom occur in addition and modify the ground state
properties. Their understanding represents one of the most
challenging and complex problems in the physics of CMR compounds
\cite{Ima98,Ram97}.

We need to generalize the interaction terms which appeared in the
undoped case to the present situation, taking into account the
constraint which follows from large $U$ that prevents the
occupancy of any Mn ion by more that one $e_g$ electron.
Therefore, we restrict the space to Mn$^{3+}$ and Mn$^{4+}$
configurations. Furthermore, at low hole concentrations the doped
holes are trapped and form polarons, as follows from the {\it
insulating\/} magnetic phases found at low doping in the phase
diagrams of Fig. \ref{fig:expphd}. Therefore, a polaronic model
was developed in Ref. \cite{Sces99}, with the low-energy effective
model Hamiltonian given by:
\begin{equation}
{\cal H}=H_J(e_g)+H_J(t_{2g})+H_{\rm JT}+H_{\tau}+H_{pol}.
\label{mantJ}
\end{equation}
It includes superexchange terms due to $e_g$ and $t_{2g}$
electrons [$H_J(e_g)$ and $H_J(t_{2g})$], orbital interactions
($H_{\rm JT}$), the
 crystal-field splitting ($H_{\tau}$)
induced by the JT effect, and the polaronic energy ($H_{pol}$)
which contributes at finite doping with new FM interactions due to
the effective processes involving the excitations around a
Mn$^{4+}$ ion created by a hole.

The superexchange between total spins $S=2$ is generalized with
respect to the undoped case by including the constraint on the
$e_g$ occupation to,
\begin{eqnarray}
H_J(e_g)&=&\frac{1}{4}J\sum_{\langle
ij\rangle}f_i^{}f_i^{\dagger}\left\{
 - \frac{8}{5} \frac{U}{\varepsilon(^6\!A_1)}
   \left(\frac{1}{4}\vec{S}_i\cdot\vec{S}_j+\frac{3}{2}\right)
   {\cal P}_{\langle ij\rangle}^{\zeta\xi}\right. \nonumber \\
& & \left. \hskip 1cm +\left[\frac{U}{\varepsilon(^4\!E)}
   + \frac{3}{5}\frac{U}{\varepsilon(^4\!A_1)} \right]
   \left(\frac{1}{4}\vec{S}_i\cdot\vec{S}_j-1\right)
   {\cal P}_{\langle ij\rangle}^{\zeta\xi}\right.           \nonumber \\
& &\left. \hskip 1cm +\left[ \frac{U}{\varepsilon(^4\!E)}
   + \frac{U}{\varepsilon(^4\!A_2)} \right]
   \left(\frac{1}{4}\vec{S}_i\cdot\vec{S}_j-1\right)
   {\cal P}_{\langle ij\rangle}^{\zeta\zeta}\right\}f_j^{}f_j^{\dagger},
\label{H_Je}
\end{eqnarray}
where $J=t^2/U$, and $t$ is the hopping between $3z^2-r^2$
orbitals along the $c$-axis. The excitation energies are given by
Eqs. (\ref{excd51})--(\ref{excd54}), and the orbital projection
operators have been introduced in Sec. III. The hole operators
$f_i^{\dagger}$ and $f_i$ guarantee that this superexchange term
contributes only if both sites carry a single $e_g$ electron
($f_if_i^{\dagger}=1$), and the orbital operators ${\cal
P}_{\langle ij\rangle}^{\zeta\xi}$ and ${\cal P}_{\langle
ij\rangle}^{\zeta\zeta}$ decide about the strength of a particular
superexchange contribution.

The $t_{2g}$-superexchange is isotropic in leading order,
\begin{eqnarray}
H_J(t_{2g})&=&\sum_{\langle ij\rangle}\left\{
       J_t \left(\frac{1}{4}\vec{S}_i\cdot\vec{S}_j-1\right)
       f_i^{}f_i^{\dagger}f_j^{}f_j^{\dagger}
 +{\hat J}_t\left(\frac{4}{9}\vec{S}_i\cdot\vec{S}_j-1\right)
         f_i^{\dagger}f_i^{}f_j^{\dagger}f_j^{}\right.   \nonumber \\
& &\left.\hskip .5cm +{\bar J}_t
\left(\frac{1}{3}\vec{S}_i\cdot\vec{S}_j-1\right)
         \left( f_i^{\dagger}f_i^{}f_j^{}f_j^{\dagger}
               +f_i^{}f_i^{\dagger}f_j^{\dagger}f_j^{}\right)\right\},
\label{H_Jt}
\end{eqnarray}
with the AF superexchange constants $J_t$, ${\hat J}_t$, and
${\bar J}_t$ obtained from the hopping of $t_{2g}$ electrons for
the pairs of Mn$^{3+}$--Mn$^{3+}$, Mn$^{4+}$--Mn$^{4+}$, and
Mn$^{3+}$--Mn$^{4+}$ ions, respectively. The spin operators
$\vec{S}_i$ correspond to spins $S=2$ and $S=3/2$ of Mn$^{3+}$ and
Mn$^{4+}$ ions, when the number of holes at site $i$ is either
$f_i^{\dagger}f_i^{}=0$ or $f_i^{\dagger}f_i^{}=1$, respectively.
The JT term (\ref{jtterm}) leads to static distortions
\cite{Mil96} which induce intersite orbital interactions
($\propto\kappa$) between the $e_g$ orbitals and is rewritten in a
similar way,
\begin{equation}
H_{\rm JT}=\kappa\sum_{\langle ij\rangle}f_i^{}f_i^{\dagger}\left(
        {\cal P}_{\langle ij\rangle}^{\zeta\zeta}
      -2{\cal P}_{\langle ij\rangle}^{\zeta\xi  }
       +{\cal P}_{\langle ij\rangle}^{\xi  \xi  }\right)f_j^{}f_j^{\dagger}.
\end{equation}

The problem of a Mott insulator doped by a low number of holes is
one of the fascinating problems in the field of strongly
correlated electrons. In the $t$-$J$ model the motion of a hole
added to an AF background is hindered by the string effect
\cite{Shr88}, and the free propagation disappears. In manganites
the situation is even more complex due to the presence of lattice
and orbital degrees of freedom, which are in first place
responsible for hole localization at lower doping concentrations.
In fact, in an orbitally degenerate Mott-Hubbard system there
exists a direct coupling between holes and orbitals due to the
polarization of $e_g$ orbitals in the neighborhood of a hole
\cite{Kil99}. This coupling might be strong enough to form a bound
state of a hole with the surrounding orbitals, leading to a
dressing of hole by orbital excitations \cite{vdB00} (Sec. VII.B),
and yielding a strong reduction of the bandwidth. Such
orbital-hole bound states lead to an exponential suppression of
the bandwidth which makes the system unstable towards hole
localization. Here we assume that the holes loose most of their
kinetic energy due to the lattice distortions of the breathing
mode type (Fig. \ref{fig:jtmodes}), and only the effective
processes survive which excite a hole to its nearest neighbors
\cite{Sces99}.
\begin{figure}
\centerline{\epsfig{figure=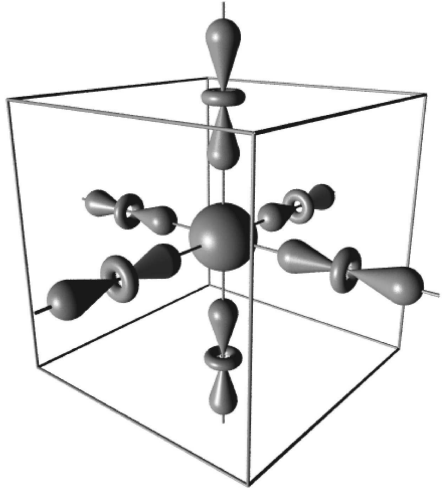,width=6cm}}
\smallskip
\caption {Orbital polaron in the strong coupling limit. Six
$|\zeta\rangle$ occupied
 $e_g$ orbitals at Mn$^{3+}$ ions point towards a central hole at a
 Mn$^{4+}$ site
 (after Ref. \protect\cite{Kil99}).}
\label{fig:polaron}
\end{figure}
The cubic symmetry of perovskite manganites is locally broken
close to a hole which appears as the removal of the $e_g$
degeneracy on the sites adjacent to a doped hole. The origin of
this splitting is in the distortions of the oxygen ions that point
towards the empty site, and by the electrostatic potential between
positive hole and negative electrons. The occupied $e_g$ orbitals
around a hole are likely to be modified, and if any other orbital
interactions were absent, an {\it orbital polaron\/} bound state
would be formed around a hole, with occupied $|\zeta\rangle$
orbitals oriented along the bonds towards the hole site (Fig.
\ref{fig:polaron}).

In the insulating regime, the localized holes are randomly
distributed and form locally lattice polarons with energy $E_p$.
The new interactions originate from the DE mechanism which induces
a FM superexchange for Mn$^{4+}$--Mn$^{3+}$ pairs $\propto
J_p=\bar{J}_p/(1+E_p/2J_H)$ (with $\bar{J}_p=t^2/2E_p$)
\cite{Sces99},
\begin{equation}
H_{pol}= -E_p\sum_{i}f_i^{\dagger}f_i^{}
 -\sum_{\langle ij\rangle} f_i^{\dagger}f_i^{}
  \left[ \bar{J}_p +
  \frac{1}{8}J_p \left(\vec{S}_i\cdot\vec{S}_j-3\right) \right]
  P_{j\zeta} f_j^{}f_j^{\dagger}.
\label{H_pol}
\end{equation}
Similar to the undoped case, the exchange terms do depend now on
the orientation of the occupied $e_g$ orbital on the Mn$^{3+}$ ion
by the projection operator $P_{j\zeta}$ (\ref{projind}), with the
largest FM contribution when $\langle P_{j\zeta}\rangle=1$, i.e.,
for the directional $3z^2-r^2$-type orbitals $|\zeta\rangle$
oriented along the bonds towards the Mn$^{4+}$ ion (Fig.
\ref{fig:polaron}). The classical phase diagram can be now
investigated in a similar fashion as in Sec \ref{sec:somd9}.B,
assuming that the holes are localized and randomly distributed
over the lattice. This assumption corresponds to the dilute limit
and describes well the features of polaronic phase in the low
doping regime $x=1-n<0.16$.

\begin{figure}
\centerline{\epsfig{figure=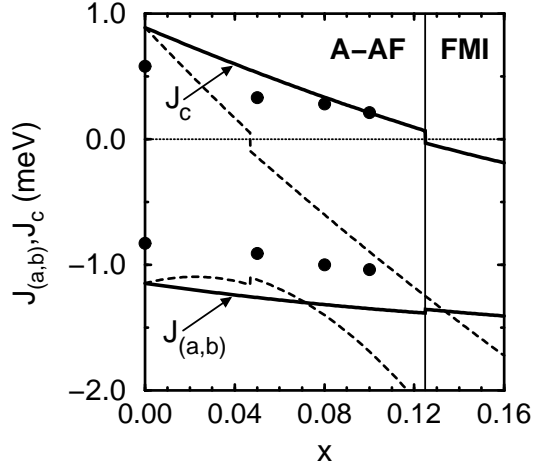,width=8.5 cm}}
\smallskip
\caption {Exchange interactions $J_{(a,b)}$ and $J_c$ in the A-AF
and FMI ground state
    of La$_{1-x}$Ca$_x$MnO$_3$ ($J_H/U=0.095$, $\kappa=0.5J$, $E_p=0.67$ eV) for
    rigid orbital order (full lines) and with $|\zeta\rangle$-type  occupied
    $e_g$ orbitals around Mn$^{4+}$ ions as in Fig. \protect\ref{fig:polaron}
    (dashed lines). The experimental data points \protect\cite{Mou99} are shown
    by filled circles (after Ref. \protect\cite{Sces99}).
} \label{fig:jpol}
\end{figure}

We have seen that for realistic parameters taken from spectroscopy
the A-AF observed in LaMnO$_3$ at $x=0$ is reproduced (Sec. V.A).
Instead of studying the phase diagram as a function of parameters,
we address here a simpler question: How the anisotropic exchange
constants $J_{(a,b)}=-1.15$ and $J_c=0.88$ meV found in LaMnO$_3$
\cite{Fei99} change as a function of doping. Localized polarons in
doped systems stabilize locally FM order around Mn$^{4+}$ defects
as $J_p$ is the largest (FM) exchange element, with $J_p\simeq 4J$
taking $E_p\simeq 0.67$ eV \cite{Sces99}, which might provide a
natural explanation of a gradual magnetic transition {\it within
the insulating phase\/} by the DE mechanism. Indeed, if one
assumes that the Mn$^{4+}$ are randomly distributed and that the
orbital ordering on the Mn$^{3+}$--Mn$^{3+}$ bonds remains
unchanged, one finds that the AF coupling $J_c$ is gradually
weakened with increasing doping $x$, while the FM interaction
$|J_{(a,b)}|$ increases much slower (Fig. \ref{fig:jpol}), in good
agreement with recent experiments \cite{Mou99}. However, a good
agreement with the experimental points was obtained assuming that
the occupied orbitals around a hole {\it are not modified\/} to
$|\zeta\rangle$ orbitals shown in Fig. \ref{fig:polaron}. In
contrast, if the directional orbitals $|\zeta\rangle$ are assumed,
a much faster transition to the FM phase follows (Fig.
\ref{fig:jpol}) which does not agree with experiment. This
suggests that the structure of polaron is rather rigid and
determined primarily by lattice distortions and local JT effect
rather than by the electronic interactions. It might be that some
corrections to the perturbative treatment which leads to Eq.
(\ref{H_pol}) are needed, but the present result suggests that the
polarons realized in La$_{1-x}$Sr$_x$MnO$_3$ are not orbital
polarons of Fig. \ref{fig:polaron}.
A better understanding of polarons is very important in order to
get more insight into the observed insulator-metal transition from
the insulating to metallic FM phase with increasing doping. It is
experimentally observed that La$_{1-x}$Ca$_x$MnO$_3$ is metallic
with FM LRO for hole concentrations in the range of $0.17<x<0.5$.
This metallic state can be turned into an insulating one by
increasing the temperature above the Curie temperature T$_C$, or
by decreasing the hole concentration below x$_{crit}\simeq 0.17$.

The character of the first transition, from a metallic ferromagnet
to an insulating paramagnetic state, can be addressed at first
sight in the framework of the interplay between the DE mechanism
and the lattice-polaron formation \cite{Mil99}. Indeed, in this
case the crossover from metallic to insulating state is controlled
by the ratio $\lambda=E_b/E_{kin}$ of polaron binding energy $E_b$
to the average kinetic energy $E_{kin}$ of the charge carriers.
The formation of a bound state with the lattice distortion is
favored only if the loss in kinetic energy is balanced by a gain
in the binding energy. This condition can be reached by raising
the temperature in the DE system, since the loss of FM
correlations is accompanied by a shrinking of the bandwidth, so
that it eventually increases the value of $\lambda$ and induces a
carrier localization.

A different situation is represented by the metal-insulator
transition induced by the varying hole concentration within the FM
phase at low temperature. The total Hamiltonian in the metallic
phase has to include explicitly the kinetic energy of $e_g$
electrons determined by the symmetry of allowed hopping processes,
as presented in Sec. VI. One arrives at the effective $t$-$J$ like
model which is characterized by two energy scales for orbital and
charge excitations: $W_{orb}$ and $W_{ch}$. Assuming that the
orbitals are disordered, i.e., in an orbital liquid state
\cite{Nag97}, the former quantity describes the orbital
fluctuations determined by the value of $J$, and a fraction of the
hopping amplitude proportional to the hole density $\propto xt$
which follows from strong correlations (see Sec. VI.B). The
quantity $W_{ch}$ determines the rate of the charge fluctuations
and the amount of free kinetic energy, determined the bandwidth of
$6t$, which corresponds to the uncorrelated electrons.

If the orbitals order, the charge fluctuations are reduced, i.e.,
the kinetic energy is lowered below its value in a correlated
metal. Since the holes can move coherently even within an
antiferro-type orbital arrangement \cite{vdB00}, it has been
argued \cite{Kil99} that the reduction of the kinetic energy is
only of $\sim 5\%$ including the incoherent processes, while it
amounts to $\sim 30\%$ without these contributions. This suggests
that orbital ordering cannot suffice to explain the observed
localization and one might expect that a more plausible mechanism
of localization involves either the formation of orbital polarons,
as proposed recently by Kilian and Khaliullin \cite{Kil99}, or the
lattice deformations.

In summary, the crossover from a free-carrier to a small-polaron
picture can be triggered in a DE system either by a reduction of
the doping concentration, or by an increase in temperature; the
former acts via an enhancement of the polaron binding energy,
while the latter by constraining the motion of holes via the DE
mechanism. An interesting suggestion that orbital polarons play a
major role in both transitions was put forward recently
\cite{Kil99}. It seems, however, that lattice contribution to
localization is more important and the question about the possible
role of orbital polarons remains open.



\section{ Electronic structure and excitations in doped manganites }

\subsection{ Double exchange in uncorrelated $e_g$ orbitals }

As we have already shown, the orbital degeneracy gives a rich
structure of superexchange which contains competing FM and AF
terms, and allows thus for the formation of anisotropic magnetic
structures, both in cuprates (Sec. III.A) and in manganites (Sec.
V.A). In this Section we will discuss a problem of {\it
double-exchange via degenerate orbitals} in doped manganites. It
is easier to consider first a simpler situation with only few
electrons in the $e_g$ band that is realistic, for example, for
Ca$_{1-y}$Sm$_y$MnO$_3$ \cite{Mai98}, where $y\ll 1$ ($y=1-x$ for
conventional notation as used in La$_{1-x}$A$_x$MnO$_3$). For
$y=0$ this system is a Mott insulator, with high-spin states
$S=3/2$ per site due to the $t_{2g}$ orbitals filled by three
electrons, while $e_g$ states are empty. Upon doping electrons
enter into the doubly degenerate $e_g$ orbitals ($E_z=0$). In
Ca$_{1-y}$Sm$_y$MnO$_3$ a canted spin structure was observed for
$0.07<x<0.12$ \cite{Mai98}. Other data show that C-AF order is
realized in Nd$_{1-x}$Sr$_x$MnO$_3$ for $0.6<x<0.8$, while A-AF
phase is stable for $0.5<x<0.6$ \cite{Aki98}. In
Pr$_{1-x}$Sr$_x$MnO$_3$ the A-AF order exists for $0.5<x<0.7$
\cite{Kaw97}.

The effective Hamiltonian which describes the low energy
properties of electron doped manganites and represents a
generalization of the DE model to degenerate orbitals may be
written in the form suggested by van den Brink and Khomskii
\cite{Bri99},
\begin{eqnarray}
H=-\sum_{\langle ij\rangle\alpha\beta\sigma} t_{ij}^{\alpha\beta}
 (c_{i \alpha \sigma}^{\dagger} c_{j\beta\sigma}^{} + H.c.)
 - J_H \sum_{i\alpha\sigma\sigma{^\prime}} {\vec S}_i\cdot
   c_{i\alpha\sigma}^{\dagger}\vec{\sigma}_{\sigma\sigma{^\prime}}^{}
   c_{i\alpha\sigma{^\prime}}^{}
 + J_{AF}\sum_{\langle ij\rangle}{\vec S}_i\cdot {\vec S}_j.
\label{eq:deeg}
\end{eqnarray}
The first term describes the kinetic energy of $e_g$ electrons
which are labeled by the site index $i$, spin index $\sigma$, and
also by the orbital index $\alpha(\beta)=1,2$, corresponding to
the local basis, e.g., to the $|x\rangle$ and $|z\rangle$
orbitals. As in the DE model, $e_g$ electrons interact by Hund's
rule exchange $\propto J_H$ with core $t_{2g}$ spins, and the
model is made more realistic by adding the superexchange between
$t_{2g}$ spins $\propto J_{AF}$ which might be determined as in
Sec. V.A. However, for the present purpose it was treated as a
parameter \cite{Bri99}. The presence of the orbital degeneracy,
together with the very particular relations between the hopping
matrix elements $t_{ij}^{\alpha\beta}$ \cite{Sla54,Zaa93}, makes
this problem and its outcome very different from the usual DE
model (Sec. II.B).

The quasiclassical approach to study the model (\ref{eq:deeg})
follows the standard route as for the DE mechanism for electrons
which move in a nondegenerate band \cite{And55}. In the first step
we have to determine the spectrum of the $e_g$-electrons ignoring
their interaction with the localized spins. This spectrum is given
by the solution of the matrix equation \cite{Tak98},
\begin{eqnarray}
|| t_{\mu\nu}- \epsilon \delta_{\mu\nu} || =0, \label{eq:ME}
\end{eqnarray}
where
\begin{eqnarray}
\label{eq:hop1} t_{11}&=& -2t_{ab}(\cos k_x + \cos k_y),  \\
\label{eq:hop2} t_{12}&=& t_{21}= -\frac{2}{\sqrt{3}} t_{ab}(\cos
k_x - \cos k_y),  \\ \label{eq:hop3} t_{22}&=& -\frac{2}{3}
t_{ab}(\cos k_x + \cos k_y) -\frac{8}{3} t_c \cos k_z.
\end{eqnarray}
Here $t_{11}$ is the dispersion due to an overlap between two
$|x\rangle$ orbitals, $t_{12}$ --- between a pair of different
orbitals, one $|x\rangle$ and one $|z\rangle$ orbital, and
$t_{22}$ between two $|z\rangle$ orbitals on neighboring sites. In
writing Eqs. (\ref{eq:hop1})--(\ref{eq:hop3}) we have taken into
account the ratios of different hopping integrals
\cite{Sla54,Zaa93}, which are determined by the symmetry of the
$e_g$ wave functions, and introduced the notation $t_{ab}$ and
$t_c$, to be defined further on. The solutions
$\epsilon_{\pm}({\vec k})$ of Eq. (\ref{eq:ME}) are:
\begin{eqnarray}
\epsilon_{\pm}( {\vec k} ) &=& -\frac{4t_{ab}}{3}(\cos k_x+\cos
k_y)-\frac{4t_c}{3}\cos k_z \nonumber \\ &\pm&
 \left\{\left[ \frac{2t_{ab}}{3}(\cos k_x + \cos k_y)-
 \frac{4t_c}{3}\cos k_z \right]^2
 + \frac{4t_{ab}^2}{3} (\cos k_x - \cos k_y)^2 \right\} ^{1/2},
\label{eq:disp}
\end{eqnarray}
These bands were obtained in Refs. \cite{Tak98} and \cite{Bri99}.

Following van den Brink and Khomskii \cite{Bri99}, we now take
into account the interaction of the $e_g$ electrons with the
magnetic background. We assume that the underlying magnetic
structure is characterized by two sublattices, with a possible
canting, so that the angle between neighboring spins in the
$(a,b)$-plane is $\theta_{ab}$ and along the $c$-direction is
$\theta_{c}$. This rather general assumption covers all magnetic
phases shown in Fig. \ref{fig:mo+oo} --AF phases: G-type
(two-sublattice structure, $\theta_{ab}=\theta_{c}=\pi$), A-type
(FM planes coupled antiferromagnetically, $\theta_{ab}=0$ and
$\theta_{c}=\pi$), and C-type structures (FM chains coupled
antiferromagnetically, $\theta_{ab}=\pi$ and $\theta_{c}=0$), and
the FM phase with $\theta_{ab}=\theta_{c}=0$. As assumed in the
nondegenerate DE model \cite{And55}, we then have the effective
hopping matrix elements determined by the spin background:
$t_{ab}=t\cos(\theta_{ab}/2)$ and $t_c=t\cos(\theta_c/2)$. Note
that here $t$ is chosen as the hopping element between two
$|x\rangle$ orbitals in $(a,b)$ planes, unlike in the other
Sections. We show below that solving this problem in the
quasiclassical approximation introduced for the nondegenerate DE
model (Sec. II.B), the energy spectrum (\ref{eq:disp}) is
renormalized by the magnetic order and, because of the
orbital-dependent hopping matrix elements in a degenerate system,
this results in an anisotropic magnetic structure.

When we dope the system by adding electrons, these go first into
states with minimal energy, in our case into the states close to
the $\Gamma$-point at ${\vec k}=0$. Let us first assume that all
doped charges go into the state with the lowest energy, which is
strictly speaking only the case for very small doping ($y\simeq
0$), and neglect for the moment the effects promoted by a finite
filling of the bands. At the $\Gamma$-point the energies are
$\epsilon_{\pm}(0)=-\frac{4}{3}(2t_{ab}+t_c)\pm\frac{4}{3}|t_{ab}-t_c|$.
Using this simplifying assumption which overestimates the kinetic
energy, the total energy per site of the system containing $y$
electrons reads:
\begin{eqnarray}
E(\theta_{ab},\theta_c)&=& \frac{J}{2}
(\cos\theta_c+2\cos\theta_{ab})+y\epsilon_-({\vec k}=0) \nonumber
\\
 &=& -\frac{3J}{2} + 2J \cos^2 (\theta_{ab}/2) +J\cos^2 (\theta_c/2)
     \nonumber \\
 &-&\frac{4}{3} y t\left[  2 \cos (\theta_{ab}/2) + \cos(\theta_c/2)
  + | \cos (\theta_{ab}/2) - \cos (\theta_c/2) |  \right].
\label{eq:energy}
\end{eqnarray}

Minimizing the energy of the doped state (\ref{eq:energy}) over
two angles $\theta_{ab}$ and $\theta_c$, one encounters two
possible situations \cite{Bri99}. If $\cos(\theta_{ab}/2)>\cos
(\theta_c/2)$, then the magnetic structure is A-type-like; in this
case the minimization of the energy with respect to the angles
$\theta_{ab}$ and $\theta_c$ gives
\begin{eqnarray}
\cos\left({\theta_{ab}\over 2}\right)=\frac{t}{J}y, \hskip 1cm
\cos\left({\theta_c\over 2}\right)=0, \label{eq:angle1}
\end{eqnarray}
and the energy of the corresponding state amounts to
\begin{eqnarray}
E^{(1)} = -\frac{3}{2}J-\frac{2t^2}{J} y^2. \label{eq:energy1}
\end{eqnarray}
Physically this state corresponds to an $(a,b)$-plane with a
canted structure, with the spins in neighboring planes being
antiparallel. If instead $\cos(\theta_{ab}/2)<\cos(\theta_c/2)$,
then the magnetic structure is C-type-like and one finds for
$\theta_{ab}$ and $\theta_c$:
\begin{eqnarray}
\cos\left({\theta_{ab}\over 2}\right)=\frac{t}{3J}y, \hskip 1cm
\cos\left({\theta_c\over 2}\right)=\frac{4t}{3J}y.
\label{eq:angle2}
\end{eqnarray}
The energy $E^{(2)}$ of this state is exactly equal to that of the
A-type state $E^{(1)}$. In this situation we have a canted
structure in all three directions, with the spin correlations in
the $c$-direction being closer to the FM state. Thus, in the
lowest-order approximation in $J$, the two solutions
(\ref{eq:angle1}) and (\ref{eq:angle2}) are degenerate. We note
that these phases have in general canted spin structures, and are
therefore different from the magnetic structures shown in Fig.
\ref{fig:mo+oo} in a broad range of parameters. One can easily
show that in this approximation we would have degenerate solutions
up to a concentration $y_c=3J/4t$, beyond which the A-type
solution becomes energetically favorable. In fact, this solution
never evolves into a FM state -- for $y>J/t$ the canting angle
$\theta_{ab}=0$ in Eq. (\ref{eq:angle1}) and the basal plane
becomes FM, while the moments of the neighboring planes are
opposite to each other, i.e., one finds pure A-type
antiferromagnetism. In contrast to that, the C-type-like solution
(\ref{eq:angle2}) would give with increasing $y$ first the state
with completely polarized FM chains, but with the magnetic moments
of neighboring chains pointing in the directions which differ by a
certain angle $\theta_{ab}$, and finally, at $y=3J/t$, also the
angle $\theta_{ab}=0$ and one finds an isotropic FM state.
However, the A-type solution has a lower energy in this regime of
parameters; thus one never finds a FM state in this approximation,
contrary to experiment.

With increasing electron doping the higher lying band states will
be filled which will modify the above picture. We report the same
calculation as above, but taking into account that the $e_g$
electrons gradually fill the available band states \cite{Bri99}.
One finds that at very low doping concentrations the A-type
solution given by Eq. (\ref{eq:angle1}) and the C-type solution
given by Eq. (\ref{eq:angle2}) are indeed the magnetic structures
of lowest energy, but the degeneracy of these states is lifted and
the A-type solution has always a somewhat lower energy than the
C-type solution. This has a simple physical reason. In the A-type
structure the dispersion of the bands is strictly 2D, so that the
density of states (DOS) at the band edge is finite. For the C-type
solution of Eq. (\ref{eq:angle2}), the bands have a highly
anisotropic but already 3D character, leading typically to a
vanishing DOS at the band edge. Therefore, the A-type magnetic
structure is stabilized as it has a larger DOS at the band edge.
At a somewhat higher doping level, however, the quasi-1D peak in
the DOS close to the band edge starts to play a role and can cause
the transition to a C-type state.

\begin{figure}
\centerline{\psfig{figure=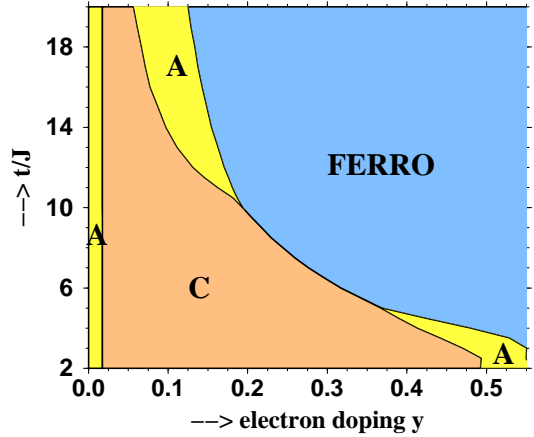,width=7cm}}
\smallskip
\caption {Phase diagram of the DE model with degenerate $e_g$
bands. Depending on the electron doping concentration and the
ratio of the $e_g$ bandwidth and the $t_{2g}$ superexchange one
finds either AF A-type (A), or C-type (C), or FM (FERRO) order
(after Ref. \protect\cite{Kho99}).}
 \label{fig:phase}
\end{figure}

The complete phase diagram of DE model with degenerate $e_g$
orbitals (\ref{eq:deeg}) is presented in Fig. \ref{fig:phase}. As
we have discussed above, the sequence of phases follows from the
modulation of the DOS by the DE mechanism. The phase diagram has
some similarity to that obtained by Maezono, Ishihara and Nagaosa
\cite{Mae98}, and qualitatively reproduces a transition from the
FM to C-AF phase, as obtained in Nd$_{1-x}$Sr$_x$MnO$_3$ for
$x>0.5$ ($y<0.5$) \cite{Aki98}. Experimentally, the phases found
for $y<0.5$ are insulating (Fig. \ref{fig:expphd}), and the
present model gives a C-phase which is susceptible to disorder and
likely to become insulating in this range of parameters for
$t/J_{AF}\simeq 0.25$. However, for realistic values of parameters
$t/J_{AF}\simeq 0.01$ (Table 2), it would give a FM phase except
for a narrow range of low doping $y<0.1$ (Fig. \ref{fig:phase}).
Thus, the model (\ref{eq:deeg}) is not complete and we believe
that lattice distortions play an important role, changing the
balance between the kinetic energy and magnetic interactions and
leading to a different phase diagram.
\subsection{ Spin-waves in a metallic phase }

Unfortunately, the phase diagram shown in Fig. \ref{fig:phase}
does not allow to conclude anything about the favored magnetic
phases near the filling of one $e_g$ electron per atom, i.e., in
hole doped La$_{1-x}$Sr$_x$MnO$_3$ with $x<0.5$, and other related
compounds. In this case the $e_g$ electrons are strongly
correlated and one cannot use the DE for degenerate orbitals as
discussed in the previous Section. Due to strong on-site Coulomb
interactions, the motion of $e_g$ electrons is then allowed only
in the restricted space, without creating double occupancies which
leads to the superexchange as explained in Sec. V.A. If a
half-filled system is doped, the motion of holes in the $e_g$ band
becomes possible when it is accompanied by a backflow of electrons
which carry spin and orbital index. Thus, one has to consider a
hopping problem which resembles the $t$-$J$ model. In order to
make such a model realistic for doped manganites, the degeneracy
of $e_g$ orbitals and the previously derived form of superexchange
in the doped system (\ref{H_Je}) and (\ref{H_Jt}) have to be
included.

The problem of DE and the resulting phase diagram in the regime of
hole doping belongs to the unsolved problems. The approximate
solutions were presented by several groups
\cite{Mae98,Kil99,Ish00}, but either the models were
oversimplified, or only certain aspects of the phase diagram were
treated. We shall discuss this problem in Sec. VIII; here we
analyse only the FM metallic phase and show how the spin-waves
follow from DE and superexchange magnetic interactions. Thus we
consider an effective $t$-$J$ model for doped
La$_{1-x}$A$_x$MnO$_3$ compounds in the FM regime \cite{Ole99},
\begin{equation}
{\cal H}=H_t+H_J(e_g)+H_J(t_{2g}), \label{mantj}
\end{equation}
where $H_t$ describes the correlated hopping in the $e_g$-band
(\ref{hkin}), with the operators
$\hat{c}_{i\alpha\sigma}^{\dagger}$ which act in the projected
space without double occupancies in $e_g$ orbitals, and $H_J(e_g)$
and $H_J(t_{2g})$ stand for the SE terms (\ref{egterm}) and
(\ref{t2gterm}), respectively.

The spectrum of magnetic excitations in FM phase is very important
for a better understanding of the physics of manganites.
Experimentally the spin-waves were measured by Perring {\it et
al.\/} in La$_{0.7}$Pb$_{0.3}$MnO$_3$ \cite{Per96} along the main
high-symmetry directions, and an isotropic dispersion was found.
Similar observation were made by Endoh {\it et al.\/} in
La$_{1-x}$Sr$_{x}$MnO$_3$ \cite{End97}. They found out that unlike
in localized FM systems, the energy of the spin-wave near the zone
boundary is not approximately equal to $k_BT_C$, where $T_C$ is
the Curie temperature, but is significantly larger. This can be
interpreted as an experimental proof that the ferromagnetism in
doped manganites is itinerant. However, the stiffness constant $D$
at low momenta is almost constant for different compounds at
$x\simeq 0.3$ doping \cite{Fer98}. This universality is puzzling
and suggests that a common mechanism is responsible in first place
for the Goldstone modes and the spin dynamics at low momenta
${\vec q}$, while some other processes might be responsible for a
much broader spectrum of the observed Curie temperatures $T_C$.
First we will address the origin of spin-waves and show that the
stiffness constant is approximately determined by the DE mechanism
in a {\it correlated\/} $e_g$ band, $D\simeq J_{\text{DE}}S$,
where $J_{\text{DE}}$ is the effective exchange constant which
couples spins $S\simeq 2$. The corrections will come from
superexchange which operates as well in doped systems and {\it
counteracts\/} the DE mechanism.

In order to construct the spin excitations, we separate first the
spin dynamics from charge and orbital dynamics. In reality all
these processes which involve different degrees of freedom are
coupled and are described by fermion operators
$\hat{c}_{i\alpha\sigma}^{\dagger}$ in $H_t$. We represent such
electron operators by fermion operators
$\{f_{ix}^{\dagger},f_{iz}^{\dagger}\}$ which carry an orbital
index, and by a Schwinger boson operator, either
$a_{i\uparrow}^{\dagger}$ or $a_{i\downarrow}^{\dagger}$,
depending on the spin $\sigma$ of the moving electron,
\begin{equation}
\hat{c}_{i\alpha\sigma}^{\dagger}=
f_{i\alpha}^{\dagger}a_{i\sigma}^{\dagger}. \label{electron}
\end{equation}
As in Sec. V.A, we use a local basis for the occupied $e_g$
orbitals: $|x\rangle\equiv |x^2-y^2\rangle$ and $|z\rangle\equiv
|3z^2-r^2\rangle$. The decomposition (\ref{electron}) allows to
simplify the MF analysis in the large-$U$ limit, where the Hilbert
space contains only two kind of stated: Mn$^{3+}$ $(d^4)$ ions if
a single $e_g$ electron is present, and Mn$^{4+}$ $(d^3)$ ions if
an $e_g$ electron was removed. Formally this can be written using
the configurations \cite{pm99}:
\begin{equation}
|i\theta,M\rangle_4=f_{ix}^{\dagger}b_{ix}^{\dagger}|M\rangle,
\hskip .7cm
|i\epsilon,M\rangle_4=f_{iz}^{\dagger}b_{iz}^{\dagger}|M\rangle,
\hskip .7cm |i,m\rangle_3=b_{i0}^{\dagger}|m\rangle,
\label{allstates}
\end{equation}
with $M$ being the component of spin $S=2$ for Mn$^{3+}$ sites,
and $m$ being the component of the core $t_{2g}$-spin $S=3/2$ for
Mn$^{4+}$ sites. The spin part is described by Schwinger bosons in
a standard way \cite{Aue94}:
$|M\rangle=(a_{i\uparrow}^{\dagger})^{S+M}(a_{i\downarrow})^{S-M}|0)$,
where $|0)$ is the vacuum state, and similar for $|m\rangle$. The
operators $b_{ix}^{\dagger}$ and $b_{iz}^{\dagger}$ in Eqs.
(\ref{allstates}) are slave boson operators which carry the
orbital index and are introduced in order to restrict the physical
space in analogy to the Kotliar-Ruckenstein bosons in the Hubbard
model \cite{Kot86}. The empty boson $b_{i0}^{\dagger}$ counts
Mn$^{4+}$ ions and has a similar meaning. Using these boson
operators we implement the local constraints,
\begin{eqnarray}
 a_{i  \uparrow}^{\dagger}a_{i  \uparrow}^{}
+a_{i\downarrow}^{\dagger}a_{i\downarrow}^{}+b_{i0}^{\dagger}b_{i0}^{}&=&2S,
\label{constrs}                                     \\
 b_{ix}^{\dagger}b_{ix}^{}+b_{iz}^{\dagger}b_{iz}^{}
+b_{i0}^{\dagger}b_{i0}^{}&=&1. \label{constro}
\end{eqnarray}
They restrict the physical space to contain no double occupancy in
the $e_g$ orbitals. In addition, the number of fermions is equal
to the number of bosons for each orbital,
$f_{i\alpha}^{\dagger}f_{i\alpha}^{}=b_{i\alpha}^{\dagger}b_{i\alpha}^{}$,
$\alpha=x,z$.

As a next step, we rewrite the hopping Hamiltonian $H_t$ assuming
the FM metallic phase and derive the effective fermion problem
with the constraints replaced by the operator expressions
$z_{i\alpha}$ which contain the bosons used in Eq. (\ref{constro})
that accompany individual hopping processes in Eq. (\ref{hkin})
\cite{pm99},
\begin{eqnarray}
H_t&=&-\tilde{t}\sum_{\langle ij\rangle\perp,\sigma}
a_{i\sigma}^{\dagger}z_{iz}^{\dagger}f_{iz}^{\dagger}
                                     f_{jz}^{}z_{jz}^{}a_{j\sigma}^{}
                                                             \nonumber \\
&-&\case{1}{4}\tilde{t}\sum_{\langle ij\rangle\parallel,\sigma}
a_{i\sigma}^{\dagger}\left[3z_{ix}^{\dagger}f_{ix}^{\dagger}f_{jx}^{}z_{jx}^{}
+z_{iz}^{\dagger}f_{iz}^{\dagger}f_{jz}^{}z_{jz}^{}\right.
\nonumber \\ & &\hskip 1.5cm \left.\pm\sqrt{3}\left(
 z_{ix}^{\dagger}f_{ix}^{\dagger}f_{jz}^{}z_{jz}^{}
+z_{iz}^{\dagger}f_{iz}^{\dagger}f_{jx}^{}z_{jx}^{}\right)\right]a_{j\sigma}^{},
\label{seraht}
\end{eqnarray}
where the renormalized hopping $\tilde{t}$ is introduced in order
to compensate for the extra factors resulting from the Schwinger
boson operators, $\sim a_{i\sigma}^{\dagger}a_{j\sigma}^{}$, when
the kinetic energy is determined from $H_t$. The terms
$z_{i\alpha}$ contain the orbital bosons $b_{i\alpha}^{\dagger}$
and an empty boson $b_{i0}^{\dagger}$, and are similar to the
respective factors $z_{i\sigma}$ used in the spin problem
\cite{Kot86}. The first sum includes the bonds $\langle
ij\rangle\perp$ along the $c$-axis, while the second sum includes
the bonds $\langle ij\rangle\parallel$ within the $(a,b)$ planes.
The MF theory is constructed by averaging over the orbital and
empty bosons, and one finds the simpler form of the hopping
Hamiltonian,
\begin{eqnarray}
H_t&=&-\tilde{t}\sum_{\langle ij\rangle\perp,\sigma}
a_{i\sigma}^{\dagger}q_{iz}^*f_{iz}^{\dagger}f_{jz}^{}q_{jz}^{}a_{j\sigma}^{}
                                                             \nonumber \\
& &-\case{1}{4}\tilde{t}\sum_{\langle ij\rangle\parallel,\sigma}
a_{i\sigma}^{\dagger}\left[3q_{ix}^*f_{ix}^{\dagger}f_{jx}^{}q_{jx}^{}
+q_{iz}^*f_{iz}^{\dagger}f_{jz}^{}q_{jz}^{}\right.   \nonumber \\
& &\hskip 1.5cm \left.\pm\sqrt{3}\left(
 q_{ix}^*f_{ix}^{\dagger}f_{jz}^{}q_{jz}^{}
+q_{iz}^*f_{iz}^{\dagger}f_{jx}^{}q_{jx}^{}\right)\right]a_{j\sigma}^{},
\label{htsb}
\end{eqnarray}
where the renormalization factors at site $i$ ($x=1-n$),
\begin{equation}
q_{ix}=\left(\frac{2x}{1+x+(1-x)\cos 2\theta_i}\right)^{1/2},
\hskip .7cm q_{iz}=\left(\frac{2x}{1+x-(1-x)\cos
2\theta_i}\right)^{1/2}, \label{qsb}
\end{equation}
depend on the angle $\theta_i$ which defines the occupied orbital
state (\ref{mixing}). In the FM state we assume that the orbitals
are homogeneous, and the angles correspond to an average orbital
state, $\theta_i=\theta$.

Now we consider the Schwinger boson operators in Eq. (\ref{htsb}).
At low temperatures the magnetic moment of FM metallic manganites
is almost fully saturated. It is therefore reasonable to expand
Eq. (\ref{htsb}) around a FM ground state. Technically this is
done by condensing the spin-up Schwinger bosons
$a_{i\uparrow}^{}\simeq\sqrt{2\bar{S}}$ (if the magnetic moments
point upwards), and by treating the spin-down Schwinger bosons in
leading order of $1/\bar{S}$ expansion to describe spin-wave
excitations around this ground state. Assuming that the spins are
pointing upwards in the MF state, we derive from the constraint
(\ref{constrs}) the following expansion of the
$a_{i\uparrow}^{\dagger}$-bosons \cite{pm99},
\begin{equation}
a_{i\uparrow}^{}=\sqrt{2\bar{S}-a_{i\downarrow}^{\dagger}a_{i\downarrow}^{}}
\simeq\sqrt{2\bar{S}}\left(
1-\frac{1}{4\bar{S}}a^{\dagger}_{i\downarrow}
a_{i\downarrow}^{}\right), \label{saddlesch}
\end{equation}
where the effective spin $2\bar{S}=2S-x$ is defined using the MF
state in Eq. (\ref{constrs}), with $\langle
b_{i0}^{\dagger}b_{i0}^{}\rangle=x$. We see that the extra factor
generated by Schwinger bosons is equal $2\bar{S}$, and thus one
has to use $\tilde{t}=t/2\bar{S}$ in Eq. (\ref{seraht}). The final
form of the hopping Hamiltonian in the correlated $e_g$ band is
obtained after replacing the Schwinger bosons by
Holstein-Primakoff boson operators $a_i$. As the expansion
(\ref{saddlesch}) is performed around the FM state, we use
therefore $a_{i\downarrow}\equiv a_i$ to study the fluctuations,
\begin{equation}
\sum_{\sigma}a_{i\sigma}^{\dagger}a_{j\sigma}^{}\simeq
 2\bar{S}-\frac{1}{2}\left(
  a_i^{\dagger}a_i^{}+a_j^{\dagger}a_j^{}-2a_i^{\dagger}a_j^{}\right).
\label{saddleexp}
\end{equation}
This expansion leads to the following form of the kinetic energy
term,
\begin{equation}
H_t = -\sum_{\langle ij \rangle\alpha\beta}t_{ij}^{\alpha\beta}
q_{i\alpha}^{}q_{j\beta}^{}f_{i\alpha}^{\dagger}f_{j\beta}^{}
+\frac{1}{4\bar{S}}\sum_{\langle ij \rangle\alpha\beta}
t_{ij}^{\alpha\beta}q_{i\alpha}^{}q_{j\beta}^{}
f_{i\alpha}^{\dagger}f_{j\beta}^{}\Big(
  a_i^{\dagger}a_i^{}+a_i^{\dagger}a_{i}^{}-2a_i^{\dagger}a_j^{}\Big).
\label{htsbexp}
\end{equation}

\begin{figure}
      \epsfysize=50mm
      \centerline{\epsffile{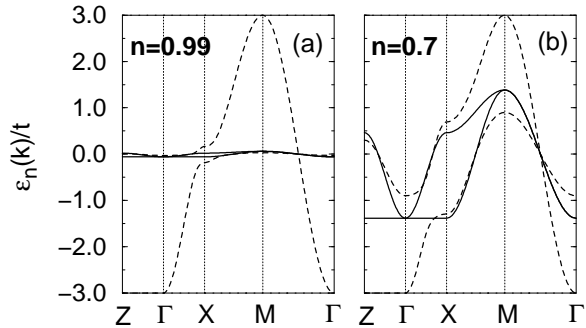}}
\caption {Correlated band structure as obtained for two
 densities of $e_g$ electrons: (a) $n=0.99$, and (b) $n=0.70$, in the
 FM disordered state with the same electron density in both $e_g$ orbitals,
 $\langle n_x\rangle=\langle n_z\rangle=n/2$ (full lines), and for the
 fully polarized orbital liquid state \protect\cite{Nag97} with
 $\langle n_x\rangle=n$ (dashed lines) (after Ref. \protect\cite{Fei00}). }
\label{fig:coba}
\end{figure}

Consider first the zeroth-order Hamiltonian,
\begin{equation}
H_t^{(0)} = -\sum_{\langle ij
\rangle\alpha\beta}t_{ij}^{\alpha\beta}
q_{i\alpha}^{}q_{j\beta}^{}f_{i\alpha}^{\dagger}f_{j\beta}^{}.
\label{htsbbands}
\end{equation}
It describes the orbital model with correlated fermions; the
hopping amplitudes $\propto
t_{ij}^{\alpha\beta}q_{i\alpha}^{}q_{j\beta}^{}$ depend on the
actual shape of occupied $e_g$ orbitals \cite{Fei00}. This problem
was studied qualitatively by van Veenendaal and Fedro \cite{Vee99}
who pointed out that the asymmetry in the magnetic phase diagrams
of doped manganites between $x>0.5$ and $x<0.5$ doping follows
from the essentially uncorrelated electrons in the former, and
strongly correlated electrons that avoid each other in the latter
case. The correlations cause strong band narrowing in the regime
of $x\to 0$ due to the reduction of the hopping elements by the
$q_{ix}$ and $q_{iz}$ factors in Eq. (\ref{qsb}). One finds that
the correlated band structure is drastically reduced in the
disordered states with $\cos 2\theta=0$, while smaller reductions
are found when the occupied orbitals are close to either
$|x\rangle$ or $|z\rangle$ states (Fig. \ref{fig:coba}). By this
mechanism one finds a tendency towards an {\it orbital liquid\/}
state as observed by Ishihara, Yamanaka and Nagaosa \cite{Nag97}.
Such a symmetry breaking in the orbital space happens easily in a
2D case, where the kinetic energy drives the system into the phase
with $|x\rangle$ orbitals occupied \cite{Mac99}, while there are
also other possibilities to stabilize a disordered state in three
dimensions \cite{Fei00}, and thus the orbital liquid of this type
is unlikely.

\begin{figure}
\centerline{\epsfig{figure=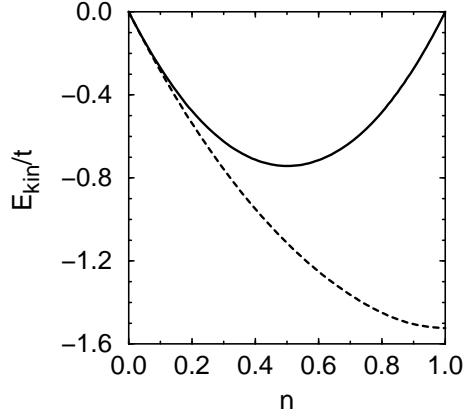,width=7cm}} \caption {Kinetic
energy as a function of $e_g$ band filling $n$ for the correlated
(\protect\ref{htsbbands}) (solid line) and uncorrelated
 (\protect\ref{eq:deeg}) electrons (dashed line)
 (after Ref. \protect\cite{Fei00}).
} \label{fig:ke}
\end{figure}

The orbital model (\ref{htsbbands}) in the limit of $U\to\infty$
is equivalent to the spinless fermion problem. The band structure
collapses at $n=1$ to a line at $\omega=0$, and the kinetic energy
vanishes. In analogy to the Hubbard model in $U\to\infty$ limit,
the kinetic energy, $E_{\rm kin}=\langle H_t^{(0)}\rangle$, is
symmetric with respect to the filling of $n=0.5$, i.e., the system
gains kinetic energy at increasing hole concentration in the
regime of $0<x<0.5$ ($1.0>n>0.5$) (Fig. \ref{fig:ke}). In
contrast, free $e_g$ electrons would give a metallic behavior with
the largest kinetic energy near $n=1$. This demonstrates that the
correlations due to the orbital degree of freedom have to be
explicitly included in order to reproduce the insulating state in
the limit of $x\to 0$.
After studying DE in degenerate $e_g$ orbitals (Sec. VI.A), it
might be expected that this mechanism produces in first instance
anisotropic magnetic phases, such as A-AF and C-AF states.
However, these solutions are {\it de facto} stabilized really by
the superexchange interactions in Ref. \cite{Bri99}. However, we
have shown in Sec. V.A (see Table 2) that the realistic
superexchange constants are much smaller that the hopping element
$t$, and thus the magnetic properties in the highly doped regime
are determined primarily by the DE term (\ref{htsbbands}).
Consequently, it is allowed to assume that the ground state is FM,
and to expand around this state which leads to the DE part of the
spin dynamics described by the first order term in Eq.
(\ref{htsbexp}) when the orbital and charge dynamics are averaged
out,
\begin{equation}
H_t^{(1)}\simeq\frac{1}{4\bar{S}}\sum_{\langle
ij\rangle\alpha\beta}
t_{ij}^{\alpha\beta}q_{i\alpha}^{}q_{j\beta}^{} \langle
f_{i\alpha}^{\dagger}f_{j\beta}^{}\rangle\Big(
  a_i^{\dagger}a_i^{}+a_i^{\dagger}a_{i}^{}-2a_i^{\dagger}a_j^{}\Big).
\label{htsbspins}
\end{equation}
Using the renormalization factors for the correlated band
structure (\ref{qsb}), this averaging yields the magnon dispersion
due to the DE mechanism \cite{pm99},
\begin{equation}
\omega_{\vec q}=\frac{t}{2\bar{S}}\left(2[1-\gamma_+({\vec
q})]R_{ab} +[1-\gamma_z({\vec q})]R_c\right), \label{fmswrr}
\end{equation}
where $\gamma_+({\vec q})$ and $\gamma_z({\vec q})$ are defined by
Eqs. (\ref{gammap}) and (\ref{gammaz}), respectively, and $t=0.40$
eV is the largest local hopping element between two $3z^2-r^2$
orbitals along the $c$-axis \cite{Fei99}. The lattice sums
$R_{ab}$ and $R_c$ can be easily derived and depend on the average
occupancy of $|x\rangle$ and $|z\rangle$ orbitals, $n_x$ and
$n_z$.

For a disordered state represented by the orbitals with complex
coefficients \cite{Fei00}, and $\cos 2\theta=n_x-n_z=0$, the
Gutzwiller renormalization factors (\ref{qsb}) become equal,
$q_{ix}=q_{iz}=q$, and it is convenient to express the result as,
\begin{equation}
\omega_{\vec q}=zJ_{\rm DE}\bar{S}(1-\gamma_{\vec q}),
\label{fmsw}
\end{equation}
with $\gamma_{\vec q}$ defined as in Eq.
(\ref{defDispRelFermicubic}), $z=6$, and the effective FM exchange
constant determined by the kinetic energy of correlated electrons,
\begin{equation}
J_{\rm DE}=\frac{tq^2}{2\bar{S}^2}\,\frac{1}{N}\sum_{{\vec k}\nu}
\langle f_{{\vec k}\nu}^{\dagger}f_{{\vec k}\nu}^{}\rangle\cos
k_{\alpha}. \label{J_DE}
\end{equation}
The summation in Eq. (\ref{J_DE}) runs over the occupied band
states in each subband $\nu$, and the band structure of degenerate
$e_g$ orbitals (\ref{eq:disp}) is renormalized by $q^2=2x/(1+x)$.
It gives the same result for any cubic direction $\alpha=a,b,c$,
so only one component $\propto\cos k_{\alpha}$ was included. Thus,
one finds isotropic spin-waves with the spin-wave stiffness
constant $D=J_{\rm DE}\bar{S}$. The qualitative result of the
Kondo lattice model with a nondegenerate band \cite{Fur96} is
therefore reproduced in the correlated $e_g$ band. The advantage
is that the present approach \cite{pm99} is more straightforward
and may be directly used to study the corrections of the DE result
due to the superexchange terms (\ref{H_Je}) and (\ref{H_Jt}). Both
superexchange terms may be expanded using Schwinger bosons around
the FM state and lead to an isotropic reduction of the effective
FM DE interactions, if a disordered orbital state with
$\langle{\cal P}_{\langle ij\rangle}^{\zeta\xi}\rangle=1/2$ and
$\langle{\cal P}_{\langle ij\rangle}^{\zeta\zeta}\rangle=1/2$ is
considered. Under these circumstances, the $e_g$ superexchange
(\ref{H_Je}) is isotropic and weakly AF, taking the parameters
$U=7.3$ eV and $J_H=0.69$ eV, as given in Sec. V.A. The second AF
term comes from the $t_{2g}$ superexchange (\ref{H_Jt}), with the
constants $J_t=2.1$ meV, ${\hat J}_t=4.6$ meV, and ${\bar
J}_t=5.5$ meV for the pairs of Mn$^{3+}$--Mn$^{3+}$,
Mn$^{4+}$--Mn$^{4+}$, and Mn$^{3+}$--Mn$^{4+}$ ions, respectively
(see Sec. V.B).
\begin{figure}
\centerline{\epsfig{figure=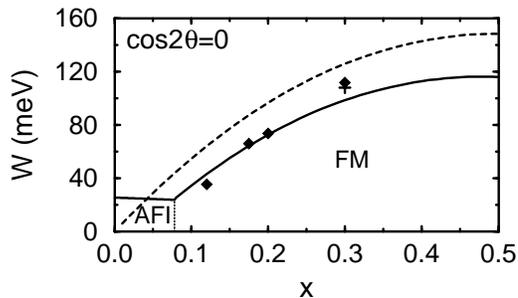,width=8cm}} \caption {Width $W$
of the magnon dispersion in FM manganites La$_{1-x}$A$_x$MnO$_3$
 as a function of doping $x$, as obtained including only the DE mechanism
(\protect\ref{fmsw}) (dashed line), and both DE and superexchange
 contributions (full line). In the AF insulating (AFI) phase at $x<0.08$ only
 the anisotropic superexchange interactions contribute. Experimental
 points correspond to:
 La$_{1-x}$Sr$_x$MnO$_3$ \protect\cite{End97} (diamonds) and
 La$_{0.7}$Pb$_{0.3}$MnO$_3$ \protect\cite{Per96} (cross)
 (after Ref. \protect\cite{pm99}).}
 \label{fig:sw}
\end{figure}
If hole doping $x$ increases, first the A-AF insulating (AFI)
state is modified as reported in Sec. V.B. In this regime the
superexchange dominates over the DE term, and a different
expansion around the A-AF phase with the polaronic FM
superexchange (\ref{H_pol}) has to be used to derive the
spin-waves (Sec. V.B). We present the numerical result for the
total magnon width $W$ obtained by these different approaches for
the AFI and FM phases in Fig. \ref{fig:sw}. The theory predicts an
observed increase of the magnon width $W$ with increasing doping
$x$ \cite{pm99} due to the DE which dominates in the metallic
regime of $x>0.08$. The DE contribution to $W$ vanishes in the
$x\to 0$ limit, in contrast to the unphysical result of band
structure calculations that ignore electron correlations, and give
the largest FM interactions at $x=0$ \cite{Sol99}, precisely at
the point where the A-AF ordering in insulating LaMnO$_3$ is
observed.

\begin{figure}
\centerline{\epsfig{figure=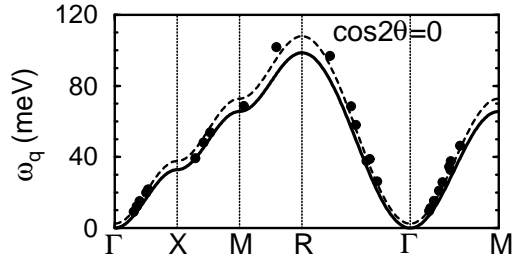,width=8cm}}
\smallskip
\caption {Magnon dispersion $\omega_{\bf q}$ as obtained at
$x=0.3$ doping using
 DE and superexchange contributions (heavy line);
 parameters as in Ref. \protect\cite{Fei99}. Experimental
 data for La$_{0.7}$Pb$_{0.3}$MnO$_3$ (circles and dashed line) are
 reproduced from Ref. \protect\cite{Per96} (after Ref.
 \protect\cite{pm99}).}
\label{fig:perring}
\end{figure}

The magnon dispersion is {\it isotropic\/} if the orbitals are
disordered. We emphasize that orbital ordering leads instead to
anisotropic magnon dispersion, and therefore such states as
obtained recently by Okamoto, Ishihara and Maekawa \cite{Ish00}
would lead to anisotropic spin-wave dispersions not only in the
A-AF phase, but also in the FM regime. The quality of the model
may be best appreciated by an excellent agreement with the
experimental results for $x=0.3$ doped manganite
La$_{0.7}$Pb$_{0.3}$MnO$_3$ \cite{Per96}. The exchange
interactions found in Ref. \cite{pm99} at $x=0.3$:
$J_{ab}\bar{S}=J_c\bar{S}=8.24$ meV are isotropic and reproduce
very well the experimental points (Fig. \ref{fig:perring}). In
conclusion, the magnon dispersion derived from DE for degenerate
$e_g$ orbitals supplemented by smaller superexchange terms agrees
well with the experimental findings in FM metallic manganites
\cite{Per96,End97}.

\subsection{Magnon softening in ferromagnetic manganites}

One of the puzzling features is nonuniversality of the magnetic
and transport properties of FM manganites at doping $x\simeq 0.30$
\cite{End97}. As we have shown in Sec. VI.B, the spin-wave
dispersion which follows from the DE model for correlated $e_g$
electrons (\ref{fmsw}) is isotropic in all three cubic directions
and of the nearest-neighbor Heisenberg type. Recently, unexpected
deviations from this dispersion with a peculiar softening of the
magnon spectrum close to the magnetic zone boundary has
experimentally been observed in compounds with low values of
$T_C$. Quite prominent in this respect are measurements of the
spin dynamics of the FM manganese oxide
Pr$_{0.63}$Sr$_{0.37}$MnO$_3$ \cite{Hwa98}. While exhibiting
conventional Heisenberg behavior at small momenta, the magnetic
excitations soften at the boundary of the BZ. Assuming the magnon
dispersion to be of Heisenberg type with an exchange constant
$J_{eff}$, the dispersion is quadratic in the regime of ${\vec
q}\simeq 0$, $\omega_{\vec q}\simeq Dq^2$, with the spin-wave
stiffness $D\propto J_{eff}$. Since the latter also controls the
Curie temperature $T_C\propto J_{eff}$, the ratio of $D$ and $T_C$
is expected to be a universal constant. Manganites, however,
exhibit a pronounced deviation from this behavior: $D/T_C$
increases significantly as one goes from compounds with high to
compounds with low values of $T_C$ \cite{Fer98}. This feature
together with the above magnon softening at the magnetic zone
boundary indicate that the DE model does not suffice and some
specific feature of magnetism in manganites has not yet been
identified.

Here we report shortly an interesting recent proposal by
Khaliullin and Kilian \cite{Kha00} that charge and coupled
orbital-lattice fluctuations are responsible for the unusual
magnon softening. The starting point is not the $t$-$J$ model for
manganites (\ref{mantj}), but a Kondo lattice model with orbital
degeneracy which is a generalization of the DE model
(\ref{kondol}) and takes into account the correlated nature of
$e_g$ electrons. We begin by analyzing the DE part,
\begin{equation}
H = -\sum_{\langle ij \rangle\alpha\beta\sigma}
t_{ij}^{\alpha\beta}\Big(
\hat{c}^{\dagger}_{i\alpha\sigma}\hat{c}^{}_{j\beta\sigma}+H.c.\Big)
-J_H\sum_{i\alpha\sigma\sigma{^\prime}} {\vec S}_i\cdot
   c_{i\alpha\sigma}^{\dagger}\vec{\sigma}_{\sigma\sigma{^\prime}}^{}
   c_{i\alpha\sigma{^\prime}}^{} ,
\label{HHP}
\end{equation}
where the constrained operators
$\hat{c}^{\dagger}_{i\alpha\sigma}$ act in the projected space
without double occupancies in the $e_g$ orbitals, as in Eq.
(\ref{mantj}). Due to the strong Hund's coupling $\propto J_H$,
core spins ${\vec S}_i$ and itinerant $e_g$ spins ${\vec s}_i$ are
not independent of each other; rather a high-spin state with total
on-site spin $S=S_t+\frac{1}{2}$, where $S_t=3/2$, is formed. This
unification of band and local spin subspaces suggests to decompose
the $e_g$ electron into its spin and orbital/charge components. As
we have shown in Sec. VI.B following Ref. \cite{pm99}, the $e_g$
spin can then be absorbed into the total spin, allowing an
independent treatment of spin and orbital/charge degrees of
freedom. The formal procedure which allows for this separation
scheme is realized by introducing Schwinger bosons $d_{i\uparrow}$
and $d_{i\downarrow}$ to describe the $e_g$ spin subspace,
\begin{equation}
s_i^+ = d^{\dagger}_{i\uparrow} d_{i\downarrow}^{}, \hskip .7cm
s_i^- = d^{\dagger}_{i\downarrow} d_{i\uparrow}^{}, \hskip .7cm
s_i^z = \case{1}{2}(d^{\dagger}_{i\uparrow  }d_{i\uparrow  }^{}-
                    d^{\dagger}_{i\downarrow}d_{i\downarrow}^{}),
\label{schweg}
\end{equation}
as well as Schwinger bosons $a^{\dagger}_{i\uparrow}$ and
$a^{\dagger}_{i\downarrow}$ to model the {\it total\/} on-site
spin $S=2$,
\begin{equation}
S_i^+ = a^{\dagger}_{i\uparrow} a_{i\downarrow}^{}, \hskip .7cm
S_i^- = a^{\dagger}_{i\downarrow} a_{i\uparrow}^{}, \hskip .7cm
S_i^z = \case{1}{2}(a^{\dagger}_{i\uparrow  }a_{i\uparrow  }^{}-
                    a^{\dagger}_{i\downarrow}a_{i\downarrow}^{}).
\label{schwt2g}
\end{equation}
These auxiliary particles are subject to the following constraints
that depend on the $e_g$ occupation number $n_i$:
\begin{eqnarray}
d^{\dagger}_{i\uparrow}d_{i\uparrow}^{}
+d^{\dagger}_{i\downarrow}d_{i\downarrow}^{} &=& n_i, \label{CON1}
\\ a^{\dagger}_{i\uparrow}a_{i\uparrow}^{}
+a^{\dagger}_{i\downarrow}a_{i\downarrow}^{} &=& 2S-1+n_i.
\label{CON2}
\end{eqnarray}
This is in fact a different way of writing the constraints
(\ref{constrs}) and (\ref{constro}). By construction, one assumes
in both approaches that high-spin states $S=2$ are realized at the
sites occupied by $e_g$ electrons which corresponds to the large
$J_H$ limit. The creation and annihilation operators for $e_g$
electrons can then be expressed in terms of spinless fermions
$f_{i\alpha}^{\dagger}$ which carry charge and orbital pseudospin
and Schwinger bosons which carry spin in analogy to Eq.
(\ref{electron}),
\begin{equation}
\hat{c}_{i\alpha\sigma}^{\dagger}=
\hat{c}_{i\alpha}^{\dagger}d_{i\sigma}^{\dagger}. \label{hatf}
\end{equation}
The Bose operators are subject to the constraint (\ref{CON1}) that
enforces the operators $d_{i\sigma}$ and $d^{\dagger}_{i\sigma}$
to act only in the projected Hilbert space with one or zero
Schwinger bosons, respectively. Our aim is to absorb the $e_g$
spin into the total spin, which requires to map the $e_g$
operators $d_{i\sigma}$ onto the operators $a_{i\sigma}$ for the
total spin. It is easy to obtain the following mapping
\begin{equation}
d_{i\sigma} = \frac{1}{\sqrt{2S}} a_{i\sigma}. \label{twosch}
\end{equation}

The kinetic energy in Hamiltonian (\ref{HHP}) now describes the
simultaneous transfer of pairs of spinless fermions and Schwinger
bosons, and can be rewritten in the form analogous to Eq.
(\ref{htsb}). However, one can still use the fermion operators
which carry the orbital index, unlike in Sec. VI.B, where orbital
bosons were introduced as well in Eq. (\ref{htsb}). One finds,
\begin{equation}
H_t = -\frac{1}{2S}\sum_{\langle ij \rangle\alpha\beta\sigma}
t_{ij}^{\alpha\beta}\left(\hat{c}^{\dagger}_{i\alpha}\hat{c}_{j\beta}^{}
a^{\dagger}_{i\sigma}a_{j\sigma}^{}+H.c.\right), \label{HTB}
\end{equation}
Thus, the above consideration provides a formal proof that the
above construction with {\it two\/} Schwinger bosons $d_{i\sigma}$
and $a_{i\sigma}$ \cite{Kha00} is completely equivalent to
introducing only {\it one\/} type of Schwinger bosons referring to
the total spins $S=2$ \cite{Ole00}, as shown in Sec. V.B. One
finds therefore the same expansion of the kinetic energy as given
by Eq. (\ref{htsbexp}), except that the Kotliar-Ruckenstein bosons
have not yet been introduced in the present case (\ref{HTB}), and
one may therefore study the renormalization of magnons by quantum
fluctuations beyond the MF theory.

To study the propagation of the magnetic excitations in hole-doped
DE systems, we now derive first the correct spin operator taking
into account the fact that the total on-site spin depends on
whether a hole or an $e_g$ electron is present at that site. The
spin quantum number is $S-\frac{1}{2}$ in the former and $S$ in
the latter case. In general, a spin excitation is created by the
operator $S_i^+$. Expressing this operator in terms of Schwinger
bosons $S^+_i = a^{\dagger}_{i\uparrow}a_{i\downarrow}$, and next
condensing $a_{i\uparrow}$ and mapping $a_{i\downarrow}$ onto the
magnon annihilation operator $a_i$ as in Sec. V.B, the following
representation is obtained,
\[
S^+_i = \left\{
\begin{array}{ll}
\displaystyle \sqrt{2S} \; a_i, & \text{for sites with $e_g$
electron},\\ \displaystyle \sqrt{2S-1} \; a_i, & \text{for sites
with hole}.
\end{array}\right.
\]
Assuming $S$ to be the ``natural'' spin number of the system, the
magnon operator $a_i$ hence has to be rescaled by a factor
$[(2S-1)/(2S)]^{1/2}$ when being applied to hole sites,
\[
A_i = \left\{
\begin{array}{ll}
\displaystyle a_i, & \text{for sites with $e_g$ electron},\\
\displaystyle \sqrt{(2S-1)/(2S)} \; a_i, & \text{for sites with
hole}.
\end{array}\right.
\]
The general magnon operator $A_i$ that automatically probes the
presence of an $e_g$ electron can finally be written as
\begin{equation}
A_i = a_i \Bigg[n_i + \sqrt{\frac{2S-1}{2S}} (1-n_i)\Bigg] \approx
a_i-\frac{1}{4S} (1-n_i)\;b_i, \label{trueboson}
\end{equation}
where $n_i$ is the number operator of $e_g$ electrons at site $i$.
It turns out that $A_i$ represents the true Goldstone operator of
hole-doped DE systems. Its composite character comprises local and
itinerant spin features which is a consequence of the fact that
static core and mobile $e_g$ electrons together build the total
on-site spin. While the itinerant part of $A_i$ is of order $1/S$
only, it nevertheless is of crucial importance to ensure
consistency of the spin dynamics with the Goldstone theorem, i.e.,
to yield an excitation mode whose energy vanishes at zero momentum
\cite{Kha00}.

Having derived the correct magnon operators $A_i$ for doped DE
systems (\ref{trueboson}) allows to study the energies of magnetic
excitations, and their renormalization caused by the coupling of
magnons to other quasiparticles present in the correlated $e_g$
band, and the coupling to the lattice. If such processes are
neglected, one finds the magnon dispersion determined primarily by
the DE in a strongly correlated $e_g$ band, as shown in Sec. VI.B.
If, however, the fermion and orbital variables are not averaged
out, dynamical processes become possible which dress the magnons
and result in finite selfenergy $\Sigma(\omega,{\vec q})$.
Therefore, the magnon spectrum in an interacting system
$\tilde{\omega}_{\vec p}$ contains the many-body correction
expressed by the magnon selfenergy,
\begin{equation}
\tilde{\omega}_{\vec q} = \omega_{\vec
q}+\text{Re}[\Sigma(\omega_{\vec q},{\vec q})]. \label{WTI}
\end{equation}
The MF magnon dispersion $\omega_{\vec q}$ is of conventional
nearest-neighbor Heisenberg form (\ref{fmsw}), and we have seen in
Sec. VI.B that it gives the spin-wave stiffness constant
$D=J_{\text{DE}}S$.

Apart from electron dynamics in the correlated $e_g$ orbitals
which is treated by the projected operators
$\hat{c}^{\dagger}_{i\alpha}$, virtual charge-transfer processes
across the Hubbard gap contribute to the superexchange. At low and
intermediate doping levels the superexchange due to the hopping of
$e_g$ electrons is of importance, and this contributes with a FM
component as the high-spin state gives the only nonvanishing
magnetic term, if electrons are polarized. These superexchange
processes $\propto J_{\text{SE}}$ establish an intersite
interaction, which in the limit of a strong Hund's coupling could
be written in the form \cite{Kha00},
\begin{equation}
H_J = - J_{\text{SE}} \sum_{\langle ij \rangle}
\left(\case{1}{4}-\tau_i^{\alpha}\tau_j^{\alpha}\right)
\left[{\vec S}_i{\vec S}_j + S(S+1)\right]n_i n_j. \label{HJS}
\end{equation}
Thus one might naively expect that the stiffness constant $D$ is
increased by the superexchange $J_{\text{SE}}$ which thus
amplifies the DE effect $\propto J_{\text{DE}}$ in the correlated
$e_g$ band, resulting in $D\propto J_{\text{DE}}+J_{\text{SE}}$
\cite{Kha00}. While the FM term is the only term which contributes
in the ground state, the situation is different in the excited
states and the above simplified picture is incorrect. It has been
shown in Sec. VI.B that the DE effect is {\it reduced by
superexchange interactions\/}. The reason is twofold: (i) a few AF
interactions are generated by the effective processes which
involve $e_g$ electrons (\ref{egterm}), and they have to be
included for the realistic parameters with finite $J_H$ (and not
using $J_H=\infty$), and (ii) the processes which involve $t_{2g}$
electrons contribute as well with AF superexchange terms
(\ref{H_Jt}). Therefore, the AF superexchange due to both $e_g$
and $t_{2g}$ electrons always dominates over a single FM $e_g$
contribution, and thus the total superexchange decreases the
stiffness constant and {\it de facto counteracts\/} the DE in the
FM manganites.

\begin{figure}
\begin{minipage}[b]{0.32\linewidth}
\centering \epsfig{file=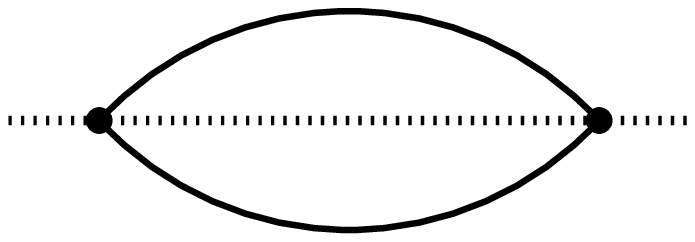,width=\linewidth}\\[3pt] (a)
\end{minipage}
\hfill
\begin{minipage}[b]{0.32\linewidth}
\centering \epsfig{file=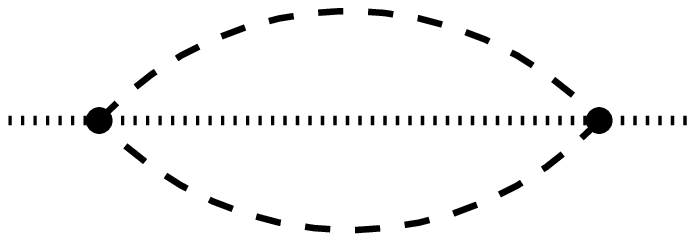,width=\linewidth}\\[3pt] (b)
\end{minipage}
\hfill
\begin{minipage}[b]{0.32\linewidth}
\centering \epsfig{file=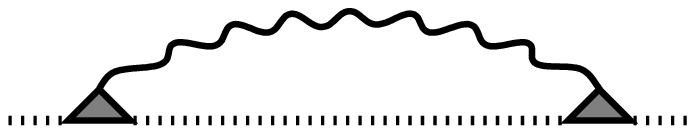,width=\linewidth}\\[3pt] (c)
\end{minipage}\\[6pt]
\caption {Magnon selfenergies describing the effect of magnon
scattering on (a) orbital fluctuations, (b) charge fluctuations,
and (c) phonons. Solid, dashed, dotted, and wiggled lines denote
orbiton, holon, magnon, and phonon propagators, respectively
(after Ref. \protect\cite{Kha00}). } \label{fig:SCO}
\end{figure}

Fortunately, the above incorrect interpretation of Ref.
\cite{Kha00} is of quantitative nature and one may still use the
same expansion of the DE processes around the FM state in order to
study the consequences of magnon interactions with orbital
excitations and with the lattice. Using the slave formalism
similar to that introduced in Sec. VI.B, Khaliullin and Kilian
derived the effective processes which describe the coupling of
magnons to charge and orbital fluctuations separately
\cite{Kha00}. These processes involve always scattering of either
charge (fermionic) or orbital (bosonic) states on the magnons, and
lead in lowest order to the contributions to the magnon selfenergy
$\Sigma(\omega,{\vec q})$ shown in Figs. \ref{fig:SCO}(a) and
\ref{fig:SCO}(b).

The softening of magnons at the zone boundary was observed in
Pr$_{0.63}$Sr$_{0.37}$MnO$_3$, the compound which has a lower
value of $T_C$ \cite{Hwa98}, while no softening was found earlier
in metallic La$_{0.7}$Pb$_{0.3}$MnO$_3$ \cite{Per96}. The
compounds with lower values of $T_C$ are worse metals and become
in some cases insulating \cite{Fer98}, as for instance
Pr$_{1-x}$Ca$_x$MnO$_3$. This suggests that the lattice degrees of
freedom are likely to play an important role in the magnon
softening. The crystal dynamics is given by the Hamiltonian,
\begin{equation}
H_{\rm ph}=\frac{1}{2}K\sum_i{\vec Q}_i^2+K_1\sum_{\langle
ij\rangle} Q_i^{\alpha}Q_j^{\alpha}+\frac{1}{2M}\sum_i{\vec
P}_i^2, \label{phononkk}
\end{equation}
where ${\vec Q}_i=(Q_{2i},Q_{3i})$,
$Q_i^{a(b)}=(Q_{3i}\pm\sqrt{3}Q_{2i})/2$, $Q_i^c=Q_{3i}$, and
$Q_{2i}$ and $Q_{3i}$ are JT phonons of Fig. \ref{fig:jtmodes}.
The coupling of the spin-waves to phonons depends on the ratio
$k_1=K_1/K$.

\begin{figure}
\centering \epsfig{file=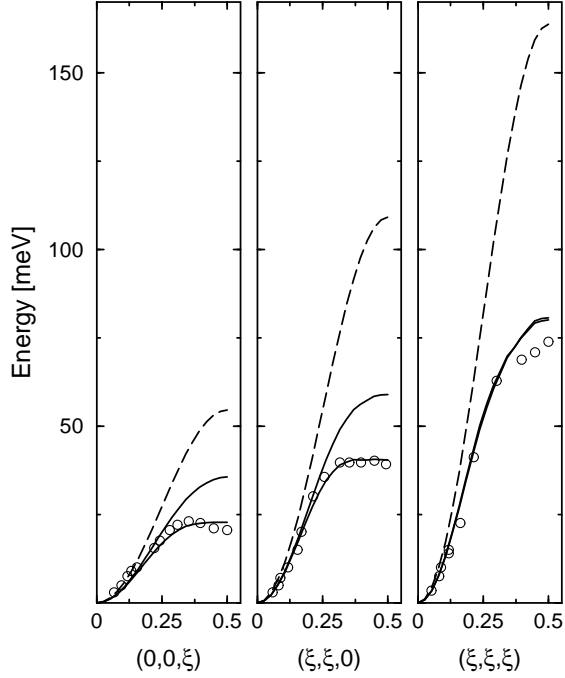,width=0.5\linewidth}\\[6pt]
\caption {Magnon dispersion along $(0,0,\xi)$, $(\xi,\xi,0)$, and
$(\xi,\xi,\xi)$ directions, where $\xi=0.5$ is at the cubic zone
boundary. Experimental data from Ref. \protect\cite{Hwa98} are
indicated by circles, the MF dispersion $\omega_{\vec q}$ is
marked by dashed lines. Solid lines represent the theoretical
result for the dispersion $\tilde{\omega}_{\vec q}$ defined by Eq.
(\protect\ref{WTI}); it includes charge, orbital, and lattice
effects. The upper curve is obtained for dispersionless phonons
with $k_1=0$, the lower one is a fit to the experimental data with
$k_1=-0.33$ corresponding to ferrotype orbital-lattice
correlations (after Ref. \protect\cite{Kha00}).}
 \label{fig:MRN}
\end{figure}

One of the central results of Ref. \cite{Kha00} is that the
coupling between spins and phonons is indirect -- it is mediated
via the orbital channel. Orbital fluctuations couple to the spins
be the DE term (\ref{HTB}), while the coupling of orbitals to
phonons (\ref{hjth1}) admixes low phononic frequencies into
orbital fluctuations. The corresponding effective spin-phonon
interaction is of the form,
\begin{equation}
H_{\rm s-ph} = \sum_{{\vec p}{\vec q}\lambda}g_{{\vec p}{\vec
q}}^{\lambda} (b_{{\vec q}\lambda}^{\dagger}+b_{-{\vec
q}\lambda}^{}) A_{\vec p}^{\dagger}A_{{\vec p}+{\vec q}}^{},
\label{sph}
\end{equation}
where $b_{{\vec q}\lambda}^{\dagger}$ are the phonon creation
operators for the mode $\lambda=1,2$ with momentum ${\vec q}$,
$A_{\vec p}^{\dagger}$ are the Fourier transforms of the boson
operators $A_i$ (\ref{trueboson}), and $g_{{\vec p}{\vec
q}}^{\lambda}$ are the respective coupling constants which depend
on the frequency of the involved phonon mode.
The corresponding diagram which contributes to the magnon
selfenergy is shown in Fig. \ref{fig:SCO}(c). Using perturbation
theory, different processes shown in Fig. \ref{fig:SCO} lead to
the expressions which contain summations over internal momenta.
Such summations were performed numerically using a Monte-Carlo
algorithm in Ref. \cite{Kha00} and give the result shown by solid
lines in Fig. \ref{fig:MRN}. We note that the bare MF dispersion
$\omega_{\vec p}$ is somewhat overestimated for the chosen
parameters, as the contributions coming from the AF superexchange
were neglected \cite{Kha00}. Therefore, the good agreement with
experiment claimed for the stiffness constant $D$ using the
spectroscopic value of $t=0.4$ eV \cite{Boc96} is fortuitous; in
fact it implies that the value of $t$ is somewhat different in
Pr$_{0.63}$Sr$_{0.37}$MnO$_3$ from the above value which gives a
very good agreement for the stiffness constant in
La$_{0.7}$Pb$_{0.3}$MnO$_3$ when the superexchange contributions
are correctly included \cite{pm99}. It is clear, however, that the
experimental magnon dispersion can be reproduced by Eq.
(\ref{WTI}) in the parameter space of the model which includes the
coupling to the lattice (\ref{sph}). In the actual calculation,
with the result shown in Fig. \ref{fig:MRN} \cite{Kha00}, the
phonon contribution was determined using:
$E_{\text{JT}}a_0^2\equiv (g_2a_0)^2/2K=0.004$ eV, $\omega_0=0.08$
eV \cite{Oki95}, and $\Gamma=0.04$ eV. More details may be found
in Ref. \cite{Kha00}.

As the main result, a pronounced softening of magnons at large
momenta can be reproduced in the theory which treats the coupling
of spin waves to the fluctuations of the orbital and lattice
degrees of freedom \cite{Kha00}. Interestingly, charge
fluctuations are found to play only a minor role. This follows
qualitatively from the energy scales -- the spectral density of
charge fluctuations $\propto U$ lies well above the magnon band.
In contrast, orbital and lattice fluctuations have rather low
characteristic frequencies ($\propto xt$ and $\propto\omega_0$,
respectively) and hence may couple stronger to spin-waves. The
precise mechanism of this coupling is not yet completely
understood, however. The numerical study of Ref. \cite{Kha00}
suggests that the presence of JT phonons amplifies the magnon
softening which follows in first instance from orbital
fluctuations. The softening at the zone boundary, which occurs
simultaneously with practically unaffected spin dynamics at small
momenta that enters the spin-wave stiffness $D$ \cite{Fer98},
indicates that the instability towards an orbital-lattice ordered
state is responsible for this phenomenon. The unusual magnon
dispersion experimentally observed in low-$T_C$ manganites
\cite{Hwa98} can hence be understood as a precursor effect of
orbital-lattice ordering.
%


\section{ Orbital degrees of freedom in a ferromagnet }

\subsection{ Orbital excitations }

While the electronic interactions in cuprates might stabilize a
spin liquid in particular situations (Sec. IV), the spin-orbital
model in manganites is in the opposite limit. Large Hund's rule
interactions $\propto J_H$ stabilize the orbital ordered states at
the filling of one $e_g$ electron per site and at low doping, and
FM planes stagger in the A-AF phase. Thus, the spin degree of
freedom can be integrated out and one is left with orbital
dynamics. In the undoped case one finds an anisotropic pseudospin
model with an interesting behavior, as the pseudospin quantum
number is not conserved \cite{vdB99}.

The proper understanding of pure orbital excitations is very
important, both for fundamental reasons, and as a starting point
to consider weakly doped FM $(a,b)$ planes of
La$_{1-x}$A$_x$MnO$_3$. Therefore, we consider an idealized
uniform FM phase. In this case the only superexchange channel
which contributes is the effective interaction via the high-spin
state, and one finds the effective Hamiltonian which describes
$e_g$ electrons in a cubic crystal at strong-coupling
\cite{Fei97,Ish97,Shi97},
\begin{equation}
{\cal H} = H_J(e_g) + H_{\tau}. \label{somfff}
\end{equation}
It consists of the superexchange part $H_J(e_g)$, and the orbital
splitting term due to crystal-field term $H_{\tau}$ (\ref{htau})
which is connected with the uniaxial pressure and was introduced
before in Sec. III.A. An example of such a model is the
superexchange interaction in the FM state of LaMnO$_3$ which
originates from $d_i^4d_j^4\rightleftharpoons d_i^3d_j^5$
excitations into a high-spin $d_j^5$ state $|^6\!A_1\rangle$ (with
$d_i^4\equiv t_{2g}^3e_g$, $d_i^3\equiv t_{2g}^3$, and
$d_i^5\equiv t_{2g}^3e_g^2$). It gives the effective Hamiltonian
with orbital interactions,
\begin{equation}
H_J(e_g) = - \frac{t^2}{\varepsilon(^6\!A_1)}
        \sum_{\langle ij\rangle} {\cal P}_{\langle ij\rangle}^{\zeta\xi},
\label{generic}
\end{equation}
where $t$ is the hopping element between the directional
$3z^2-r^2$ orbitals along the $c$-axis, and $\varepsilon(^6\!A_1)$
is the excitation energy \cite{Fei99}. The orbital degrees of
freedom are described by the projection operators ${\cal
P}_{\langle ij\rangle}^{\zeta\xi}$ (\ref{projbond1}) which select
a pair of orbitals $|\zeta\rangle$ and $|\xi\rangle$, being
parallel and orthogonal to the directions of the considered bond
$\langle ij\rangle$ in a cubic lattice.

\begin{figure}
\centerline{\epsfig{figure=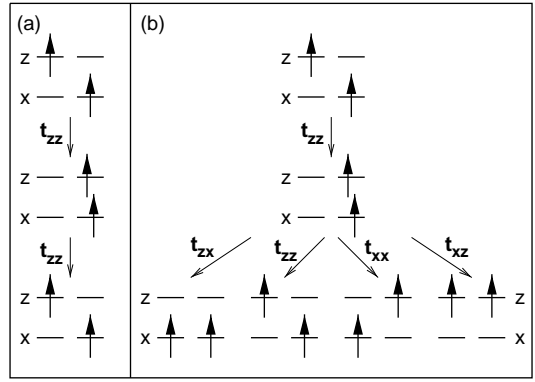,width=7cm}}
\smallskip
\caption{Schematic representation of the virtual
$d_i^4d_j^4\rightarrow d_i^3d_j^5$
 excitations in LaMnO$_3$ for the starting FM configuration
 $d_{iz\uparrow}^{\dagger}d_{jx\uparrow}^{\dagger}|0\rangle$ which
involve
 the high-spin $|^6\!A_1\rangle$ state and generate effective orbital
 superexchange interactions (\protect\ref{eq:hamorb}):
 (a) for a bond along the $c$-axis, $(ij)\perp$;
 (b) for a bond within the $(a,b)$ plane, $(ij)\parallel$
 (after Ref. \protect\cite{vdB99}).}
\label{fig:ose}
\end{figure}

The Hamiltonian (\ref{generic}) has cubic symmetry and may be
written using any reference basis in the $e_g$ subspace. For the
conventional choice of $|x\rangle$ and $|z\rangle$ orbitals, the
above projection operators ${\cal P}_{\langle
ij\rangle}^{\zeta\xi}$ are represented by the orbital operators
$\tau_i^{\alpha}$, with $\alpha=a,b,c$ for three cubic axes,
defined by Eqs. (\ref{orbop0}) and (\ref{orbop1}). We replace them
by {\it pseudospin operators\/} $T^x_i=\frac{1}{2}\sigma^x_i$ and
$T^z_i=\frac{1}{2}\sigma^z_i$, where $\sigma^x_i$ and $\sigma^z_i$
are the Pauli matrices. It is convenient to use the prefactor
$J=t^2/\varepsilon(^6\!A_1)$ in Eq. (\ref{generic}) as the energy
unit for the superexchange interaction [note that this definition
of $J$ is different from that used in Sec. V.A by a factor of
$U/(U-5J_H)\simeq 2$]. Thus, one finds a pseudospin Hamiltonian,
\begin{equation}
\label{eq:hamorb} H_J(e_g) = \case{1}{2}J \sum_{\langle
ij\rangle\parallel} \left[
        T^z_iT^z_j+3T^x_iT^x_i\mp\sqrt{3}( T^x_iT^z_j+T^z_iT^x_j )\right]
           \nonumber \\
     +  2J \sum_{\langle ij\rangle\perp} T^z_i T^z_j ,
\end{equation}
where the prefactor of the mixed term $\propto\sqrt{3}$ is
negative in the $a$-direction and positive in the $b$-direction,
and the meaning of the pseudospin components
$\mid\uparrow\rangle=|x\rangle$ and
$\mid\downarrow\rangle=|z\rangle$ is the same as in Sec. III.A. We
choose the same convention as in Eqs. (\ref{hpara}) and
(\ref{hperp}) that the bonds labeled as $\langle
ij\rangle\!\!\parallel$ ($\langle ij\rangle\!\!\perp$) connect
nearest-neighbor sites within $(a,b)$ planes (along the $c$-axis).
By construction, the superexchange interaction occurs only between
the pairs of ions with singly occupied {\it orthogonal\/} $e_g$
orbitals $|\zeta\rangle$ and $|\xi\rangle$ at two nearest-neighbor
sites (two orthogonal $e_g$ orbitals are then singly occupied in
the intermediate high-spin excited states). The virtual
excitations which lead to the interactions described by Eq.
(\ref{eq:hamorb}) are shown in Fig. \ref{fig:ose}. Here we
neglected a trivial constant term which gives the energy of $-J/2$
per bond, i.e., $-3J/2$ per site in a 3D system. We emphasize that
the SU(2) symmetry is explicitly broken in $H_J$, and the
interaction depends only on two pseudospin operators, $T^x_i$ and
$T^z_i$. Moreover, there is an interesting balance of
symmetry-breaking $\sim T^z_iT^z_j$ and fluctuating $\sim
T^x_iT^x_j$ terms, with Ising-like $c$-bonds, and more quantum
fluctuations on the bonds within the $(a,b)$ planes, assuming the
symmetry breaking with $\langle T^z_i\rangle\neq 0$. However, the
overall coefficients of both types of terms are equal to $3J$
which shows that the symmetry breaking may happen in any spatial
direction and will give {\it the same classical energy\/}.
However, the models in lower dimension will have different
properties and will be more classical. In fact, the 1D model has
only Ising interactions, if the orbitals $|\xi\rangle$ and
$|\zeta\rangle$ are chosen as a basis.
Before analyzing the excitation spectra of the Hamiltonian
(\ref{eq:hamorb}), one has to determine first the classical ground
state of the system. The classical configurations which minimize
the interaction terms (\ref{eq:hamorb}) are characterized by the
two-sublattice pseudospin order, with two angles describing
orientations of pseudospins, one at each sublattice. As usually,
the classical ground state is obtained by minimizing the energy
with respect to these two rotation angles, i.e., by choosing the
optimal orbitals.

Let us consider first the 3D case at orbital degeneracy $E_z=0$.
The superexchange interaction (\ref{generic}) seems to induce the
alternation of {\it orthogonal orbitals\/} in the ground state in
all three directions, which would be equivalent to the {\it
two-sublattice {\rm G-AF} order in the pseudospin space\/}.
Indeed, this configuration gives the lowest energy on the MF level
for individual directions, as the virtual transitions represented
in Fig. \ref{fig:ose} give the largest contribution, if the
hopping involves one occupied and one unoccupied orbital of the
same type (e.g., either directional or planar with respect to the
direction of the bond $\langle ij\rangle$). However, it is not
possible to realize a G-AF state in a 3D lattice, as the orbitals
that are orthogonal in $(a,b)$ planes are not orthogonal along the
$c$-direction.

In order to investigate the ground state on the classical level,
we perform a uniform rotation of $\{|z\rangle,|x\rangle\}$
orbitals at each site,
\begin{equation}
\left( \begin{array}{c}
 |i\bar{\mu}\rangle   \\
 |i\bar{\nu}\rangle
\end{array} \right) =
\left(\begin{array}{cc}
 \ \ \cos\theta &  \sin\theta \\
-\sin\theta &  \cos\theta
\end{array} \right)
\left( \begin{array}{c}
 |iz\rangle   \\
 |ix\rangle
\end{array} \right) ,
\label{newstateimn}
\end{equation}
and generate the new orthogonal orbitals, $|i\bar{\mu}\rangle$ and
$|i\bar{\nu}\rangle$, which are used to determine the energy as a
function of $\theta$. The rotation (\ref{newstateimn}) leads to
the following transformation of the pseudospin operators,
\begin{eqnarray}
T^x_i &\rightarrow&  T^x_i \cos 2\theta - T^z_i \sin 2\theta,
\nonumber\\ T^z_i &\rightarrow&  T^x_i \sin 2\theta + T^z_i \cos
2\theta, \label{eq:rot}
\end{eqnarray}
and the interaction Hamiltonian $H_J$ is then transformed into,
\begin{equation}
{\cal H}^\theta = H_\parallel^\theta + H_\perp^\theta,
\label{hrot}
\end{equation}
\begin{eqnarray}
\label{hrotpara} H_\parallel^\theta &=&  \case{1}{2}J
\sum_{\langle ij\rangle\parallel}
  \left[\right. (2+\cos 4\theta\mp \sqrt{3}\sin 4\theta ) T^x_i T^x_j
  +(2-\cos 4\theta\pm\sqrt{3}\sin 4\theta )T^z_i T^z_j
                                                          \nonumber \\
 & & \hskip 1.1cm -(\sin 4\theta\pm\sqrt{3}\cos 4\theta )
                      (T^x_iT^z_j + T^z_iT^x_j) \left. \right],
                      \\ \nonumber
                      \\
H_\perp^\theta &=& J\sum_{\langle ij\rangle\perp} \left[ \right.
      (1-\cos 4\theta) \ T^x_i T^x_j + (1+\cos 4\theta) \ T^z_i T^z_j
    + \sin 4\theta (T^x_iT^z_j + T^x_iT^z_j)\left. \right].
\label{hrotperp}
\end{eqnarray}
The Hamiltonian given by Eqs. (\ref{hrotpara}) and
(\ref{hrotperp}) has the symmetry of the cubic lattice, but {\em
surprisingly one finds the full rotational symmetry\/} of the
present interacting problem on the classical level at orbital
degeneracy. The occupied orbitals are $|i\bar{\mu}\rangle$ and
$|i\bar{\nu}\rangle$ on $A$ and $B$ sublattice, respectively, and
classically the lowest energy is $E_{\rm MF}=-3J/4$ per site,
independent of the rotation angle $\theta$, as long as the
occupied orbitals are staggered.
\begin{figure}
\centerline{\epsfig{figure=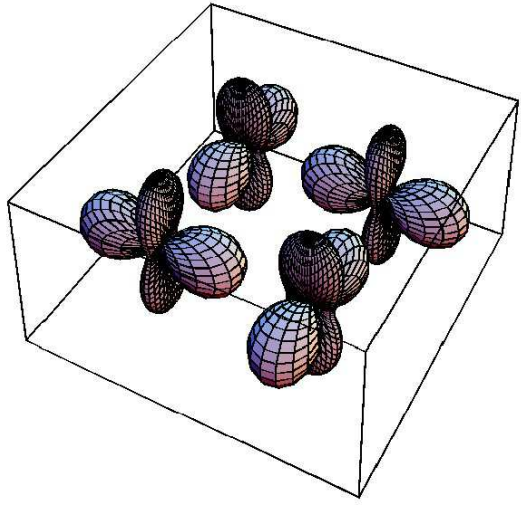,width=7cm}}
\smallskip
\caption { Alternating orbital order in FM cubic LaMnO$_3$ and in
a 2D model of
 a FM $(a,b)$ plane in LaMnO$_3$:
$(|x\rangle + |z\rangle )/\sqrt{2}$ an $(|x\rangle - |z\rangle
)/\sqrt{2}$ as found at $E_z\to 0$ (after Ref.
\protect\cite{vdB99}). } \label{fig:oo}
\end{figure}
A finite orbital field $E_z\neq 0$ breaks the rotational symmetry
on the classical level. It acts along the $c$-axis, and it is
therefore easy to show that the ground state in the limit of
$E_z\to 0$ is realized by the alternating occupied orbitals being
symmetric/antisymmetric linear combinations of $|z\rangle$ and
$|x\rangle$ orbitals, i.e., the occupied states correspond to the
rotated orbitals (\ref{newstateimn}) $|i\bar{\mu}\rangle$ and
$|i\bar{\nu}\rangle$ on the two sublattices with an angle
$\theta=\pi/4$, shown in Fig. \ref{fig:oo}. In particular, this
state is different from the alternating directional orbitals,
$3x^2-r^2$ and $3y^2-r^2$, which might have been naively expected.
It follows in the limit of degenerate orbitals from the
'orbital-flop' phase, in analogy to a spin-flop phase for the
Heisenberg antiferromagnet at finite magnetic field. With
increasing (decreasing) $E_z$ the orbitals tilt out of the state
shown in Fig. \ref{fig:oo}, and approach $|x\rangle$ $(|z\rangle)$
orbitals, respectively, which may be interpreted as an increasing
FM component of the orbital polarization in the pseudospin model.

To describe the tilting of pseudospins due to the crystal-field
$\propto E_z$ we make two different transformations
(\ref{newstateimn}) at both sublattices, rotating the orbitals by
an angle $\theta=\case{\pi}{4}-\phi$ on sublattice $A$,
\begin{equation}
\left( \begin{array}{c}
 |i\mu\rangle   \\
 |i\nu\rangle
\end{array} \right) \! = \!
\left(\begin{array}{cc}
 \ \ \cos(\case{\pi}{4}-\phi) & \sin(\case{\pi}{4}-\phi) \\
    -\sin(\case{\pi}{4}-\phi) & \cos(\case{\pi}{4}-\phi)
\end{array} \right)\!
\left( \begin{array}{c}
 |iz\rangle   \\
 |ix\rangle
\end{array} \right) ,
\label{flopi}
\end{equation}
and using a similar transformation to Eq. (\ref{flopi}) with an
angle $\theta=\case{\pi}{4}+\phi$ on sublattice $B$, so that the
relative angle between the {\it occupied\/} orbitals
$|i\mu\rangle$ ($i\in A$) and $|j\nu\rangle$ ($j\in B$) is
$\case{\pi}{2}-2\phi$, and decreases with increasing $\phi$. The
operators $T^x_i$ and $T^z_i$ may be now transformed as in Eqs.
(\ref{eq:rot}) using the actual rotations by
$\theta=\case{\pi}{4}\pm\phi$, as given in Eq. (\ref{flopi}). As
before, the orbitals $|i\mu\rangle$ and $|j\nu\rangle$ are
occupied on two sublattices, $i\in A$ and $j\in B$, respectively,
and the orbital order in the classical state is described by the
transformed operators $\langle T^z_i\rangle=-1/2$ and $\langle
T^z_j\rangle=+1/2$, respectively.

Using the new operators $T^x_i$ and $T^z_i$, the transformed
Hamiltonian (\ref{somfff}) takes the form,
\begin{equation}
\label{hflop} {\cal
H}^\phi=H_\parallel^\phi+H_\perp^\phi+H_{\tau}^\phi,
\end{equation}
\begin{eqnarray}
H_\parallel^\phi &=& \frac{1}{2}J\sum_{\langle
ij\rangle\parallel}\left[
         (2\cos 4\phi -1) T^x_i T^x_j
        +(2\cos 4\phi +1) T^z_i T^z_j \right. \nonumber \\
    & & \hskip 1.1cm +2\sin 4\phi(T^x_i T^z_j-T^z_i T^x_j)
    \pm\! \sqrt{3} (T^x_i T^z_j+T^z_i T^x_j)\left.\right],
\label{hfloppara}
\end{eqnarray}
\begin{eqnarray}
H_\perp^\phi = J\sum_{\langle ij\rangle\perp}\left[ \right.
            (\cos 4\phi +1)T^x_iT^x_j +(\cos 4\phi -1)T^z_iT^z_j
     -\sin 4\phi( T^x_i T^z_j - T^z_i T^x_j )\left.\right],
\label{hflopperp}
\end{eqnarray}
\begin{equation}
H_{\tau}^\phi = E_z\sum_i (\lambda_i\sin 2\phi T^z_i - \cos 2\phi
T^x_i), \label{hflopez}
\end{equation}
where $\lambda_i=-1$ for $i\in A$ and $\lambda_i=1$ for $i\in B$.
The energy of the classical ground state is given by,
\begin{equation}
E_{\rm 3D}^{\rm MF}=-\frac{3}{4}J\cos 4\phi
                    +\frac{1}{2}E_z\sin 2\phi,
\label{emf3d}
\end{equation}
and is minimized by,
\begin{equation}
\sin 2\phi=-\frac{E_z}{6J}. \label{cos3d}
\end{equation}
The above result (\ref{cos3d}) is valid for $|E_z|\leq 6J$;
otherwise one of the initial orbitals (either $|x\rangle$ or
$|z\rangle$) is occupied at each site, and the state is fully
polarized ($\sin 2\phi=\pm 1$).

In contrast, in the 2D case the cubic symmetry is explicitly
broken, and the classical state is of a spin-flop type (Fig.
\ref{fig:oo}). It corresponds to alternatingly occupied orbitals
on the two sublattices in the plane, with the orbitals given by
$\theta=\pi/4$ in Eq. (\ref{newstateimn}) at $E_z=0$. A finite
value of $E_z$ tilts the orbitals out of the planar $|x\rangle$
orbitals by an angle $\phi$, and the Hamiltonian reduces to
\begin{eqnarray}
{\cal H}^\phi_{2D}=H_\parallel^\phi+H_{\tau}^\phi,
\end{eqnarray}
as there is no bond in the $c$-direction. The classical energy is
\begin{equation}
E_{\rm 2D}^{\rm MF}=-\frac{1}{4}J(2\cos 4\phi+1)
                    +\frac{1}{2}E_z\sin 2\phi.
\label{emf2d}
\end{equation}
Therefore, one finds the same energy of $-3J/4$ as in a 3D model
at orbital degeneracy. This demonstrates a particular {\it
frustration of orbital superexchange interactions\/}, where the
orbital energy cannot be gained from the third direction once the
orbitals have been optimized with respect to the other two. The
bonds along the third direction only allow for restoring the
rotational symmetry in the 3D model on the classical level by
rotating the orthogonal orbitals in an arbitrary way. The energy
(\ref{emf2d}) is minimized by,
\begin{equation}
\sin 2\phi=-\frac{E_z}{4J}, \label{cos2d}
\end{equation}
if $|E_z|\leq 4J$; otherwise $\sin 2\phi=\pm 1$. Interestingly,
the value of the field at which the orbitals are fully polarized
is reduced by one third from the value obtained in three
dimensions (\ref{cos3d}). This shows that although the orbital
exchange energy can be gained in a 3D model only on the bonds
along two directions in the alternating (orbital-flop) phase
either at or close to $E_z=0$, one has to counteract the
superexchange on the bonds in all three directions when the field
is applied.

The superexchange in the orbital subspace is AF and one may map
the orbital terms in the Hamiltonian (\ref{hflop}) onto a spin
problem in order to treat the elementary excitations within the
LSW theory. It is convenient to derive the excitations for the
spin-flop phase induced by an orbital-field starting from the
rotated Hamiltonian (\ref{hflop}). Here we choose the
Holstein-Primakoff transformation \cite{Faz99,Aue94} for localized
pseudospin operators ($T=1/2$),
\begin{equation}
T_i^+ = \bar{a}^{\dag}_i(1-\bar{a}_i^{\dag}\bar{a}_i)^{1/2},
\hskip .7cm T_i^- =  (1-\bar{a}_i^{\dag}\bar{a}_i)^{1/2}\bar{a}_i,
\hskip .7cm T_i^z = \bar{a}^{\dag}_i\bar{a}_i-\frac{1}{2},
\label{hpi}
\end{equation}
for $i\in A$ sublattice and
\begin{equation}
T_j^+ = (1-\bar{b}_j^{\dag}\bar{b}_j)^{1/2}\bar{b}_j,
\hskip .7cm T_j^- =
\bar{b}^{\dag}_j(1-\bar{b}_j^{\dag}\bar{b}_j)^{1/2}, \hskip .7cm
T_j^z = \frac{1}{2}-\bar{b}^{\dag}_j\bar{b}_j, \label{hpj}
\end{equation}
for $j\in B$ sublattice. In the harmonic approximation the terms
$\sim T_i^zT_j^x$ do not contribute to the boson Hamiltonian as
they give only odd numbers of boson operators. Therefore, the
phase dependence in the terms $\propto\pm\sqrt{3}$ is lost in the
LSW approximation.

After performing a Fourier transformation to
$\{\bar{a}^{\dag}_{\vec k},\bar{b}^{\dag}_{\vec k}\}$ operators,
the Hamiltonian may be further simplified by using the symmetry in
${\vec k}$-space and introducing new boson operators,
\begin{equation}
a_{\vec k}=\frac{1}{\sqrt{2}}(\bar{a}_{\vec k}-\bar{b}_{\vec k})
,\hskip 1cm b_{\vec k}=\frac{1}{\sqrt{2}}(\bar{a}_{\vec
k}+\bar{b}_{\vec k}) . \label{eq:branch}
\end{equation}
which leads to the effective orbital Hamiltonian of the form,
\begin{eqnarray}
\nonumber H_{\rm LSW} &=& J \sum_{\vec k} \left[ A_{\vec
k}a^{\dag}_{\vec k}a_{\vec k}^{} + \frac{1}{2} B_{\vec
k}(a^{\dag}_{\vec k}a^{\dag}_{-{\vec k}} + a^{}_{-{\vec
k}}a^{}_{\vec k}) \right] \\ & &+ J \sum_{\vec k} \left[ A_{\vec
k}b^{\dag}_{\vec k}b_{\vec k}^{} - \frac{1}{2} B_{\vec
k}(b^{\dag}_{\vec k}b^{\dag}_{-{\vec k}} + b^{}_{-{\vec
k}}b^{}_{\vec k}) \right] , \label{hfou}
\end{eqnarray}
where the coefficients $A_{\vec k}$ and $B_{\vec k}$ depend on
angle $\phi$,
\begin{eqnarray}
A_{\vec k}&=&3-B_{\vec k},    \\ B_{\vec
k}&=&\frac{1}{2}\left[(2\cos 4\phi-1)\gamma_+({\vec k})
          + (\cos 4 \phi+1)\gamma_z({\vec k}) \right] ,
\label{eq:gamma}
\end{eqnarray}
and the ${\vec k}$-dependence is given by $\gamma_+({\vec k})$,
and by $\gamma_z({\vec k})$, defined by Eqs. (\ref{gammap}) and
(\ref{gammaz}), respectively. After a Bogoliubov transformation,
\begin{equation}
\label{Bog1} a_{\vec k}=u_{\vec k}\alpha_{\vec k}^{}+v_{\vec
k}\alpha^{\dagger}_{-{\vec k}}, \hskip 1cm b_{\vec k}=u_{\vec
k}\beta_{\vec k}^{}+v_{\vec k}\beta^{\dagger}_{-{\vec k}},
\end{equation}
with the coefficients
\begin{eqnarray}
u_{\vec k}=\sqrt{\frac{A_{\vec k}}{2\zeta_{\vec k}}+\frac{1}{2}},\
\ \ v_{\vec k}=-{\rm sgn}(B_{\vec k}) \sqrt{\frac{A_{\vec
k}}{2\zeta_{\vec k}}-\frac{1}{2}}, \label{uvlsw}
\end{eqnarray}
where $\zeta_{\vec k}=\sqrt{A_{\vec k}^2-B_{\vec k}^2}$, the
Hamiltonian (\ref{hfou}) is diagonalized and takes the following
form,
\begin{equation}
H_{\rm LSW}=\sum_{\vec k}\left[ \omega^-_{\vec
k}(\phi)\alpha^{\dag}_{\vec k}\alpha_{\vec k} + \omega^+_{\vec
k}(\phi) \beta^{\dag}_{\vec k} \beta_{\vec k}\right] . \label{Hnu}
\end{equation}
The orbital-wave dispersion is given by
\begin{eqnarray}
\omega_{\vec k}^{\pm}(\phi)=3J\left\{ 1 \pm \case{1}{3}\left[
 (2\cos4\phi-1)\gamma_+({\vec k})
 + (\cos4\phi+1)\gamma_z({\vec k})\right]\right\}^{1/2}.
\label{eq:omega_phi}
\end{eqnarray}
The orbital excitation spectrum consists of two branches like, for
instance, in an anisotropic Heisenberg model \cite{Faz99}.
\begin{figure}
\centerline{\epsfig{figure=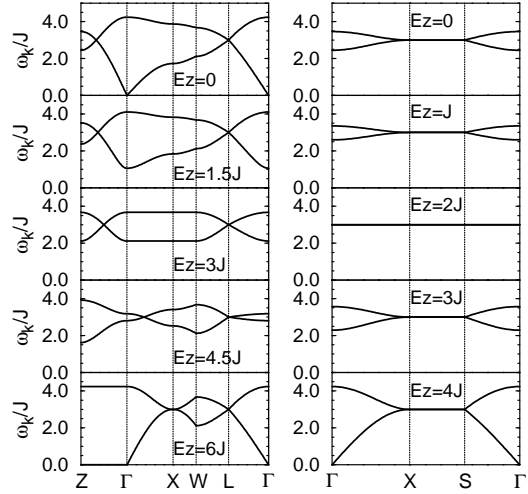,width=7cm}}
\smallskip
\caption {Orbital-wave excitations as obtained for different
values of
 the crystal-field splitting $E_z$ for a 3D (left) and 2D (right) orbital
 superexchange model (\protect{\ref{somfff}}). The result shown for a 3D
 system at $E_z=0$ was obtained for the orbitals rotated by $\theta=\pi/4$
 in Eqs. (\protect{\ref{newstateimn}}), and corresponds to the $E_z\to 0$
 limit of the orbital-flop phase (after Ref. \protect\cite{vdB99}).
} \label{fig:ow}
\end{figure}
The dependence on the field $E_z$ is implicitly contained in the
above relations via the angle $\phi$, as determined by Eq.
(\ref{cos3d}) and for the 3D model. The orbital-wave dispersion
for a 2D system, can easily be obtained from Eq.
(\ref{eq:omega_phi}) by setting $\gamma_z({\vec k})=0$ and
selecting $\phi$ according to Eq. (\ref{cos2d}). First we discuss
the excitation spectra given by Eq. (\ref{eq:omega_phi}) shown in
Fig. \ref{fig:ow} for different crystal-field splittings $E_z$ for
the 3D system, along different high-symmetry directions of the
$fcc$ BZ appropriate for the alternating orbital order. Most
interestingly, a gapless orbital-wave excitation is found for the
3D system at orbital degeneracy. Obviously, this is due to the
fact that the classical ground state energy is independent of the
rotation angle $\theta$ at $E_z=0$. At first glance, however, one
does not expect such a gapless mode, as the Hamiltonian
(\ref{eq:hamorb}) does not obey a continuous SU(2) symmetry. The
cubic symmetry of the model, however, is restored if one includes
the quantum fluctuations, as shown in Fig. \ref{fig:theta}. Note
that the quantum corrections found in the 3D orbital model
(\ref{eq:hamorb}) are somewhat smaller than those for the 3D
Heisenberg antiferromagnet. They do depend on the rotation angle
$\theta$, as the orbital-wave dispersion does.

As a special case, the orbital-wave dispersion for a quasi-2D
situation in a 3D case can easily be obtained from Eq.
(\ref{eq:omega_phi}) by eliminating the term
$\propto\gamma_z({\vec k})$ and assuming $\phi=\pi/4$,
\begin{equation}
\label{omegaquasi2d} \omega_{\vec
k}^{\pm}(\phi=\pi/4)=3J\sqrt{1\pm\gamma_+(\vec k)}.
\end{equation}
In a 3D system it applies to the ground state given by alternating
$|x\rangle$ and $|z\rangle$ orbitals on the two sublattices, and
one finds the largest quantum corrections, as this dispersion has
a line of nodes along the $\Gamma-Z$ direction, i.e.,
$\omega^-_{(0,0,q)}=0$ for $0<q<\pi$. In higher order spin-wave
theory, however, it might very well be that a gap opens in the
excitation spectrum. We expect, however, that this gap, if it
arises, is small, with its size being self-consistently determined
by quantum fluctuations.

For the 2D system the situation at orbital degeneracy is quite
different (see Fig. \ref{fig:ow}). The lack of interactions along
the $c$-axis breaks the symmetry of the model already at $E_z=0$,
opens a gap in the excitation spectrum,
\begin{equation}
\label{omega2d} \omega_{\vec
k}^{\pm}(\phi=0)=3J\sqrt{1\pm\case{1}{3}\gamma_+(\vec k)},
\end{equation}
and suppresses quantum fluctuations. Thus, one encounters an
interesting example of a more classical behavior in lower
dimension. In fact, the 1D model (\ref{eq:hamorb}) is classical as
only Ising interactions are left and the modes are dispersionless
(local mode at $\omega_k=2J$).

At increasing the orbital field $|E_z|$, the 2D and 3D system
resemble each other, with a large gap in the excitation spectrum
at $|E_z|=2J$ ($|E_z|=3J$) in the 2D (3D) system (Fig.
\ref{fig:ow}). At larger $|E_z|$ the gap gradually closes when an
orbital field which compensates the energy loss due to the orbital
superexchange between identical (FM) orbitals is approached. At
this value of the field ($E_z=4J$ and $E_z=6J$ in a 2D and 3D
model, respectively), the full dispersion of the orbital waves is
recovered, the spectrum is gapless (Fig. \ref{fig:ow}), and the
quantum fluctuations reach a maximal value. We would like to
emphasize that this behavior is qualitatively different from the
Heisenberg antiferromagnet, both in two and three dimensions,
where the anomalous terms $\propto T_i^+T_j^+$ and $\propto
T_i^-T_j^-$ are absent which results in the conserved total spin
$T^z=\sum_iT_i^z$, and quantum fluctuations vanish at the
crossover from the spin-flop to FM phase.
\begin{figure}
\centerline{\epsfig{figure=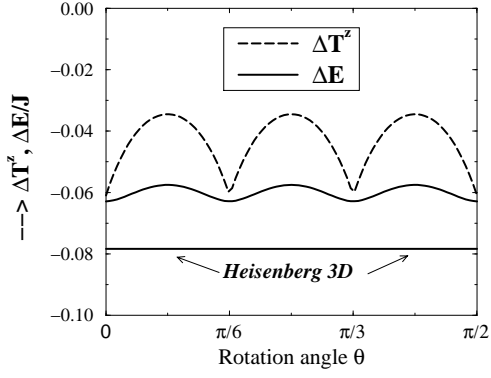,width=7cm}}
\smallskip
\caption {Quantum corrections for the 3D orbital model
(\protect\ref{somfff}) as
 functions of rotation angle $\theta$ (\protect{\ref{newstateimn}}) for:
 the order parameter $\Delta T^z$ (full lines), and the ground-state
 energy $\Delta E/J$ (dashed lines)
 (after Ref. \protect\cite{vdB99}). }
\label{fig:theta}
\end{figure}
It is instructive to make a comparison between the analytic
approximations of the LSW theory and the exact diagonalization
(ED) of 2D finite clusters using the finite temperature
diagonalization method \cite{Jak94,Jak00}. As in the mean-field
approach, one finds also a unique ground state for finite clusters
at $E_z=0$ by ED. Making the rotation of basis (\ref{newstateimn})
is however still useful in the ED method as it gives more physical
insight into the obtained correlation functions which become
simpler and more transparent when calculated within an optimized
basis. Moreover, they offer a simple tool to compare the results
obtained by ED with those of the analytic approach. We shall
present below the results obtained with $4\times 4$ clusters;
similar results were also found for 10-site clusters \cite{Hor99}.

\begin{figure}
\centerline{\epsfig{figure=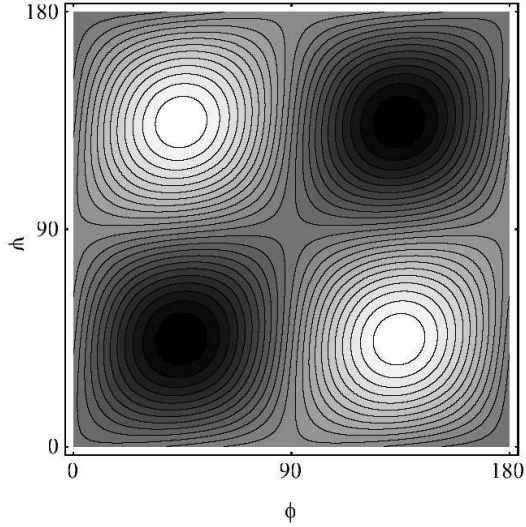,width=7cm}}
\smallskip
\caption { Contour plot of the {\em rotated\/} nearest-neighbor
orbital correlation
 function $\langle {\tilde T}^z_i {\tilde T}^z_{i+R}\rangle$ as function of
 the angles $\phi$ and $\psi$ for a 16-site planar cluster with $E_z=0$ and
 $T=0.1J$. White regions correspond to positive (FM) and black areas to
 negative (AF) orbital correlations, i.e.,
 $\langle {\tilde T}^z_i {\tilde T}^z_{i+R}\rangle>0.24$ ($<-0.24$),
 respectively. They are separated by 25 contour lines chosen with the
 step of 0.02 in the interval [-0.24,0.24] (after Ref. \protect\cite{vdB99}.}
\label{fig:contour}
\end{figure}

Let us look at nearest-neighbor correlation function in the ground
state $\langle \tilde{T}^z_i\tilde{T}^z_{i+R}\rangle$, where the
operators with a tilde refer to a rotated basis,
\begin{eqnarray}
\tilde{T}^z_i     &=& \cos 2\phi T^z_i     + \sin 2\phi T^x_i
\nonumber \\ \tilde{T}^z_{i+R} &=& \cos 2\psi T^z_{i+R} + \sin
2\psi T^x_{i+R}, \label{ttilde}
\end{eqnarray}
so that the correlation function depends on two angles: $\phi$ and
$\psi$. In Fig. \ref{fig:contour} the intersite orbital
correlation in the ground state is shown as a contour-plot. The
intensity of the grey scale changes from positive to negative
values of the correlation function, $\langle
\tilde{T}^z_i\tilde{T}^z_{i+R}\rangle$. One finds that the
neighbor correlations have their largest value if the orbitals are
rotated by $\phi=\pi/4$ and $\psi=3\pi/4$ (or $\phi=3\pi/4$ and
$\psi=\pi/4$), i.e., under this rotation of the basis states the
system looks like a ferromagnet, indicating that the occupied
$(|x\rangle+|z\rangle)/\sqrt{2}$ and
$(|x\rangle-|z\rangle)/\sqrt{2}$ orbitals are staggered in a 2D
model, as shown in Fig. \ref{fig:oo}. Note that quantum
fluctuations are small as in the ground state and one finds
$\langle\tilde{T}^z_i\tilde{T}^z_{i+R}\rangle\simeq 0.246$. One
finds that although the symmetry is not globally broken in a
finite system, the short-range order resembles that found in the
symmetry-broken state with orbital LRO. This demonstrates at the
same time the advantage of the rotated basis in the ED study,
because in the original unrotated basis one finds instead $\langle
T^z_i T^z_{i+R}\rangle\simeq 0$, which might lead in a naive
interpretation to a large overestimation of quantum fluctuations.
\begin{figure}[h]
\centerline{\epsfig{figure=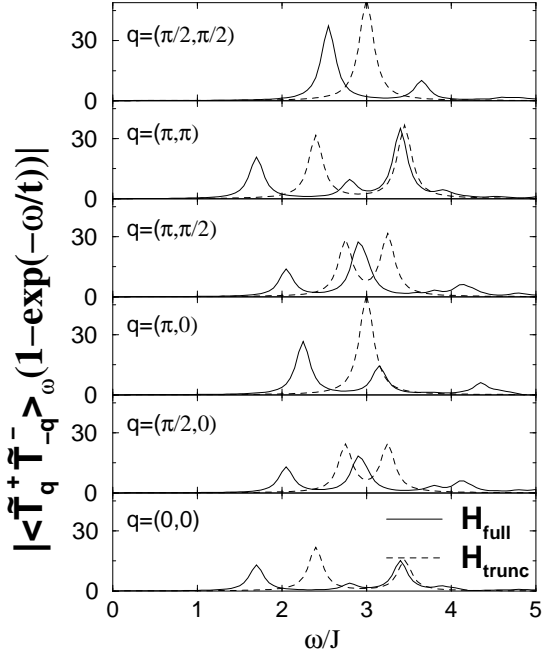,width=7cm}} \caption
{Transverse response function ${\tilde T}^{+-}_{\bf q}(\omega)$
for the
 rotated orbitals as a function of frequency $\omega$ for different momenta
 at low temperature. Calculations were performed for a 16-site 2D cluster at
 $E_z=0$ and $T=0.1J$ for:
 (a) the 2D orbital model given by Eq. (\ref{eq:hamorb}) ($H_{\rm full}$,
     full lines), and
 (b) neglecting the mixed terms $\propto T^x_iT^z_j$ ($H_{\rm trunc}$, dashed
    lines). The spectra are broadened by $\Gamma=0.1J$
 (after Ref. \protect\cite{vdB99}).}
\label{fig:transrot}
\end{figure}
The ground-state correlations
$\langle\tilde{T}^z_i\tilde{T}^z_{i+R}\rangle$ found by ED are in
excellent agreement with the results of LSW theory. By comparing
the results obtained at $T=0.1J$, $0.2J$ and $0.5J$, it has been
found in Ref. \cite{vdB99} that the calculated low-temperature
correlation functions are almost identical in this temperature
range, and thus the values shown in Fig. \ref{fig:contour} for
$T=0.1J$ are representative for the ground state. They demonstrate
an instability of the system towards the symmetry-broken state.

In order to verify the accuracy of the LSW approach for finding
the excitation spectrum, we discuss the results for the dynamical
orbital response functions in the case of orbital degeneracy. The
transverse response function for the orbital excitations evaluated
with respect to the {\it rotated\/} local quantization axes
(\ref{ttilde}) is defined as follows,
\begin{equation}
{\tilde T}^{+-}_{\vec
q}(\omega)=\frac{1}{2\pi}\int^{\infty}_{-\infty} dt\; \langle
{\tilde T}^+_{\vec q} {\tilde T}^-_{-{\vec q}}(t)\rangle\;
\exp(-i\omega t). \label{tomega}
\end{equation}
As we have already mentioned, the LSW approximation does not allow
to investigate the consequences of the coupling of single
excitonic excitations to the order parameter, represented by the
terms $\propto T^x_iT^z_j$. Therefore, strictly speaking the LSW
approach corresponds to the truncated Hamiltonian when such terms
are not included. The ED gives then a double-peak structure in the
response function ${\tilde T}^{+-}_{\vec q}(\omega)$ which agrees
well with the dispersion of two modes found in the LSW approach
(Fig. \ref{fig:transrot}). However, if the terms $\propto
T^x_iT^z_j$ are included, one finds different structures -- the
lowest energy excitation moves to lower energies, and satellite
structures appear which describe the incoherent processes in
orbital dynamics.

\begin{figure}
\centerline{\epsfig{figure=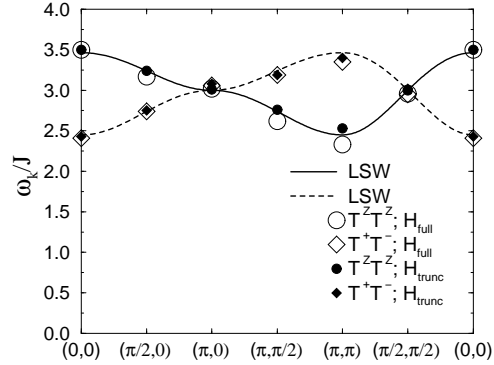,width=7cm}}
\smallskip
\caption {Dispersion of orbital waves along the main directions in
the 2D BZ. Results
 for the first moment calculated for a $4\times 4$ cluster with the full
 ($H_{\rm full}$ and truncated ($H_{\rm trunc}$ Hamiltonian (empty and
full
 symbols) are compared with the dispersions for the two modes found using
the
 LSW theory (solid and dashed line) (after Ref. \protect\cite{vdB99}).}
\label{fig:ed+lsw}
\end{figure}

In spite of some additional incoherent processes in the spectra,
the first moment of ${\tilde T}^{+-}_{\vec q}(\omega)$ determined
by ED agrees very well with the dispersion found within the LSW
theory [as determined from Eq. (\ref{omega2d})] (Fig.
\ref{fig:ed+lsw}). The values of the first moments are only
slightly changed when instead the truncated Hamiltonian (without
the processes $\propto T^x_iT^z_j$) is used in a numerical
approach. This comparison demonstrates that the LSW approach
captures the leading term in the orbital dynamics and may be thus
used to investigate the consequences of orbital excitations on the
hole dynamics, as presented in the next Section.
\subsection{ Hole propagation in an orbital ordered state }

As an interesting example of the consequence of orbital degrees of
freedom in the hole excitation spectra, the problem of hole
propagation in an orbital ordered 2D FM planes, as found in
LaMnO$_3$, has been considered recently by van den Brink, Horsch
and Ole\'s \cite{vdB00}. The main idea is that the hole which
moves in an orbital ordered plane [Fig. \ref{fig:hop}(a)] dresses
by orbital excitations, and a polaron is formed. Its properties
will depend on the actual parameters and it is interesting to
investigate: (i) whether a quasiparticle (QP) state will form
under these circumstances in
    analogy with the spin problem --
    a hole moving in a quantum antiferromagnet \cite{Mar91}, and
(ii) how much the spectral function changes when the orbital
degeneracy is
     removed.

Let us consider the simplest situation and assume that the $e_g$
orbitals are degenerate, $E_z=0$. If only few holes are doped to a
FM plane as in the A-AF phase of La$_{1-x}$Ca$_x$MnO$_3$ with
$x\ll 1$, the ground state is determined in first instance by a 2D
version of the orbital model (\ref{eq:hamorb}),
\begin{equation}
\label{H_J2D} H_J = \case{1}{2}J\sum_{\langle ij\rangle}
  \left[T^z_i T^z_j + 3T^x_i T^x_j\mp\sqrt{3} (T^x_iT^z_j+T^z_i T^x_j)\right],
\end{equation}
where the bonds $\langle ij\rangle$ connect nearest neighbors in a
single $(a,b)$ plane, and the AF superexchange $J$ implies that
the orbitals order. In order to see the consequences of this
ordered state for the kinetic energy of doped holes, it is
convenient to transform the hopping term (\ref{hkin}) along the
bonds $\langle ij\rangle$ in the $(a,b)$ FM planes to the basis
defined by Eqs. (\ref{flopi}) at $\phi=0$ which corresponds to the
symmetry-broken ground state with the orthogonal orbitals
$|i\mu\rangle$ and $|j\nu\rangle$ occupied on the two sublattices
(Fig. \ref{fig:oo}). Let us introduce the new fermion (hole)
operators which correspond to the occupied orbitals $|i0\rangle$
as $f_{i0}^{\dagger}$, and the operators which correspond to the
excited states $|i1\rangle$ as $f_{i1}^{\dagger}$. Thus, the
operator $f_{i0}^{\dagger}$ create a hole in the orbital
$|i\mu\rangle$ for $i\in A$ and $|i\nu\rangle$ for $i\in B$,
respectively. The transformed kinetic energy is \cite{vdB00},
\begin{eqnarray}
\label{eq:ht01} H_t&=&\case{1}{4}t\!\sum_{\langle ij\rangle}\left[
  f_{i0}^{\dagger}f_{j0}^{}+f_{i1}^{\dagger}f_{j1}^{}
 +2(f_{i1}^{\dagger}f_{j0}^{}+f_{i0}^{\dagger}f_{j1}^{}) \right. \nonumber \\
& & \hskip .8cm \left. \pm\sqrt{3}\left(
   f_{i1}^{\dagger}f_{j0}^{}-f_{i0}^{\dagger}f_{j1}^{}\right)+H.c.\right].
\end{eqnarray}
Together with the usual crystal-field splitting term $H_{\tau}$
(\ref{htau}), Eqs. (\ref{H_J2D}) and (\ref{eq:ht01}) define the
{\it orbital $t$-$J$ model\/},
\begin{equation}
{\cal H}=H_t+H_J+H_{\tau}. \label{orbtj}
\end{equation}

The first interesting observation is that the hole motion is not
completely suppressed by the orbital ordering, unlike in the spin
$t$-$J$ model. The processes $\propto f_{i0}^{\dagger}f_{j0}^{}$
concern the occupied orbitals and thus a hole may always
interchange with an electron without disturbing the orbital
ordering [Fig. \ref{fig:hop}(b)]. These processes lead to a band
in the limit of $U\to\infty$, where the constraint of no double
occupancy is implemented with the slave-boson operators $b_{i0}$
and $b_{i1}$, standing for the orbital flavors which accompany the
hole operators $h_i^{\dagger}$ according to the prescription:
\begin{equation}
\label{fbh} f_{i0}^{\dagger}=b_{i0}h_i^{\dagger}, \hskip 1cm
f_{i1}^{\dagger}=b_{i1}h_i^{\dagger}.
\end{equation}
In the orbital ordered state the $b_{i0}$ bosons are condensed,
$b_{i0}=1$, which leads after a Fourier transformation from the
hole operators $h_i^{\dagger}$ to $h_{\vec k}^{\dagger}$ to a free
hole propagation in the lower Hubbard band,
\begin{equation}
\label{bandinf} H_h=\sum_{\vec k}\varepsilon^0_{\vec
k}(\phi)h_{\vec k}^{\dagger}h_{\vec k}^{},
\end{equation}
with a dispersion determined by the orbital order via
\begin{equation}
\label{freehole} \varepsilon^0_{\vec k}(\phi)=(-2\sin
2\phi+1)t\gamma_+({\vec k}),
\end{equation}
where $\gamma_+({\vec k})$ is defined by Eq. (\ref{gammap}). The
angle $\phi$ (\ref{flopi}) depends on the optimal orbitals
(\ref{cos2d}): $\phi=0$ at the orbital degeneracy ($E_z=0$), while
$\phi\neq 0$ if the orbital degeneracy is removed by a finite
field $E_z\neq 0$. Note that the largest dispersion is found for
$|x\rangle$ orbitals at $\phi=-\pi/4$, while the dispersion
vanishes at $\phi=\pi/8$ due to the conflicting phases of
$|x\rangle$ and $|z\rangle$ orbitals.
\begin{figure}
\centerline{\epsfig{figure=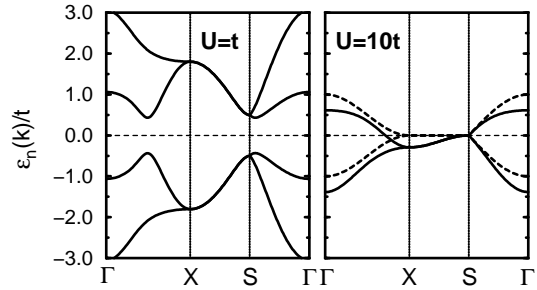,width=7cm}}
\smallskip
\caption {Dispersion relations in the reduced BZ [$X=(\pi,0)$,
$S=(\pi/2,\pi/2)$] in the mean-field approximation which simulates
the treatment of Coulomb repulsion in LDA+U as obtained for: $U=t$
(left), and in the lower Hubbard band centered at $\omega=0$ for
$U=10t$ (right, full lines). In the right panel the dispersion
obtained in the $U\rightarrow\infty$ limit is shown by dashed
lines for comparison (after Ref. \protect\cite{vdB00}). }
\label{fig:lda+u}
\end{figure}
If the other processes which couple the moving hole to the orbital
excitations could be neglected, the band (\ref{bandinf}) would
give a coherent spectral function of a hole. Furthermore, an
electron doped at $n=1$ would propagate in the upper Hubbard band
by a similar dispersion, i.e., due to the term $\propto
f_{i1}^{\dagger}f_{i1}^{}$ in Eq. (\ref{eq:ht01}). These subbands
are separated by the Coulomb interaction $U$ which acts between
two $e_g$ states, in this case between the occupied and unoccupied
orbital states, 
\begin{equation}
\label{uterm} H_U=U\sum_i
f_{i0}^{\dagger}f_{i0}^{}f_{i1}^{\dagger}f_{i1}^{}.
\end{equation}
We recall, however, that the interorbital Coulomb interaction is
invariant with respect to the choice of orbital basis only if the
pair-hopping terms in Eq. (\ref{hint}) are included \cite{Ole83}.
In the present case with the occupied $|i0\rangle$ orbitals at
$n=1$, we may take $f_{i0}^{\dagger}f_{i0}^{}=1$, and the Coulomb
term reduces to a local potential which acts on the {\it
unoccupied\/} states $|i1\rangle$. When the hopping Hamiltonian
(\ref{eq:ht01}) is supplemented by this potential $\propto U$, one
finds that the bands are separated by a gap which opens at $U=0$
(Fig. \ref{fig:lda+u}). The bands change drastically as a function
of $U$ \cite{vdB00}: if $U$ is small, the bands resemble the
uncorrelated problem (\ref{eq:disp}) of Sec. VI.A, while the shape
of the lower Hubbard band becomes close to the $U\to\infty$ limit
(\ref{bandinf}) already at $U/t\approx 10$ which may be taken a
representative value for manganites \cite{Fei99}. These changes of
the bands between the small and large $U$ regime simulate the
effect of the local potentials which act on the unoccupied states
in the LDA+U method \cite{Ani91}.
\begin{figure}
\centerline{\epsfig{figure=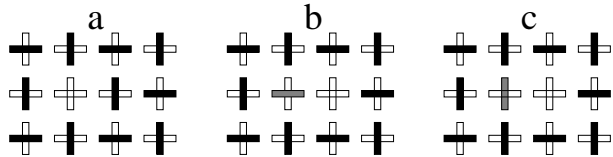,width=8cm}}
\vspace{0.15in} \caption{A single hole added in an orbital-ordered
ground state (a);
 the occupied (empty) orbitals $|\mu\rangle$ and $|\nu\rangle$ are shown
as filled (empty) rectangles. The hole can move either without
disturbing the orbital order (b), or by creating orbital
excitations (c) (after Ref. \protect\cite{vdB00}).}

  \label{fig:hop}
\end{figure}
Thus, one finds a free hole band shown in Fig. \ref{fig:lda+u} and
the question is how this band changes when the hole starts to
dress by orbital excitations of Sec.VII.A [Fig. \ref{fig:hop}(c)].
It may be expected that the free hole dispersion (\ref{bandinf})
is drastically modified by the hole-orbiton coupling which allows
the hole propagation by frustrating the orbital order, just as the
AF order is locally disturbed in the $t$-$J$ model, and the
'string' of excited bonds is created on the path of the hole
\cite{Shr88}. This problem was recently investigated in the
orbital model by van den Brink, Horsch, and Ole\'s \cite{vdB00}
who have shown that the interaction between the hole and the
orbitals is so strong in the manganites that propagating holes are
dressed with many orbital excitations and form polarons that have
large mass, small bandwidth and low QP weight.

The orbital background is described by a local constraint for
$T=1/2$ pseudospins,
\begin{equation}
\label{constb}
b_{i0}^{\dagger}b_{i0}^{}+b_{i1}^{\dagger}b_{i1}^{}=2T,
\end{equation}
where $b_{i0}^{\dagger}$ and $b_{i1}^{\dagger}$ are boson
operators which refer to the occupied and empty state at site $i$
(\ref{fbh}). By making the lowest order expansion of the
constraint around the ground state with orbital ordering, one
recovers the linear orbital-wave (LOW) approximation analyzed in
Sec. VII.A. Having chosen the occupied orbitals as
$b_{i0}^{\dagger}b_{i0}^{}\simeq 1$ bosons at every site, the
expansion is the same for both sublattices and reads,
\begin{equation}
\label{txtz} T_i^x=\case{1}{2}(b_{i}^{}+b_{i}^{\dagger}), \hskip
1cm T_i^z=          T-b_{i}^{\dagger}b_{i}^{},
\end{equation}
with $b_{i}\equiv b_{i1}$ playing the role of a Holstein-Primakoff
boson. It corresponds to the spin problem with the spins rotated
by $\pi$ at one of the sublattices \cite{Faz99,Aue94}. The
resulting effective boson Hamiltonian is diagonalized by a Fourier
and Bogoliubov transformation (Sec. VII.A). One finds therefore an
interacting problem in LOW approximation,
\begin{equation}
\label{strongcoupling} {\cal H}_{\rm LOW}=H_h+H_o+H_{ho},
\end{equation}
where the orbital waves (orbitons) for a 2D model are given by,
\begin{equation}
\label{magnons} H_o=\sum_{\vec k} \omega_{\vec
k}(\phi)\alpha^{\dag}_{\vec k}\alpha^{}_{\vec k},
\end{equation}
with
\begin{equation}
\label{omega2dphi} \omega_{\vec k}(\phi)= 3J\left[ 1
+\case{1}{3}(2\cos4\phi-1)\gamma_+({\vec k})\right]^{1/2},
\end{equation}
standing for the orbiton dispersion. The single mode written now
for convenience in the full BZ in Eq. (\ref{magnons}) is
equivalent to two branches of orbital excitations obtained in the
folded zone in Sec. VII.A \cite{vdB99}. The orbital excitations
depend sensitively on the orbital splitting $E_z$. At orbital
degeneracy ($E_z=0$) one finds a maximum of $\omega_{\vec
k}(\phi)$ at the $\Gamma=(0,0)$ point and a weak dispersion $\sim
J$ [see also Eq. (\ref{omega2d})]. In contrast, for $E_z=\pm 2J$
orbital excitations are dispersionless, and $\omega_{\vec k}=3J$.
The remaining part of Eq. (\ref{strongcoupling}) describes the
hole-orbiton interaction [Fig. \ref{fig:hop}(c)],
\begin{eqnarray}
\label{orbpol} H_{ho}=t\sum_{{\vec k},{\vec q}}h_{{\vec k}-{\vec
q}}^{\dagger}h_{\vec k}^{}
  \left[M_{{\vec k},{\vec q}}\alpha_{\vec q}^{\dagger} \right.
 +\left. N_{{\vec k},{\vec q}}\alpha_{{\vec q}+{\vec
Q}}^{\dagger}+H.c.\right],
\end{eqnarray}
where ${\vec Q}=(\pi,\pi)$, the vertex functions are:
\begin{eqnarray}
\label{vertexm} M_{{\vec k},{\vec q}}&=&2\cos 2\phi\left[u_{\vec
q}\gamma_+({\vec k}-{\vec q})
                                        +v_{\vec q}\gamma_+({\vec k})\right],
                    \\    \label{vertexn}
N_{{\vec k},{\vec q}}&=&  -\sqrt{3}\left[u_{\vec q}\gamma_-({\vec
k}-{\vec q})
                                        -v_{\vec q}\gamma_-({\vec k})\right],
\end{eqnarray}
and $\gamma_-({\vec k})=\gamma_+({k_x,k_y+\pi})$ is defined by Eq.
(\ref{gammam}).

In this way, the orbital $t$-$J$ model (\ref{orbtj}) leads to an
effective Hamiltonian (\ref{strongcoupling}), describing a
many-body problem due to the hole-orbiton coupling term $H_{ho}$.
Although the analytic structure and the form of Eq.
(\ref{strongcoupling}) resembles the usual $t$-$J$ model written
in the slave fermion formalism \cite{Mar91}, there are important
differences. First of all, a hole may propagate freely in the
orbital model by the term (\ref{freehole}), as it would be also
the case in the quantum antiferromagnet with further-neighbor
hopping. Second, the orbital waves do not obey the nesting
symmetry, i.e., $\omega_{{\vec k}+{\vec Q}}(\phi)\neq \omega_{\vec
k}(\phi)$, where ${\vec Q}=(\pi,\pi)$, and are more classical than
the spin waves, with a finite gap in the excitation spectrum
(\ref{omega2dphi}). Furthermore, the hole-orbiton interaction
(\ref{orbpol}) has a richer analytic structure than that of the
$t$-$J$ model, as the scattering processes which conserve the
momentum modulo ${\vec Q}$ due to a new vertex $\propto N_{{\vec
k},{\vec q}}$ (\ref{vertexn}). Finally, an important feature is
also that both hole dispersion $\varepsilon^0_{\vec k}(\phi)$ and
orbiton dispersion $\omega_{\vec k}(\phi)$ depend on the
crystal-field splitting $E_z$, and thus the analytic structure is
richer. As a special case, at $E_z=-2J$ both modes are
dispersionless, and the model orbital becomes equivalent to a hole
which moves in a classical antiferromagnet described by the Ising
model, with no quantum fluctuations \cite{Kan89}.
\begin{figure}
\centerline{\epsfig{figure=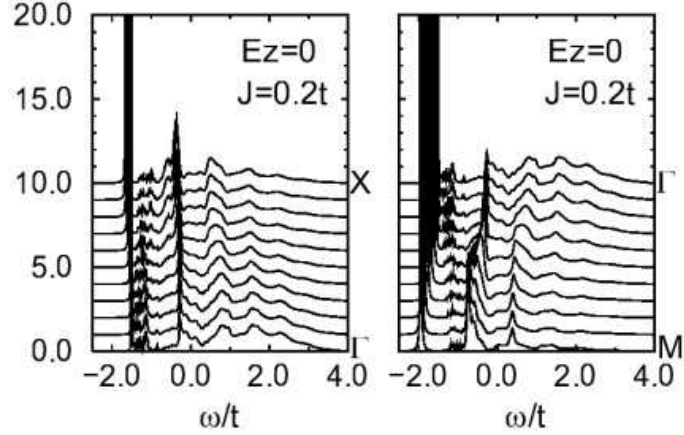,width=9cm}}
\smallskip
\caption {Spectral functions for the orbital $t$-$J$ model as
obtained in two
 high-symmetry directions of the 2D BZ: $\Gamma-X$ (left) and $M-\Gamma$
 (right) for $J=0.2t$ and $E_z=0$ [$\Gamma=(0,0)$, $X=(\pi,0)$,
 $M=(\pi,\pi)$] (after Ref. \protect\cite{vdB00}).}
\label{fig:sf0}
\end{figure}
The many-body problem obtained for a hole propagating in an
orbital ordered background (\ref{strongcoupling}) may be solved
using the self-consistent Born approximation (SCBA)
\cite{Sch89,Kane89}. This method gives results of high quality and
compares favorably with ED for the hole which moves in a quantum
antiferromagnet \cite{Mar91}. Treating ${\cal H}_{\rm LOW}$ in the
SCBA, one finds the selfenergy \cite{vdB00},
\begin{eqnarray}
\label{selfen} \Sigma({\vec k},\omega)&=& t^2\sum_{\vec q}
\left[M^2_{{\vec k},{\vec k}-{\vec q}} G({\vec k}-{\vec
q},\omega-\omega_{\vec q}) + N^2_{{\vec k},{\vec k}-{\vec q}}
G({\vec k}-{\vec q},\omega-\omega_{{\vec q}+{\vec Q}})\right].
\end{eqnarray}
Here $G({\vec k},\omega)$ stands for the hole Green function which
obeys the Dyson equation,
\begin{equation}
\label{dyson} G^{-1}({\vec k},\omega)= \omega-\varepsilon^0_{\vec
k}(\phi)-\Sigma({\vec k},\omega).
\end{equation}
Eqs. (\ref{selfen}) and (\ref{dyson}) represent a closed set of
equations which has to be solved numerically by iteration on a
lattice. We discuss below some representative results obtained
recently by van den Brink {\it et al.} \cite{vdB00}.
\begin{figure}
 \centerline{\epsfig{figure=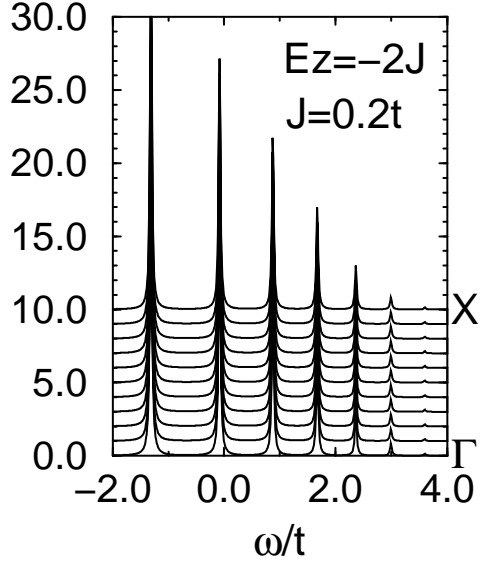,width=9cm}}
\smallskip
\caption {Ladder spectrum for the orbital $t$-$J$ model obtained
for $E_z=-2J$ and $J=0.2t$ (after Ref. \protect\cite{vdB00}). }
\label{fig:sf-2}
\end{figure}
First of all, the hole spectral function is drastically changed by
the coupling processes to the orbital excitations which is
particularly strong when the orbital superexchange $J$ is much
lower than the hopping $t$. Examples of rather complex spectra are
shown in Figs. \ref{fig:sf0}--\ref{fig:sf+2} for $J/t=0.1$ which
corresponds to the realistic parameters for manganites
\cite{Fei99}. As in the $t$-$J$ model, a QP state appears at the
threshold, and this results from a strong dressing of a hole by a
cloud of orbital excitations. The incoherent part extends over the
scale of $\sim 6t$ which corresponds to the full dispersion in the
$e_g$ band (\ref{eq:disp}).

The energy and momentum dependence of the incoherent spectra is
markedly different from the spin $t$-$J$ problem. At orbital
degeneracy (Fig. \ref{fig:sf0}) particular satellite structures
are obtained with a rather weak ${\vec k}$-dependence. Their
origin may be understood by looking at the hole spectrum found at
$E_z=-2J$ which corresponds to the dispersionless orbiton spectrum
(\ref{omega2dphi}) and to the vanishing hole dispersion
(\ref{freehole}) due to the conflicting phases in the $e_g$
electron hopping. In this case a ladder spectrum of the $t$-$J^z$
model is reproduced \cite{Kan89,Mar91} (Fig. \ref{fig:sf-2}), but
the values of $t$ and $J^z$ have to scaled for the orbital model
and the present case corresponds to a larger ratio $J^z/t\simeq
0.3$ in the spin problem \cite{vdB00}. Such ladder features at
decreasing distances result in the pronounced satellites observed
still in the spectra at orbital degeneracy. In contrast, with
increasing value of $E_z$, the incoherent part changes to a rather
smooth curve dominated by a broad maximum that corresponds to the
free dispersion which is still visible but strongly damped by the
hole scattering on the orbital excitations. An example of such
spectra is shown in Fig. \ref{fig:sf+2}.
\begin{figure}
\centerline{\epsfig{figure=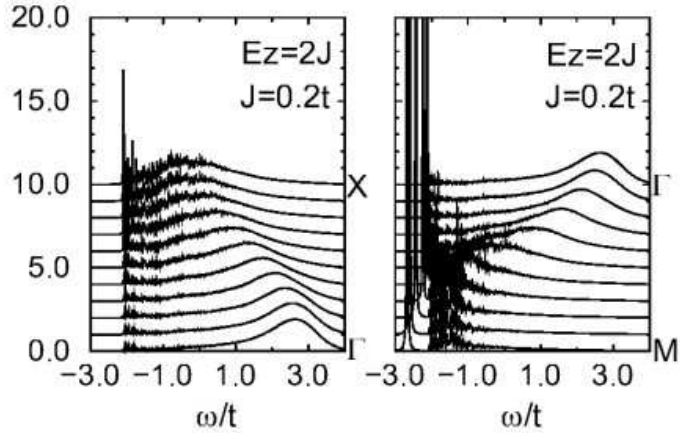,width=9cm}}
\smallskip
\caption {The same as in Fig. \protect\ref{fig:sf0}, but for
$E_z=+2J$
 (after Ref. \protect\cite{vdB00}).}
\label{fig:sf+2}
\end{figure}
The QP states have a narrow dispersion $\propto J$, as in the
$t$-$J$ model (Fig. \ref{fig:mstar}). Note that the QP's are well
defined and have a large weight $a_M$ close to the minimum of the
QP band at the $M=(\pi,\pi)$ point, while they are weaker at
higher energies. As in the $t$-$J$ model \cite{Mar91}, at the
maximum of the QP band, found here at the $\Gamma=(0,0)$ point,
the QP has the lowest spectral weight $a_{\Gamma}$. Note that the
QP minimum is here determined by the minimum of free dispersion
(Fig. \ref{fig:lda+u}) rather than by the enhanced quantum
fluctuations at the boundary of the folded BZ, as encountered in
the spin $t$-$J$ model.

The effective mass $m^*$ which may be defined be the momentum
dependence of the QP energy \cite{vdB00}, and the QP bandwidth
$W^*$ increase first linearly with increasing $J$ in the range of
$J/t<0.3$, while they approach the free values in the
weak-coupling regime of $J/t>1$. However, an additional dependence
on the orbital splitting makes the QP dispersion rather narrow
close to the orbital degeneracy ($E_z=0$) and at $E_z<0$, while it
broadens up when $E_z>0$ and the uniform phase with $|x\rangle$
orbitals occupied is approached. This has interesting consequences
for the experimental situation and suggests that the photoemission
spectra of manganites should have a strong dependence on the
deviations from the cubic symmetry which remove the orbital
degeneracy $(E_z\neq 0$). Furthermore, the dependence on the
doping might be very interesting in these situations when the
orbital ordering changes, as for instance in the layered compounds
\cite{Mac99}.
\begin{figure}[t]
\centerline{\epsfig{figure=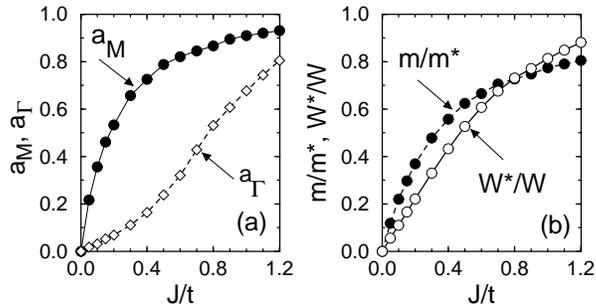,width=8cm}}
\smallskip
\caption {Quasiparticle properties at $E_z=0$ as functions of
$J/t$:
 (a) weights of the two QP bands at the $\Gamma$ point:
     $a_{M}$ (filled circles), and $a_{\Gamma}$ (open squares);
 (b) inverse effective mass $m/m^*$ (full circles) and
     the width of QP band $W^*/W$ (empty circles), normalized
     by the LDA+U values $m$ and $W$, respectively
     (after Ref. \protect\cite{vdB00}).}
\label{fig:mstar}
\end{figure}


\section{ Phase diagrams of manganites -- open problems }

\subsection{Magnetic, orbital, and charge ordering in CE-phase}

Over the last few years much attention has been focused on the
interplay between charge and orbital ordering occurring in
half-doped manganites ($x=0.5$). A direct evidence of the charge
ordered (CO) state in half-doped manganite has been provided by
the electron diffraction for La$_{0.5}$Ca$_{0.5}$MnO$_3$
\cite{Cheo96}. Similar observations have also been reported for
Pr$_{0.5}$Sr$_{0.5}$MnO$_3$ \cite{Tom95}
Nd$_{0.5}$Sr$_{0.5}$MnO$_3$ \cite{Kuw95}, and for
Pr$_{1-x}$Ca$_{x}$MnO$_3$ with $x=0.4$ and 0.5 \cite{Zim99}. This
CO state is characterized by an alternating arrangement of
Mn$^{3+}$ and Mn$^{4+}$ ions in $(a,b)$ planes and the charge
stacking in $c$-direction. In CO state these systems show an
insulating behavior with a very peculiar form of AF spin ordering.
The observed magnetic structure is the so-called magnetic CE phase
and consists of quasi-1D FM zigzag chains coupled
antiferromagnetically in both directions. In addition, the
occupied orbitals at Mn$^{3+}$ positions show in these systems
$d_{3x^2-r^2}$/$d_{3y^2-r^2}$ orbital ordering, staggered along
the FM chains.

The insulating CO state can be transformed into a metallic FM
state either by doping, or by applying an external magnetic field
\cite{Ima98}. Other interesting observations were done by studying
Pr$_{1-x}$(Ca$_{1-y}$Sr$_y$)$_x$MnO$_3$ crystals with controlled
one-electron bandwidth. As already mentioned above, at half-doping
Pr$_{0.5}$Ca$_{0.5}$MnO$_3$ has a CO CE-type insulating state.
However, by substitution Ca with Sr leading to the increase of the
carrier bandwidth, one induces the collapse of the CO insulating
state, and the A-type metallic state with $d_{x^2-y^2}$ orbital
ordering is realized in Pr$_{0.5}$Sr$_{0.5}$MnO$_3$ \cite{Kaw97}.
The coexistence of the A-type spin ordered and CE-type spin/charge
ordered states has been detected in the bilayer
LaSr$_2$Mn$_2$O$_7$ \cite{Kub98} and in 3D
Nd$_{0.5}$Sr$_{0.5}$MnO$_3$ \cite{Kuw95}. These results indicate
the competition between the metallic A-type (uniform)
$d_{x^2-y^2}$ orbital ordering, and the insulating CE-type
$d_{3x^2-r^2}/d_{3y^2-r^2}$ orbital ordering at half-doping and
demonstrate the importance of the coupling between magnetic,
charge and orbital ordering in these compounds.
\begin{figure}[t]
      \epsfysize=51mm
      \centerline{\epsffile{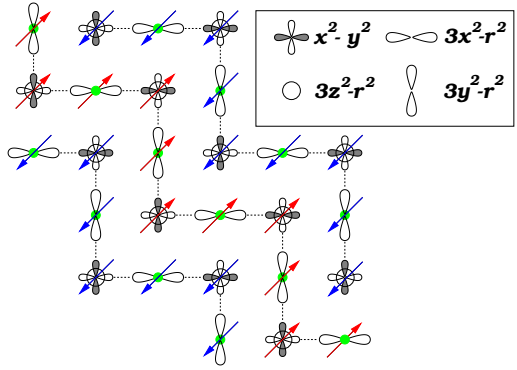}}
\caption {View of the CE phase in the $(a,b)$ plane. The $x^2-y^2$
orbitals at corner sites have positive (white) and negative (grey)
lobes, while the phases of the other orbitals are positive. The
grey dots at the bridge sites represent a charge surplus (after
Ref. \protect\cite{Kho99}).} \label{fig:ce_phase}
\end{figure}
Being experimentally well established, theoretically, however, the
nature of the CO state in half-doped manganites and the origin of
the unconventional zigzag magnetic structure remain challenging
open problems. As shown in Fig. \ref{fig:ce_phase}, the CE-phase
unit cell contains two geometrically inequivalent sites
\cite{Kho99}, so-called bridge and corner sites. Note that the
specific choice of basis orbitals shown in Fig. \ref{fig:ce_phase}
is motivated by the convenience of this basis for the calculation
of electronic structure. The expectation values of actual
observables such as the charge distribution and the type of
occupied orbitals are, of course, independent of the choice of
basis Wannier orbitals.

\begin{figure}[t]
      \epsfxsize=50mm
      \centerline{\epsffile{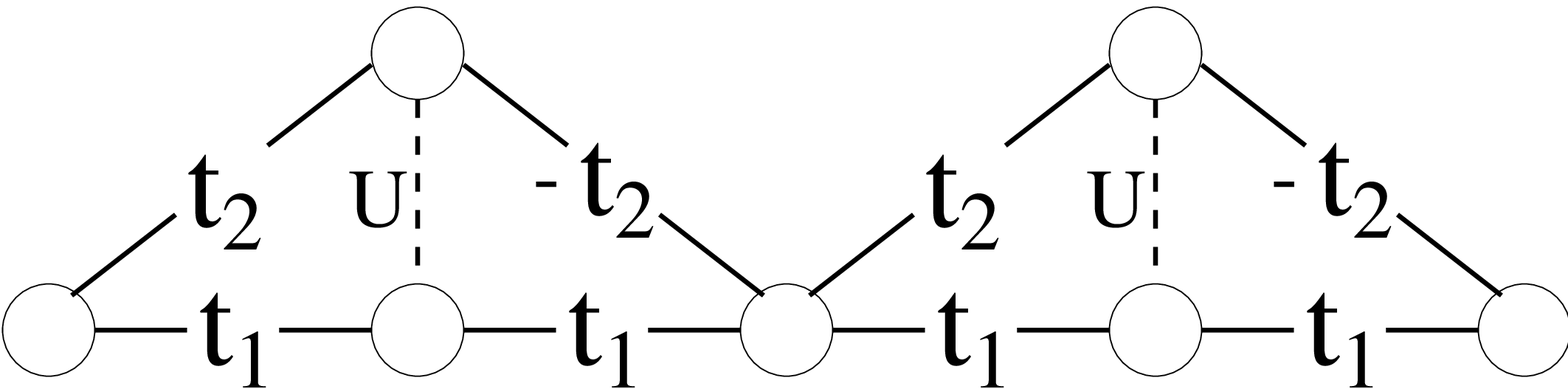}}
      \epsfxsize=85.7mm
      \centerline{\epsffile{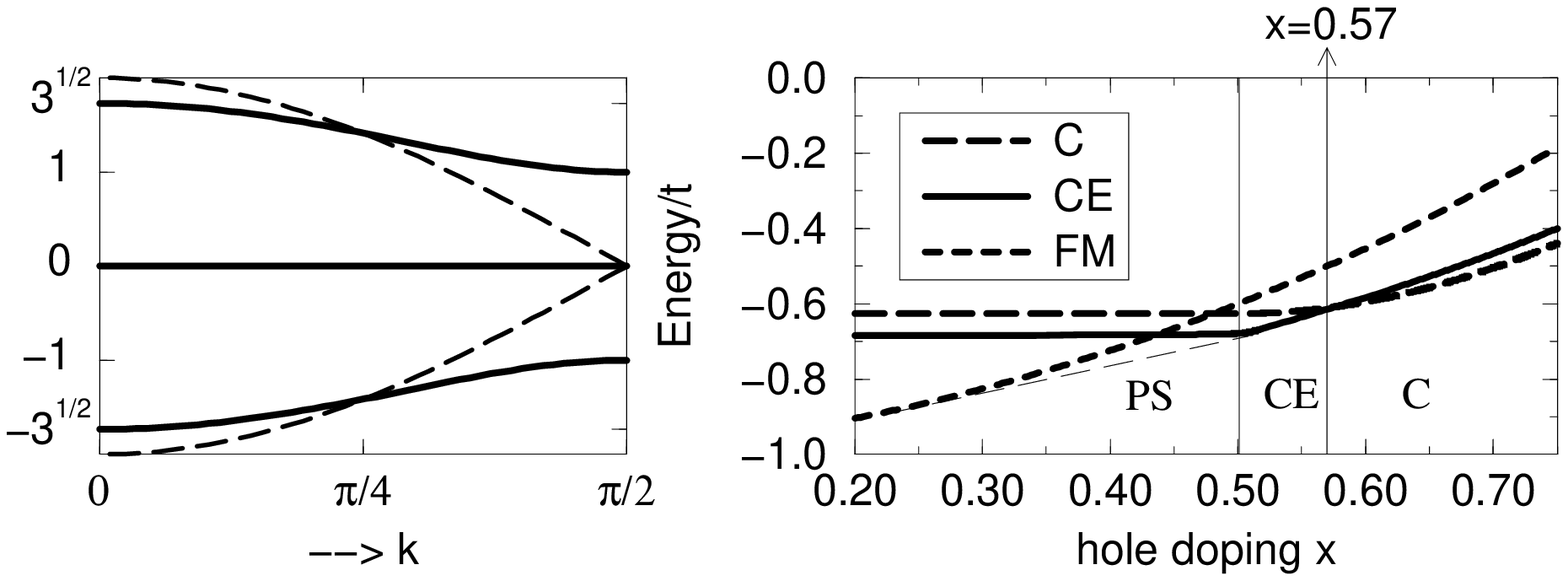}}
\caption {Top: topology of the hopping in a zigzag chain, where
$t_1=t/2$,
 $t_2=t\sqrt{3}/2$, and $U$ is the Coulomb interaction (\protect\ref{ouru}).
 Bottom left: electron dispersion in the zigzag chain of the CE phase for
 $U=0$, and electron dispersion in a C phase (dashed line). Bottom right:
total energy per site for the CE, C and FM phase for $t/J=5$. The
Maxwell construction in the phase separated (PS) region is shown
by the thin dashed line (after Ref. \protect\cite{Kho99}).}
\label{fig:zigzag}
\end{figure}

It is straightforward to solve the band structure problem at no
electron interactions \cite{Solce}. The hopping elements follow
from the Slater-Koster rules \cite{Zaa93}, and are shown in Fig.
\ref{fig:zigzag}. The kinetic energy $H_t$ is given by the same
expression as in Eq. (\ref{hkin}), with the hopping integrals
connecting now the orbitals along a single FM chain of the CE
phase. We note that the unit cell consists of four atoms, but the
electronic problem is simplified by a particular choice of orbital
phases \cite{Kho99}. The orbitals perpendicular to the chain
direction at the bridge positions $|\xi\rangle$ are decoupled from
the directional $|\zeta\rangle$ orbitals and from $|x\rangle$ and
$|z\rangle$ orbitals at corner sites -- hence the bands are
obtained by the solution of a $3\times 3$ matrix. One finds two
bands with energies $\epsilon_{\pm}(k)=\pm t\sqrt{2-\cos 2k}$,
where $k$ is the wave vector ($0<k\leq\pi/2$), and two
nondispersive bands at zero energy. In Fig. \ref{fig:zigzag} we
reproduce the bands reported in Ref. \cite{Kho99}, where a
different convention was used and the gap was found at $k=\pi/2$;
the conventional definition of momentum $k$, however, leads with
the alternating phases of $x^2-y^2$ orbitals at corner sites to
the gap at $k=0$.

At $x=0.5$ the $\epsilon_{-}(k)$ band is fully occupied, and all
other bands are empty. The system is insulating as the occupied
and empty bands are separated by a gap $\Delta=t$ at $k=0$. In the
C and CE phase the FM chains are decoupled due to the DE mechanism
(Sec. II.B), and the kinetic energy is supplemented by the
magnetic energy, equal for both structures. However, the opening
of the gap at the Fermi-energy in the CE phase lowers its kinetic
energy if the lowest band is filled, i.e., the system is
half-doped. This mechanism is equivalent to the situation known
from the lattice-Peierls problem, where the opening of a gap
stabilizes the ground state with a lattice deformation. In the
half-doped manganites, however, the gap is a direct consequence of
the symmetry of the $e_g$ wave functions and is therefore a very
robust feature.
In the insulating state at $x=0.5$ the average occupancy of the
$|\zeta\rangle$ orbital at the bridge site ($3x^2-r^2$ or
$3y^2-r^2$) is $n_b=\langle n_{m\zeta}\rangle=1/2$, while the
orthogonal $|\xi\rangle$ orbitals at bridge positions are empty.
The charge is thus uniformly distributed, with the average
occupancy at the corner sites equal to $n_c=\langle
n_{ix}+n_{iz}\rangle=1/2$, and the ratio between $|x\rangle$ and
$|z\rangle$ occupancy of $n_x:n_z=\sqrt{3}:1$. This ratio reflects
simply the ratio of hopping amplitudes between the bridge
$|\zeta\rangle$ orbital and both orbitals at the corner positions.
This symmetry in charge distribution is removed when the Coulomb
interaction $U$ between $e_g$ electrons occupying different
orbitals at the same site is taken into account. Using the above
notation we write the Coulomb interaction as follows,
\begin{equation}
H_U= U\left(\sum_{m\in B}n_{m\zeta}n_{m\xi}+\sum_{i\in
C}n_{ix}n_{iz}\right), \label{ouru}
\end{equation}
where $n_{i\alpha}=c^{\dagger}_{i\alpha}c_{i\alpha}^{}$ are the
respective electron number operators, and the summations run over
bridge ($m\in B$) and corner ($i\in C$) positions of zigzag
chains. For the $e_g$ electrons in the manganites $U\approx 10t$,
so that the system is strongly correlated. We have verified using
a HF approximation \cite{Cuo00} that the interactions have a very
different effect on the corner and bridge-sites. On the bridge
positions the $|\xi\rangle$ orbitals are always empty, so that the
Coulomb repulsion is ineffective. In contrast, both orbitals
($|x\rangle$ and $|z\rangle$) are partially occupied on the corner
sites, and thus the total electron density $n_c$ decreases with
increasing $U$. The same effect was found by ED of finite clusters
and with the Gutzwiller projection method by van den Brink,
Khaliullin, and Khomskii \cite{Kho99}.

The phase diagram at different doping and $U=0$ in depicted in
Fig. \ref{fig:zigzag}. We observe that CE-phase is stable only in
the nearest vicinity of half-doping. For $x>1/2$ the holes that
are doped into the lower $\epsilon_-$-band efficiently suppress
the CO state. In this doping range the CE phase becomes unstable
with respect to the C phase, and the kinetic energy of the C phase
is lower for $x>0.57$ (Fig. \ref{fig:zigzag}). For $x<1/2$ the
energy per site of the CE phase is constant because the extra
electrons are doped in the nondispersive bands at zero energy;
this is reflected by a kink obtained in the energy as a function
of doping at $x=1/2$. For lower hole-doping (higher electron
concentration) the homogeneous FM phase is more stable, as
expected.

\begin{figure}[t]
      \centerline{\hbox{
      \epsfysize=7cm
      \epsfxsize=7cm
      \epsffile{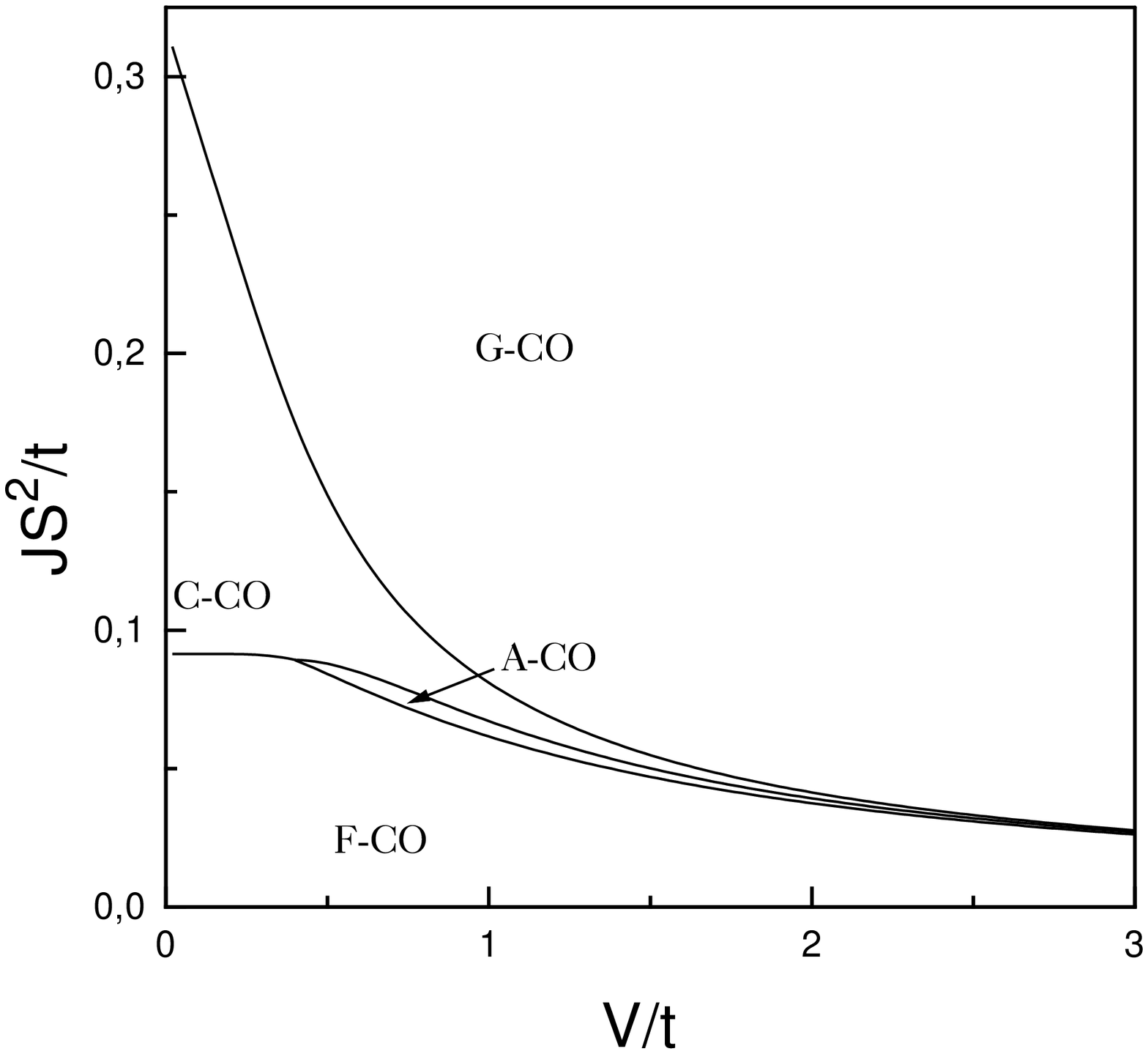}
      \epsfysize=7cm
      \epsfxsize=7cm
      \epsffile{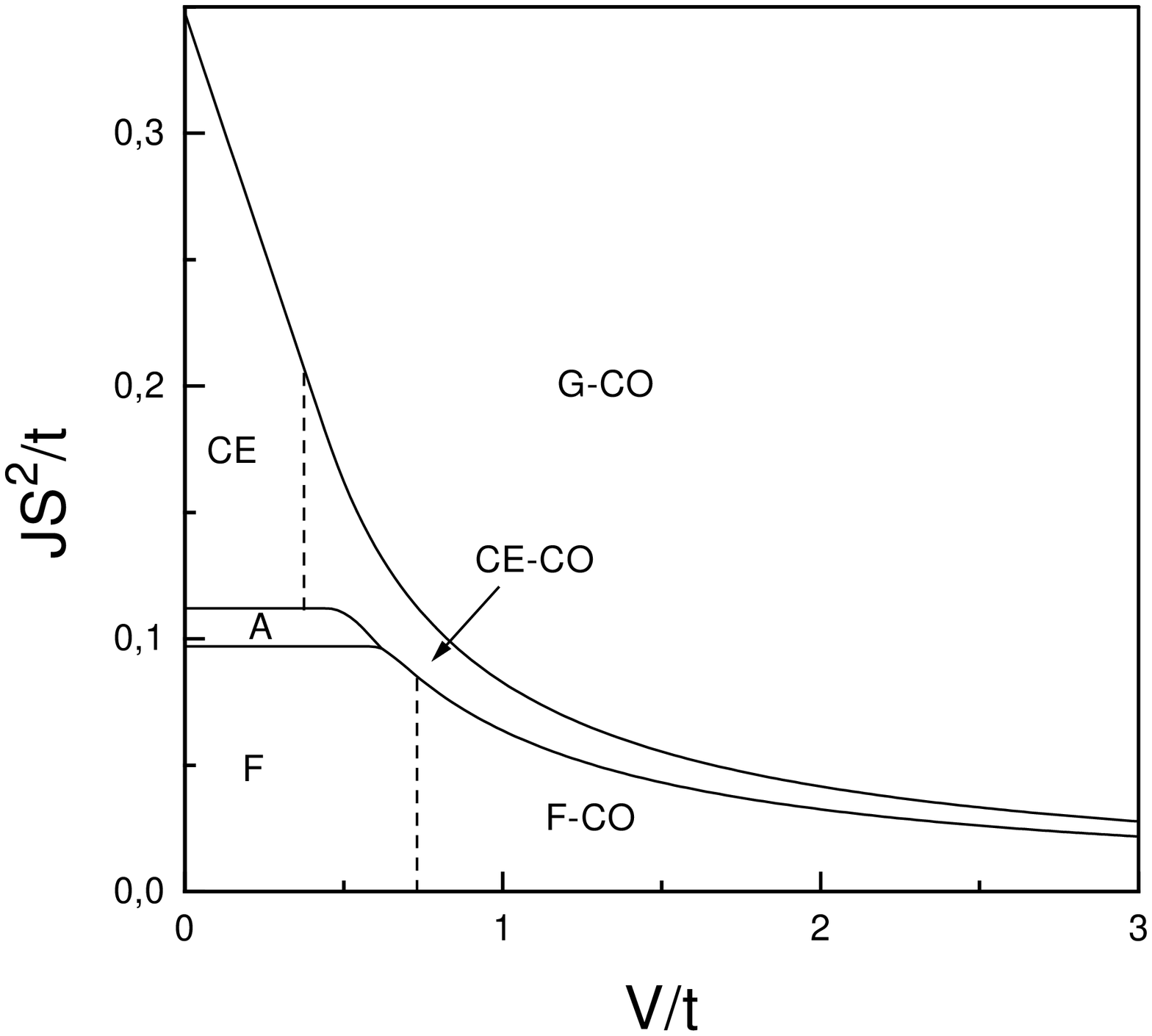}}}
\smallskip
\caption{
 Phase diagram of the one-orbital model (left) two-orbital model
 (\protect\ref{m1}) (right) for different values of $JS^2/t$ ($J=J_{AF}$) and
 $V/t$ parameters. Here A, C, and G stand for different magnetic ordering:
 A-AF, C-AF, G-AF and FM, respectively, and CO is charge ordering. The dashed
 line in the two-orbital case separates
 the uniform phase from the CO states in corresponding magnetic phases
 (after Ref. \protect\cite{Jac00}).}
\label{fig:onetwo} \label{oneorb}
\end{figure}

The interest in the origin of CO is also motivated by recent
experimental data. The charge order functions are markedly
stronger than the orbital fluctuations, both below and above the
magnetic transition \cite{Zim99}. Thus it appears that the
transition is driven by CO fluctuations, and the orbital state
follows. Recognizing that the on-site Coulomb interaction between
the orbitals at corner sites (\ref{ouru}) leads to the CO state,
we point out that another possibility to stabilize this state
follows from the long-range Coulomb repulsion. Thus we consider
the two-orbital FM Kondo lattice model (\ref{eq:deeg}) filled by
$n=0.5$ electron per site,
\begin{equation}
H=-\sum_{ij\alpha\beta\sigma}t_{ij}^{\alpha\beta}
c_{i\alpha\sigma}^{\dagger}c_{j\beta\sigma}^{}
-J_H\sum_{i\alpha\sigma\sigma{^\prime}} {\vec S}_i\cdot
   c_{i\alpha\sigma}^{\dagger}\vec{\sigma}_{\sigma\sigma{^\prime}}^{}
   c_{i\alpha\sigma{^\prime}}^{}
+J_{AF}\sum_{\langle ij\rangle}{\vec S}_i\cdot{\vec S}_j +
V\sum_{\langle ij\rangle}n_{i}n_{j}, \label{m1}
\end{equation}
which is extended by the intersite Coulomb repulsion $V$. On-site
Coulomb interaction is not included, however, so the previous
mechanism is absent. Studying the model (\ref{m1}) within the MF
approximation, a competition between different types of magnetic
ordering was established \cite{Jac00}. Let us compare the phase
diagrams obtained for a nondegenerate Kondo lattice model
(\ref{kondol}) extended by a similar intersite Coulomb interaction
term $\propto V$, and for the two-orbital model (\ref{m1}). In the
case of the one-orbital model the MF theory predicts a 'continuous
increase of the CO due to the intersite Coulomb interaction $V$.
Depending on the value of the intersite Coulomb interaction $V$
and on superexchange coupling $J_{AF}$, different types of spin
ordering shown in Fig. \ref{fig:onetwo} (A-AF, C-AF, G-AF, and FM)
coexist with the CO in the ground state of the one-orbital model.
In contrast, in the two-orbital model the transition to the CO
states occurs only at a finite critical value of $V$; thus
magnetic states are obtained either with or without CO, depending
on the value of $V$.
The presence of orbital degeneracy with the peculiar anisotropic
hopping amplitudes between $e_g$ orbitals introduces a new
magnetic state which was absent in the nondegenerate model,
CE-type spin ordering (Fig. \ref{fig:onetwo}). In contrast to the
nondegenerate case, the C-AF ordering is never found within the
model (\ref{m1}) due to its instability against the effective
"dimerization" and the onset of the zigzag FM order. The
alternation of the FM bonds in $a(b)$ directions leads to the
alternation of the hopping amplitude. As a result the bare band is
split into bonding and antibonding states and the "dimerization"
gap opens on the Fermi surface at half-doping. The CE-type spin
ordered state is accompanied with $d_{3x^2-r^2}/d_{3y^2-r^2}$
orbital ordering that fits naturally to the topology of the zigzag
structure as it was discussed above. The CE phase with magnetic
and charge ordering as shown in Fig. \ref{fig:ce_phase} also wins
over the charge disordered A-AF ordering in the regime of
intermediate parameters. One may speculate that the competition
between A-AF charge disordered and CE CO states indicates that
parameters of the system are close to the critical values.
Therefore, the small change of the bandwidth or the coupling to
the lattice might stabilize one or the other state.

We have to conclude that these answers concerning the stability
and the properties of the CE phase have to be treated as rather
preliminary. The insulating behavior which follows from a
topological phase factor in the hopping is certainly an
interesting observation, and the CE phase seems to be indeed a
particular type of order driven by the degeneracy of $e_g$
orbitals. The on-site and intersite Coulomb interactions are
likely to help each other in stabilizing the CO states
\cite{Mut99}. Although the intersite interaction stabilizes the CO
state in a 2D lattice, it favors the same charge alternation in
the third direction, contrary to the structure of the CE phase.
This qualitative trend may by reversed only by other interactions.
However, the precise role of different Coulomb interactions and of
the JT effect in the stability of the CE phase \cite{Miz97} are
not understood at present and have to be established by future
studies.

\subsection{ Stripes in manganites }

As we have discussed above, one of the unique aspects of physics
of manganites is the unusually strong interaction between charge
carriers and lattice degrees of freedom, due to which markedly
distinct types of charge, orbital and magnetic ordering are
observed in different doping regimes. The strong electron-phonon
coupling, which can be tuned by varying the electronic doping,
electronic bandwidth and disorder, gives rise to a complex
phenomenology, in which crystallographic structure, magnetic
structure and transport properties are intimately interrelated.
Below a certain temperature $T_{CO}$, electronic carriers become
localized onto specific sites, which display LRO throughout the
crystal structure (CO states). Moreover, the filled $e_g$
($3z^2-r^2$-like) orbitals at Mn$^{3+}$ ions and the associated
lattice distortions (elongated Mn--O bonds) also develop LRO
(orbital ordering). Finally, the magnetic exchange interactions
between neighboring Mn ions, mediated by oxygen ions, become
strongly anisotropic which gives rise to complex magnetic ordering
in the stable structures. Historically, magnetic ordering was the
first to be experimentally investigated in manganites
\cite{Wol55}, where the magnetic structures of a series of
manganese perovskites with general formula
La$_{1-x}$Ca$_{x}$MnO$_{3}$ were studied using neutron powder
diffraction. It was not until recently, however, that the
crystallographic superstructure of this compound was
experimentally observed by electron diffraction \cite{Rad97}. The
curious static patterns in the spin, charge, and orbital densities
observed in manganites are currently attracting much attention
\cite{Che98,Che99}.
\begin{figure}[tbp]
\centerline{\epsfig{figure=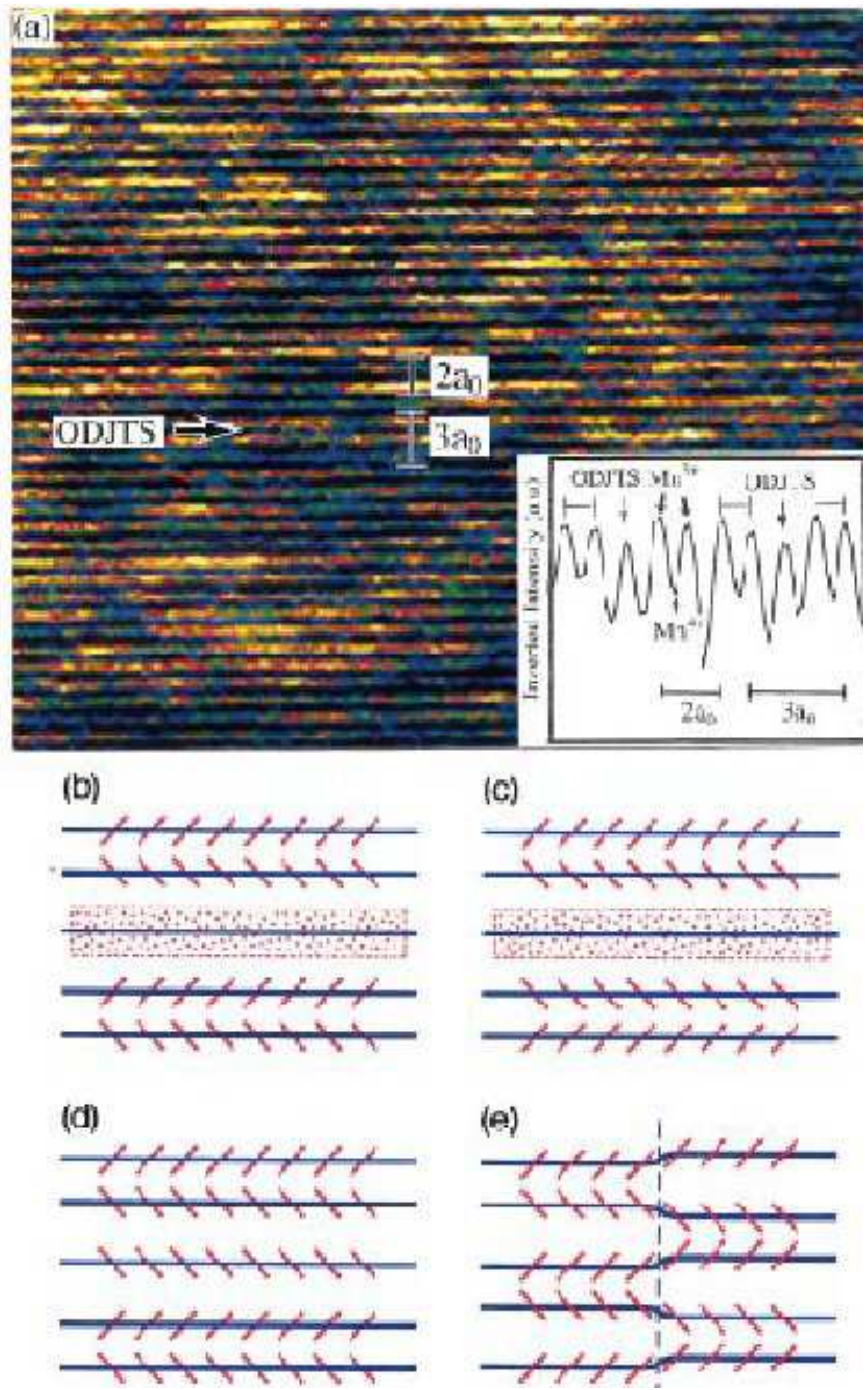,width=11.5cm}}
\smallskip
\caption {(a) High resolution lattice image in the FM phase at 206
K showing a mixture of incommensurate charge-ordered and FM charge
disordered microdomains. The arrow stands for an unpaired JT
stripes (JTS) between paired ones. An inverted intensity profile
of this configuration is plotted in the inset of (a). (b) and (c)
show in-phase and out-of-phase configurations, respectively, of
the paired JTS surrounding the orbital disordered JTS. The lines
represent Mn$^{3+}$ JTS while the slanted slashes (random dotted
strip) stand for $d_{z^2}$ orbital ordering (disordering). (b) and
(c) go into residual discommensuration (d) and antiphase boundary
(e), respectively, in the incommensurate phase when complete
orbital order is realized (after Ref. \protect\cite{Che98}). }
\label{fig:mori}
\end{figure}
It is very interesting, that the charge and orbital ordering (COO)
observed experimentally is always concomitant with the stripe-AF
(S-AF) phase. At $x=1/2$, the COO, known as a CE phase (Sec.
VII.B), has been confirmed by the synchrotron X-ray diffraction
experiment \cite{Murce98}. At $x=2/3$, however, two different COO
patterns have been reported: (i) the {\it bistripe\/} (BS)
structure \cite{Che98} (Fig. \ref{fig:mori}), in which the main
building block of the COO pattern at $x=1/2$ persists even at
$x=2/3$, and (ii) the {\it Wigner-crystal\/} (WC) structure
\cite{Rad99}, in which the COO occurs with the maximized distance
between $e_g$ electrons. The appearance of these two different
structures suggests that: (i) the corresponding energies are very
close to each other and the ground state has (quasi-)degeneracy,
and (ii) the conversion between them is prohibited by a large
energy barrier. Under these circumstances, it is of limited
relevance to investigate which of the two states has a lower
energy using some model Hamiltonian. Rather than attempt to
investigate which one gives the ground state for particular
parameters, we present {\it in extenso\/} the recent ideas of
Hotta {\it et al.\/} \cite{Hot00} on the origin of the near BS-WC
degeneracy.

Let us discuss the importance of the role of the 1D conducting
zigzag paths in the $(a,b)$ basal plane, and considered
parallel-transport of an $e_g$ electron along these paths through
the JT centers composed of MnO$_6$ octahedra. The transport
invokes the Berry-phase connection and we can introduce the
``winding number'' $w$ as a direct consequence of topological
invariance which should be conserved irrespective of the details
of $H$. Consider $e_g$ electrons coupled both to localized
$t_{2g}$ spins with the Hund's rule coupling $J_H$, and to JT
distortions of the MnO$_6$ octahedra. Since $J_H$ is the largest
characteristic energy among those considered here, it is taken to
be infinite for simplicity. This implies that the spin of each
$e_g$ electron at a Mn site aligns completely in parallel with the
direction of the $t_{2g}$ spin $S=3/2$ at the same site. Thus, the
spin degrees of freedom are effectively lost for the $e_g$
electrons, and the spin index will be dropped hereafter. Since it
is known experimentally that the $t_{2g}$ spins are antiparallel
along the $c$-axis, we can assume that the $e_g$ electrons can
move only in the $(a,b)$ plane due to the DE mechanism.

The above situation is well described by the Hamiltonian
\cite{Hot00},
\begin{eqnarray}
 H &=& - \! \sum_{ij\alpha\beta}
 t^{\alpha\beta}_{ij} c_{i\alpha}^{\dag} c_{j\beta}^{}
 \!+\! J_{AF}\sum_{\langle ij\rangle}{\vec S}_i\cdot {\vec S}_j \nonumber \\
 &+& \!E_{\rm JT} \! \sum_i \Bigl[
 2\sum_{\alpha\beta}c_{i\alpha}^{\dagger}
 (Q_{2i}^{}\sigma_i^x+Q_{3i}^{}\sigma_i^z)_{\alpha\beta}^{}c_{i\beta}^{}
+(Q_{2i}^2+Q_{3i}^2) \Bigr], \label{stripeham}
\end{eqnarray}
with the same notation as used in Eq. (\ref{hamhotta}). The second
term $\propto J_{AF}$ represents the AF coupling between
nearest-neighbor classical $t_{2g}$ spins which are normalized for
convenience to $|{\vec S}_i|=1$. The third term is controlled by
the JT energy $E_{\rm JT}$ and describes the coupling of an $e_g$
electron with the $(x^2$$-$$y^2)$- and $(3z^2$$-$$r^2)$-type
(dimensionless) JT modes, given by $Q_{2i}$ and $Q_{3i}$ (Fig.
\ref{fig:jtmodes}), respectively.

Intuitively, it can be understood that the competition between
kinetic and magnetic energies can produce an S-AF state. The
ground state of the system described by Eq. (\ref{stripeham}) with
$J_{AF}=0$ is a 2D FM metal and optimizes the kinetic energy of
$e_g$ electrons, while it becomes a 2D AF insulator at $J_{AF}\geq
t$ to exploit the magnetic energy of the $t_{2g}$ spins. For
smaller but nonzero values of $J_{AF}$, a competition between
these states occurs which results in a mixture of FM and AF
states. In this S-AF state, a 1D conducting path can be defined by
connecting nearest-neighbor sites with parallel $t_{2g}$ spins. A
path with a large stabilization energy is needed to construct a
stable 2D structure. Thus, Hotta {\it et al.\/} \cite{Hot00}
concentrated themselves on specifying the shapes of (quasi-)stable
1D paths in the S-AF manifold.

Let us start with the case of $E_{\rm JT}$=$0$ and no electron
correlation, which allows us to illustrate the importance of
topology in the present problem. In a 2D FM metal, the
kinetic-energy gain is reduced by $2J_{AF}$, due to the loss of
magnetic energy per site. In contrast, in a 2D AF insulator, the
magnetic energy gain per site is $2J_{AF}$. In S-AF states, the
optimized periodicity $M$ for a 1D path along the $a$-axis
direction is numerically found to be given by $M$=$2/n$
\cite{Hot00}, in agreement with Ref. \cite{Koi98}, where $n(=1-x)$
is the $e_g$-electron number per site. The S-AF structure with
zigzag path is nothing but the well-known CE phase, and the
analysis of Ref. \cite{Hot00} predicts that this state is very
stable at $x$=1/2. However, the same analysis for $x\ge 2/3$ does
not lead to a zigzag path but instead to a straight line as the
optimized structure, which disagrees with experiment. Thus, it is
necessary to find a quantity other than the energy to discuss the
possible preferred paths that may arise from a full calculation,
including nonzero $E_{\rm JT}$ and Coulomb interactions.
Reconsidering the results at $x$=$1/2$ led Hotta {\it et al.\/}
\cite{Hot00} to the idea that the number of vertices along the
path, $N_{\rm v}$, may provide the key difference among paths. A
confirmation of this idea is provided by the calculation of
energies for the $2^6$ and $2^8$ paths at $x=2/3$ ($M$=6) and
$3/4$ ($M$=8), respectively.

In the next step one may include the JT distortions, $E_{\rm
JT}\neq 0$ in Eq. (\ref{stripeham}). By writing the JT modes in
polar coordinates as $Q_{2i}=Q_i\sin\theta_i$ and
$Q_{3i}=Q_i\cos\theta_i$, ``phase-dressed'' fermion operators,
${\tilde c}_{ix}$ and ${\tilde c}_{iz}$, are introduced as:
\begin{eqnarray}
{\tilde c}_{ix}&=& e^{i\theta_i/2}
 [ c_{ix}\cos(\theta_i/2) + c_{iz}\sin(\theta_i/2)],                 \\
{\tilde c}_{iz}&=& e^{i\theta_i/2}
 [-c_{ix}\sin(\theta_i/2) + c_{iz}\cos(\theta_i/2)],
\end{eqnarray}
with $e^{i\theta_i/2}$ representing the molecular Aharonov-Bohm
effect. The amplitude $Q_{i}$ is determined by a MF approximation
\cite{Hot00}, while the phases $\theta_i$'s are interrelated
through the Berry-phase connection to provide the winding number
$w$ along the 1D path as $w=\oint_c d{r}\cdot\nabla\theta
/(2\pi)$, where $c$ forms a closed loop for the periodic lattice
boundary conditions. Mathematically, the winding number $w$ is
proven to be an integer \cite{Hot98}. In this system, it may be
decomposed into two terms as $w=w_{\rm g}+w_{\rm t}$. The former,
$w_{\rm g}$, is the geometric term which becomes $w_{\rm g}=0$ or
1, corresponding to the periodic or antiperiodic boundary
condition in the $e_g$ electron wave function. It may be shown
that the kinetic energy is lower with $w_{\rm g}$=$0$ than that
with $w_{\rm g}$=$1$ for $x\ge 1/2$. Thus, $w_{\rm g}$ is taken as
zero hereafter.
\begin{figure}[b]
\centerline{\epsfig{figure=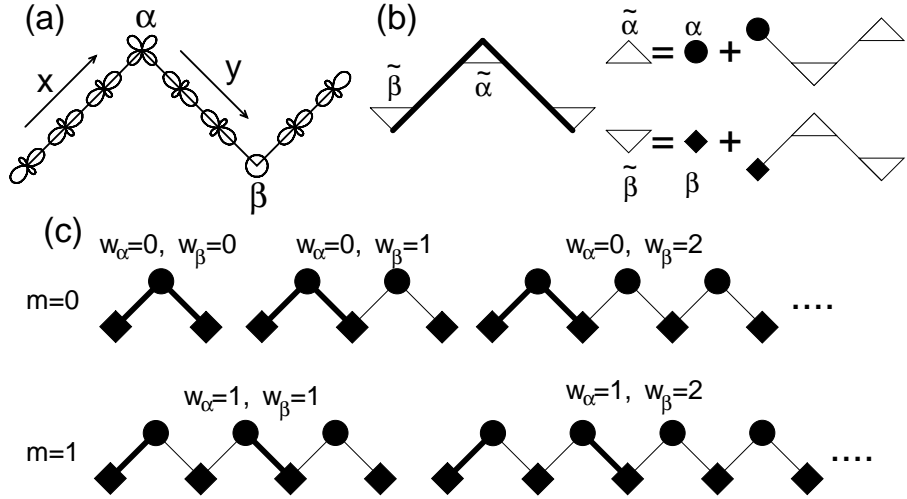,width=13cm}}
\smallskip
\caption {(a) A typical building block for a 1D path for an $e_g$
electron with JT distortions. (b) General structure of the
lowest-energy-state path and the renormalization scheme for the
vertices $\alpha$ and $\beta$. The thick (thin) line denotes the
straight-line part with (without) an $e_g$ electron localized on
it. The solid circle and diamond denote, respectively, the bare
vertices, $\alpha$ and $\beta$, while open up-and down-triangles
indicate the renormalized vertices, ${\tilde\alpha}$ and ${\tilde
\beta}$. Note that the periodicity of the 1D path is given by
$M=2/n=2/(1-x)$. (c) Groups of 1D paths derived from mother states
with $m=0$ and 1. Paths in the first column corresponding to the
mother WC structures with $w$=$2m$+$1$, which produce daughter
states with $w=2m+2, 2m+3, \cdots$ (after Ref.
\protect\cite{Hot00}). } \label{fig:hotta2}
\end{figure}
To show that only the number of vertices $N_{\rm v}$ along a given
path determines the topological term $w_{\rm t}$, let us consider
the transfer of a single $e_g$ electron along the path shown in
Fig. \ref{fig:hotta2}(a). On the straight line part in either $a$-
or $b$-direction, the phase is fixed at $\theta_a$=$2\pi/3$
($\theta_b$=$4\pi/3$), because the $e_g$-electron orbital is
polarized along the direction of the bridge bonds. This effect may
be called an {\it orbital double-exchange\/} in the sense that the
orbitals align along the actual chain direction to maximize the
kinetic energy, similarly as the FM alignment of $t_{2g}$ spins in
induced by the usual DE mechanism (Sec. II.B). Thus, $w_{\rm t}$
does not change when one of the bridge positions is passed. In
contrast, when the electron passes by one of the vertex positions,
the phase changes from  $\theta_a$ to $\theta_b$ (from $\theta_b$
to $\theta_a$), indicating that the electron picks up a phase
change of $2\pi/3$ ($4\pi/3$). Since these two vertices appear in
pairs, $w_{\rm t}$(=$w$) is evaluated as $w_{\rm t}=(N_{\rm
v}/2)(2\pi/3+4\pi/3)/(2\pi)=N_{\rm v}/2$. The phases at the corner
positions are assigned as an average of the phases sandwiching
those vertices, $\theta_{\alpha}=\pi$ and $\theta_{\beta}=0$, to
keep $w_{\rm g}$ invariant. Then, the phases are determined at all
the sites once $\theta_x$, $\theta_y$, $\theta_{\alpha}$, and
$\theta_{\beta}$ are known.

Now we include the cooperative JT effect, important ingredient to
determine COO patterns in the actual manganites. Although its
microscopic treatment is involved, we can treat it
phenomenologically as a constraint for macroscopic distortions
\cite{Hot00}, energetically penalizing $w=0$ and $M/2$ paths. In
fact, it is numerically found that $w=1, 2, \cdots, M/2-1$ paths
constitute the lowest-energy band and they can be regarded as
degenerate, since its bandwidth is about $0.01t$, much smaller
than the interband energy difference ($\sim 0.1t$). Summarizing,
the cooperative JT effect gives us two rules for the localization
of $e_g$ electrons \cite{Hot00}:
 (i) electrons never localize at vertices;
(ii) electrons localize pairwise -- any electron that localizes on
one of the segments in the $a$-direction has a partner that
localizes on one of the segments in the $b$-direction.

Applying these rules, we obtain a general structure for the
lowest-energy path. Important features are the {\it renormalized
vertices\/}, ${\tilde\alpha}$ and ${\tilde\beta}$, abbreviated
notations to represent the set of straight-line parts that are not
occupied by $e_g$ electrons. The winding number assigned to
${\tilde\alpha}$ (${\tilde\beta}$) is $1/3$+$w_{\alpha}$
($2/3$+$w_{\beta}$), where the number of vertices included in
${\tilde \alpha}$ (${\tilde \beta}$) is $1$+$2w_{\alpha}$
($1$+$2w_{\beta}$). Thus, the lowest-energy path is labeled by the
nonnegative integers $w_{\alpha}$ and $w_{\beta}$, leading to a
total winding number $w$=$1$+$w_{\alpha}$+$w_{\beta}$. Although
the topological argument does not determine the precise position
at which an $e_g$ electron localizes in space, it is enough to
regard a charged straight-line part as a {\it quasi-charge\/}.
Since the quasi-charges align at equal distance in the WC
structure, the corresponding path is labeled by
$w_{\alpha}$=$w_{\beta}=m$, with $m$ a nonnegative integer. By
increasing $w_{\beta}$ keeping $w_{\alpha}$ fixed, we can produce
any non-WC-structure paths with $w_{\alpha}=m$ and $w_{\beta}=m+1,
m+2, \cdots$ [Fig. \ref{fig:hotta2}(c)]. In this way, the WC
structure with $w=2m+1$ can be considered the {\it mother state\/}
for all non-WC-structure paths with $w=2m+2, 2m+3, \cdots$,
referred to as the {\it daughter states\/}. The states belonging
to different $m$'s are labeled by the same $w$, but a large energy
barrier exists for the conversion among them, since an $e_g$
electron must be moved through a vertex in such a process. Thus,
the state characterized by $w$ in the group with $m$, once formed,
it cannot decay, even if it is not the lowest-energy state.

Note that the topological argument works irrespective of the
details of the Hamiltonian $H$, since $w$ is a conserved quantity.
However, it cannot single out the true ground state, since the
quantitative discussion on the ground state energy depends on the
choice of $H$ and on the approximations employed. In fact, either
the BS or WC structure can be stable, but in view of the small
energy difference, their relative energy will likely change
whenever a new ingredient is added to $H$. It may be expected that
these phases are rather sensitive to the Coulomb interactions
which will play an important role in stabilizing one of these two
structures.

\begin{figure}[t]
\centerline{\epsfig{figure=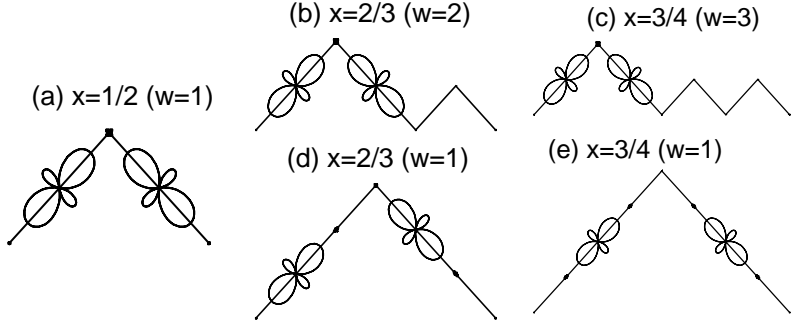,width=14cm}} \vspace{0.15in}
\caption {(a) Path with $w=1$ at $x=1/2$ for $E_{\rm JT}=2t$. At
each site, the
   orbital shape is shown with its size in proportion to the orbital density.
(b) The BS-structure path with $w=2$ at $x=2/3$. (c) The
BS-structure path with $w=3$ at $x=3/4$. (d) The WC-structure path
with $w=1$ at $x=2/3$. (e) The WC-structure path with $w=1$ at
$x=3/4$ (after Ref. \protect\cite{Hot00}). } \label{fig:hotta3}
\end{figure}

Now we analyze following Hotta {\it et al.\/} \cite{Hot00} the
charge and orbital arrangement in La$_{1-x}$Ca$_x$MnO$_3$
($x>0.5$), in which the experimental appearance of the BS
structure provides key information to specify the 1D path. Since
the quasi-charges exist in a contiguous way in the BS structure,
its path is produced from the mother state with $m=0$ [see Fig.
\ref{fig:hotta2}(c)]. In particular, the COO pattern in the
shortest 1D path is uniquely determined as shown in Figs.
\ref{fig:hotta3}. At $x=1/2$, the path is characterized by $w=1$
which is the basic mother state with $m=0$. The COO pattern shown
in Fig. \ref{fig:hotta3}(a) leads to the CE-type AF state
\cite{Murce98}. The other paths with $w=2$ and 3 are nothing but
the BS structures experimentally observed at $x=2/3$ and 3/4
\cite{Che98}.

It may be assumed that the long-range Coulomb interaction $V$
destabilizes the BS structure and transforms it to the WC
structure, but this is not the case; for the BS $\rightarrow$ WC
conversion with the help of $V$, an $e_g$ electron must be on the
vertex in the path with $w$=$2$ or $3$ [see Figs.
\ref{fig:hotta3}(b) and \ref{fig:hotta3}(c)]. This is against rule
(i) and thus, the BS structure, once formed, is stable due to the
topological condition, even including a weak repulsion $V$. In the
group of $m=0$, the WC structure appears only in the path with
$w=1$. Thus, the WC-structure paths with $w=1$ at $x=2/3$ and 3/4
are obtained by simple extension of the straight-line part in the
path at $x=1/2$ [see Figs. \ref{fig:hotta3}(d) and
\ref{fig:hotta3}(e)]. The detailed charge distribution inside the
quasi-charge segment is determined by a self-consistent
calculation with the JT effect, leading to the WC structure. Even
if the non-WC structure occurs for $w=1$, it is unstable in the
sense that it is easily converted to the WC structure, because no
energy barrier exists for an $e_g$ electron shift along the
straight-line part.
The above topological analysis shows that:
 (i) the WC structure is made of $w_{\rm WC}$=$1$ zigzag paths, and
(ii) the BS structure contains a shorter-period zigzag path with
     $w_{\rm BS}=M/2-1=x/(1-x)$.
Note that on the BS path, the less-distorted Mn$^{4+}$ sites
occupy all the vertices ($N_{\rm v}$ equals the number of
Mn$^{4+}$ ions), while the heavily distorted Mn$^{3+}$ sites
appear in pairs (the number of Mn$^{3+}$ ions equal to $2$). Thus,
$w_{\rm BS}$ is rewritten as
\begin{equation}
\label{w2} w_{\rm BS} = {N_{\rm v} \over 2} ={{\rm
Number~of~Mn}^{4+}~{\rm ions} \over {\rm Number~of~Mn}^{3+}~{\rm
ions}} ={x \over 1-x}.
\end{equation}
Since $w_{\rm BS}$ is an integer, we can predict that at specific
values of $x$[=$w_{\rm BS}$/($1$+$w_{\rm BS}$)], such as $1/2$,
$2/3$, $3/4$, {\it etcetera\/}, nontrivial charge and orbital
arrangement will be stabilized in agreement with the experimental
observation \cite{Che98}.

\subsection{ Orbital ordering and phase separation }

Finally, we turn to the competition between different magnetic
phases in doped manganites. There is no controversy on the fact
that ferromagnetism at large doping is promoted by the DE
mechanism -- the kinetic energy of $e_g$ electrons is maximized
when the $t_{2g}$ spins and their own spins are aligned. Within
this scenario, one gets a natural explanation of the FM metallic
phase. When an $e_g$ electron moves in some region, it does not
pay an energy $J_H$ if all the $t_{2g}$ spins in its neighborhood
are parallel. The hole-spin scattering is reduced in this way, and
one gains the kinetic energy. As the carrier concentration
increases, the FM polarization clouds around the holes start to
overlap and the ground state becomes metallic, with $e_g$
electrons moving in the correlated degenerate band, inducing the
saturated FM ordering.
\begin{figure}[t]
\centerline{\epsfig{figure=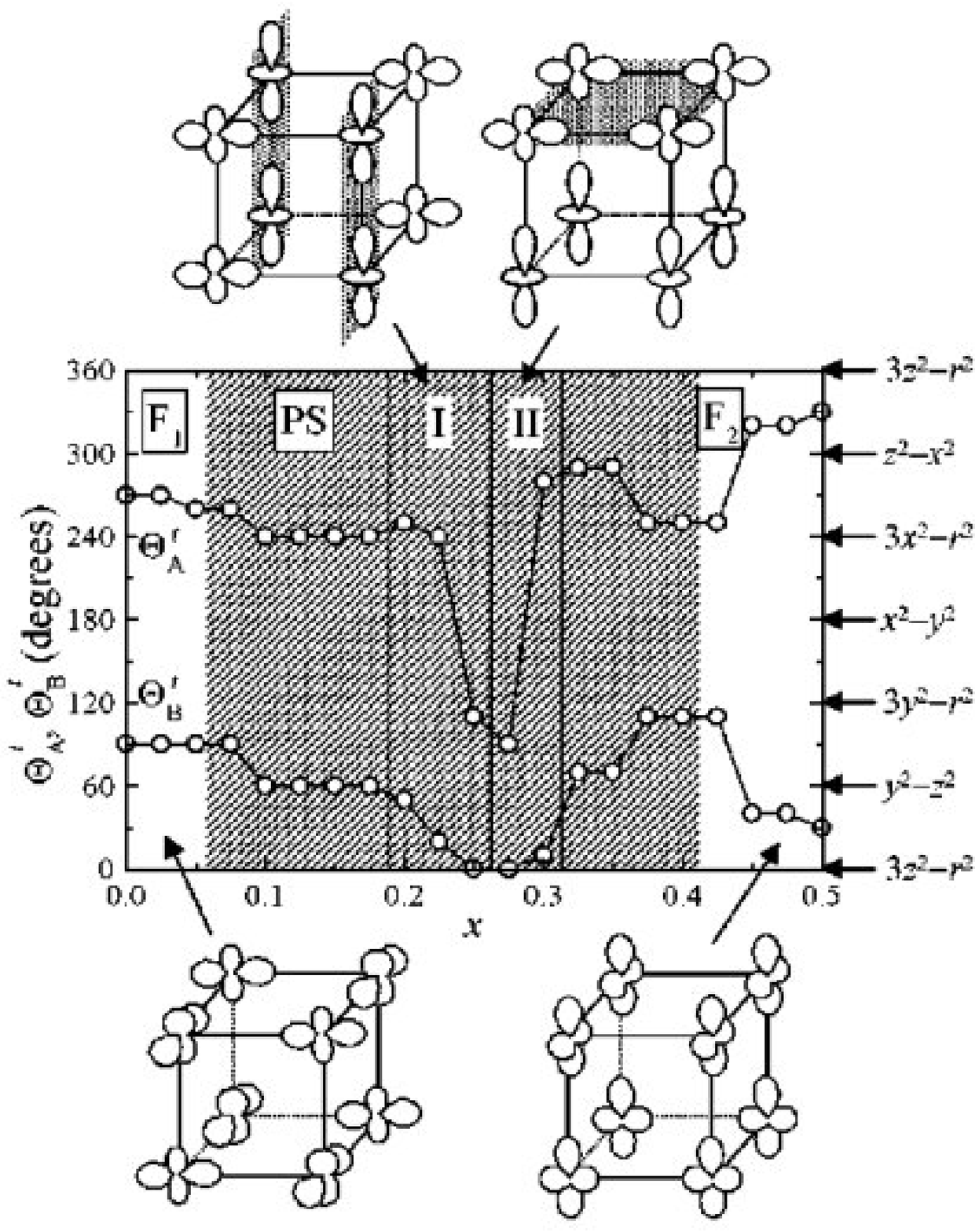,width=13cm}}
\vspace{0.15in} \caption {A sequential change of orbital states as
a function of hole concentration $x$ (after Ref.
\protect\cite{Ish00}). $\Theta_{A(B)}^t$ is the angle in the
orbital space in the $A(B)$ orbital sublattice. Note that these
angles are related to the angle used in Eq.
(\protect\ref{newstate}) by $\theta_{A(B)}=2\Theta_{A(B)}^t$. The
schematic orbital states are shown in phase-$I$ and phase-$II$;
the dotted areas show the regions where the hole concentration is
rich.} \label{fig:oodoping}
\end{figure}
The understanding of the role played by the orbital degrees of
freedom in the FM metallic phase is not complete, however. As
suggested by the studies of DE both in the electron and hole
doping regime \cite{Bri99,Ish00,Mae98}, one possibility is that
the orbital ordering of some type is stabilized in particular
doping regimes. The transitions between different types of orbital
ordering are promoted by the kinetic energy which is optimized by
either planar $|x\rangle$-type or directional $|z\rangle$-type
orbitals, while the kinetic energy in the phases with orbital
ordering and occupied orbitals that are linear combinations of
$|x\rangle$ and $|z\rangle$ states as in Eqs. (\ref{orbmoffa}) is
lower \cite{Kha00,Fei00}. This explains the mechanism of an
orbital liquid discovered by Nagaosa {\it et al.\/} \cite{Nag97},
and confirmed by numerical studies \cite{Mac99}.

It has been found that due to a competition between the
superexchange which favors orbital ordering and DE which favors a
uniform phase, the magnetic phase diagrams obtained from the
models which include the degeneracy of $e_g$ orbitals have orbital
ordering also in the FM phase \cite{Ish00,Mae98}. An example is
shown in Fig. \ref{fig:oodoping}, where all FM states have some
kind of orbital ordering, modified under increasing doping. Such
states are in qualitative agreement with the experimental
observations at low doping, where the orbital ordering was
observed in La$_{0.88}$Sr$_{0.12}$MnO$_3$ \cite{End99}, and may be
concluded from the anisotropic exchange interactions found in
La$_{0.85}$Ca$_{0.15}$MnO$_3$ \cite{Mou00}. The charge is
uniformly distributed in this doping regime, while the orbital
ordering occurs only in low temperatures in
La$_{0.88}$Sr$_{0.12}$MnO$_3$, with a FM phase with disordered
orbitals in the intermediate temperatures \cite{End99}. At higher
doping, the transition to the orbital liquid state is indeed
realized in double-layered manganites, where it is consistent with
the observed lattice deformation along $c$ axis and allows to
explain the observed the observed spin ordering and its anisotropy
\cite{Nag00}

In contrast, two phases (I and II) found for $0.18<x<0.32$ doping
(Fig. \ref{fig:oodoping}) are characterized by the doped holes
concentrated either in the regions of directional (I), or planar
(II) orbitals, separated by the 1D or 2D structures with few holes
and the orbitals being closer to the undoped situation. Whether or
not such a competition between the phases with differently ordered
orbitals really happens close to $x\sim 0.25$ is still
controversial at the moment -- it might be that the phase diagram
of Okamoto, Ishihara, and Maekawa \cite{Ish00} is closer to the
situation found in the insulating rather than metallic FM
manganites. In fact, this competition suggests that there are also
other ways of gaining the kinetic energy in the metallic phase, as
for instance realized in uniform orbital phases with complex
coefficients of $e_g$ orbitals \cite{Fei00}, and such states might
be better candidates in the metallic phase.
\begin{figure}[t]
\centerline{\epsfig{figure=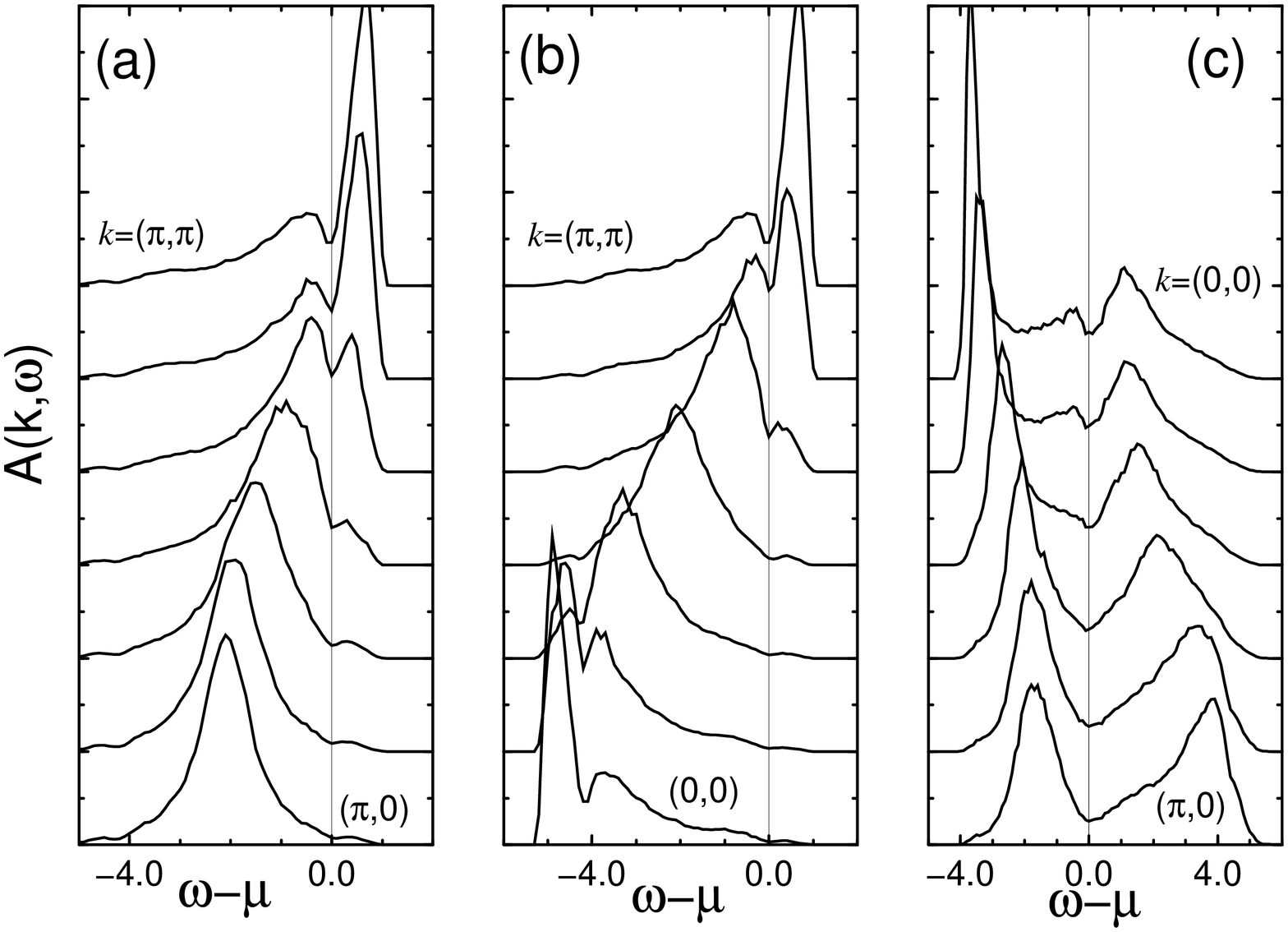,width=8cm,angle=270}}
\vspace{0.2in} \caption {Spectral functions $A({\vec k},\omega)$
at $J_H=\infty$
 for a 2D one-orbital case with $T=1/30$, $\langle n\rangle\sim 0.92$,
 on a $12\times 12$ cluster for: (a) along $(\pi,0)$ to $(\pi,\pi)$;
 (b) along $(0,0)$ to $(\pi,\pi)$. Part (c) shows the results for the 2D
 two-orbital model with $T=1/10$, $\lambda=1.5$, $\langle n\rangle\sim 0.70$,
 on a $10\times 10$ cluster along $(0,0)$ to $(\pi,0)$
 (after Ref. \protect\cite{Mor99}).}
\label{fig:pseudogap}
\end{figure}
The experimental phase diagrams of manganites (Fig.
\ref{fig:expphd}) are typically richer than those obtained from
model calculations. For instance, the phase diagram of Fig.
\ref{fig:oodoping} shows several different types of orbital
ordering, but the magnetic state is FM in the whole range of
doping $0<x<0.5$. Instead, the A-AF order is observed at small
doping \cite{Mou99}, and CE-phase (Sec. VIII.A) at doping $x\simeq
0.5$. It is rather difficult to reproduce this complex behavior in
the theoretical models. However, the interplay of charge, spin,
lattice and orbital degrees of freedom belongs to the generic
features of this class of compounds, and the challenge in the
theory is still to describe these various energy contributions
with comparable and sufficient accuracy.

The competition between the fully polarized FM state and other
types of order which occurs as temperature and electron
concentration are varied is a central problem in the physics of
manganite perovskites. This problem has attracted a lot of
interest recently, especially due to the unusual experimental
results obtained in the manganese oxides \cite{Ram97,cheong}. For
instance, above the Curie temperature $T_C$ and for a wide range
of densities, several manganites show an insulating behavior of
not completely clear origin that contributes to the large
magnetoresistance effects. The low temperature phases have complex
structures, not fully understood within the DE scheme, that
includes different phases with AF and CO ordering, orbital
ordering, FM insulating phase, and tendencies towards the
formation of charge inhomogeneities even in the FM phase.
Moreover, many experimental results show the occurrence of charge
inhomogeneities in macroscopic form or in small clusters of one
phase embedded into another. It turns out the metallic FM phase
has regions where FM clusters coexist with another phase in a
range of temperature and concentration \cite{Sch95,Ram96}, either
in Sr- and Ca-doped manganese oxides.

In this context, De Gennes suggested rather early that the
competition between the AF superexchange and the DE results in the
canting of the AF state \cite{deG60}, that is the angle $\theta$
between the spins from different sublattices becomes smaller than
$\pi$. The canting angle grows with the concentration of charge
carriers, which might explain the increase of magnetization upon
doping observed in La$_{1-x}$Ca$_x$MnO$_3$. Already rather long
ago arguments were given against the stability of canted ground
state \cite{Nag79,Nag96}. In the De Gennes approach the local
spins were treated classically. It was found that quantum
corrections stabilize the AF state and the canted state appears
only above a certain concentration of charge carriers
\cite{Nag96}. By this mechanism the canted state might also
disappear -- in fact it was not observed in
La$_{1-x}$Ca$_x$MnO$_3$ at increasing doping $0<x<0.15$
\cite{Mou99}.

A more fundamental problem however is that at partial filling of
the conduction band, the homogeneous ground state might be
unstable against phase separation (PS). Experimentally the PS was
recently observed by Babushkina {\it et al.} \cite{Bab99}, and it
is also obtained in theory. For instance, the phase diagram of
Fig. \ref{fig:oodoping} contains several regions with PS, either
between different orbital states, or between differently doped
phases. Another possibility is the PS between the regions with
different magnetic ordering which accompanies differently doped
regimes of the sample. This implies that many experimental data on
doped manganites should be reinterpreted taking into account the
inhomogeneity of the ground state. In particular, the charge
transport and the metal-insulator transition should be more
appropriately described in terms of percolation rather than by the
properties of the pure states.

The problem of the nature of the PS has been studied extensively
by means of QMC techniques within the FM Kondo model with one and
two orbitals, including the JT effect for taking into account the
occurrence of orbital order, though without fully considering the
consequences of large Coulomb interactions \cite{Yun98,Hu98}.
Several unexpected results have been found in that study. In
particular, either for the one- or two-orbital model, when
calculating the density of $e_g$ electrons as a function of
varying chemical potential, one finds that some densities are
unstable \cite{Yun98,Hu98}; that is, the density is changing
discontinuously at particular values of the chemical potential.
Other calculations done in the canonical ensemble where the
density is kept fixed, showed that the ground state is not
homogeneous, while being constituted by separate regions with
values of densities corresponding to the unstable regime. This
phenomenon is usually referred to as {\it phase separation\/},
such as the familiar liquid-vapor coexistence in the phase diagram
of water where it is known that the compressibility becomes
negative negative. It is also similar to the stripe instability in
the cuprates \cite{Zaa98}, but the present phenomenon is more
difficult to study as it happens in a larger space which involves
the orbitals and lattice.

In the regime of large Hund's coupling $J_H$ and intermediate
values of the JT coupling, PS occurs both for small and large hole
densities. In particular at small $e_g$ densities, the PS appears
between an electron-undoped AF state and a metallic uniform
orbital-ordered FM state. At small hole concentrations, the latter
phase coexists with an insulating staggered orbital ordered FM
phase. In the QMC results \cite{Yun98,Hu98}, the PS manifests
itself as the occurrence of a macroscopic separation of two phases
with different charge distribution. Actually, this possibility
should be prevented by long-range Coulomb interactions, which is
usually neglected. In fact, even taking into account polarization
effects and screening, a complete separation would lead to a huge
loss of energy. These considerations suggest that two large
regions should split into many smaller pieces in order to
distribute the charge more uniformly and reduce the energy loss in
the Coulomb channel.

If the PS involves just single carriers (doped holes) with their
local environment that has been distorted by the presence of a
hole, such states are referred to as polarons. The distortion can
be either in the magnetic channel (spin polaron), or in the
phononic channel (lattice polaron), or in the orbital channel
(orbital polaron), or may be even a combination of these effects.
Furthermore, the case where such regions of reduced electron
density involve more than one carrier are referred to as {\it
clusters\/} or {\it droplets\/}. The competition between the
long-range Coulomb repulsion and the magnetic interaction would
determine the size and the shape of the resulting clusters. The
stable state, obtained when the extended Coulomb interactions are
included, is considered as a charge inhomogeneous state which has
to be distinguished from the metastable state that appears in a
standard first order transition.
\begin{figure}[b]
\centerline{\epsfig{figure=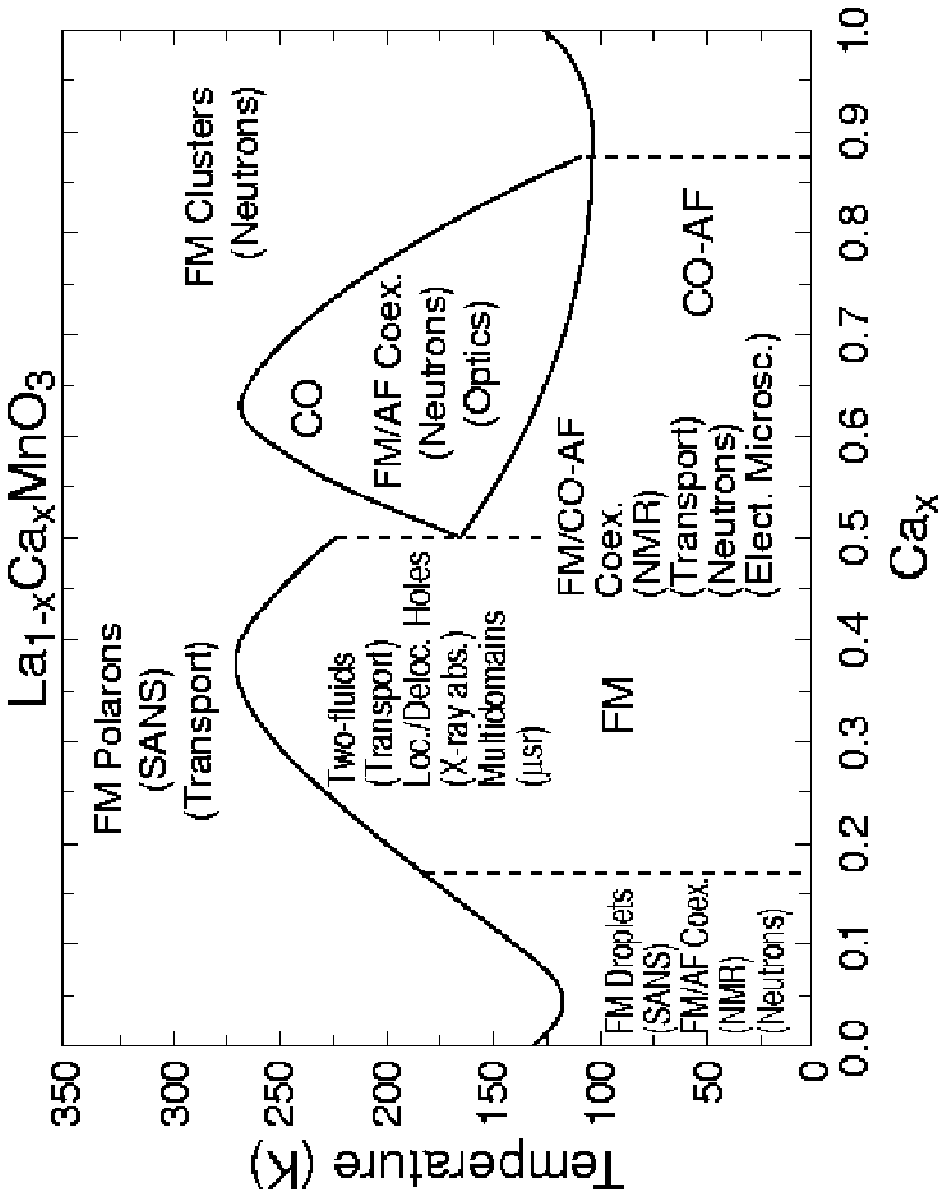,width=8cm,angle=270}}
\smallskip
\caption {A schematic version of the phase diagram for
La$_{1-x}$Ca$_x$MnO$_3$ \protect\cite{Sch95}. The abbreviations
Coex., Loc., Deloc., abs., $\mu$SR, and Elect. Microsc., stand for
coexistence, localized, delocalized, absorption, muon spin
relaxation, and electron microscopy, respectively (after Ref.
\protect\cite{Adriana}). } \label{fig:ps}
\end{figure}
A large-scale PS is expected if the competing phases have
approximately the same density, as it happens experimentally at
$x=0.5$ which is a singular point in the phase diagrams (Sec.
VII.B). It is of fundamental importance to investigate how the
properties of the ordered FM state are influenced by the vicinity
of a PS regime, especially in connection to the CMR phenomenon.
One of the consequences of this proximity to PS state is observed
as an effective increase of the compressibility in the FM phase
which implies the occurrence of strong charge fluctuations. This
result implies that even within the FM phase, which is uniform
when time averaged, there is a dynamical tendency toward cluster
formation. This effect is expected to influence the transport
property. Calculations done on finite size clusters show that the
resistivity behaves as observed experimentally, in other words it
is insulating at small $x$ and rapidly decreases when $x$
increases, and a metallic state is approached.

A very interesting observation was made recently by Moreo, Yunoki
and Dagotto that the PS is responsible for a pseudogap formation
in the spectral function $A({\vec k},\omega)$ calculated using
Monte-Carlo techniques for the model of manganites which includes
the coupling to JT phonons \cite{Mor99}. The pseudogaps was found
to be a robust feature at all momenta along the Fermi surface and
occurs due to the existence of FM clusters coexisting with the AF
ordering. A particularly pronounced pseudogap was found in the
model with doubly degenerate $e_g$ orbitals which demonstrates
again that the orbital degrees of freedom play an important role
in manganites. The results have striking similarity with the angle
resolved photoemission studies of La$_{1.2}$Sr$_{1.8}$Mn$_2$O$_7$
\cite{Des98} which suggests that microscopic inhomogeneities exist
in this system, both above and below $T_C$.

The PS scenario is likely to play an important role in the low and
intermediate doping regime, where several mechanisms have been
proposed to explain the CMR effects in manganese oxides. It adds
to the simple DE idea the possibility of charge inhomogeneities as
the main effect competing with ferromagnetism. This picture is
different from that proposed by Millis {\it et al.\/}
\cite{Mil95}, where the interplay between spin disorder and the
formation of lattice polarons via JT-coupling is fundamental for
explaining the metal-insulator transition. In this respect, though
the JT-coupling turns out to be important in both scenarios, in
the PS scheme a state given by independent local lattice polarons
is a special case of a more general situation where droplets of
various shape and size might form. Still, the dynamical
fluctuation of charge inhomogeneities increase as T$_C$ decreases,
thus explaining the CMR effect at the boundary of the FM phase.
Other theories emphasize the electronic localization effect
induced by off-diagonal disorder in the spin correlations
\cite{Mul96,Var96,Cal98}, or diagonal disorder due to chemical
substitution.

Summarizing, the generic phase diagram of manganites shown in Fig.
\ref{fig:ps} represents still a very challenging and interesting
problem in the field of correlated electrons in degenerate $e_g$
bands. There are many open questions related to the understanding
of the insulating state of manganese oxides, especially in the
framework of tendencies towards charge inhomogeneous state.
Analytical techniques beyond the local MF approximations are
required to reproduce the essential physics of the nonuniform
charge phase and to investigate the role of orbital ordering in
doped manganites. More detailed analysis is needed to understand
the shape and the size of droplets in such inhomogeneous states,
and the crossover from the PS regime to polaronic regime which has
been observed experimentally \cite{Uru95,cheong,Fuj98}.

\section{ Summary and conclusions }

We have reported a systematic analysis of the consequences of {\it
orbital degrees of freedom\/} in a class of correlated MHI: copper
and manganese oxides. These systems are characterized by several
different interactions which involve three different types of
degrees of freedom in the undoped compounds: spins, orbitals, and
lattice. If such systems are doped, one has to include in addition
charge dynamics. The main question we have been dealing with
through this report was: What is the role of the orbital degrees
of freedom in presence of strong electron correlations and/or
lattice effects, and what kind of {\it magnetic and orbital
ordering} is promoted by them?

The essence of the problem posed by the superexchange interactions
at orbital degeneracy is captured by the spin-orbital model
(\ref{somcu})--(\ref{htau}) for the cuprates, defined for the
$d^9$ ions as in KCuF$_3$. This model is of particular interest as
it combines only two aspects of a more complex problem: magnetic
(spin) and orbital degrees of freedom, which are discussed at
integral filling, i.e., in the absence of doping, and without
taking seriously into account the JT effect. The model Hamiltonian
has already been proposed long ago \cite{Kug73}, but its full
consequences have been appreciated only recently
\cite{Fei97,Ole00}.

The Hamiltonian derived for the cuprates from the spectroscopic
information about the spectra of excited states has been first
considered on the classical level, where we have shown in Sec. III
that it gives a particular frustration of spin and orbital
interactions. This is best represented by a singular point at the
origin of the phase diagram for the $d^9$ spin-orbital model (Fig.
\ref{fig:3d}) which is highly degenerate at the classical level,
so that many axes separating different classical phases should
emerge from it. In this respect the classical phase diagram is
incomplete: the quantum mechanics is likely to take over in this
regime and decide about the actual ordered or disordered state. As
a main result of this part, we have shown that enhanced quantum
fluctuations close to the transition lines between different
classical phases destroy classical states (Sec. IV). Unlike in the
frustrated spin models, the Gaussian fluctuations around classical
states involve the orbital sector, probably yielding novel
spin-orbital liquids in the form of generalized (R)VB states
\cite{Fei97}. Such states turn out to be very good variational
wave functions, and turn out to be exceptionally stable in the
$d^9$ model with respect to the classical phases.

Whether the phase diagram presented in Fig. \ref{fig:rpa} is the
qualitatively correct picture of the quantum disorder realized
next to the orbital degeneracy, is still an open question. Another
possibility would be an ordered state which results by the
order-out-of disorder mechanism. Unfortunately, even the physics
at the above multicritical point of the phase diagram of the $d^9$
model is not captured by the SU(4) model which, instead, turns out
to be an idealized generalization of the Kugel-Khomskii model,
putting the spin and orbital variables on the same footing.
However, this highly symmetric problem is not easier to
investigate except for the 1D case, and thus the final answer to
the problem of the ground state of the spin-orbital $d^9$ model in
the vicinity of the multicritical point is still an open issue.

The physical properties of a system described by the spin-orbital
model (\ref{somcu})--(\ref{htau}) are mainly determined by the
nature of the collective excitations. In addition to the usual
magnon dispersion known from the spin systems, we have to consider
the pure orbital (or excitonic) excitations, and the mixed modes
which involve simultaneously both spin and orbital excitation
(spin-and-orbital modes). These modes couple to the magnons and
may thus have measurable consequences, predicted as an anomalous
spin response detected in neutron scattering experiments. It is
thus challenging for the experimentalists to investigate carefully
any peculiarities which are expected to arise in the spin spectra
of the MHI with orbital degeneracy.

The above ideas on the spin liquid near orbital degeneracy and on
the role played by the orbital fluctuations should be verified by
future experiments. It is not easy to find compounds whose
microscopic parameters live in the region close to this peculiar
point in the phase diagram, and a progress in {\it material
science} in needed. Furthermore, though close to this region in
the sector of the electronic interactions, the cooperative and
local JT effects convey to remove the orbital degeneracy
stabilizing a particular spin pattern, as for example, it is
realized in NaNiO$_2$.
In this respect, an interesting situation seems to happen in the
case of LiNiO$_2$, where these competing lattice-induced phenomena
that promote ordered states are absent, and a quantum critical
state characterized by power-law behavior of the spin correlation
functions appears instead.

The spin-orbital model (\ref{full}) derived for the $d^4$ ions as
in LaMnO$_3$ \cite{Fei99} has many common features with the
simpler model in the $d^9$ case (Sec. V). The spin interactions
are different as they concern large spins $S=2$ of Mn$^{3+}$ ions,
and more excited states of Mn$^{2+}$ ions are involved, even when
the $e_g$ part of superexchange (\ref{egterm}) is considered
alone. As a consequence of larger spins, the model is more
classical and the disordered states are not likely taking the
realistic situation in manganites, with the strong JT effect which
drives the system away from the multicritical point of frustrated
interactions. The classical phases in the $d^4$ case are similar
to those found for the $d^9$ model, but quantitatively the regions
of their stability are different. Therefore, the A-AF phase
obtained by a particular interrelation between the magnetic and
orbital ordering is obtained in the $d^4$ case in a natural way as
a consequence of the $e_g$ part of superexchange interactions --
this phase is robust and agrees with experimentally observed
ordering in manganites. An important finding in this respect is
that the electronic mechanism {\it alone\/} is responsible of the
A-AF ordering, and that the JT interaction is only changing this
state quantitatively and tuning a somewhat different orbital
ordering. Furthermore, when a FM $(a,b)$ plane of the A-AF phase
is considered, the spins are integrated out and one finds the same
orbital part of the superexchange Hamiltonian discussed in Sec.
VII.

The analysis of the $d^9$ and $d^4$ models may be treated as the
first step on the way towards the understanding of one of the most
challenging problems in the modern solid state physics: How the
delicate balance of magnetic and orbital interactions is affected
by doping in the manganese oxides, and which physical mechanisms
are responsible for various types of ordering observed in the CMR
materials? We have analyzed this issue in an extensive way in
Secs. V-VIII. Firstly, we have discussed the low doping regime of
the manganese oxides which is simpler because the lattice helps to
restrict the doped holes to particular sites, and the model deals
in principle still only with the same degrees of freedom as in
undoped LaMnO$_3$: spin, orbital, and lattice, but they are
modified by the added charges. This part of the phase diagram is
dominated by many competing effects both in the magnetic and
orbital channel. The central issue is the presence of an {\it
orbital polaron\/} regime which mediates the crossover from the
A-AF to the FM insulating phase. The DE mechanism stabilizes the
polarons and their localization by new FM effective interactions
that give rise to the transition to the FM insulating phase as the
hole concentration is increased (Sec. V.B). The binding energy of
these orbital-hole bound state depends on the scale of
fluctuations that involve orbitals, and therefore on the
concentration of doped holes. Different configurations of the
polarons may be stabilized by the lattice effects; the polarons
are separated  from each other \cite{Ahn98} and may induce
interesting orbital ordering in their neighborhood \cite{Miz00},
but finally start to overlap. Thus, when the hole concentration
increases, the orbital fluctuations also do, breaking eventually
the localization process and favoring instead a different kind of
orbital (ordered or disordered) state in a FM metallic phase. This
problem is at heart of the CMR phenomenon.

Naively, the transition to the FM metallic phase occurs due to the
DE mechanism. This picture is oversimplified -- it ignores the
orbital degrees of freedom which are of importance in the doped
regime and lead to several types of magnetic ordering when the DE
model is generalized to the case of degenerate $e_g$ orbitals
(Sec. VI). We would like to emphasize that the DE via degenerate
orbitals has very interesting special features that are very
different from the conventional nondegenerate situation. The
orbital degeneracy leads in general to the formation of
anisotropic magnetic structures that depend on the actual doping
concentration, and include the layered magnetic A-type structures,
predominantly the $x^2-y^2$ orbitals, and chain-like structures of
the C-type. The stability of such states will be influenced by
lattice distortions -- in this case a compression of the MnO$_6$
octahedra along one of the cubic directions. One can show,
however, that due to the anharmonicity the JT distortion always
leads to a local elongation. If strong enough along the $c$-axis,
this tendency would favor a structure in which the $3z^2-r^2$
orbitals are occupied, in this case lowering the energy of C-type
structures. These considerations demonstrate the importance of the
JT effect. Cooperative JT coupling between the individual MnO$_6$
centers in the crystal leads to the simultaneous ordering of the
octahedral distortions and the electronic orbitals. Part of the
electronic energy is thereby lost and therefore an accurate
description of the orbital state in manganites is possible only
when the lattice effects are {\it explicitly} included in the
model.

Unlike the DE Hamiltonian for nondegenerate orbitals, the
realistic DE model which includes the degeneracy of $e_g$
electrons cannot be considered without the strong Coulomb
correlations. Electronic correlations are very important for the
understanding CMR manganites and cannot be neglected. Very
different results are obtained when local Coulomb correlations are
not included -- for instance: the tendency towards the FM metallic
phase is strongest at the filling of one $e_g$ electron per site
\cite{Sol99}, and the band gap practically vanishes without JT
distortions \cite{Pop00}, while in reality the gap is primarily
due to large on-site Coulomb interactions \cite{Var96} (Sec. VII).
The main result of the DE model without Coulomb interactions is
that it can explain FM metallic phase only for doping $x>0.5$,
where, however, the localized G-AF phases are typically found
(Fig. \ref{fig:expphd}). At lower values of $x$, the suppression
of double occupations by the local Coulomb repulsion becomes more
and more important, and leads to a crossover from DE to
superexchange. First it reduces somewhat the effective FM
interactions which follow from the DE mechanism, leading to the
reduction of the magnon bandwidth \cite{pm99} and of the Curie
temperature $T_C$ \cite{Hel00} with decreasing $x$, and eventually
the superexchange between the Mn$^{3+}$ ions becomes more
important and stabilizes the A-AF ordering coexisting with the
orbital ordering.

Recently it became clear that the nature of FM states may be
studied by analyzing the spectrum of magnetic excitations. In
recent neutron scattering experiments \cite{Per96,End97,Hwa98} the
spin wave dispersion throughout the BZ was measured in various
manganites in the FM phase. It has been found that the spin waves
are clearly resolved at low temperature, and the stiffness
constant shows a universal behavior. This universal behavior and
also the higher-energy magnons in the metallic phase are well
described by the DE model with degenerate $e_g$ orbitals,
supplemented by smaller SE terms (Sec. VI.B). At higher
temperature, however, different charge dynamics in metallic and
localized compounds manifest themselves in the heavy damping of
high-frequency spin waves even below T$_c$. Therefore, the simple
DE model does not describe well the observed spectra in such
situations when the doped holes localize, high-frequency magnons
soften, and the value of $T_C$ is reduced. We have discussed that
orbital and charge fluctuations are responsible for a strong
modulation of the exchange bonds, leading to a softening of the
magnon excitation spectrum close to the BZ boundary (Sec. VI.C).
The presence of JT phonons further enhances this effect. This
peculiar interplay between DE physics and orbital-lattice dynamics
becomes dominant close to the instability towards an
orbital-lattice ordered state. The unusual magnon dispersion
experimentally observed in low-$T_C$ FM manganites can hence be
considered as a precursor effect of orbital-lattice ordering. A
complete understanding of these complex phenomena is still an open
problem \cite{MN00} and may be achieved only by taking into
account all relevant degrees of freedom, including the JT effect
which is still present in the metallic phase \cite{Despi}.

The proper understanding of pure orbital excitations is also very
important, both in the undoped regime, and for the FM phase where
such excitations are responsible for the orbital dynamics.
Starting from the uniform FM phase, where the spin operators can
be integrated out, we considered the interaction which occurs only
between the pairs of ions with singly occupied staggered $e_g$
orbitals at nearest neighbor sites (Sec. VII). A gapless
orbital-wave excitation is found for the 3D system at orbital
degeneracy, a peculiarity related to the invariance of the
classical ground state energy with respect to the orbital
rotations by angle $\theta$ at $E_z=0$. This rotational invariance
is broken in the 2D systems by the lack of interactions along the
$c$-axis, and the 2D orbital model is more classical, with a gap
in the excitation spectrum and suppressed quantum fluctuations,
unlike in spin systems which are more quantum when one comes to a
lower dimension. At increasing orbital field $E_z$, however, the
system resembles again the behavior of a 3D system, with finite
quantum fluctuations even when the orbitals are aligned which
follows from the nonconservation of the orbital quantum number in
the subspace of $e_g$ orbitals.

Another open problem is the mechanism of stability of the CE phase
in half-doped manganites with $x=0.5$. The insulating CO state has
been observed in almost all such compounds at half-doping,
accompanied by a peculiar magnetic structure: 1D FM zigzag chains
coupled antiferromagnetically within the $(a,b)$ planes, and along
the $c$-axis. In addition, these systems show
$d_{3x^2-r^2}$/$d_{3y^2-r^2}$ orbital ordering along the FM
chains. The is no doubt that the degeneracy of $e_g$ orbitals
plays here once again a crucial role. It follows both from a
different behavior of the effective DE models with nondegenerate
and degenerate orbitals, and from the specific mechanism of an
insulating behavior discussed in Sec. VIII.A. The stability of
different magnetic states is changed by orbital degeneracy, and
for instance the C-type spin ordering is never achieved in the
two-orbital case due to its instability against the effective {\it
dimerization\/} and formation of the zigzag FM order. The
alternating $d_{3x^2-r^2}/d_{3y^2-r^2}$ orbital ordering along the
FM zigzag chain follows naturally the topology of the hopping in
the CE phase. However, the role of the JT effect played in this
phase and the magnetic and orbital excitation spectra are not yet
understood.

The competition between different phases represents a particularly
challenging problem. It results in a PS between AF and FM domains
which together with the tendency to form inhomogeneous structures,
seems to be a generic property of systems with strongly correlated
electrons. Depending on the specific situation tuned by the
strength of the coupling to the lattice and the doping level, the
instability of homogeneous state and the tendency to PS may result
in a formation of different structures: either random,
percolation-like networks or regular structures, e.g. stripes,
Wigner crystals, and the like (Sec. VIII). Many properties of such
phase-separated states differ markedly from those of homogeneous
states, and the possibility of PS has to serve as a starting point
for the future explanation of certain anomalous phenomena observed
experimentally in manganites. The PS scenario should lead to
certain extensions of the DE model for degenerate $e_g$ orbitals
by the ideas of charge inhomogeneities as the main effect
competing with ferromagnetism, and by considering the insulating
properties above the Curie temperature as a consequence of the
formation of dynamical clusters. How the PS scenario occurs is
still an open problem and it is related to the understanding of
the insulating states in manganites and other materials with
strongly correlated electron bands. Analytical techniques beyond
the local MF approximations are required to get the essential
physics of the non-uniform CO phase under control. In experiment,
the photoemission spectroscopy is expected to be of particular
importance, as such features as the pseudogap \cite{Mor99}, and
the changes of the spectra between the compounds with different
number of adjacent magnetic layers, lattice distortions, and
doping levels, would help to characterize the orbital and magnetic
ordering in the underlying phases.

Summarizing, the realistic model for manganites has to include the
superexchange of spin-orbital type, the coupling to the lattice
due to the cooperative and local JT effect, and some form of the
DE in the correlated $e_g$ orbitals, either localized or
itinerant. Several theoretical models have already been proposed,
but in most cases they neglect at least one of the important
aspects of the problem (as, for instance, either $e_g$ orbital
degeneracy, or coupling to the lattice, or electron correlations),
and then use some parameters to fit such properties as the type of
magnetic ordering, the stiffness constant, or the types of
occupied orbitals to the experiment --- this is usually successful
as the parameter space is large enough. While the model proposed
by Feiner and Ole\'s \cite{Fei99} can certainly still be improved,
it not only includes all the essential aspects mentioned above,
but also takes the parameters from spectroscopy, except for the
orbital interaction which follows from the JT effect and was fixed
by experiment. In this way, the parameters which determine the
magnetic transitions are all fixed and the predictions of this
model may by confronted with experiment. The quality of this model
may be appreciated by looking at Table 3. The N\'eel temperatures
of the Mott-Hubbard insulators: LaMnO$_3$ and CaMnO$_3$, of the
weakly doped polaronic La$_{0.92}$Ca$_{0.08}$MnO$_3$ compound, as
well as the Curie temperature of the metallic FM
La$_{0.7}$Pb$_{0.3}$MnO$_3$, are all obtained within 30 \% to the
experimental value. Of course, there is still a lot of room for
improvement and the understanding is far from complete, in
particular at finite temperature. Future progress may be expected
by including the JT effect in a better way, and by a more careful
analysis of the orbital ordered and disordered phases in various
doping regimes.

We believe that the present report makes the role played by the
correlated $e_g$ orbitals in the MHI more transparent, and
demonstrates that they have to be included to obtain qualitatively
correct answers in still open problems, such as: the stability of
different phases including the CE phase and stripe phases, the
mechanism of PS, the metal-insulator transition at finite
temperature, and the mechanism of the CMR itself.

\begin{table}
\caption{ Magnetic transition temperatures for the A-AF and G-AF
states in undoped compounds: LaMnO$_3$, CaMnO$_3$, and in
polaronic A-AF phase in La$_{0.92}$Ca$_{0.08}$MnO$_3$ ($T_N$), and
in the metallic FM La$_{0.7}$Pb$_{0.3}$MnO$_3$ ($T_C$), as
obtained from the spin-orbital model in doped manganites presented
in Refs. \protect\cite{Fei99,pm99}, including a reduction factor
due to quantum fluctuations (theory), compared with the
experimental values (exp). The physical mechanisms which
contribute to the stability of these different magnetic phases are
indicated by '+'; DE effect was included either by polaronic (P)
or by itinerant (I) mechanism. Orbital ordering stabilized by a
superposition of the $e_g$-superexchange and the JT effect above
$T_N$ plays a crucial role in driving the magnetic ordering in
LaMnO$_3$ and in La$_{0.92}$Ca$_{0.08}$MnO$_3$ into the A-AF
phase. } \vspace{0.15in}
\begin{tabular}{c|cccccccc}\hline
 &  & \multicolumn{2}{c} {superexchange} & & &
      \multicolumn{3}{c} {$T_N$ ($T_C$)} \\
compound & phase & $e_g$ part& $t_{2g}$ part& DE& JT& theory &
exp  &
 Ref. \\
\hline LaMnO$_3$                    & A-AF & + & + & --& + &  106
K & 139 K &
 \protect\cite{Mou99} \\
La$_{0.92}$Ca$_{0.08}$MnO$_3$& A-AF & + & + & P & + &   95 K & 122
K &
 \protect\cite{Mou99} \\
La$_{0.7}$Pb$_{0.3}$MnO$_3$  & FM   & + & + & I & --&  390 K & 355
K &
 \protect\cite{Hwa98} \\
CaMnO$_3$                    & G-AF & --& + & --& --&  124 K & 110
K &
 \protect\cite{Wol55}
\label{loufe}
\end{tabular}
\end{table}

\acknowledgments

A.M.O. acknowledges warmly numerous stimulating conversations and
a very friendly collaboration with Louis Felix Feiner, Peter
Horsch and Jan Zaanen, who contributed significantly to his
present understanding of this fascinating subject. It is a
pleasure to thank G. Aeppli, J. van den Brink, B. Keimer, A.
Fujimori, G. Khaliullin, D. I. Khomskii, R. Micnas, F. Moussa, G.
A. Sawatzky, T. M. Rice, H. Shiba, J. Spa\l{}ek, Y. Tokura, and W.
Weber for valuable discussions. We are grateful to Ferdinando
Mancini for creating this opportunity to discuss the spin-orbital
models and to the stuff of the International Institute for
Advanced Studies "E.R. Caianiello" in Vietri sul Mare for their
assistance and financial support. The financial support by the
Committee of Scientific Research (KBN) of Poland, Project No.~2
P03B 175 14, is acknowledged.

N.B.P. acknowledges INTAS N97-963, INTAS N-9711066 and INFM of
Salerno for the financial support and the warm hospitality.


\end{document}